\documentclass[12pt]{article}

\usepackage{times}
\usepackage{algorithm}
\usepackage{algpseudocode}
\usepackage{lscape}
\usepackage{caption, rotating}

\newcommand{\mc}{\mathcal}
\newcommand{\ul}{\underline}

\newcommand{\tr}{\textrm}
\newcommand{\tb}{\textbf}
\newcommand{\mr}{\mathrm}
\newcommand{\mbb}{\mathbb}

\newcommand{\PRE}{\mr{PRE}} 

\newcommand{\DPE}{\mr{DPE}} 

\newcommand{\RGE}{\mr{RGE}} 

\usepackage{amsmath, amssymb}
\usepackage{graphicx, subfigure}
\usepackage{multibib} 
\newcites{MM}{References (M)}
\newcites{SM}{References (SM)}

\newtheorem{definition}{Definition}
\newtheorem{theorem}{Theorem}
\newtheorem{lemma}[theorem]{Lemma}
\newtheorem{corollary}[theorem]{Corollary}

\newenvironment{sciabstract}{%
\begin{quote} 
\bf}
{\end{quote}}

\title{Clustering Edges in Directed Graphs}

\author{%
  Manohar Murthi,$^{1\ast}$ Kamal Premaratne$^{1}$ \\
  \\
  \normalsize{$^{1}$Department of Electrical and Computer Engineering, 
  University of Miami,} \\
  \normalsize{1251 Memorial Drive} \\
  \normalsize{Coral Gables, Florida 33146, USA} \\
  \\
  \normalsize{$^\ast$To whom correspondence should be addressed; 
  E-mail: mmurthi@miami.edu.}}

\date{}

\begin{document} 


\baselineskip24pt


\maketitle 


\begin{sciabstract}
  How do vertices exert influence in graph data? We develop a framework for edge clustering, a new method for exploratory data analysis that reveals how both vertices and edges collaboratively accomplish directed influence in graphs, especially for directed graphs. In contrast to the ubiquitous vertex clustering which groups vertices, edge clustering groups edges. Edges sharing a functional affinity are assigned to the same group and form an influence subgraph cluster.  With a complexity comparable to that of vertex clustering, this framework presents three different methods for edge spectral clustering that reveal important influence subgraphs in graph data, with each method providing different insight into directed influence processes. We present several diverse examples demonstrating the potential for widespread application of edge clustering in scientific research.
\end{sciabstract}


New exploratory data analysis methods that reveal how vertices and edges collaboratively achieve directed influence in graph data are warranted. For this purpose, we develop the Flow Laplacian Edge Clustering framework that complements vertex spectral clustering, which pervades graph data analysis. In \emph{vertex clustering,} labels are assigned to \emph{vertices} based on a graph cuts-based cost expressed as a matrix quadratic form of a vertex Laplacian matrix \cite{{VonLuxburg2007SC}}.  Vertices sharing a label form a cluster that reveals vertex similarity or affinity. To understand how vertices and edges collaboratively achieve directed influence, especially in directed graphs (digraphs), we introduce \emph{edge clustering} in which one instead assigns labels to \emph{edges}. Edge labels are assigned by minimizing a matrix quadratic form involving our new Flow Laplacians that capture different types of directed influence specified as edge affinities.  Our framework produces subgraph clusters that not only reveal groups of edges sharing a functional affinity but also provide insight into collaborative vertex and edge directed influence processes, engendering new ways to explore graph data.

Our Flow Laplacian Edge Clustering Framework considers three philosophically different approaches to describing the functional affinity between directed edge pairs (see Fig.~\ref{fig:OverallFlowDiagram}), leading to three different edge clustering methods with three types of directed influence: 

\tb{(a)}~\emph{Producer-Receptor-Emphasizing (PRE)} clustering yields clusters where two edges that are both conveying influence to a vertex share a functional affinity and belong together in one group (edges (1,2) in Fig.~\ref{fig:OverallFlowDiagram}A); similarly two edges that are both conveying influence from a vertex share an affinity, and belong together in another group (edges (3,4) in Fig.~\ref{fig:OverallFlowDiagram}A). This yields a subgraph cluster in which certain vertices are mainly producers with outgoing directed edges, and the other vertices are mainly receptors with incoming directed edges. 

\tb{(b)}~\emph{Directed-Path-Emphasizing (DPE)} clustering generates clusters where edges that form a directed path share an affinity and belong together in one group (edges (1,3) in Fig.~\ref{fig:OverallFlowDiagram}A are in one group while edges (2,4) are in another; alternatively edges (1,4) in one group and edges (2,3) in another). This yields subgraph clusters that emphasize long directed paths of influence.

\tb{(c)}~\emph{Region-Emphasizing (RGE)} clustering yields clusters where all edges that share a common vertex share an affinity irrespective of their directionality (edges (1,2,3,4) in Fig.~\ref{fig:OverallFlowDiagram}A are in one group). This yields subgraph clusters that identify distinct partitioned regions of concentrated flows, a method suitable for both directed and undirected graphs.

We capture edge functional affinity via a weighted edge clustering cost expressed as a matrix quadratic, which leads to a graph cuts-based interpretation (See SM). To grasp this, suppose we are clustering $M$ edges into $K$ groups by assigning a label to each edge.  Then we can relate the edge clustering cost to a weighted sum of the form $\displaystyle\sum_{\ell_p, \ell_q\in\mc{E}} \phi_{pq} (\ell_p - \psi_{pq} \ell_q)^2$ where $\ell_p$ and $\ell_q$ are the labels of edges $e_p$ and $e_q$ respectively, $\psi_{pq} \in \{0, -1, +1\}$ determines the particular edge functional affinity, and $\phi_{pq} \geq 0$ captures the relative importance of the vertex shared by edges $e_p$ and $e_q$; it could be user specified or selected based on graph attributes, e.g., proportional to the sum of the absolute weights of the edges connected to the vertex. By expressing this cost as a matrix quadratic, we get a $M \times M$ positive semi-definite (psd) Laplacian matrix for each functional affinity, i.e., $\ul{L}_{\PRE}$ for PRE clustering, $\ul{L}_{\DPE}$ for DPE clustering, and $\ul{L}_{\RGE}$ for RGE clustering. For each of these Laplacians one can interpret clustering as signed cuts in a certain \emph{dual graph}. When $\phi_{pq}$ is appropriately chosen, we can write these three Laplacians in terms of the edge Laplacian matrix $\ul{L}_e$ \cite{Schaub2018GlobalSIP} which is less well-known than its vertex Laplacian counterpart, consequently allowing us to connect $\ul{L}_e$ to edge clustering (see SM).

To counter cluster size imbalance, we employ `normalization' for which we define the volume of an \emph{edge} to account for the edge weight, directionality, and the relative importances of its two end-vertices. By defining the normalized clustering cost as a quadratic form associated with a normalized Laplacian and relaxing the optimization, we obtain a simple algorithm for clustering $M$ edges:

\tb{(1)}~Select the edge clustering method (PRE, DPE, or RGE); 

\tb{(2)}~Choose $K$, the number of clusters desired; 

\tb{(3)}~Compute the (PRE, DPE, or RGE) normalized Laplacian; 

\tb{(4)}~Use the eigenvectors associated with its $K$ minimum eigenvalues to compute the $M \times K$ row-normalized eigenvector matrix; 

\tb{(5)}~Apply a clustering algorithm (e.g., k-means++) to its $M$ rows to obtain $K$ edge clusters. 

Fig.~\ref{fig:OverallFlowDiagram} summarizes this Flow Laplacian framework. Given the digraph's weighted adjacency matrix, the number of desired clusters $K$, and the clustering method that emphasizes a desired edge functional affinity, it produces edge clusters as its output. Figs.~\ref{fig:OverallFlowDiagram}C and \ref{fig:OverallFlowDiagram}D, where the results for the $K=2$ case are shown for two simple synthetic digraphs, illustrate the principal differences in the three clustering methods.  

Our framework allows one to focus attention on the subgraphs whose edges are collectively achieving a specified type of flow catalyzed by edge affinities. In RGE clustering, each subgraph cluster covers a region of concentrated flows among relatively important vertices. These subgraphs can be combined like puzzle pieces to re-constitute the entire graph. The `coupling’ vertices that `straddle' multiple clusters have lower relative importance in terms of their connectivity and average edge weights.  In DPE clustering, the subgraph clusters can be viewed as being assembled by linking together directed edges to connect vertices with larger relative importances. In PRE clustering, the subgraph clusters can be viewed as being formed by linking together `star’ networks where vertices with outgoing edges are connected to vertices with incoming edges, with the center vertices having relatively larger importance.  

\begin{figure}[H]
  \centering
  \subfigure[Roads near Piazza Mazzini in Rome, Italy.]{%
    \includegraphics[width=3.10in, height=2.50in]{%
      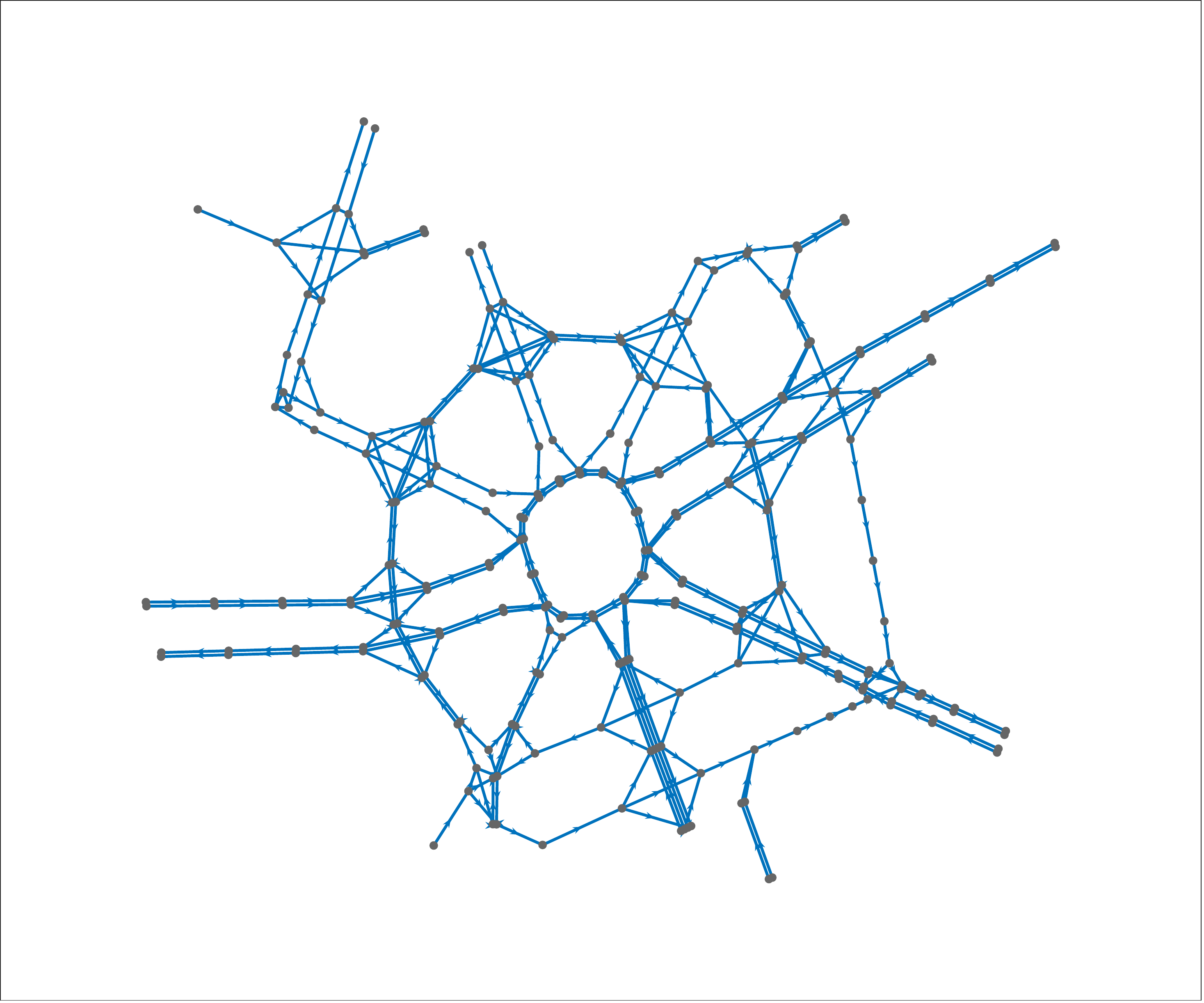}}
  \subfigure[PRE edge clusters.]{%
    \includegraphics[width=3.10in, height=2.50in]{%
      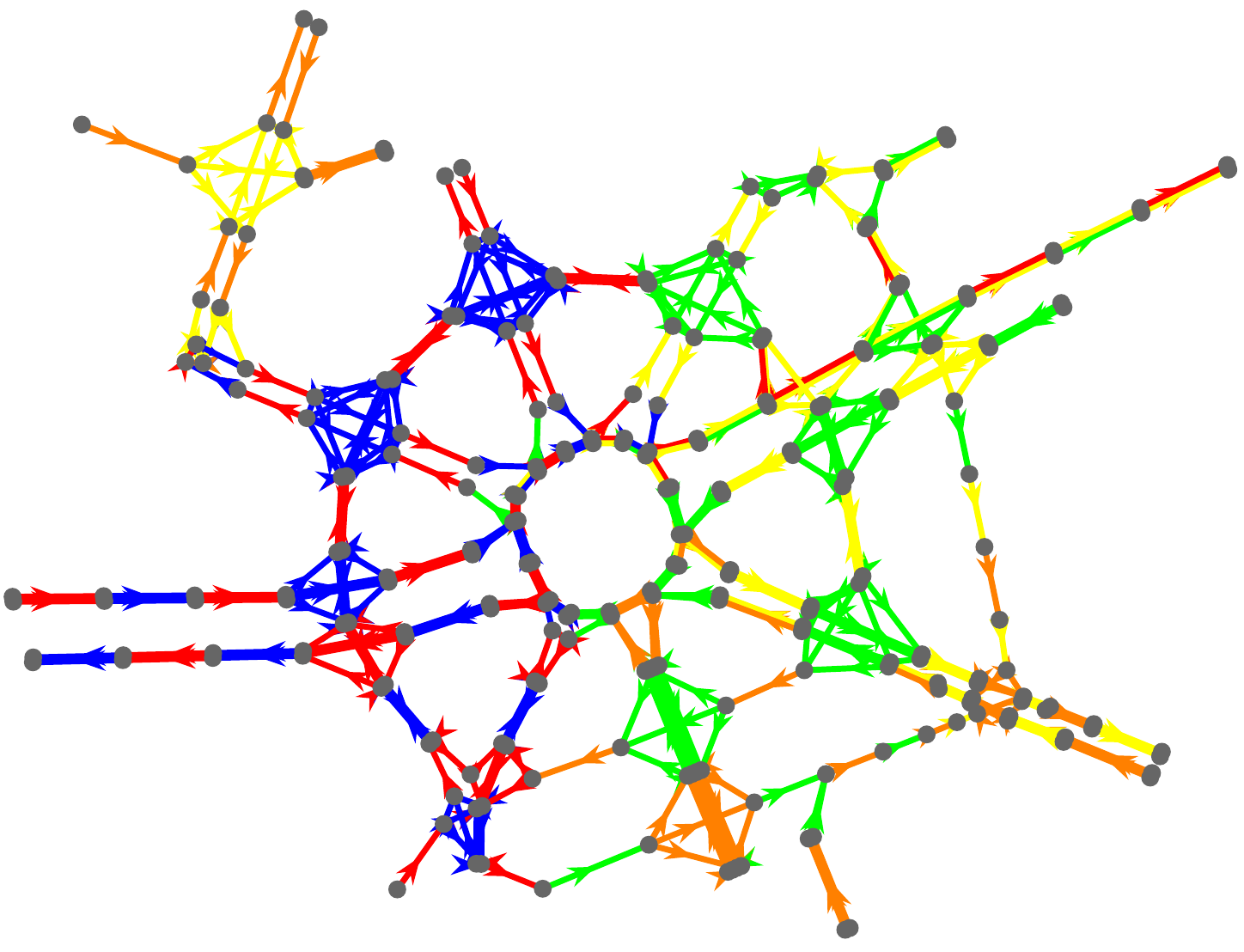}} \\   
  \subfigure[DPE edge clusters.]{%
    \includegraphics[width=3.10in, height=2.50in]{%
      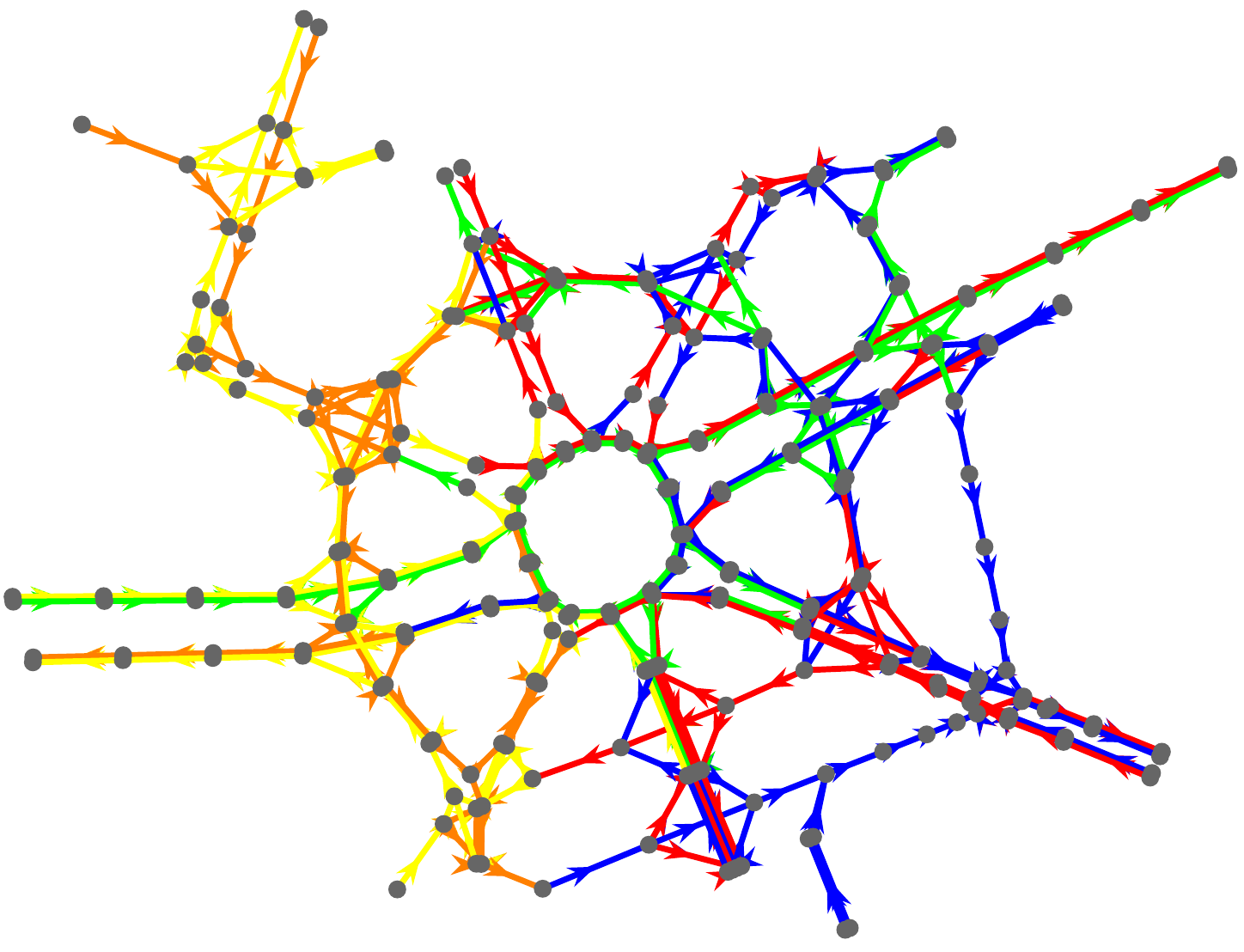}} 
  \subfigure[RGE edge clusters.]{%
    \includegraphics[width=3.10in, height=2.50in]{%
      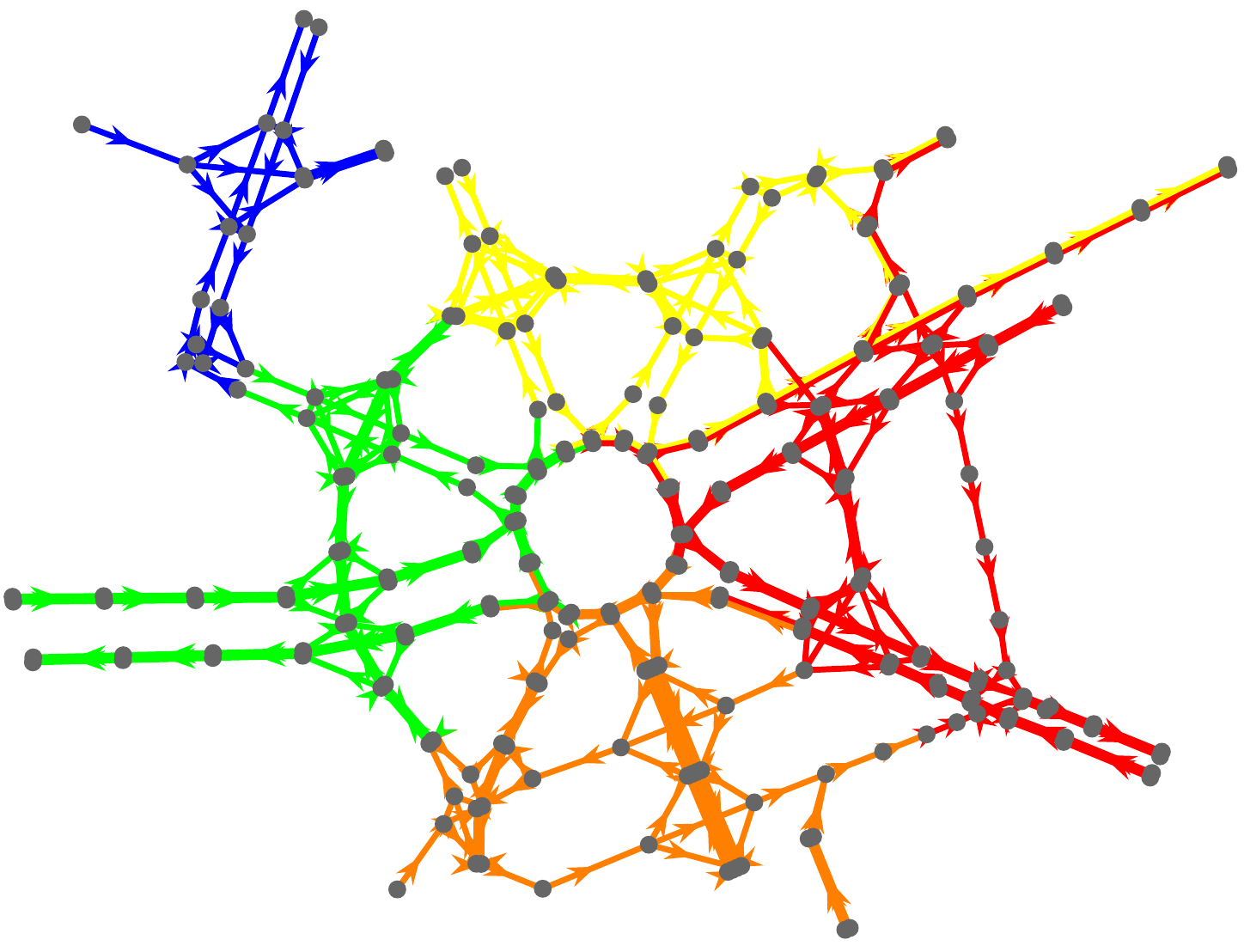}}
  \caption{Roads Near Piazza Mazzini in Rome, Italy. 
  \tb{(a)}~Vertices in this unweighted planar digraph represent waypoints; edges represent road segments; edge direction represents traffic flow \cite{Sardellitti2017_arXiv}. Edge clustering yields useful insight without having to rely on waypoint geographic coordinates. Consider the $K = 5$ case.   
  \tb{(b)}~PRE clustering highlights diamond-like edge cluster motifs identifying complex interchanges that can be points of concern for pedestrians and locations for driver confusion.
  \tb{(c)}~DPE clustering highlights subgraphs of long directed paths (highlighted in the same color). Paths can be used to head quickly to the center or avoid the center, and also highlight potential bottlenecks in the case of traffic accidents. 
  \tb{(d)}~RGE clustering partitions the digraph into distinct regions, each region featuring concentrated flows among its member waypoints. The waypoints here straddling multiple clusters may have fewer connections to roads and may indicate potential bottlenecks between different regions. By focusing on a particular RGE cluster, one obtains an organized approach to traffic planning.}
\label{fig:PiazzaMazzini}
\end{figure}

We see these general patterns writ large on real-world graphs. Consider Fig.~\ref{fig:PiazzaMazzini} which shows roads near Piazza Mazzini in Rome, Italy \cite{Sardellitti2017_arXiv}. This planar digraph makes it easier to visualize our edge clustering results. With $K=5$, PRE clustering extracts diamond-like edge cluster motifs; DPE clustering produces subgraphs of long connected road links; RGE clustering partitions the graph into $5$ regions that are spatially distinct.  We observe these same patterns for the simpler $K=2$ case (see SM).

\begin{figure}[H]
  \centering
  \includegraphics[width=\textwidth]{%
      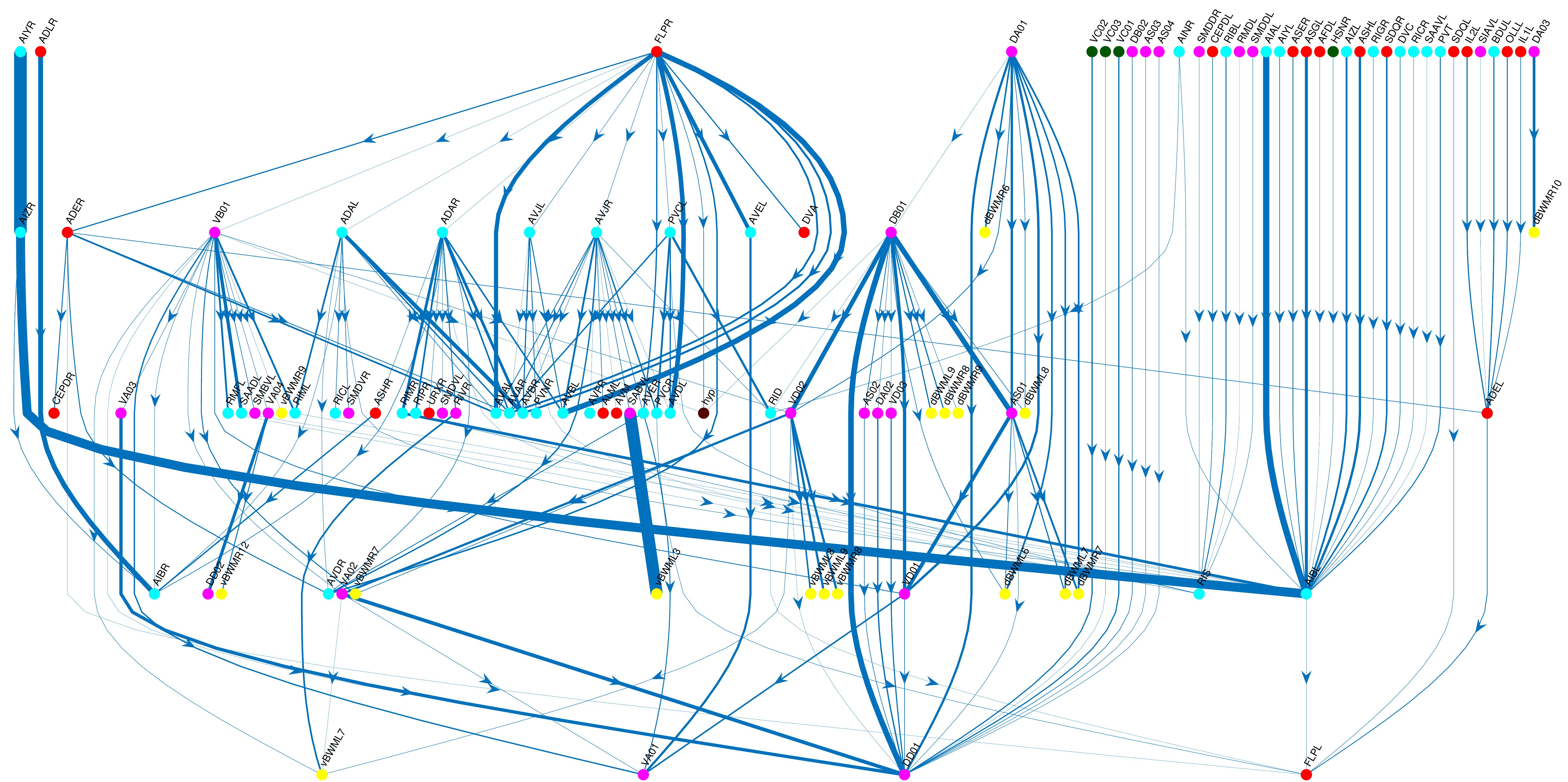}
  \caption{\emph{Caenorhabditis elegans} Connectome cluster example. Digraph vertices represent neurons; edges represent synaptic connections; edge thickness represents synaptic strength; vertex colors represent neuronal functionalities \cite{Cook2019Nature}. In contrast to vertex clustering, edge clustering  subgraphs link together neurons with very different functionalities (e.g., sensory neurons, interneurons, motor neurons, and muscles), providing a tool for scientists to explore the role of different neurons. Using only the weighted adjacency matrix, DPE clustering partitions the connectome into $K$ subgraphs that emphasize directed synaptic paths linking neurons with large weighted degree.  The subgraph cluster example above (110 neurons, 183 synaptic edges, obtained with $K=30$) features a number of directed paths emanating from the FLPR sensory neuron. It highlights circuit elements associated with anterior harsh touch response and backward locomotion.  For example, we see interconnections among the FLP, ADE, and BDU neurons associated with anterior harsh touch response.  We also see direct and indirect connections from these neurons to the AVE, AVD, AVA, AIB, RIM interneurons and VA motor neurons associated with backward locomotion. The prominent topological role of the ADA neurons, whose functions are not well understood, suggest that they are important in anterior harsh touch response and backward locomotion. The FLPR and FLPL sensory neurons appear in only two DPE clusters; the counterpart cluster highlighting directed paths emanating from FLPL is in the SM, as well as PRE and RGE clustering examples that highlight interpretable subgraphs that hint at additional neuron functionality based on the graph topology.}
\label{fig:CElegans_DPE_K30_N110_M183}
\end{figure}

\begin{figure}[H]
  \centering
  \subfigure[]{%
  \includegraphics[width=0.48\textwidth, height=1.5in]{%
      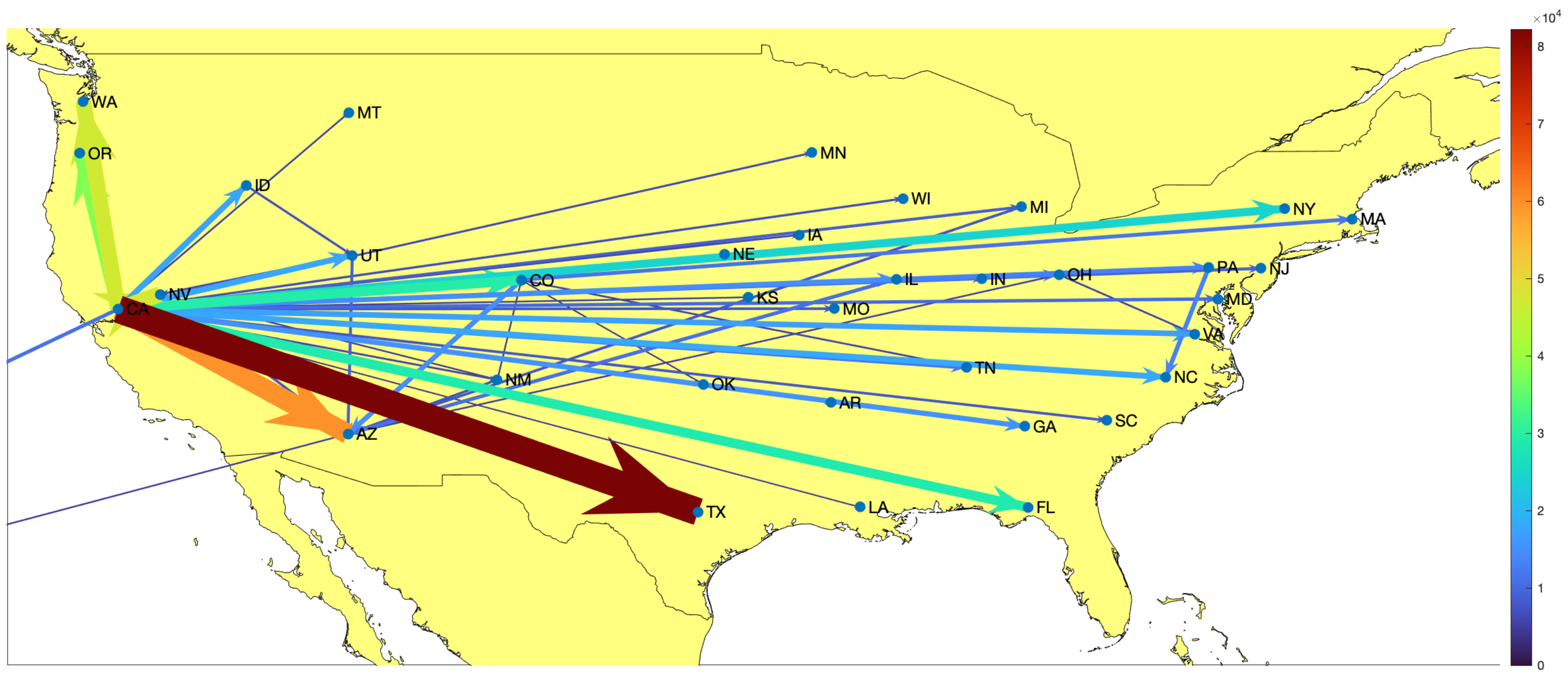}} 
  \subfigure[]{%
  \includegraphics[width=0.48\textwidth, height=1.5in]{%
      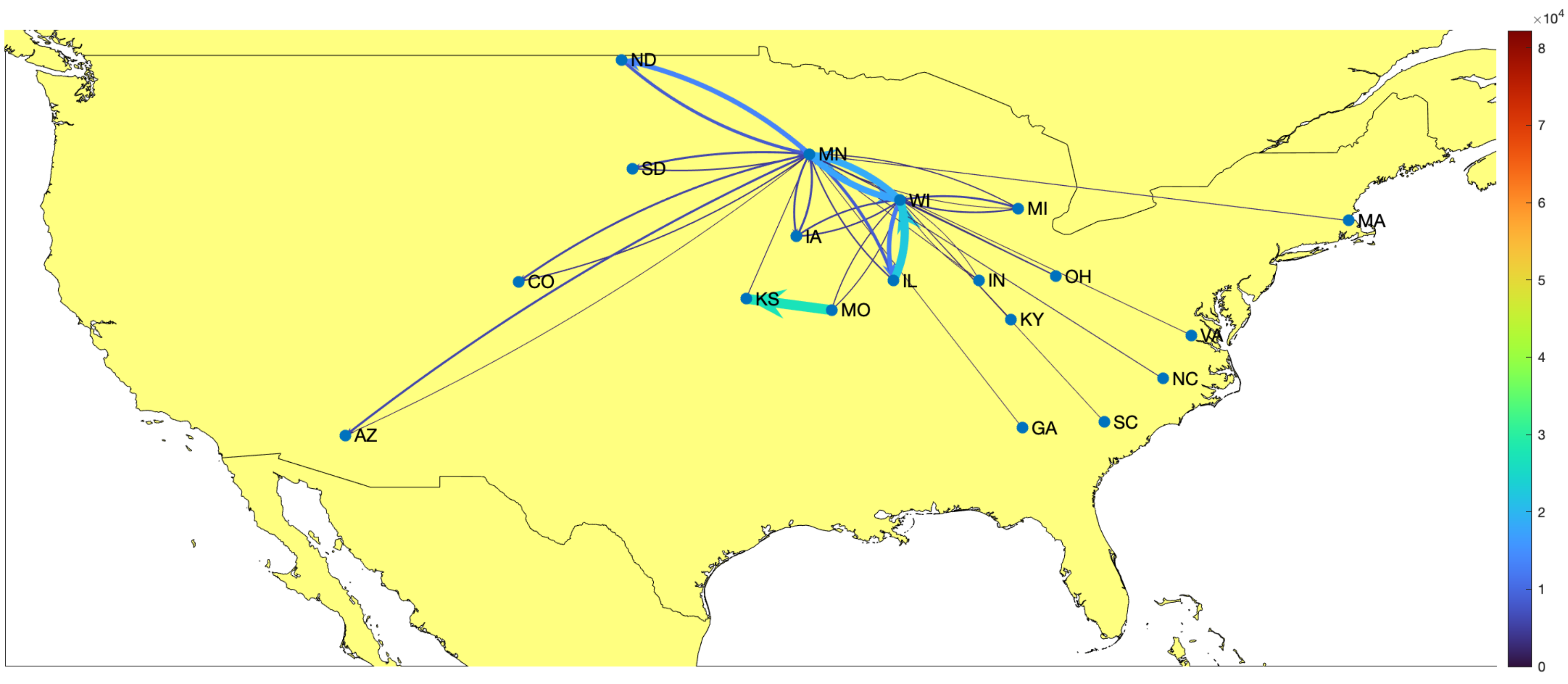}} \\
  \subfigure[]{%
  \includegraphics[width=3.00in, height=2.50in]{%
      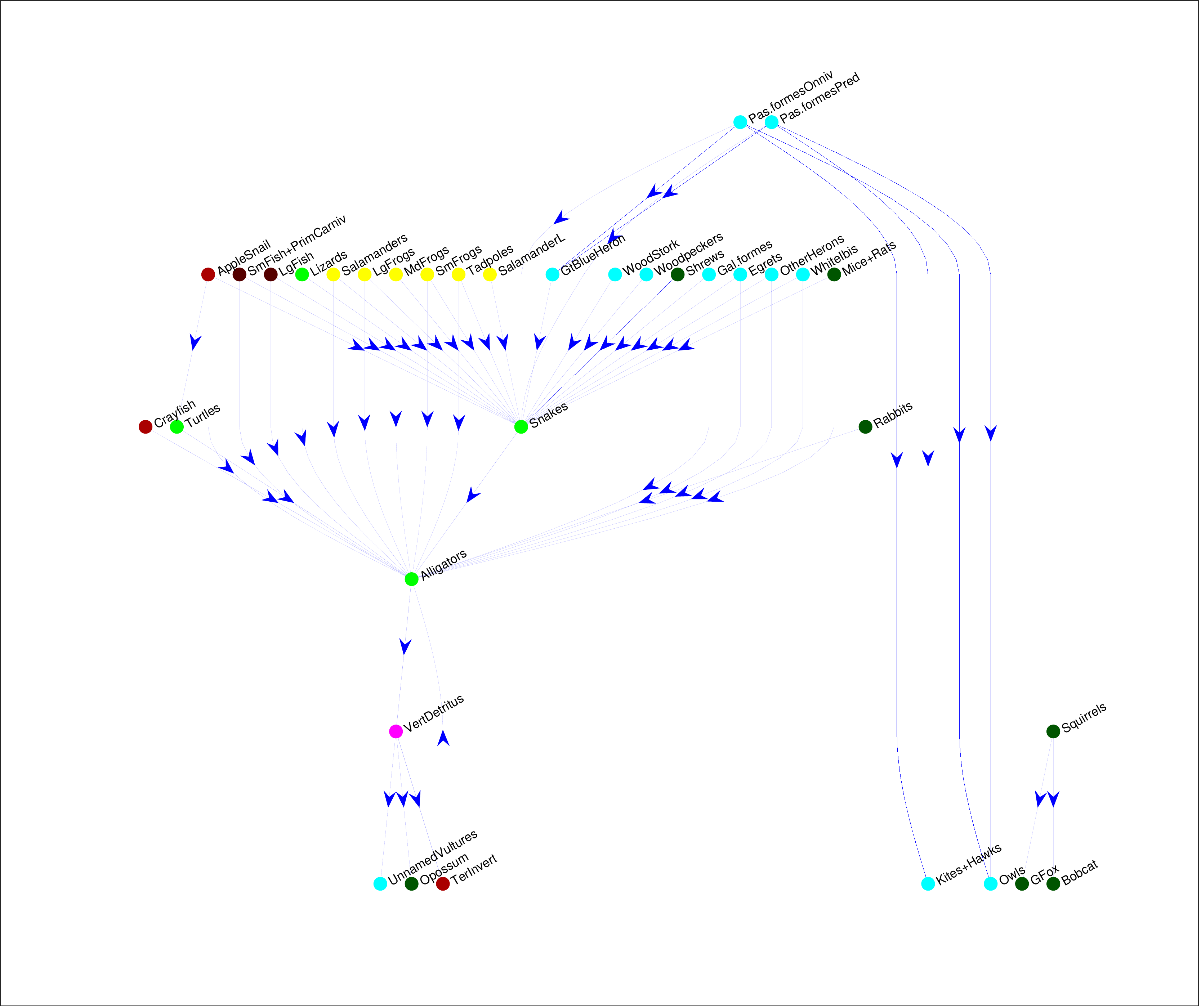}}
  \subfigure[]{%
  \includegraphics[width=3.00in, height=2.50in]{%
      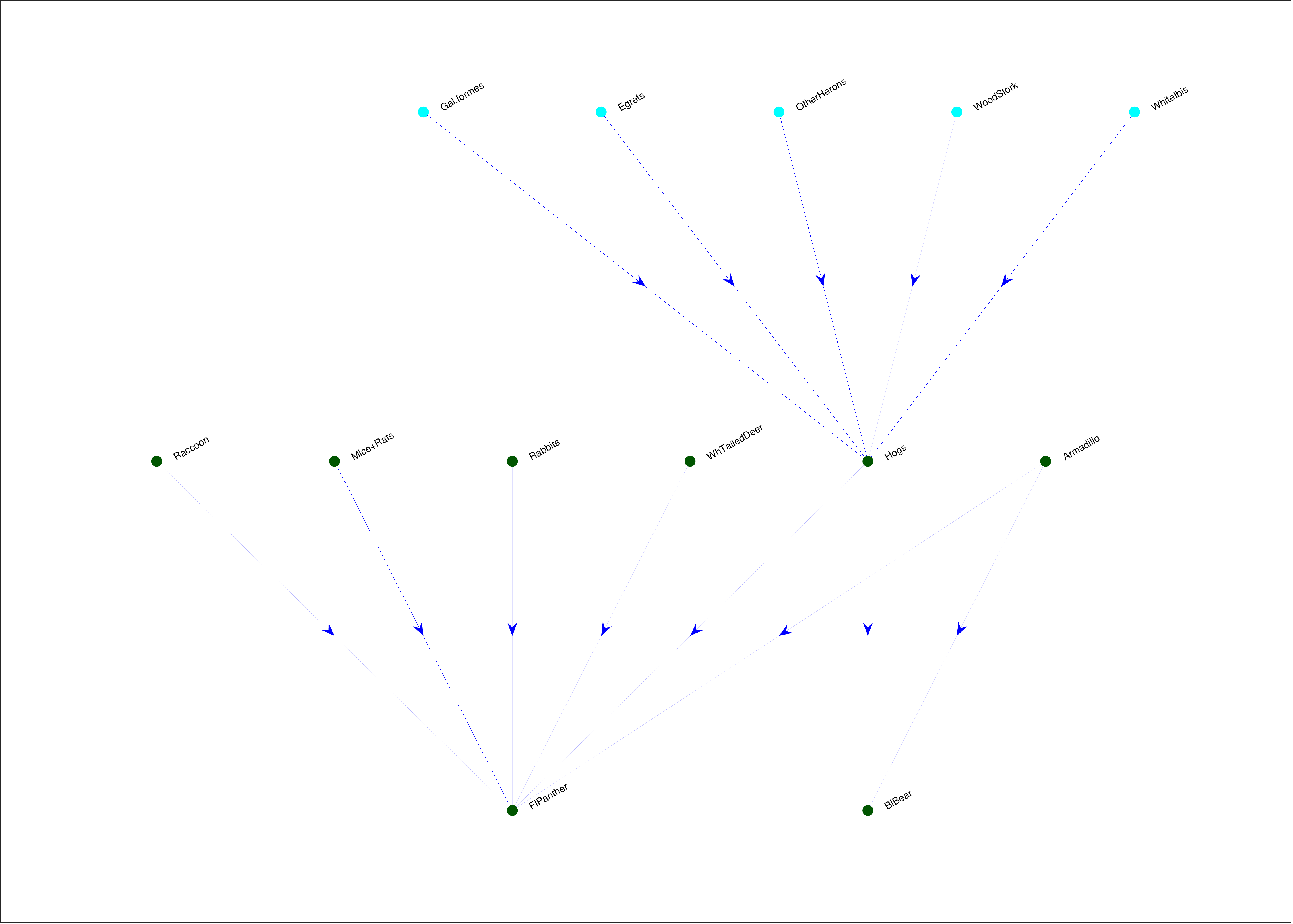}}
  \caption{2019 US Inter-State Migration Data and Florida Bay Cypress Wetlands Food Web. \\
  \emph{Top:} Migration flow data \cite{InterStateMigration_USCB2019} can be represented as a weighted digraph with edge direction representing the migratory movement. Edge clustering reveals meaningful subgraphs highlighting different types of migratory movement. For instance, the PRE cluster ($K = 6$) in \tb{(a)} confirms California as a large producer of migration, e.g., to Texas (edge color and thickness represent volume);  The other 5 clusters show migratory movement mainly involving CA, TX, NY, and FL (see SM).
  The RGE-based cluster ($K = 6$) in \tb{(b)} reveals a 'hidden' pocket of migration mainly among midwestern states centered on Minnesota; the remaining 5 clusters also reveal interesting patterns (see SM). For clarity, only edges with volumes at least 5\% of the maximum are shown. 
  \emph{Bottom:} Cypress wetlands food web represented as a weighted digraph \cite{Bondavalli2000JB}. Each vertex (color indicates one of 9 organism compartments) represents an organism; each directed edge (width is  proportional to edge weight) reflects biomass exchange. 
  DPE clustering yields meaningful subgraph clusters of focused direct and indirect carbon transfer, allowing visualization of trophic levels. The DPE edge cluster ($K = 10$) in \tb{(c)} highlights carbon transfer to-and-from alligators. 
  RGE clustering yields clusters of concentrated carbon flow among organisms. RGE cluster ($K = 10$) in \tb{(d)} highlights food sources of larger birds which in turn constitute the food source for hogs and then for black bears and the endangered Florida panther, revealing the important concentrated carbon flows involved. 
  Other clusters are in SM.}
\label{fig:USMigration_Cypress}
\end{figure}

Figs.~\ref{fig:CElegans_DPE_K30_N110_M183} and~\ref{fig:USMigration_Cypress} show examples of how edge clustering reveals meaningful subgraphs for various datasets. For the \emph{Caenorhabditis elegans} connectome, edge clustering yields subgraphs that link neurons with widely varying functionalities (e.g., sensory, inter, and motor neurons), in contrast to vertex clustering which groups similar neurons. The edge clusters suggest circuits associated with different types of behavior and provide clues for investigating additional neuronal functionality. For the US migration dataset, our edge clustering results not only confirm known patterns of migration but also reveal patterns that are otherwise not readily apparent (e.g., a `pocket' of migration among midwestern states). For the food web, edge clustering provides methods for better understanding trophic levels and for discerning carbon flows (e.g., those associated with endangered species). Additional examples on a variety of directed and undirected graphs are in the SM. For any desired number of clusters $K$, each of the three edge clustering methods reveal different insight into directed flow processes in graphs.

Our Flow Laplacian framework offers a strategy for clustering edges for the first time, thus providing unique and diverse insights into directed influence and concentrated flows in graphs via three different methods. Researchers studying undirected or directed graphs with positive weights can readily adopt our framework to reveal subgraphs that merit further study. With the weighted adjacency matrix and the desired number of clusters being the only inputs, the framework produces subgraph clusters through modest complexity, only requiring one to compute the $K$ eigenvectors associated with the minimum eigenvalues of a $M\times M$ symmetric psd matrix, followed by row normalization and k-means++ clustering. These edge clustering methods complement the well-known vertex clustering, and offer new avenues for exploratory data analysis in numerous disciplines.


\paragraph*{Acknowledgments.}


This work is partially supported by the University of Miami U-LINK and National Science Foundation Award \#2123635.

\texttt{MATLAB} code for Flow Laplacian computation and example data used are available at 

https://github.com/kpremaratne/EdgeClustering



\bibliographystyle{IEEEtran}

\bibliography{%
  LibraryPremaratne_Books,%
  LibraryPremaratne_Conferences,%
  LibraryPremaratne_Journals,%
  LibraryPremaratne_Other,%
  LibraryPremaratne_Datasets,%
  scibib}

\newpage


\captionsetup[figure]{}

\begin{figure}[H]
  \centering
  \includegraphics[width=6.3in]{%
      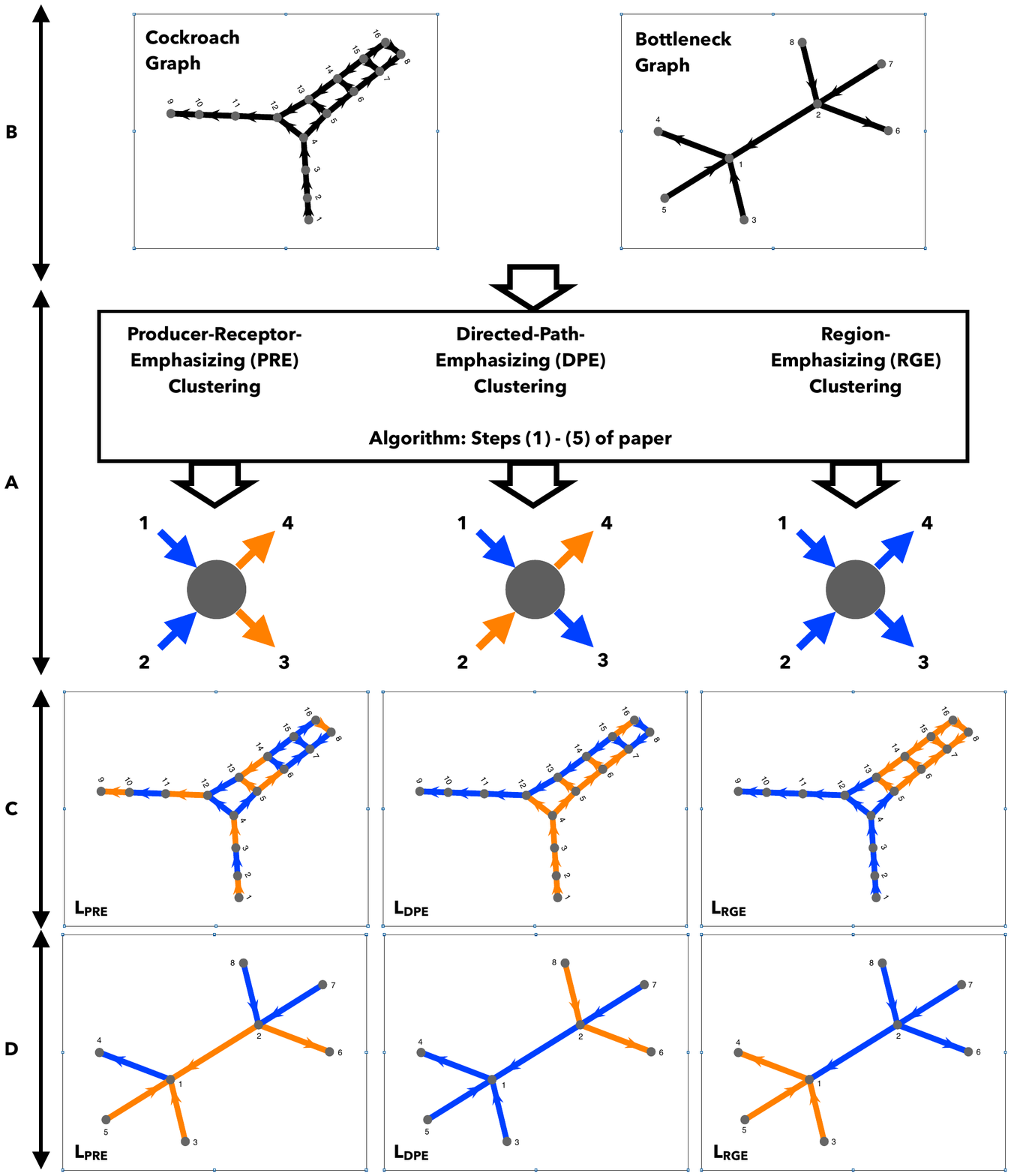}
\end{figure}
\clearpage 
\captionof{figure}{%
The Flow Laplacian Edge Clustering Framework. 
  \tb{(A)}~3 types of edge clustering: 
  \emph{Producer-Receptor-Emphasizing (PRE) clustering:} Edges conveying influence to a common vertex (edges 1,2) share a functional affinity as do edges conveying influence from a common vertex (edges 3,4). This encourages edge clusters where vertices act either mainly as producers (orange) or receptors (blue) of directed influence. 
  \emph{Directed Path-Emphasizing (DPE) clustering:} Edges along a directed path (edges (1,3) and (2,4)) share a functional affinity. This encourages clusters (blue and orange) with long directed paths. 
  \emph{Region-Emphasizing clustering (RGE) clustering:} Edges that share a common vertex (edges (1,2,3,4)) share a functional affinity. This encourages clustering of edges into localized regions. 
  \tb{(B)}~Two input unweighted digraphs: a `cockroach’ digraph and a `bottleneck’ digraph. Given the adjacency matrix and the number of clusters $K$, edges (not vertices) are labeled. 
  \tb{(C)}~Cockroach digraph edge clustering for $K=2$: PRE clustering yields an orange cluster where vertices (4,6,15) are producers, while (7,12,14) are receptors; the blue cluster has vertex 5 as a producer while 13 is a receptor. The long directed paths in the cockroach antennae have edges in alternating clusters because directed paths are not encouraged. DPE clustering yields clusters with long directed paths. RGE clustering ignores direction and partitions the edges into two regions. 
  \tb{(D)}~Bottleneck digraph clustering for $K=2$: PRE edge clustering yields an orange cluster where vertex 2 acts is a producer, while vertex 1 is mainly a receptor; in the blue cluster, these roles are reversed. DPE clustering yields clusters which feature long directed paths. RGE clustering partitions the edges into two regions.} 
\label{fig:OverallFlowDiagram}

\newpage

\vfill

\newpage

\newpage

\newpage


\setcounter{page}{1}

\renewcommand{\thesection}{S\arabic{section}}
\renewcommand{\thefigure}{S\arabic{figure}}
\renewcommand{\thetable}{S\arabic{table}}

\setlength{\parindent}{1.5em}


\begin{center}
{\Large {\bf SUPPLEMENTARY MATERIALS}}
\end{center}

{\Large Materials and Methods}


{\Large Supplementary Text} 


\section{Notation} 


\subsection{Basic Notation}


Given the set $\mc{X}$, its complement is $\overline{\mc{X}}$ and its cardinality is $\mr{card}[\mc{X}]$. The integers are $\mbb{N}$ and the real numbers are $\mbb{R}$. The $N\times 1$ column vectors and the $N\times M$ matrices with real-valued entries are denoted by $\mbb{R}^N$ and $\mbb{R}^{N\times M}$, resepectively. We will use $\mbb{N}_+$ and $\mbb{R}_+$ to denote non-negative integers and non-negative reals, respectively. Unless otherwise indicated, all vectors and matrices are taken to be real-valued. Given the real number $x$, $\mr{sgn}[x]$ denotes its sign. As usual, $\delta_{ij}$ is the Kronecker delta, i.e., $\delta_{ij} = 1$ for $i = j$, and $\delta_{ij} = 0$ otherwise.

We use $\ul{x} = \{x_i\}\in\mbb{R}^N$ to denote the $N\times 1$ column vector populated with the elements $x_i,\, i \in \{1, \ldots, N\}$, and $\ul{X} = \{x_{ij}\}\in\mbb{R}^{N\times M}$ to denote the $N\times M$ matrix populated with the elements $x_{ij},\, i \in \{1, \ldots, N\},\, j \in \{1, \ldots, M\}$. Given the matrix $\ul{X} = \{x_{ij}\}$, $\ul{X}^T$ is its transpose, $\ul{X}^H$ is its Hermitian transpose, and $|\ul{X}|$ and $\mr{sgn}[\ul{X}]$ are the matrices whose elements are the absolute values and the signs of the corresponding elements of $\ul{X}$, respectively, i.e., $|\ul{X}|=\{|x_{ij}|\}$ and $\mr{sgn}[\ul{X}] = \{\mr{sgn}[x_{ij}]\}$. $\mr{Tr}[\ul{X}]$ denotes the trace of the square matrix $\ul{X}$. We use $\ul{1}_N\in\mbb{R}^N$ to denote the $N\times 1$ column vector of all $1$\,s. For $\ul{x} \in \mbb{R}^N$, $\mr{diag}[\ul{x}]\in\mbb{R}^{N\times N}$ is the diagonal matrix with $\ul{x}$ on its main diagonal; for $\ul{X} \in \mbb{R}^{N\times N}$, $\mr{diag}[\ul{X}]\in\mbb{R}^N$ is the vector generated from the main diagonal of $\ul{X}$. By $\ul{X}\geq 0$ (or $\ul{X}>0$), we mean that all elements of the matrix $\ul{X}$ are non-negative (or positive). $\ul{X} \odot \ul{Y}$ denotes the Hadamard product (i.e., element-wise product) of matrices $\ul{X}$ and $\ul{Y}$.

The eigen-decomposition of the square symmetric matrix $\ul{X} \in \mbb{R}^{N \times N}$ is $\ul{X} = \ul{V} \ul{\Lambda} \ul{V}^H$. Here, $\ul{V} = [\ul{v}_1, \ldots, \ul{v}_N] \in \mbb{R}^{N \times N}$ is the matrix of eigenvectors $\ul{v}_i,\, i \in \{1, \ldots, N\}$; $\ul{\Lambda} = \mr{diag}[\ul{\lambda}] \in \mbb{R}^{N \times N}$, where $\ul{\lambda} = [\lambda_1, \ldots, \lambda_N]^T$, is the diagonal matrix of eigenvalues $\{\lambda_1, \ldots, \lambda_N\}$; and $\{\ul{v}_i, \lambda_i\}$ denotes the $i$-th unit eigenpair of $\ul{X}$. 

Table~\ref{tab:notation} summarizes this notation.


\subsection{Graph-Related Definitions}


We use $\mc{G}=\mc{G}(\mc{V}, \mc{E})$ to denote a digraph with $N$ vertices and $M$ directed edges. Here, $\mc{V}=\{v_1, \ldots, v_N\}$ is the set of $N$ vertices and $\mc{E}=\{e_{ij}=(v_j {\to}\, v_i)\}$, in which $e_{ij}=(v_j {\to}\, v_i)$ is the directed edge \emph{from} $v_j\in\mc{V}$ \emph{to} $v_i\in\mc{V}$, is the set of $M$ edges. Often, we will enumerate the edges in $\mc{E}$ as $\{e_1, \ldots, e_M\}$ too. If the edge $e_q,\, q \in \{1, \ldots, M\}$, corresponds to $e_{k_q,\ell_q}=(v_{\ell_q} {\to}\, v_{k_q}),\, k_q, \ell_q \in \{1, \ldots, N\}$, we will identify this correspondence as $e_q\cong e_{k_q,\ell_q}$, i.e., $e_q$ and $e_{k_q,\ell_q} = (v_{\ell_q} {\to}\, v_{k_q}\}$ is the same edge. The enumeration scheme which pairs a particular edge $e_q$ with $e_{k_q,\ell_q}$ is assumed fixed. 

The \emph{unweighted adjacency matrix} of $\mc{G}=\mc{G}(\mc{V}, \mc{E})$ is $\ul{A}=\{a_{ij}\}\in\mbb{R}^{N\times N}$ in which $a_{ij}=1$ if $e_{ij}=(v_j {\to}\, v_i)\in\mc{E}$ and $a_{ij}=0$ otherwise, i.e., if vertex $j$ has a directed edge to vertex $i$ then the $(i, j)$-th entry of $\ul{A}$ is one, otherwise it is zero. By a \emph{vertex function,} we refer to a real-valued function $\nu: \mc{V} \mapsto \mbb{R}: v_i \mapsto \nu_i \equiv \nu(v_i)$ defined on the vertices $v_i\in\mc{V}$ of the digraph $\mc{G}$; $\ul{\nu} = [\nu_1, \ldots, \nu_N]^T\in\mbb{R}^N$ is the corresponding \emph{vertex vector.} In a weighted digraph, we also have an \emph{edge weight function} defined on the edges $e_{ij}\in\mc{E}$. We then have a \emph{weighted adjacency matrix} $\ul{W}=\{w_{ij}\}\in\mbb{R}^{N\times N}$ where $w_{ij}$ denotes the weight associated with edge $e_{ij}$. We define $\ul{w} = [w_1, \ldots, w_M]^T\in\mbb{R}^M$ as the \emph{edge weight vector} where $w_q$ is the edge weight associated with edge $e_q\cong e_{k_q,\ell_q}$, i.e., using the same notation used for edges, $w_q\cong w_{k_q,\ell_q}$ refers to the weight associated with the edge $e_q\cong e_{k_q,\ell_q}$. We use $\ul{w}\cong\ul{W}$ to identify this relationship for all edges. Whenever necessary, we will use $\mc{G}(\mc{V}, \mc{E})[\ul{\nu}, \ul{w}]$ to explicitly identify the vertex and edge weight vectors associated with the digraph $\mc{G}$. 

The \emph{unweighted incidence matrix} is $\ul{B} = \{b_{ip}\}\in\mbb{R}^{N\times M}$ in which $b_{ip} = +1$ if edge $e_p$ corresponding to weight $w_p$, has vertex $v_i$ as its source vertex, $-1$ if vertex $v_i$ is its destination vertex, and $0$ otherwise. In this manner, the $(i, q)$-th element $b_{iq}$ of $\ul{B}$ encodes the relationship between edge $e_q\cong e_{k_q, \ell_q}$ and the source-destination vertex pair $(v_{\ell_q} {\to}\, v_{k_q})$ associated with it: for each edge $e_q\cong e_{k_q,\ell_q}$, the element $b_{iq} = +1$ if vertex $v_i$ is the source vertex $v_{\ell_q}$ of edge $e_q$ and $b_{iq} = -1$ if vertex $v_i$ is the destination vertex $v_{k_q}$ of edge $e_q$. So, $b_{iq} = (\delta_{i,\ell_q} - \delta_{i,k_q})$. 

For the vertex $v_i\in\mc{V}$, we use $\sigma_i \in \mbb{N}_+$ to denote the total (integer) number of edges it is connected to, and we refer to it as the \emph{social particpation} of vertex $v_i$; $d_{out,i} = \displaystyle\sum_j w_{ji}$ is its weighted out-degree and $d_{in,i} = \displaystyle\sum_j w_{ij}$ is its weighted in-degree. We also use $\langle d_{out,i}\rangle = \displaystyle\sum_j |w_{ji}|$ and $\langle d_{in,i}\rangle = \displaystyle\sum_j |w_{ij}|$ to denote the sum of the absolute values of outgoing and incoming edge weights of vertex $v_i$, respectively. The vectors $\ul{d}_{out} = \ul{W}\,\ul{1}_N\in\mbb{R}^N$ and $\ul{d}_{in} = \ul{W}^T\ul{1}_N\in\mbb{R}^N$ are the \emph{out-degree vector} and \emph{in-degree vector,} respectively; the matrices $\ul{D}_{out} = \mr{diag}[\ul{d}_{out}]\in\mbb{R}^{N\times N}$ and $\ul{D}_{in} = \mr{diag}[\ul{d}_{in}]\in\mbb{R}^{N\times N}$ whose diagonals are populated with the corresponding degree vectors are the \emph{out-degree matrix} and \emph{in-degree matrix,} respectively.

Table~\ref{tab:notation2} summarizes this notation. Additional notation introduced in the remainder of this work is summarized in Tables~\ref{tab:notation3} and \ref{tab:notation4}.

Proofs of all our claims (lemmas and theorems) appear in Section~\ref{sec:Proofs}.


\section{Edge Clustering} 


Suppose the $M$ digraph \emph{edges} are to be partitioned into the $K$ clusters, say, $\{\Sigma^{(1)}, \ldots, \Sigma^{(K)}\}$, where $\displaystyle \bigcup_{k=1}^K \Sigma^{(k)} = \mc{E}$ and $\Sigma^{(k)}\cap\Sigma^{(\ell)} = \emptyset$, for $k\neq\ell$. As in earlier \emph{vertex} clustering work \cite{Gallier2019_Book, Shi2000ToPAMI, VonLuxburg2007SC}, we now consider a `partition' matrix $\ul{X}=\{x_p^{(k)}\}\in\mbb{R}^{M\times K}$ of the type
\begin{equation}
  \ul{X} 
    =\begin{bmatrix}
       \ul{x}^{(1)} & \ldots & \ul{x}^{(k)} & \ldots & \ul{x}^{(K)}
     \end{bmatrix}
    =\begin{bmatrix}
       \ul{x}_1^T \\
       \vdots \\
       \ul{x}_p^T \\
       \vdots \\
       \ul{x}_M^T
     \end{bmatrix}.
  \label{eq:X}
\end{equation}
The $(p,k)$-th element $x_p^{(k)}$ of $\ul{X}$, which corresponds to the edge $e_p\in\mc{E}$ and the cluster $\Sigma^{(k)}$, is 
\begin{equation}
  x_p^{(k)}
    =\begin{cases}
       \alpha^{(k)},
         & \tr{when $e_p\in\Sigma^{(k)}$}; \\
       0,
         & \tr{when $e_p\in\overline{\Sigma}^{(k)}$},
     \end{cases}
  \label{eq:Indicator}
\end{equation}
for some non-zero real numbers $\alpha^{(k)},\, k\in\{1, \ldots, K\}$. Here $\overline{\Sigma}^{(k)}$ is of course the complement of $\Sigma^{(k)}$ in $\mc{E}$, i.e., $\overline{\Sigma}^{(k)} = \mc{E}\setminus\Sigma^{(k)}$. One may view the $k$-th column $\ul{x}^{(k)}\in\mbb{R}^M$ of $\ul{X}$ as an indicator vector for edge membership in the cluster $\Sigma^{(k)}$ in that each non-zero entry of this $k$-th column indicates that the corresponding edge is a member of cluster $\Sigma^{(k)}$. Let us use $\mc{X}$ to denote all partition matrices in $\mbb{R}^{M\times K}$ of the type in \eqref{eq:X}, i.e., for all $p \in \{1, \ldots, M\}$ and for all $k \in \{1, \ldots, K\}$, 
\begin{equation}
  \mc{X}
    =\{\ul{X} = [\ul{x}^{(1)}, \ldots, \ul{x}^{(K)}] \in \mbb{R}^{M \times K}    \mid 
    \ul{x}^{(k)} 
      = \alpha^{(k)} 
        [z_1^{(k)}, \ldots, z_M^{(k)}]^T,\;\;
        z_p^{(k)} \in \{0, 1\}
     \}.
  \label{eq:setX}
\end{equation}

Given that a partition matrix $\ul{X} \in \mc{X}$ is meant to specify the edges that belong to each cluster, similar to work in \cite{Gallier2019_Book, VonLuxburg2007SC}, we insist that $\ul{X}$ satisfies the following requirements:
\begin{itemize}
  \item[\tb{[P1]}] No edge belongs to multiple clusters: 
To capture this, we enforce the constraint that $\ul{x}^{(k_1)^T} \ul{x}^{(k_2)} = 0,\; k_1 \neq k_2$, i.e., the $K$ columns of $\ul{X}$ must be pairwise orthogonal. This requires $\ul{X}^T \ul{X}$ to be a diagonal matrix of the form \cite{Gallier2019_Book}
\begin{equation}
  \ul{X}^T \ul{X} 
    =\mr{diag}[M^{(1)} \alpha^{(1)^2}, \ldots, M^{(K)} \alpha^{(K)^2}],
  \label{eq:XTX}
\end{equation}
where $M^{(k)} = |\Sigma^{(k)}|$, the cardinality of cluster $\Sigma^{(k)}$ with of course $\displaystyle\sum_{k=1}^K M^{(k)} = M$, and $\alpha^{(k)} \neq 0,\, \forall k \in \{1, \ldots, K\}$.
  \item[\tb{[P2]}] No cluster of edges is empty: 
To capture this, we enforce the constraint that, for each $k \in \{1, \ldots, K\}$, $x_p^{(k)} = \alpha^{(k)} \neq 0$, for at least one $p \in \{1, \ldots, M\}$; equivalently, the cluster cardinality $M^{(k)} > 0$ \cite{Gallier2019_Book}.
\end{itemize}

Another feature that we desire is that each edge must be assigned to a cluster (i.e., no edge can be without a group membership). To capture this, as pointed out in \cite{Gallier2019_Book}, one may enforce the additional constraint $\ul{X}\, (\ul{X}^T \ul{X})^{-1} \ul{X}^T \ul{1}_M = \ul{1}_M$. However, the relaxation scheme we later employ to choose the solution $\ul{X}$ involves the application of a clustering algorthm (such as, k-means) to the rows of a certain $M\times K$ matrix, of which each row corresponds to a digraph edge (see Section~\ref{subsec:OptimizationProblem}). Since this process forces each edge to be assigned to a cluster, we proceed with only the constraints in [P1] and [P2]. 

Assigning edges to clusters then involves the selection of a partition matrix $\ul{X} \in \mc{X}$ that satisfies the constraint $\ul{X}^T \ul{X} = \mr{diag}\left[M^{(1)} \alpha^{(1)^2}, \ldots, M^{(K)} \alpha^{(K)^2}\right],\, \alpha^{(k)} \neq 0,\, M^{(k)} > 0$, while minimizing an appropriate `cost'  


\subsection{A Cost Function for Edge Clustering} 


Consider $x_p^{(k)}$ and $x_q^{(k)}$, the scalar indicator variables for whether the digraph edges $e_p$ and $e_q$ are members of cluster $\Sigma^{(k)}$, respectively. To compare the membership statuses of edges $e_p$ and $e_q$ in the cluster $\Sigma^{(k)}$, we use the following non-negative cost:

\begin{definition}
\label{def:Cost_k_pq} 
The cost associated with the edge pair $\{e_p, e_q\}$ in cluster $\Sigma^{(k)},\, k \in \{1, \ldots, K\}$, of the digraph $\mc{G}(\mc{V}, \mc{E})$ is
\[
  \tr{Cost}^{(k)}(e_p, e_q)
    =\frac{1}{2} 
     \phi_{pq}\,
     (x_p^{(k)} - \psi_{pq}\,x_q^{(k)})^2.
\]
Here, both the parameters $\psi_{pq}$ and $\phi_{pq}$ are real and symmetric in that $\psi_{pq} = \psi_{qp}$ and $\phi_{pq} = \phi_{qp}$ and, in addition, $\phi_{pq}$ takes the form 
\[
  \phi_{pq}
    =\begin{cases}
       \geq 0,
         & \tr{when $\{e_p, e_q\}$ share a common vertex and $p \neq q$}; \\
       0,
         & \tr{otherwise}. 
     \end{cases}
\]
\end{definition}

As we later show in Section~\ref{subsec:Psi}, $\psi_{pq}$ promotes a specified directed edge `functional affinity' in the clustering results by modifying how the indicator variable labels of the edge pair $\{e_p, e_q\}$ are compared; and $\phi_{pq}$ is the weight associated with the squared pairwise edge label comparison. The matrices $\ul{\Psi} = \{\psi_{pq}\}\in\mbb{R}^{M\times M}$ and $\ul{\Phi} = \{\phi_{pq}\}\in\mbb{R}^{M\times M}$ generated from the entries $\psi_{pq}$ and $\phi_{pq}$, respectively, are of course symmetric; in addition, $\ul{\Phi}$ is non-negative and has a zero diagonal. We defer a more detailed discussion regarding the design of these real scalar parameters $\psi_{pq}$ and $\phi_{pq}$ until later (see Sections \ref{subsec:Psi} and \ref{subsec:Phi}). 

With Definition~\ref{def:Cost_k_pq} in place, one may express the total cost associated with the edge pair $\{e_p, e_q\}$ by summing over the costs associated with the edge pair $\{e_p, e_q\}$ for all the clusters: 
\begin{equation}
  \tr{Cost}(e_p, e_q)
    =\sum_{k=1}^K
     \tr{Cost}^{(k)}(e_p, e_q)
    =\frac{1}{2} 
     \sum_{k=1}^K \phi_{pq}
     (x_p^{(k)} - \psi_{pq}\,x_q^{(k)})^2.
  \label{eq:edgepair_cost}
\end{equation}
On the other hand, the total cost associated with cluster $\Sigma^{(k)}$ can be obtained by summing the costs associated with corresponding indicator variable $k$ for all edge pairs:
\begin{equation}
  \tr{Cost}^{(k)}(\mc{G}) 
    =\sum_{p=1}^M \sum_{q=1}^M \tr{Cost}^{(k)}(e_p, e_q).
  \label{eq:cluster_cost}
\end{equation}

Let us now generate the diagonal matrix $\ul{D}_{\ul{\Psi}\ul{\Phi}} = \mr{diag}[\ul{d}_{\ul{\Psi}\ul{\Phi}}] \in \mbb{R}^{M\times M}$ whose diagonal is populated with the entries in the column vector $\ul{d}_{\ul{\Psi}\ul{\Phi}}\in\mbb{R}^M$ that is defined as
\begin{equation}
  \ul{d}_{\ul{\Psi}\ul{\Phi}}
    =\begin{bmatrix}
       d_{\ul{\Psi}\ul{\Phi}}(1), & d_{\ul{\Psi}\ul{\Phi}}(2), & \ldots, & d_{\ul{\Psi}\ul{\Phi}}(M)
     \end{bmatrix}^T,
  \;\;\tr{with}\;\;
  d_{\ul{\Psi}\ul{\Phi}}(p)
    =\frac{1}{2} \sum_{q=1}^M \phi_{pq}(1+\psi_{pq}^2).
  \label{eq:D_Psi_Phi}
\end{equation}
Then, $\tr{Cost}^{(k)}(\mc{G})$ in \eqref{eq:cluster_cost} associated with cluster $\Sigma^{(k)}$ can be expressed as a matrix quadratic form. 

\begin{lemma}
\label{lem:Cost_k_MQF}
The total cost $\tr{Cost}^{(k)}(\mc{G})$ associated with cluster $\Sigma^{(k)},\, k \in \{1, \ldots, K\}$, can be expressed as the matrix quadratic form 
\[
  \tr{Cost}^{(k)}(\mc{G})
    =\ul{x}^{(k)^T}(\ul{D}_{\ul{\Psi}\ul{\Phi}} - (\ul{\Psi}\odot\ul{\Phi}))\,\ul{x}^{(k)}.
\]
Here, $\ul{\Psi}\odot\ul{\Phi}\in\mbb{R}^{M\times M}$ refers to the Hadamard product (i.e., the element-wise product) of $\ul{\Psi}$ and $\ul{\Phi}$ and $\ul{x}^{(k)}$ is of course the $k$-th column of the partition matrix $\ul{X}$.
\end{lemma}

We now address how the entries of $\ul{\Psi}$ and $\ul{\Phi}$ are selected. 


\subsection{Entries of $\ul{\Psi}$ and Edge Functional Affinities}
\label{subsec:Psi}


The method by which we select the entries of the symmetric matrix $\ul{\Psi} = \{\psi_{pq}\}$ allows us to emphasize different edge functional affinities. Three edge functional affinities are of particular interest to us in our Flow Laplacian Edge Clustering framework.


\paragraph{(a)~Producer-Receptor-Emphasizing (PRE) Edge Functional Affinity}

Here we encourage edge clusters wherein edges that are both pointing to a common vertex are assigned the same edge label and hence grouped within the same cluster because of their similar functionality (viz., conveying influence to a common vertex). Similarly, we also encourage edge clusters wherein edges that are both pointing away from a common vertex are assigned the same edge label and hence grouped within the same cluster label, again because of their similar functionality (viz., conveying influence away from a common vertex). Accordingly, we select $\ul{\Psi}_{\PRE} = \{(\ul{\Psi}_{\PRE})_{pq}\}$ as follows: 
\begin{equation}
  (\ul{\Psi}_{\PRE})_{pq}
    =\begin{cases}
       +1,
         & \tr{when $e_p$ and $e_q$ both point to a common vertex}, \\
         & \tr{or when $e_p$ and $e_q$ both point away from a common vertex}; \\
       -1,
         & \tr{when $e_p$ points to and $e_q$ points away from a common vertex}, \\
         & \tr{or when $e_p$ points away from and $e_q$ points to a common vertex}; \\
       0,
         & \tr{when $e_p$ and $e_q$ do not share a common vertex}; \\
       +1,
         & \tr{when $p=q$}.
     \end{cases}
  \label{eq:Psi_PRE}
\end{equation}
With $\ul{\Phi}$ as in Definition~\ref{def:Cost_k_pq}, one may now express the Hadamard product $\ul{\Psi} \odot \ul{\Phi}_{\PRE}$ as
\begin{equation}
  (\ul{\Psi} \odot \ul{\Phi}_{\PRE})_{pq}
    =\begin{cases}
       +\phi_{pq},
         & \tr{when $e_p$ and $e_q$ both point to a common vertex}, \\
         & \tr{or when $e_p$ and $e_q$ both point away from a common vertex}; \\
       -\phi_{pq},
         & \tr{when $e_p$ points to and $e_q$ points away from a common vertex}, \\
         & \tr{or when $e_p$ points away from and $e_q$ points to a common vertex}; \\
       0,
         & \tr{when $e_p$ and $e_q$ do not share a common vertex}, \\
         & \tr{or when $p = q$}.
     \end{cases}
  \label{eq:Phi_Psi}
\end{equation}

Note that the diagonal elements of $\ul{\Psi}_{\PRE}$ in fact can be arbitrarily selected because, with $\phi_{pp} = 0$, they do not affect the Hadamard product $\ul{\Psi} \odot \ul{\Phi}_{\PRE}$ and hence the cost $\tr{Cost}^{(k)}(\mc{G})$ in Lemma~\ref{lem:Cost_k_MQF}. However, we choose $(\ul{\Psi}_{\PRE})_{pp} = +1$ in \eqref{eq:Psi_PRE} because it allows us to express $d_{\ul{\Psi}\ul{\Phi}}(p)$ in Lemma~\ref{lem:Cost_k_MQF} compactly as
\begin{equation}
  d_{\ul{\Phi}}(p)
    \equiv d_{\ul{\Psi}\ul{\Phi}}(p)
    =\sum_{q=1}^M \phi_{pq},\;
     \forall p \in \{1, \ldots, M\}.
  \label{eq:sum_phi}
\end{equation}
This allows us to express $\ul{d}_{\ul{\Psi}\ul{\Phi}} \in \mbb{R}^M$ and the diagonal matrix $\ul{D}_{\ul{\Psi}\ul{\Phi}} \in \mbb{R}^{M\times M}$ in Lemma~\ref{lem:Cost_k_MQF} as 
\begin{equation}
  \ul{d}_{\ul{\Phi}}
    \equiv\ul{d}_{\ul{\Psi}\ul{\Phi}}
    =\ul{\Phi}\, \ul{1}_M
    \in \mbb{R}^M
  \;\;\tr{and}\;\;
  \ul{D}_{\ul{\Phi}}
    \equiv\ul{D}_{\ul{\Psi}\ul{\Phi}}
    =\mr{diag}[\ul{d}_{\ul{\Phi}}]
    =\mr{diag}[\ul{\Phi}\, \ul{1}_M] \in 
    \mbb{R}^{M\times M}.    
  \label{eq:D_phi}
\end{equation}

Now let us consider \eqref{eq:edgepair_cost}, the total cost $\tr{Cost}(e_p, e_q)$ associated with the edge pair $\{e_p, e_q\}$ for this PRE edge affinity case. If $e_p$ and $e_q$ do not share a common vertex, then $\phi_{pq}=0$ and so $\tr{Cost}(e_p, e_q) = 0$, meaning that the edge pair does not contribute to the cost. On the other hand, if $e_p$ and $e_q$ both point to a common vertex, then $(\ul{\Psi}_{\PRE})_{pq} = +1$ and so
\begin{equation}
  \tr{Cost}_{\PRE}(e_p, e_q) 
    =\frac{1}{2} \phi_{pq}
     \sum_{k=1}^K 
     (x_p^{(k)} - x_q^{(k)})^2.
\end{equation}
Therefore $\tr{Cost}(e_p, e_q)$ achieves its minimum value of $0$ when $x_p^{(k)} = x_q^{(k)}$, i.e., when the edges both belong to the cluster $\Sigma^{(k)}$ or $\overline{\Sigma}^{(k)}$. The situation is similar if $e_p$ and $e_q$ both point away from a common vertex. In this manner, $\ul{\Psi}_{\PRE}$ in \eqref{eq:Psi_PRE} indeed encourages the PRE edge functional affinity. As we shall see, the PRE edge functional affinity leads to an edge clustering algorithm that generates edge clusters consisting of subgraphs of directed influence in which certain vertices are mainly producers (with outoing edges conveying influence) while other vertices are mainly receptors (with incoming edges receiving influence). 


\paragraph{(b)~Directed-Path-Emphasizing (DPE) Edge Functional Affinity}

Here we encourage edge clusters wherein a pair of edges that form a directed path through a common vertex are assigned the same edge label and hence grouped within the same cluster because of their similar functionality.  That is, if edge $e_p$ is pointing to a common vertex, and edge $e_q$ is pointing away from a common vertex, then the two edges form a directed path and are encouraged to be in the same cluster.  The same holds if edge $e_p$ is pointing away from, and edge $e_q$ is pointing to the common cluster. With an argument very similar to the one used above for PRE edge functional affinity, it is clear that this DPE edge functional affinity can be realized by selecting $\ul{\Psi}_{\DPE} = \{(\ul{\Psi}_{\DPE})_{pq}\}$ as
\begin{equation}
  \ul{\Psi}_{\DPE}
    =-\ul{\Psi}_{\PRE}.
  \label{eq:Psi_DPE}
\end{equation}
To better grasp this, consider the case when $e_p$ and $e_q$ form a directed path through a common vertex.  In this case, $(\ul{\Psi}_{\DPE}) = +1$, and so the cost becomes
\begin{equation}
  \tr{Cost}_{\DPE}(e_p, e_q) 
    =\frac{1}{2} \phi_{pq}
     \sum_{k=1}^K 
     (x_p^{(k)} - x_q^{(k)})^2,
\end{equation}
which is obviously minimized when $x_p^{(k)} = x_q^{(k)}$, i.e., when the edges both belong to the cluster $\Sigma^{(k)}$ or $\overline{\Sigma}^{(k)}$. As we shall see, this DPE edge functional affinity will lead to edge clusters consisting of subgraphs featuring long directed paths linking together vertices with high relative importance. 


\paragraph{(c)~Region-Emphasizing (RGE) Edge Functional Affinity}

Here we encourage edge clusters wherein a pair of edges that share a common vertex, irrespective of the edges' directionality, are assigned the same edge label and hence grouped within the same cluster. If edges $e_p$ and $e_q$ share a common vertex, then they are encouraged to be in the same cluster, regardless of their direction. It is clear that this RGE edge functional affinity can be realized by selecting $\ul{\Psi}_{\RGE} = \{(\ul{\Psi}_{\RGE})_{pq}\}$ as
\begin{equation}
  \ul{\Psi}_{\RGE}
    =|\ul{\Psi}_{\PRE}|
    =|\ul{\Psi}_{\DPE}|.
  \label{eq:Psi_RGE}
\end{equation}
As we shall see, the RGE edge functional affinity will lead to edge clusters consisting of subgraphs of concentrated flow among vertices of higher relative importance. 

The matrices $\ul{\Psi}_{\PRE}$, $\ul{\Psi}_{\DPE}$, and $\ul{\Psi}_{\RGE}$ above can be compactly represented as follows:

\begin{lemma}
\label{lem:Psi}
$\ul{\Psi}_{\PRE}$, $\ul{\Psi}_{\DPE}$, and $\ul{\Psi}_{\RGE}$ in \eqref{eq:Psi_PRE}, \eqref{eq:Psi_DPE}, and \eqref{eq:Psi_RGE} associated with PRE, DPE, and RGE edge functional affinities can be expressed as 
\[
  \ul{\Psi}_{\PRE} 
    =\mr{sgn}[\ul{B}^T\ul{B}];\;\;
  \ul{\Psi}_{\DPE} 
    =-\ul{\Psi}_{\PRE}
    =-\mr{sgn}[\ul{B}^T\ul{B}];\;\;
  \ul{\Psi}_{\RGE} 
    =|\ul{\Psi}_{\PRE}|
    =\mr{sgn}[|\ul{B}^T\ul{B}|],
\]    
respectively. Here, $\ul{B} \in \mbb{R}^{N \times M}$ denotes the digraph's unweighted incidence matrix.
\end{lemma}

This then leads us to 

\begin{lemma} 
\label{lem:Cost_k_L}
The total cost $\tr{Cost}^{(k)}(\mc{G})$ associated with cluster $\Sigma^{(k)},\, k \in \{1, \ldots, K\}$, is
\[
  \tr{Cost}^{(k)}(\mc{G})
    =\ul{x}^{(k)^T}\ul{L}\,\ul{x}^{(k)},
\]
where
\[
  \ul{L}
    =\begin{cases}
       \ul{L}_{\PRE}
         =\ul{D}_{\ul{\Phi}} - (\ul{\Psi}_{\PRE}\odot\ul{\Phi}),
         & \tr{for PRE edge functional affinity}; \\
       \ul{L}_{\DPE}
         =\ul{D}_{\ul{\Phi}} + (\ul{\Psi}_{\PRE}\odot\ul{\Phi}),
         & \tr{for DPE edge functional affinity}; \\
       \ul{L}_{\RGE}
         =\ul{D}_{\ul{\Phi}} - |\ul{\Psi}_{\PRE}\odot\ul{\Phi}|,
         & \tr{for RGE edge functional affinity}.
     \end{cases}
\]
Here, $\ul{D}_{\ul{\Phi}} = \mr{diag}[\ul{\Phi}\,\ul{1}_M]$ as in \eqref{eq:D_phi} and the Hadamard product $\ul{\Psi}_{\PRE}\odot\ul{\Phi}$ is as in \eqref{eq:Phi_Psi}.
\end{lemma}

We refer to $\ul{L}_{\PRE}, \ul{L}_{\DPE}$ and $\ul{L}_{\RGE}$ as \emph{Flow Laplacians,} as they allow one to express clustering costs as a matrix quadratic function, similar to the case of vertex clustering. These Flow Laplacians are at the heart of the three different edge clustering methods, each centered on a different edge functional affinity.

Table~\ref{tab:parameters} summarizes the parameters associated with the three edge functional affinities when $\ul{\Phi}$ is as in Definition~\ref{def:Cost_k_pq}.


\subsection{Entries of $\ul{\Phi}$ and Pairwise Edge Label Comparison}
\label{subsec:Phi}


The matrix $\ul{\Phi} = \{\phi_{pq}\}$ specifies the weighting of the squared edge label comparisons. Recall that $\phi_{pq} \ge 0$, and is identically zero if edges $e_p$ and $e_q$ do not share a common vertex, or if $p=q$.  We therefore focus on designing the $\phi_{pq}$ value for the case in which directed edges $e_p$ and $e_q$ share a common vertex which we shall identify as $v_i$. 

While many design choices are possible, to simplify matters, we select $\ul{\Phi} = \{\phi_{pq}\}$ as 
\begin{equation}
  \phi_{pq}
    =\begin{cases}
       \nu_i \geq 0,
         & \tr{when $\{e_p, e_q\}$ share the vertex $v_i$ and $p \neq q$}; \\
       0,
         & \tr{otherwise},
     \end{cases}
  \label{eq:phi_nu}
\end{equation}
where $\ul{\nu} = \{\nu_i\} \in \mbb{R}^N$ is a vertex vector such that $\ul{\nu} \geq 0$. We will use $\ul{\Phi}(\ul{\nu})$ to recognize this vertex function dependance.

One reasonable choice is to pick $\ul{\nu}$ so that $\nu_i$ captures the relative importance of the vertex $v_i$. This means that the $\phi_{pq}$ term weighting the pairwise edge label comparison is larger if the two edges share a relatively important common vertex, leading to a larger penalty if edges sharing an affinity are placed in separate cluster groups. In our experiments, we selected $\nu_i$ to be proportional to $\langle d_i \rangle \equiv \langle d_{out,i} \rangle + \langle d_{in,i} \rangle$, the sum of the absolute values of edge weights attached to $v_i$ (see \eqref{eq:nuvectorchoice}). For unweighted digraphs, this reduces to $\sigma_i$, the total (integer) number of edges that are connected to, or the social participation of, vertex $v_i$ (see Definition~\ref{def:VolEdge}). An added advantage that this choice offers is that it can be computed solely using the weighted adjacency matrix, and no other external information. Nevertheless, other options for designing $\phi_{pq}$ or $\ul{\nu}$ that reflect the relative importance of the common vertex of edges $p$ and $q$ as well as their relative similarities (e.g., through external side information) are possible. 


\section{Graph Cuts-Based Interpretation of Edge Clustering}


With both $\ul{\Psi}$ and $\ul{\Phi}$ specified, the three Flow Laplacians are subsequently also specified, and so one can re-focus attention on deriving the algorithms for edge clustering. However, it turns out that the three Flow Laplacians can be re-written in terms of a certain \emph{edge} Laplacian matrix, and that this connection allows us to interpret our clustering costs in terms of graph cuts.  We now unveil this connection.


\subsection{A Unified View of the Flow Laplacians Using An Edge Laplacian}


Recall that one can define the standard `vertex' Laplacian matrix of an \emph{undirected graph} as $\ul{L}_v(\ul{w}) = \ul{B}\, \mr{diag}[\ul{w}]\, \ul{B}^T \in \mbb{R}^{N\times N}$ in which $\ul{B} \in \mbb{R}^{N\times M}$ is the aforementioned unweighted incidence matrix, and $\ul{w} \in \mbb{R}^M$ is the column vector of edge weights. This suggests the following expression for an `edge' Laplacian applicable to \emph{digraphs:}

\begin{definition}[Edge Laplacian]
\label{def:Le}
The \emph{edge Laplacian} of the digraph $\mc{G}(\mc{V}, \mc{E})$ is the symmetric matrix $\ul{L}_e(\ul{\nu}) = \ul{B}^T \, \mr{diag}[\ul{\nu}]\, \ul{B} \in \mbb{R}^{M\times M}$. where $\ul{B} \in \mbb{R}^{N\times M}$ is the unweighted incidence matrix and $\ul{\nu} = \{\nu_i\} \in \mbb{R}^N$ is an arbitrary vertex vector. 
\end{definition}

This edge Laplacian, which can be used to study flows in digraphs, is a generalization of the Hodge Laplacian that is typically defined for the case of $\ul{\nu} = \ul{1}_N \in \mbb{R}^N$ \cite{Schaub2018GlobalSIP, Jia2019ACM_SIGKDD}. While the more well-known vertex Laplacian has been used to design vertex clustering algorithms, there appears to be no previous work on utilizing the edge Laplacian to cluster edges in digraphs (nor in undirected graphs). The following result exposes the reason:

\begin{theorem} 
\label{thm:DS} 
Consider the digraph edge Laplacian $\ul{L}_e(\ul{\nu}) \in \mbb{R}^{M\times M}$ in Definition~\ref{def:Le} associated with the vertex vector $\ul{\nu} \in \mbb{R}^N$. Then, for an arbitrary edge weight vector $\ul{w} \in \mbb{R}^M$, the \emph{`edge' differential} $(\ul{L}_e(\ul{\nu})\,  \ul{w})_p$ at edge $e_p \cong e_{k_p\ell_p}$ and the \emph{`edge' quadratic sum} $\ul{w}^T \ul{L}_e(\ul{\nu})\, \ul{w}$ are given by 
\[
  (\ul{L}_e(\ul{\nu})\, \ul{w})_p 
    = \nu_{\ell_p}\, d_{net,\ell_p} - \nu_{k_p}\, d_{net,k_p}
  \;\;\tr{and}\;\;
  \ul{w}^T \ul{L}_e(\ul{\nu})\, \ul{w} 
    = \sum_{v_i\in\mc{V}} \nu_i\, d_{net,i}^2,
\]
respectively. Here, $d_{out, i}$ and $d_{in, i}$ denote the weighted out-degree and the weighted in-degree of vertex $v_i\in\mc{V}$, respectively; and $d_{net,i} = d_{out,i} - d_{in,i}$.
\end{theorem}

We may view $d_{out,i}$ as the total directed influence of vertex $v_i$ on its \emph{successor vertices} and $d_{in,i}$ as the total directed influence of its \emph{predecessor vertices} on vertex $v_i$. Then $d_{net,i} = d_{out,i} - d_{in,i}$ is the \emph{net directed influence} of vertex $v_i$: if $d_{net,i}>0$ the vertex is more of a producer of directed influence; if $d_{net,i}<0$, it is more of a receptor of directed influence. Consequently the value of $d_{net,i}^2$ is larger for vertices that are either mainly producers or receptors of directed influence, and $\nu_i$ allows for vertex-dependent scaling of this squared net directed influence $d_{net,i}^2$. The quadratic form $\ul{w}^T \ul{L}_e(\ul{\nu})\, \ul{w}$, as Theorem~\ref{thm:DS} demontrates, captures the total scaled squared net directed influence for the entire digraph. Unfortunately, however, this quadratic does not allow for pairwise edge label comparisons that are needed to define edge clustering costs. To elaborate, suppose we desire to assign edges to clusters so as to minimize the total net directed flow as captued by $\ul{w}^T \ul{L}_e(\ul{\nu})\, \ul{w}$. But the cluster assignment of the edges of one vertex to minimize its net directed flow impacts the net directed flow of all its successor and predecessor vertices. 

Fortunately, as it turns out, we still can connect this edge Laplacian to edge clustering by expressing the $\ul{\Psi}$ and $\ul{\Phi}$ matrices, and hence the three Flow Laplacians, in terms of $\ul{L}_e$. To proceed, we first establish the relationship between $\ul{\Phi}$ in \eqref{eq:phi_nu} and the edge Laplacian $\ul{L}_e$ in Definition~\ref{def:Le}:

\begin{lemma}
\label{lem:Phi_Le} 
Consider the digraph edge Laplacian $\ul{L}_e(\ul{\nu}) \in \mathbb{R}^{M\times M}$ in Definition~\ref{def:Le} associated with the vertex vector $\ul{\nu} \in \mbb{R}^N$. We assume $\ul{\nu}$ has non-negative entries, i.e., $\ul{\nu} \ge 0$. Then, 
\[
  \ul{\Phi}(\ul{\nu})
    =|\ul{W}'(\ul{\nu})|,
  \;\;\tr{where}\;\;
  \ul{W}'(\ul{\nu}) 
    =\ul{L}_e(\ul{\nu}) - \mr{diag}[\mr{diag}[\ul{L}_e(\ul{\nu})]]
    \in\mbb{R}^{M\times M},
\]
where $\ul{\Phi}(\cdot) \in \mathbb{R}^{M\times M}$ is as in \eqref{eq:phi_nu}.
\end{lemma}

Thus, for a given vertex vector $\ul{\nu} \geq 0$, the matrix $\ul{\Phi}(\ul{\nu})$ in \eqref{eq:phi_nu} is the absolute value of the matrix $\ul{W}'(\ul{\nu})$ formed by first constructing the edge Laplacian $\ul{L}_e(\ul{\nu})$ in Definition~\ref{def:Le} and then removing its diagonal. 

For the strictly positive vertex vector case, i.e., for $\ul{\nu} > 0$, Lemma~\ref{lem:Phi_Le} now allows us to express the three Flow Laplacian matrices $\ul{L}_{\PRE}$, $\ul{L}_{\DPE}$, and $\ul{L}_{\RGE}$ in Lemma~\ref{lem:Cost_k_L} as follows:

\begin{theorem}
\label{thm:L_W'}
Consider the digraph edge Laplacian $\ul{L}_e(\ul{\nu}) \in \mathbb{R}^{M\times M}$ in Definition~\ref{def:Le} associated with the vertex vector $\ul{\nu} \in \mbb{R}^N$ where $\ul{\nu} > 0$. Then, $\ul{L}_{\PRE}, \ul{L}_{\DPE}, \ul{L}_{\RGE} \in \mbb{R}^{M\times M}$ in Lemma~\ref{lem:Cost_k_L} are given by 
\[
  \ul{L}_{\PRE}(\ul{\nu})
    =\ul{D}_{|\ul{W}'|}  - \ul{W}'(\ul{\nu});\;\;
  \ul{L}_{\DPE}
    =\ul{D}_{|\ul{W}'|} + \ul{W}'(\ul{\nu});\;\;
  \ul{L}_{\RGE}
    =\ul{D}_{|\ul{W}'|} - |\ul{W}'(\ul{\nu})|,
\]
where $\ul{W}'(\ul{\nu}) = \ul{L}_e(\ul{\nu}) - \mr{diag}[\mr{diag}[\ul{L}_e(\ul{\nu})]]$ and $\ul{D}_{|\ul{W}'|} \in \mbb{R}^{M\times M}$ is the diagonal matrix given by $\ul{D}_{|\ul{W}'|} = \mr{diag}[|\ul{W}'(\ul{\nu})|\, \ul{1}_M]$.
\end{theorem}

Table~\ref{tab:parameters} summarizes what Theorem~\ref{thm:L_W'} states regarding the parameters associated with the three edge functional affinities when $\ul{\Phi}(\ul{\nu})$ is chosen as in \eqref{eq:phi_nu} with $\ul{\nu} > 0$.


\subsection{Dual Graph-Based Interpertation}


\paragraph{Dual Graph} 

Noting that the edge Laplacian $\ul{L}_e(\cdot)$ is symmetric, we can view the matrix $\ul{W}^{'} = \ul{L}_e - \mr{diag}[\mr{diag}[\ul{L}_e]] \in \mbb{R}^{M\times M}$ as the adjacency matrix of an undirected, but signed, `dual' graph.

\begin{definition}[Dual Graph]
\label{def:DualGraph} 
The \emph{dual graph} of the digraph $\mc{G}(\mc{V}, \mc{E})[\ul{\nu}, \ul{w}]$, in which $\ul{\nu} \in \mbb{R}^N$ and $\ul{w} \in \mbb{R}^M$ are the vertex and edge weight vectors, respectively, is the undirected signed graph $\mc{G}'(\mc{V}', \mc{E}')[\ul{\nu}', \ul{w}']$ induced by the adjacency matrix $\ul{W}'(\ul{\nu}) = \ul{L}_e(\ul{\nu}) - \mr{diag}\mr{diag}[\ul{L}_e(\ul{\nu})]] \in \mbb{R}^{M\times M}$, where $\ul{L}_e(\ul{\nu})$ is the edge Laplacian of $\mc{G}(\mc{V}, \mc{E})[\ul{\nu}, \ul{w}]$. The vertices in $\mc{V}'$ and edges in $\mc{E}'$ are referred to as the \emph{dual vertices} and \emph{dual edges,} respectively.
\end{definition}

This notion of a dual graph $\mc{G}'(\mc{V}', \mc{E}')$ can in fact be considered a `generalization' (albeit a sign difference) of another notion that one associates with a digragh, viz., the \emph{line graph} $\mc{G}''(\mc{V}'', \mc{E}'')$ \cite{Orlin1977IM}. Edges in the original digraph become vertices in the dual graph.  Dual graph vertices are connected by a dual graph edge if they share a common vertex in the original graph. The dual graph is undirected and signed in that the dual edge weights are both positive and negative.  In fact, if the original digraph directed edges form a directed path through the common vertex $v_i$ in the original digraph, then the dual graph edge weight is $-\nu_i$; if the original digraph directed edges either both point to or both point away from $v_i$, then the dual graph edge weight is $+\nu_i$ (recall that the non-negative $\nu_i$ reflect the relative importance of the vertex $v_i$). A more in-depth examination of the dual graph appears in Section~\ref{subsec:DualGraph}. 

Consequently, we can interpet the three edge clustering methods in the original digraph $\mc{G}$ as clustering the dual vertices and as different variations on minimizing signed cuts in the signed dual graph $\mc{G}'$. To proceed, for $\ul{x}^{(k)} \in \mbb{R}^M$, the $k$-th column of a matrix $\ul{X} = [\ul{x}^{(1)}, \ldots, \ul{x}^{(K)}] \in \mc{X}$, consider the total cost associated with cluster $\Sigma^{(k)}$ in Lemma~\ref{lem:Cost_k_L}. With vertex vector $\ul{\nu} \in \mbb{R}^N,\, \ul{\nu} > 0$, substitute the expressions in Table~\ref{tab:parameters} to get the following:

\begin{lemma}
\label{lem:3Costs}
For the PRE, DPE, and RGE edge functional affinities, the total costs associated with cluster $\Sigma^{(k)},\, k \in \{1, \ldots, K\}$, are, respectively, 
\begin{alignat*}{3}
  &\tr{Cost}_{\PRE}^{(k)}(\mc{G})
    &
      &=\ul{x}^{(k)^T} \ul{L}_{\PRE}\, \ul{x}^{(k)} 
        &
          &=\frac{1}{2} 
            \sum_{p=1}^M \sum_{q=1}^M 
            |w'_{pq}|\, (x_p^{(k)} - \mr{sgn}[w'_{pq}]\, x_q^{(k)})^2; 
            \notag \\
  &\tr{Cost}_{\DPE}^{(k)}(\mc{G})
    &
      &=\ul{x}^{(k)^T} \ul{L}_{\DPE}\, \ul{x}^{(k)} 
        &
          &=\frac{1}{2} 
            \sum_{p=1}^M \sum_{q=1}^M 
            |w'_{pq}|\, (x_p^{(k)} + \mr{sgn}[w'_{pq}]\, x_q^{(k)})^2;
            \notag \\
  &\tr{Cost}_{\RGE}^{(k)}(\mc{G})
    &
      &=\ul{x}^{(k)^T} \ul{L}_{\RGE}\, \ul{x}^{(k)} 
        &
          &=\frac{1}{2} 
            \sum_{p=1}^M \sum_{q=1}^M 
            |w'_{pq}|\, (x_p^{(k)} -  x_q^{(k)})^2,
\end{alignat*}
where $W' = \{w'_{pq}\} \in \mbb{R}^{M\times M}$ is the adjacency matrix of the dual graph.
\end{lemma}

These expressions resemble vertex clustering cost functions, except that it is the edge weights of the dual graph that are being utilized. To interpet these costs in terms of cuts in the dual graph (which is signed and undirected, as shown in Lemma~\ref{lem:DualGraph}), we first need the following:

\begin{definition}
\label{def:Cuts}
For two sets of dual vertices $\mc{A}, \mc{B} \subseteq \mc{V}'$ in the dual graph $\mc{G}'(\mc{V}', \mc{E}')[\ul{\nu}', \ul{w}']$ associated with the diagraph $\mc{G}(\mc{V}, \mc{E})[\ul{\nu}, \ul{w}]$, we define the following:
\begin{itemize}
  \item[\tb{(i)}] Absolute cut between $\mc{A}$ and $\mc{B}$:
\[
  \mr{Cut}(\mathcal{A}, \mathcal{B}) 
    =\sum_{v'_p \in \mc{A}} \sum_{v'_q \in \mc{B}} 
     |w'_{pq}|.
\]
  \item[\tb{(ii)}] Sum of the positive links between $\mc{A}$ and $\mc{B}$:
\[
  \mr{Links}^+(\mc{A}, \mc{B}) 
    =\sum_{v'_p \in \mc{A}} \sum_{v'_q \in \mc{B}}
     \max\{0, +w'_{pq}\}.
\]
  \item[\tb{(iii)}] Sum of the negative links between $\mc{A}$ and $\mc{B}$:
\[
  \mr{Links}^-(\mc{A}, \mc{B}) 
    =\sum_{v'_p \in \mc{A}} \sum_{v'_q \in \mc{B}}
     \max\{0, -w'_{pq}\}.
\]
\end{itemize}
\end{definition}

It is easy to see that $\mr{Cut}(\mc{A}, \mc{\overline{A}}) = \mr{Links}^+(\mc{A}, \mc{\overline{A}}) + \mr{Links}^-(\mc{A}, \mc{\overline{A}})$, and as in \cite{Gallier2019_Book, VonLuxburg2007SC}, one may show that the three costs in Lemma~\ref{lem:3Costs} can be expressed in terms of dual graph cuts. 

\begin{lemma}
\label{lem:3costsdualgraph}
For the PRE, DPE, and RGE edge functional affinities, the total costs associated with cluster $\Sigma^{(k)},\, k \in \{1, \ldots, K\}$ in the dual graph, are, respectively, 
\begin{alignat}{3}
  &\tr{Cost}_{\PRE}^{(k)}(\mc{G})
    &
      &=\ul{x}^{(k)^T} \ul{L}_{\PRE}\, \ul{x}^{(k)}
        &
          &=\alpha^{(k)^2} 
            \left[
              \mr{Cut}(\Sigma^{(k)}, \overline{\Sigma}^{(k)}) 
                + 2\, \mr{Links}^-(\Sigma^{(k)}, \Sigma^{(k)})
            \right]; 
            \notag \\
  &\tr{Cost}_{\DPE}^{(k)}(\mc{G})
    &
      &=\ul{x}^{(k)^T} \ul{L}_{\DPE}\, \ul{x}^{(k)}
        &
          &=\alpha^{(k)^2} 
            \left[
              \mr{Cut}(\Sigma^{(k)}, \overline{\Sigma}^{(k)}) 
                + 2\, \mr{Links}^+(\Sigma^{(k)}, \Sigma^{(k)})
            \right]; 
            \notag \\
  &\tr{Cost}_{\RGE}^{(k)}(\mc{G})
    &
      &=\ul{x}^{(k)^T} \ul{L}_{\RGE}\, \ul{x}^{(k)}
        &
          &=\alpha^{(k)^2} 
            \mr{Cut}(\Sigma^{(k)}, \overline{\Sigma}^{(k)}).
  \label{eq:3Costs_cuts}
\end{alignat}
\end{lemma}

In light of Lemma~\ref{lem:3costsdualgraph}, we make the following observations:
\begin{itemize}
  \item PRE edge clustering focuses on both minimizing the absolute cut between inter-cluster dual vertices and minimizing the dual negative edge weights connecting intra-cluster dual vertices. Recall that pairs of edges forming a directed path in the original digraph leads to negative edge weights in the dual graph, meaning that intra-cluster directed paths in the original digraph are discouraged. In addition, note that minimizing the absolute cuts in the dual graph involves cutting edges with low absolute weights in the dual graph. This is equivalent to choosing `coupling vertices' in the original digraph with lower $\nu_i$, and hence relatively lower importance.  Here, coupling vertices in the original digraph refer to those vertices whose edges are in multiple clusters.
  \item DPE edge clustering focuses on minimizing the sum of the absolute cut between inter-cluster dual vertices (as PRE edge clustering does) and the dual positive edge weights connecting intra-cluster dual vertices (in contrast to what PRE clustering does). This means that in the original digraph, DPE edge clustering discourages edges that both point to, or both point away from, a common vertex from being in the same cluster; instead, it encourages clusters consisting of directed paths connecting vertices with large $\nu_i$ values in the original digraph. 
  \item RGE edge clustering focuses on minimizing the absolute cuts between inter-cluster dual vertices (as both PRE and DPE clustering do). Therefore coupling vertices are those with lower $\nu_i$ values, and hence lower relative impoartance. 
\end{itemize}

We now express the dual graph cut costs in Lemma~\ref{lem:3costsdualgraph} in the following manner (where the $\alpha^{(k)^2}$ terms are removed):

\begin{definition}[Unscaled Dual Graph Cuts]
\label{def:3unscaled_cuts}
For the PRE, DPE, and RGE edge functional affinities, the total unscaled costs associated with cluster $\Sigma^{(k)},\, k \in \{1, \ldots, K\}$, are, respectively, 
\begin{alignat*}{3}
  &\tr{UCost}_{\PRE}^{(k)}(\mc{G})
    &
      &=\frac{\ul{x}^{(k)^T} \ul{L}_{\PRE}\, \ul{x}^{(k)}}{\alpha^{(k)^2}}
        &
          &=
            \left[
              \mr{Cut}(\Sigma^{(k)}, \overline{\Sigma}^{(k)}) 
                + 2\, \mr{Links}^-(\Sigma^{(k)}, \Sigma^{(k)})
            \right]; \\
  &\tr{UCost}_{\DPE}^{(k)}(\mc{G})
    &
      &=\frac{\ul{x}^{(k)^T} \ul{L}_{\DPE}\, \ul{x}^{(k)}}{\alpha^{(k)^2}}
        &
          &= 
            \left[
              \mr{Cut}(\Sigma^{(k)}, \overline{\Sigma}^{(k)}) 
                + 2\, \mr{Links}^+(\Sigma^{(k)}, \Sigma^{(k)})
            \right]; \\
  &\tr{UCost}_{\RGE}^{(k)}(\mc{G})
    &
      &=\frac{\ul{x}^{(k)^T} \ul{L}_{\RGE}\, \ul{x}^{(k)}}{\alpha^{(k)^2}}
        &
          &= 
            \mr{Cut}(\Sigma^{(k)}, \overline{\Sigma}^{(k)}).
\end{alignat*}
\end{definition}

Edge clustering's connection to the edge Laplacian and its subsequent dual graph cuts interpretation help establish edge clustering as a counterpart and alternative to vertex clustering in graphs.  Note that one can also work with the Laplacians in Lemma~\ref{lem:Cost_k_L} to obtain a graph cuts interpretation in terms of a different dual graph defined in terms of $\ul{\Phi}$ and $\ul{\Psi}$; we do not do so for space reasons.

\section{Optimization Problem Associated With Edge Clustering}


We may now recast our edge clustering task as the following optimization problem:
\begin{multline}
  \min_{\ul{X} \in \mc{X}} \;\;
  \sum_{k=1}^K 
  \ul{x}^{(k)^T} \ul{L}\, \ul{x}^{(k)} \\
  \tr{subject to}\;\; 
  \ul{X}^T \ul{X}
    =\mr{diag}[M^{(1)} \alpha^{(1)^2}, \ldots, M^{(1)} \alpha^{(K)^2}],\,
     \alpha^{(k)} \neq 0,\, M^{(k)} > 0,
  \label{eq:optL}
\end{multline}
where we use the `generic' $\ul{L}$ to stand for $\ul{L}_{\PRE}$, $\ul{L}_{\DPE}$, or $\ul{L}_{\RGE}$. 

Direct minimization of this edge clustering cost function can lead to cluster size imbalance, a problem that is common in vertex clustering \cite{VonLuxburg2007SC, Gallier2019_Book}.  Work on vertex clustering addresses this issue by tempering the clustering cost of each cluster by its vertex cluster `volume' which takes into account the vertex degrees. This leads to the normalized vertex Laplacians. We follow along similar lines by first defining the volume of an \emph{edge} and then defining the volume of a \emph{cluster} as the sum of the volumes of its member edges. This then leads to normalized costs. 


\subsection{Volume of an Edge}


To proceed, consider the digraph edge $e_p = (v_{\ell_p} {\to}\, v_{k_p})$ with edge weight $w_p$ which is directed from vertex $v_{\ell_p}$ to vertex $v_{k_p}$. Recall that $w_p$ is equivalently $w_{k_p \ell_p}$ in the weighted adjacency matrix and $\ul{\nu} = \{\nu_i\} \in \mbb{R}^N$ denotes a vertex vector where $\nu_i$ captures the relative importance of vertex $v_i \in \mc{V}$. To simplify the notation, let us use $f_p$ to denote the volume of edge $e_p$, i.e., $f_p = \mr{Vol}(e_p)$. We postulate that $f_p$ must incorporate two main characteristics of an edge: 
  the combined importance of the two vertices $v_{\ell_p}$ and $v_{k_p}$ at the ends of the edge $e_p$; and 
  the magnitude of the edge weight $w_p$ of the edge $e_p$ and its relationship to the directionality of the flow in the graph. With these observations in mind, we propose

\begin{definition}[Volume of an Edge and Cluster] 
\label{def:VolEdge}
Consider the digraph $\mc{G}(\mc{V}, \mc{E})[\ul{\nu}, \ul{w}]$ associated with the vertex vector $\ul{\nu} \in \mbb{R}^N$ and the edge weight vector $\ul{w} \in \mbb{R}^M$.
\begin{itemize}
  \item[\tb{(i)}] The \emph{volume of the edge} $e_p=(v_{\ell_p} {\to}\, v_{k_p})$ from vertex $v_{\ell_p}$ to vertex $v_{k_p}$ in the digraph $\mc{G}$ is 
\[
  \mr{Vol}(e_p) 
    =f_p 
    =\begin{cases}
       \displaystyle\frac{|w_p|}{2}
       \left(
         \displaystyle
         \frac{\sigma_{\ell_p}\nu_{\ell_p}}{\langle d_{out, \ell_p} \rangle} 
                 + \frac{\sigma_{k_p}\nu_{k_p}}{\langle d_{in, k_p} \rangle}
       \right),
         & \tr{when $w_p \neq 0$}; \\
       0,
         & \tr{when $w_p = 0$},
     \end{cases}
\]
where $\langle d_{out, \ell_p} \rangle = \displaystyle\sum_i |w_{i\ell_p}|$ is the absolute sum of the edge weights of the outgoing edges from $v_{\ell_p}$, $\langle d_{in, k_p} \rangle = \displaystyle\sum_i |w_{k_pi}|$ is the absolute sum of the edge weights of the incoming edges to $v_{k_p}$, and $\sigma_i$ is the total integer number of edges connected to vertex $v_i$ which we refer to as its \emph{social participation.} 
  \item[\tb{(ii)}] Accordingly, the \emph{volume of a cluster} $\Sigma^{(k)} \subseteq\mc{E}$ of edges is 
\[
  \mr{Vol}(\Sigma^{(k)}) 
    =\sum_{e_p\in\Sigma^{(k)}\subseteq\mc{E}} f_p
    =\frac{\ul{x}^{(k)^T} \ul{F}\,\ul{x}^{(k)}}{\alpha^{(k)^2}},
\]
where $\ul{F} = \mr{diag}[\ul{f}] \in \mbb{R}^{M\times M}$ is the diagonal matrix whose diagonal is populated with the vector $\ul{f} = \{f_p\} \in \mbb{R}^M$ and $\ul{x}^{(k)} \in \mbb{R}^M$ denotes the aforementioned indicator vector associated with edge membership in cluster $\Sigma^{(k)}$.  The volume of a cluster of edges is the sum of the volumes of the member edges.
\end{itemize}
\end{definition}

Note how $f_p$ accounts for the combined importance of the two end-vertices via the vertex functions $\nu_{\ell_p}$ and $\nu_{k_p}$ and the social participation values $\sigma_{\ell_p}$ and $\sigma_{k_p}$ at the two edge-end vertices $v_{\ell_p}$ and $v_{k_p}$, respectively. In addition, it accounts for both the magnitude of the edge weight and the directionality of the edge via $|w_p|/\langle d_{out, \ell_p} \rangle$ and $|w_p|/\langle d_{out, k_p} \rangle$ which capture the importance of the role that the edge $e_p$ plays in conveying influence away from $v_{\ell_p}$ and toward $v_{k_p}$, respectively. See Fig.~\ref{fig:h_p}. 


\paragraph{Volume of an Edge and Flows at Its End-Vertices} 

To explore further the factors affecting edge volume, consider the directed edge $e_p \cong e_{k_p, \ell_p}$, with edge weight $w_p \cong w_{k_p, \ell_p}$, from vertex $v_{\ell_p}$ to vertex $v_{k_p}$ in an arbitrary digraph. See Fig.~\ref{fig:EdgeVolume}(a). 

Furthermore, consider the case where we set the vertex vector $\ul{\nu} = \{\nu_i\}$ to be the sum of the absolute values of edge weights attached to vertex $v_i$ (normalized with respect to the total sum of absolute values of all the digraph's edge weights), viz., 
\begin{equation}
  \nu_i 
    = \frac{\langle d_i \rangle}{\Vert\ul{w}\Vert_1}    
    = \frac{\langle d_{out,i} \rangle 
              + \langle d_{in,i} \rangle}
           {\Vert\ul{w}\Vert_1}.
  \label{eq:nuvectorchoice}          
\end{equation} 
This is exactly the choice that we have made in all our experiments. Then we can establish the following result:

\begin{lemma}
\label{lem:EdgeVol}
Consider the directed edge $e_p \cong e_{k_p, \ell_p}$, with edge weight $w_p \cong w_{k_p, \ell_p}$, from vertex $v_{\ell_p}$ to vertex $v_{k_p}$ in a digraph. With the vertex vector $\ul{\nu} = \{\nu_i\}$ chosen as in \eqref{eq:nuvectorchoice}, the following is true regarding the volume $f_p = \mr{Vol}(f_p)$ of edge $e_p$:
\[
  f_p
    =\frac{\sigma_{\ell_p}}{2}
     \cdot
     \frac{|w_p|}{\Vert\ul{w}\Vert_1}
     \cdot
     \frac{\langle d_{\ell_p}\rangle}
          {\langle d_{out, \ell_p}\rangle}
       +\frac{\sigma_{k_p}}{2} 
        \cdot
        \frac{|w_p|}{\Vert\ul{w}\Vert_1}
        \cdot
        \frac{\langle d_{k_p}\rangle}
             {\langle d_{in, k_p}\rangle}.
\]
As a consequence, the following are true:
\begin{itemize}
  \item[(i)] With the other variables fixed, $f_p$ is monotonically increasing with increasing $|w_p|$.
  \item[(ii)] With the other variables fixed, $f_p$ is monotonically decreasing with increasing $\langle d_{out, \ell_p}'\rangle$ and $\langle d_{in, k_p}'\rangle$, where $\langle d_{out, \ell_p}'\rangle = \langle d_{out, \ell_p}\rangle - |w_p|$ and $\langle d_{in, k_p}'\rangle = \langle d_{in, k_p}\rangle - |w_p|$. 
  \item[(iii)] With the other variables fixed, $f_p$ is monotonically increasing with increasing $\langle d_{in, \ell_p}\rangle$ and $\langle d_{out, k_p}\rangle$.
\end{itemize}
\end{lemma}

Lemma~\ref{lem:EdgeVol} allows us to make the following observations regarding the importance of edge $e_p$ as captured by its volume $f_p = \mr{Vol}(e_p)$:
\begin{itemize}
  \item From Lemma~\ref{lem:EdgeVol}(i): 
\begin{equation}
  0
    \leq f_p
    \leq \frac{\sigma_{\ell_p} + \sigma_{k_p}}{2},
\end{equation}
with $f_p$ attaining its minimum value $0$ when $|w_p| = 0$ and it maximum value $(\sigma_{\ell_p} + \sigma_{k_p})/2$ when $|w_p| \to \infty$. 

  \item From Lemma~\ref{lem:EdgeVol}(ii): 
  A higher $\langle d_{out, \ell_p}' \rangle$ means that the source vertex $v_{\ell_p}$ has an increasing ability to direct influence through either an increasing number of outgoing edges, or an increase in the absolute sum of the weights of the outgoing edges, or both. Similarly, a higher $\langle d_{in, k_p}' \rangle$ means that the destination vertex $v_{k_p}$ has an increasing ability to receive influence from other vertices through either an increase in the number of incoming edges, or an increase in the absolute sum of the weights of the incoming edges, or both. Therefore, as $\langle d_{out, \ell_p}'\rangle$ and/or $\langle d_{in, k_p}'\rangle$ increase, the edge $e_p$ is relatively less important in conveying influence and flow in the digraph, and so its edge volume $f_p$ decreases. 
  
  Conversely, a lower $\langle d_{out, \ell_p}' \rangle$ means that the source vertex $v_{\ell_p}$ has a reduced ability to convey influence. Similarly, a lower $\langle d_{in, k_p}' \rangle$ means that the destination vertex $v_{k_p}$ has a reduced ability to receive influence.  Therefore, as $\langle d_{out, \ell_p}'\rangle$ and/or $\langle d_{in, k_p}'\rangle$ decrease, the edge $e_p$ is relatively more important in conveying influence and flow in the digraph, and so its edge volume $f_p$ increases. 
  
  \item From Lemma~\ref{lem:EdgeVol}(iii): 
  A higher $\langle d_{in, \ell_p} \rangle$ means that the source vertex $v_{\ell_p}$ is receiving a larger flow. Similarly, a higher $\langle d_{out, k_p} \rangle$ means that the destination vertex $v_{k_p}$ is discharging a larger flow. Therefore, as $\langle d_{in, \ell_p} \rangle$ and/or $\langle d_{out, k_p} \rangle$ increase, the edge $e_p$ is relatively more important in conveying influence and flow in the digraph, and so its edge volume $f_p$ increases. 
  
  Conversely, a lower $\langle d_{in, \ell_p} \rangle$ and/or $\langle d_{out, k_p} \rangle$ means that the edge $e_p$ is relatively less important in conveying influence and flow in the digraph, and so its edge volume $e_p$ decreases.
\end{itemize}

To see how the number of edges of the two end-vertices $v_{\ell_p}$ and $v_{k_p}$ affect $f_p$, it is easiest to consider the case of a \emph{binary} digraph where all edges have weight one. 

\begin{corollary}
\label{cor:EdgeVol_binary}
Consider the directed edge $e_p \cong e_{k_p, \ell_p}$, with edge weight $w_p \cong w_{k_p, \ell_p}$, from vertex $v_{\ell_p}$ to vertex $v_{k_p}$ in a binary digraph where $w_p = 1,\, \forall p \in \{1, \ldots, M\}$. With the vertex vector $\ul{\nu} = \{\nu_i\}$ chosen as in \eqref{eq:nuvectorchoice}, the following is true regarding the volume $f_p = \mr{Vol}(f_p)$ of edge $e_p$:
\[
  f_p
    =\frac{1}{2M}
     \left(
       \frac{\sigma^2_{\ell_p}}{\sigma_{out, \ell_p}}
         +\frac{\sigma^2_{k_p}}{\sigma_{in, k_p}}
     \right)
    =\frac{1}{2M}
     \left[
       \left(
         \sigma^{1/2}_{out, \ell_p} 
           + \frac{\sigma_{in, \ell_p}}{\sigma^{1/2}_{out, \ell_p}}
       \right)^2
         +\left(
            \sigma^{1/2}_{in, k_p} 
              + \frac{\sigma_{out, k_p}}{\sigma^{1/2}_{in, k_p}}
       \right)^2
     \right],
\]
where $\sigma_{\ell_p}$ and $\sigma_{out, \ell_p}$ denote respectively the number of total and the number of outgoing edges of $v_{\ell_p}$, and $\sigma_{k_p}$ and $\sigma_{in, k_p}$ denote respectively the number of total and the number of incoming edges of $v_{k_p}$. 
\end{corollary}

Corollary~\ref{cor:EdgeVol_binary} allows us to make the following observation  regarding the importance of edge $e_p$ in a \emph{binary} graph as captured by its volume $f_p = \mr{Vol}(e_p)$:
\begin{equation}
  \frac{1}{2M}(\sigma_{out, \ell_p} + \sigma_{in, k_p})
    \leq f_p
    \leq M.
\end{equation}
Note the following for binary graphs:
\begin{itemize}
  \item The edge volume $f_p$ attains its minimum value $(\sigma_{out, \ell_p} + \sigma_{in, k_p})/2M$ when $\sigma_{\ell_p} = \sigma_{out, \ell_p}$  (and hence $\sigma_{in, \ell_p} = 0$) and $\sigma_{k_p} = \sigma_{in, kp_p}$ (and hence $\sigma_{out, k_p} = 0$), i.e., when all edges of $v_{\ell_p}$ are outgoing and when all edges of $v_{k_p}$ are incoming. This means that the edge $e_p$ is unremarkable and relatively less important in conveying flow and influence in that it is one of the many outgoing edges of $v_{\ell_p}$ and also one of the many incoming edges of $v_{k_p}$, and so its edge volume is the minimum above. We also note that, the absolute minimum value that $f_p$ can reach is $1/M$; this occurs when the end-vertices $\{v_{\ell_p}, v_{k_p}\}$ are isolated from the rest of the digraph with $e_p$ being the only directed edge between them. (See Fig.~\ref{fig:EdgeVolume}(b)). 
  \item $f_p$ attains its maximum when $\sigma_{\ell_p} = M > 1$ and $\sigma_{out, \ell_p} = 1$ (so that $\sigma_{in, \ell_p} = M - 1$) and $\sigma_{k_p} = M > 1$ and $\sigma_{in, k_p} = 1$ (so that $\sigma_{out, k_p} = M - 1$). This corresponds to the case when the vertices $\{v_{\ell_p}, v_{k_p}\}$ consitute the only vertices with $e_p$ being the only directed edge from $v_{\ell_p}$ to $v_{k_p}$ while the remaining $M-1$ directed edges are from $v_{k_p}$ to $v_{\ell_p}$. The edge $e_p$ thus acts as a bottleneck link. (See Fig.~\ref{fig:EdgeVolume}(c)).
\end{itemize}


\subsection{Normalized Costs}


We now consider normalized clustering costs to address potential cluster imbalance. Since the following development is common to all three edge functional affinities, we use the generic $\ul{L}$ to stand for $\ul{L}_{\PRE}$, $\ul{L}_{\DPE}$, or $\ul{L}_{\RGE}$, as before.

To proceed, we define the normalized cost for cluster $\Sigma^{(k)}$ as the dual graph cuts unscaled cost in Definition~\ref{def:3unscaled_cuts} divided by the volume of the cluster given by Definition~\ref{def:VolEdge}. This normalized cost can also be expressed as a ratio of two matrix quadratic functions.  

\begin{definition}[Normalized Dual Graph Costs] 
\label{def:NCuts}
The normalized cost associated with cluster $\Sigma^{(k)},\, k \in \{1, \ldots, K\}$, is
\[
  \tr{NCost}^{(k)}(\mc{G}) 
    =\frac{\tr{UCost}^{(k)}(\mc{G})}
          {\mr{Vol}(\Sigma^{(k)})} 
    =\frac{\ul{x}^{(k)^T} \ul{L}\, \ul{x}^{(k)}}
          {\ul{x}^{(k)^T} \ul{F}\, \ul{x}^{(k)}}.
\]
The total normalized cost for clustering $M$ edges into $K$ clusters is taken as the sum of the normalized costs associated with each cluster, i.e., 
\[
  \tr{NCost}(\mc{G}) 
    =\sum_{k=1}^K \tr{NCost}^{(k)}(\mc{G})
    =\sum_{k=1}^K 
     \frac{\ul{x}^{(k)^T} \ul{L}\, \ul{x}^{(k)}}
          {\ul{x}^{(k)^T} \ul{F}\, \ul{x}^{(k)}}.
\]
\end{definition}


\subsection{Normalized Cost-Based Optimization Problem}
\label{subsec:OptimizationProblem}


Correspondingly, we now attempt to solve the edge clustering task via the following normalized cost-based optimization problem:
\begin{multline}
  \min_{\ul{X} \in \mc{X}} \;\;
  \sum_{k=1}^K 
     \frac{\ul{x}^{(k)^T} \ul{L}\, \ul{x}^{(k)}}
          {\ul{x}^{(k)^T} \ul{F}\, \ul{x}^{(k)}} \\
  \tr{subject to}\;\; 
  \ul{X}^T \ul{X} 
    =\mr{diag}[M^{(1)} \alpha^{(1)^2}, \ldots, M^{(K)} \alpha^{(K)^2}],\, 
     \alpha^{(k)} \neq 0,\, M^{(k)} > 0,
  \label{eq:optL2}
\end{multline}

We can now take an approach similar to what appears in existing work on vertex clustering \cite{Gallier2019_Book, VonLuxburg2007SC, Kunegis2010SIAM_ICDM} to simplify the optimization problem in \eqref{eq:optL}. Henceforth, we assume that the edge volumes are positive, i.e.,  $f_p > 0, p=1, \cdots, M$.  One can ensure positive edge volumes by having $\ul{\nu} > 0$, i.e., the vector of vertex relative importances is strictly positive, which is a reasonable assumption (in the experiments, we set $\nu_i$ proportional to $\langle d_i\rangle = \langle d_{out,i}\rangle + \langle d_{in,i}\rangle$, the sum of the absolute values of edge weights attached to vertex $v_i$, ensuring that all entries of $\ul{\nu}$ are positive). 

With the diagonal matrix $\ul{F} \in \mbb{R}^{M \times M}$ having positive diagonal entries $f_p$, the orthogonality constraint in \eqref{eq:XTX} can be equivalently formulated as the matrix quadratic constraint 
\begin{equation}
  \ul{X}^T \ul{F}\, \ul{X}
    =\mr{diag}[\mr{Vol}[\Sigma^{(1)}]\, \alpha^{(1)^2}, \ldots, 
               \mr{Vol}[\Sigma^{(K)}]\,\alpha^{(K)^2}].
\end{equation}
By picking the indicator labels as 
\begin{equation}
  \alpha^{(k)}
    =\frac{1}{\sqrt{\mr{Vol}[\Sigma^{(k)}]}},\;
     k \in \{1, \ldots, K\},
\end{equation}
the orthogonality constraint is equivalent to 
\begin{equation}
  \ul{X}^T \ul{F}\, \ul{X}
    =\ul{I}_K,
\end{equation}
where $\ul{I}_K$ is the $K \times K$ identity matrix. Thus, the normalized cost-based optimization problem in \eqref{eq:optL2} becomes
\begin{equation}
  \min_{\ul{X} \in \mc{X}} \;\;
  \sum_{k=1}^K 
     \frac{\ul{x}^{(k)^T} \ul{L}\, \ul{x}^{(k)}}
          {\ul{x}^{(k)^T} \ul{F}\, \ul{x}^{(k)}}
  \;\;\tr{subject to}\;\; 
  \ul{X}^T \ul{F}\, \ul{X} 
    =\ul{I}_K.
  \label{eq:optL3}
\end{equation}


\subsection{Relaxed Optimization Problem} 


This problem is NP hard. To achieve a computationally more feasible approximate solution, we allow the partition matrices to be from $\mbb{R}^{M \times K}$ (instead of from $\mc{X}$) thus yielding the relaxed problem
\begin{equation}
  \min_{\ul{X} \in \mbb{R}^{M \times K}} \;
  \sum_{k=1}^K 
     \frac{\ul{x}^{(k)^T} \ul{L}\, \ul{x}^{(k)}}
          {\ul{x}^{(k)^T} \ul{F}\, \ul{x}^{(k)}}
  \;\;\tr{subject to}\;\; 
  \ul{X}^T \ul{F}\, \ul{X} 
    =\ul{I}_K.
  \label{eq:optL4}
\end{equation}
Apply the change of variables 
\begin{equation}
  \ul{Y} 
    =\ul{F}^{1/2} \ul{X}
  \iff
  \ul{X}
    =\ul{F}^{-1/2} \ul{Y},
\end{equation}
to arrive at
\begin{equation}
  \min_{\ul{Y} \in \mbb{R}^{M \times K}} 
  \mr{Tr}[\ul{Y}^T \widetilde{\ul{L}}\, \ul{Y}]
  \;\;\tr{subject to}\;\; 
  \ul{Y}^T \ul{Y} 
    =\ul{I}_K,
  \label{eq:optL5}
\end{equation}

where 
$\widetilde{\ul{L}} = \ul{F}^{-1/2} \ul{L}\, \ul{F}^{-1/2}$ is the \emph{normalized Flow Laplacian.} In particular, for the three edge functional affinities, we can now define their respective normalized Flow Laplacians:

\begin{definition}[Normalized Flow Laplacians]
\label{def:normalizedFlowLaplacians}
The three normalized Flow Laplacians are defined as
\[
  \widetilde{\ul{L}}_{\PRE} 
    =\ul{F}^{-1/2} \ul{L}_{\PRE}\, \ul{F}^{-1/2};\;\;
  \widetilde{\ul{L}}_{\DPE} 
    =\ul{F}^{-1/2} \ul{L}_{\DPE}\, \ul{F}^{-1/2};\;\;
  \widetilde{\ul{L}}_{\RGE} 
    =\ul{F}^{-1/2} \ul{L}_{\RGE}\, \ul{F}^{-1/2}.
\]
\end{definition}

Now consider the eigendecomposition of the normalized Flow Laplacian matrix: \begin{equation}
  \widetilde{\ul{L}} 
    =\widetilde{\ul{U}}\, \widetilde{\ul{\Lambda}}\, \widetilde{\ul{U}}^T,
\end{equation}
where we assume that the unit eigenpairs $\{\widetilde{\ul{u}}_p, \widetilde{\lambda}_p\},\, p \in \{1, \ldots, M\}$, of $\widetilde{\ul{L}}$ are ordered as
\begin{equation}
  \widetilde{\lambda}_1
    \leq\widetilde{\lambda}_2
    \leq\cdots 
    \leq\widetilde{\lambda}_M.
\end{equation}
Then, as a consequence of the Poincar\'e Separation Theorem, we can conclude that \cite{Horn2013_Book, Liang2013LAIA}
\begin{alignat}{2}
  &\min_{\ul{Y} \in \mbb{R}^{M \times K}} 
   \mr{Tr}[\ul{Y}^T \widetilde{\ul{L}}\, \ul{Y}]
     &
       &=\sum_{p=1}^K \widetilde{\lambda}_p,
         \;\;\tr{with}\;\;
         \notag \\
  \widehat{\ul{Y}}
    \equiv
  &\arg\min_{\ul{Y} \in \mbb{R}^{M \times K}}
   \mr{Tr}[\ul{Y}^T \widetilde{\ul{L}}\, \ul{Y}]
     &
       &=\begin{bmatrix}
           \widetilde{\ul{u}}_1 & \ldots & \widetilde{\ul{u}}_K
         \end{bmatrix}
    \in\mbb{R}^{M \times K}.
\end{alignat}
In other words, the solution $\widehat{\ul{Y}} \in \mbb{R}^{M \times K}$ to the relaxed normalized cost-based edge clustering problem in \eqref{eq:optL5} is the $M \times K$ matrix formed from the unit eigenvectors associated with the $K$ smallest eigenvalues of the normalized Laplacian $\widetilde{\ul{L}}$. 

Similar to vertex clustering \cite{Gallier2019_Book, VonLuxburg2007SC, Kunegis2010SIAM_ICDM, Riolo2014JCN}, we interpret the $p$-th row of this $\widehat{\ul{Y}}$ as the $1 \times K$ feature vector associated with the digraph edge $e_p$. These $M$ row feature vectors are then normalized one at a time such that each one has unit energy. These $M$ row normalized feature vectors are grouped into $K$ clusters by applying a clustering algorithm (e.g., \texttt{k-means++}). We have thus essentially placed the $M$ corresponding edges into $K$ groups as well, resulting in digraph edge clustering. Procedure~\ref{Pro:EdgeClustering} summarizes this scheme and provides the pseudocode for the algorithm that was used in the simulations. The pseudocode for a more general edge clustering scheme that does not utilize the edge Laplacian and instead relies upon a user-specified matrix $\ul{\Phi}$, which contains the weights for the pairwise edge comparisons in the cost function, is provided in Procedure~\ref{Pro:GenericEdgeClustering}.

Finally, we wish to make several remarks about our Flow Laplacian edge clustering framework. 


\paragraph{Complexity} 

The first step in edge clustering is to choose the type of edge clustering one wants, i.e., PRE, DPE, or RGE, since each method is based on a different approach to defining edge affinities.  Based on the edge clustering choice, one then calculates the corresponding Flow Laplacian, and then its normalized version by using the edge volumes. The main complexity incurred is the step of finding $K$ eigenvectors of the $K$ minimum eigenvalues of the $M \times M$ normalized Laplacian matrix.  


\paragraph{Symmetric PSD Property} 

All of the normalized Flow Laplacians are symmetric psd matrices, and are therefore amenable to matrix computation methods that exploit this property.


\paragraph{Constructing $\ul{\Phi}$} 

In order to calculate the chosen normalized Laplacian, one must specify $\ul{\Phi}$, the symmetric matrix with non-negative values that capture the scaling associated with the squared pairwise edge comparisons.  The values must be consistent with Definition~\ref{def:Cost_k_pq}. One approach is to set these scaling values to the relative importance, $\nu_i$, of the common vertex $v_i$ associated with the two edges being compared.   If one has external side information regarding the relative importances of vertices, then one can utilize those values.  In the absence of external side information, we set $\nu_i$ proportional to the weighted social scaling, i.e., the sum of the absolute values of the edge weights of all edges connected to vertex $v_i.$ One advantage of this choice is that it can be directly calculated from the weighted adjacency matrix. Moreover, this $\nu_i > 0$ choice allows one to calculate the various Flow Laplacians as functions of the edge Laplacian $\ul{L}_e.$  Nevertheless, other choices are certainly possible. 


\paragraph{Handling Signed Weights} 

The Flow Laplacian framework currently assumes all directed edges have positive weights. If one wishes to calculate edge clusters for graphs with signed edges (both positive and negative edge weights), then one should calculate the normalized Flow Laplacian using the absolute values of the edge weights.  This means that the edge clustering will ignore the signs of the edges. In order to take into account the sign of an edge, one must further consider the meaning of the negative sign (e.g., antipathy, inhibitory signal, etc.), and subsequently re-formulate the edge affinity descriptions and basic costs. This is a subject for future research. 


\paragraph{Handling Undirected Graphs} 

For clustering undirected graphs, the RGE edge clustering method is the most appropriate choice since its edge affinity ignores directions and encourages all edges that share a common vertex to be in the same cluster. To proceed, one treats the weights $w_{ij}$ and $w_{ji}$, which are identical, as distinct directed edges, one in each direction. And so $M$, the number of directed edges clustered by the RGE clustering method is double the number of original undirected edges. One could derive an RGE algorithm solely for handling undirected edges, but this would lead to additional paper length and notational burden, and so we do not include it here. It is also important to note that the RGE edge clustering results for undirected graphs are not the same as vertex clustering results. Besides the simple fact that clustering edges is not the same as clustering vertices, we must also note that the RGE edge clustering method has coupling vertices $v_i$ (which are connected to at least two edge clusters) that have a lower $\nu_i$ value, which means lower relative importance. In contrast, vertex clustering has coupling edges that are generally of lower edge weight.


\paragraph{Versatility With Digraphs} 

For digraphs, the Flow Laplacian edge clustering framework offers capabilities not always possible with vertex clustering. For example, the Chung Laplacian \cite{Chung2005AC} for digraphs assumes strongly connected graphs, or requires modifications of the random walk framework.  In contrast, the Flow Laplacian framework can handle any digraph. In fact, the Flow Laplacian edge clustering framework allows for digraphs with self-edges and multi-edges. The reason is that the $\ul{\Psi}$ and $\ul{\Phi}$ matrices describe relationships between edges, meaning that a self-edge or a multi-edge simply adds to the list of edges. One can also define the edge Laplacian directly through edge lists (see Lemma~\ref{lem:LeEntries}). 

\paragraph{Mixed Methods}
In principle, one can design mixed edge clustering algorithms that combine two edge affinities through a positively weighted linear combination of suitable Laplacian matrices.  For example, one can mix $\ul{L}_{PRE}$ and $\ul{L}_{RGE}$ by multiplying them by positive scalars and then adding them to obtain a mixed Laplacian that can be used for a mixed edge clustering algorithm. Or one can mix $\ul{L}_{DPE}$ and $\ul{L}_{RGE}$ as well.  One should not mix $\ul{L}_{PRE}$ and $\ul{L}_{DPE}$ as they have opposing objectives.  We do not pursue this subject here.

\paragraph{Edge Clustering Complements Vertex Clustering}
The Flow Laplacian edge clustering framework provides methods to cluster graph edges, and consequently provides new avenues for exploratory data analysis by revealing subgraph clusters geared towards emphasizing different types of directed influence and concentrated flows among vertices, with the user being able to define vertex relative importance.  The edge clusters feature subgraphs that often link together vertices with widely varying functions, and consequently illustrate how vertices with different types of functions work together to achieve directed influence and concentrated flows, depending upon the specified edge clustering method.  In contrast, vertex clustering is mainly focused on grouping similar vertices together, meaning that vertices with widely different functions are not usually clustered together. Therefore, the edge clustering methods complement the well-known vertex clustering methods and reveal new insight into graph data.  Since clustering edges is fundamentally different from clustering vertices, we do not provide comparisons with vertex clustering. 


\section{Edge Clustering Results}


We now document the results of several experiments applying our Flow Laplacian Edge Clustering Framework to both synthetic and real-world graphs. For synthetic graphs, we emphasize simple digraphs that allow for intuitive explanations and visualizations of the three different edge clustering methods. For real-world graphs, we examine a wide variety of directed and undirected graphs to illustrate how the different edge clustering methods yield different subgraphs indicating different types of directed influence.

All our experiments are conducted with the vertex vector $\ul{\nu} = \{\nu_i\}$ chosen as in \eqref{eq:nuvectorchoice}, i.e., $\nu_i$ is the sum of the absolute values of edge weights attached to vertex $v_i$ (normalized with respect to the total sum of absolute values of all the digraph's edge weights). For unweighted directed graphs, the corresponding edge weights are set to ones. The edge clustering algorithm used is described in Procedure~\ref{Pro:EdgeClustering}. 


\subsection{Synthetic Digraphs}


For the synthetic graphs, we focus on simple digraphs to illustrate the main differences between the different types of edge clustering methods and their results. 

\paragraph{10-Vertex/21-Edge Digraph from \cite{Benson2016Science}}

Consider the edge clustering results for a simple 10 vertex and 21 edge unweighted digraph in \cite{Benson2016Science} for which we clustered the edges into $K=2$ groups. See Fig.~\ref{fig:Cockroach_2Clusters_SocialScaling}(a), (b), (c). 

For PRE clustering (a), the orange cluster consists of 9 edges while the blue cluster is 12 edges. We see that in this PRE blue cluster vertex 1 and vertex 8 are producers while vertex 2 is a receptor. In the PRE orange cluster, vertex 2, along with vertex 9, are now producers, while vertex 1 is a now a receptor. Vertex 2 has the largest $\nu$ value (integer number of connected edges in this unweighted graph case), and so that is why its outgoing and incoming edges are separated into separate clusters; its producing edges (outgoing edges) are in the organge cluster, while its receptor edges (incoming edges) are in the blue cluster.  While Vertex 1 and Vertex 6 have the same $\nu$ values, there are only 2 choices for group membership, and so vertex 6 is neither a pure producer nor receptor in the two clusters. PRE clustering for this simple unweighted digraph here places edges in two groups such that each group/cluster has concentrated directed flows of production and reception based on the PRE edge affinity.

For DPE clustering (b) that emphasizes edge affinities focused on long directed paths, the blue cluster has 10 edges while the orange cluster has 11. Both clusters have length 4 directed paths, each featuring vertex 2 which has the largest $\nu$ value. The blue length 4 directed path is the path involving vertices 9, 8, 2, 1, and 6; the orange path involve vertices 8, 9, 7, 6, and 2.

For RGE clustering (c), the edges are partitioned with vertices 6, 7, and 8 serving as coupling vertices that have edges in both clusters. Each cluster has edges that have an affinity for concentrated flows regardless of direction. The blue cluster includes the edges of the large $\nu$ vertices 2 and 1. 

\paragraph{7-Vertex/9-Edge Digraph from \cite{Lai2010PhysicaA}}

The edge clustering results for this simple 7 vertex and 9 edge unweighted digraph appear in Fig.~\ref{fig:Cockroach_2Clusters_SocialScaling} (d), (e), (f).  

For PRE clustering (d), the four edges of the blue cluster consists of the edges connected to vertex 4.  This is to be expected, as it has the largest $\nu$ value. It consists of only outgoing edges, so its role as a producer is captured in the blue cluster. The orange cluster shows vertices 1 and 5 as producers and vertex 6 as a receptor.  

For DPE clustering (e), the orange cluster has a length 3 path involving vertices 4, 5, 7, and 6, while the blue cluster has a length 2 path involving vertices 4, 1, and 2. Both these paths feature vertex 4, which is to be expected given the large social participation of vertex 4 and the emphasis DPE places on generating clusters with directed paths. 

RGE clustering (f) partitions the graph into two clusters, with the edges connected to vertex 4 all in the blue cluster.


\subsection{Real Digraphs}


\paragraph{Piazza Mazzini Square Street Map \cite{Sardellitti2017_arXiv}} 

This is a 236-vertex/349-edge planar unweighted digraph representing the road network in the vicinity of Piazza Mazzini in Rome, Italy. While the geographic locations of the digraph vertices are provided, this information is not utilized in our edge clustering algorithms. Earlier, in Fig.~\ref{fig:PiazzaMazzini}, we showed our results with $K=5$. As we now see in Fig.~\ref{fig:PiazzaMazzini_2Clusters}, the same general patterns can be observed when our edge clustering algorithms are used with $K=2$. The left column has the clustering results without the geographic coordinates, while the right column has the results overlaid on the geographic locations. PRE clusters feature diamond like motifs based on its edge affinity of encouraging outgoing edges to be in one cluster, and incoming edges in another cluster, meaning that certain traffic junctions are acting more as producers of traffic, and as others as receptors of traffic. These indicate complex interchanges or points of driver confusion. DPE clusters feature long directed paths on which one can drive.  These paths can be used to drive to the center, or avoid the center, and point to potentials for bottlenecks in the case of accidents. RGE clusters partition the graph into different sections of concentrated flows, with the coupling vertices/waypoints consisting of traffic junctions that have lower relative $\nu$ which means fewer connected edges in this case. So we can view each cluster in RGE clustering as a subgraph in which one can have concentrated traffic flows in either direction. The three methods guide the formation of groups of edges sharing specified respective edge affinities. 


\paragraph{Connectome of \emph{Caenorhabditis elegans} \cite{Cook2019Nature, WormWiring_Dataset}}

The \emph{C elegans} connectome is often studied to connect neuron functions to organism behavior. Our focus is on the hermaphrodite chemical synapse network. We reiterate that all edge clustering is computed solely using properties of the weighted adjacency matrix of the connectome.  The vertex (neuron) relative importances $\nu$ were set proportional to the vertex weighted absolute degree (i.e., \eqref{eq:nuvectorchoice}). For \emph{C elegans,} the weighted absolute degree of a neuron is the sum of all of its edge weights, since the chemical synapse weights are all positive. When applied to the \emph{C elegans} connectome, vertex clustering places neurons in different groups.  In contrast, edge clustering produces edge cluster subgraphs that link together neurons possessing widely varying functions.  These subgraph clusters suggest circuits associated with different organism functions and behaviors, and allow for a topological basis for investigating or understanding functions of neurons or groups of neurons. While a directed synapse edge can be a member of only one edge cluster, a neuron can possibly be a member of multiple edge clusters, allowing one to understand different facets of neuron functionality based on its role in different subgraph circuits. 

We now view some of the results. All three edge clustering methods were run with $K=30$ edge clusters. Each method offers different types of insight into the \emph{C elegans} connectome, as we shall see.  Much of the information regarding the known functions of neurons was obtained from \cite{WormAtlas}. 

\emph{DPE clustering:} 
DPE clustering produces subgraph clusters that emphasize long directed paths of synaptic links between neurons with large $\nu$. In Fig.~\ref{fig:CElegans_DPE_K30_N110_M183}, we have already seen that DPE clustering with $K=30$ produced a cluster that showed circuit elements associated with anterior harsh touch response and backward locomotion, with a prominent role for the FLPR sensory neuron.  Noting that the FLPR and FLPL neurons only appear in two of the 30 DPE clusters, we provide the other cluster in Fig.~\ref{fig:CElegans_DPE_K30_N123_M198}.  This subgraph highlights directed paths emanating from the FLPL sensory neuron and also features feedforward circuit elements associated with anterior harsh touch response and backward locomotion.  Note that feedback circuit elements are produced as well (e.g., feedback directed paths from muscles, motor, or interneurons to sensory neurons). Therefore, these clusters help one better understand the feedback and feedforward chains of synaptic connections associated with anterior harsh touch response and backward locomotion. DPE clustering allows researchers to investigate and understand the functions of neurons or groups of neurons and associated edges by focusing on subgraphs emphasizing directed synaptic paths among neurons with large $\nu$.

\emph{PRE clustering:} 
PRE clustering allows one to focus on neurons in their roles mainly as either producers or receptors of synaptic connections.  That is, a PRE subgraph cluster consists mainly of connected star graphs in which neurons with larger relative importance (as measured by $\nu$) are either mainly producing synapses (with outgoing edges) or mainly receiving synapses (with incoming edges), but generally not both. Neurons that are neither strictly producers or receptors in a cluster generally allow connections between the producers and receptors.  Therefore PRE clustering allows one to focus on a particular neuron as a producer, and see which neurons it is mainly influencing from a graph perspective and the effects on the rest of the circuit; alternatively, one can focus on a neuron as a receptor, and see which other neurons mainly influence it. Since each edge can belong to only one cluster, a synaptic edge from neuron A to neuron B is either in a cluster in which the role of A as a producer of synaptic influence is emphasized, or it is in a cluster in which the role of B as a receptor of synaptic influence is emphasized, assuming both neuron A and B have relatively larger $\nu$. Given that neurons have many outgoing and incoming edges, PRE allows one to focus on groups of neurons and synapses in a subgraph circuit in which neurons are either mainly producers of synaptic influence or receptors of influence. 

As examples of PRE clustering of \emph{C elegans,} consider Figs~\ref{fig:CElegans_PRE_K30_N85_M122} and \ref{fig:CElegans_PRE_K30_N158_M325_trimmed_N103_M179}. In the first subgraph Fig~\ref{fig:CElegans_PRE_K30_N85_M122}, we see AIZR and AIBR as mainly receptors, and AIAR, AIYR, and HSNL as mainly producers.  In contrast, in the second subgraph Fig.~\ref{fig:CElegans_PRE_K30_N158_M325_trimmed_N103_M179}, we see AIZR and AIBR mainly as producers while AIAR, AIYR and HSNL are mainly receptors. It is important to note that HSNL is a major receptor in another subgraph cluster not pictured. These two subgraphs are associated with sensory integration and elements of egg-laying. Given the prominent roles of AIA, AIY and AIZ and AIB, these subgraphs can provide insight into how these four neurons inhibit turns and promote turns.  Given the importance of HSNL in controlling egg-laying and the production of serotonin, we can view, purely from a topological perspective, which outgoing synaptic edges are most important in its role as a producer.  Note that each neuron (vertex) can be in many edge clusters. For example, there is a large weight outgoing edge from AIZR to the command neuron AVEL that is not in the cluster in Fig.~\ref{fig:CElegans_PRE_K30_N158_M325_trimmed_N103_M179}.  This edge is then a member of another cluster and is in another subgraph, e.g., a cluster that emphasizes AVEL as a receptor. So AIZR can potentially be in many different clusters highlighting its different roles in different circuits. PRE allows one to focus on which edges and associated groups of neurons are important to a particular neuron in its role as either a producer or receptor of influence, and the effects on associated organism function. 

\emph{RGE clustering:} 
RGE clustering allows us to focus on subgraph circuits among groups of neurons linked by concentrated synaptic connections regardless of directions. Therefore each cluster reveals a subgraph of concentrated synaptic connections among neurons with larger relative importance as captured by $\nu$ (recall that $\nu$ is the sum of the absolute values of the weights of all edges connected to a neuron).  Therefore each subgraph cluster is a circuit of concentrated activity that connects to organism function. Consider the case of neurons for which we have limited information regarding its role or function. We can investigate the functions of these neurons by examining their connections to other neurons and circuit functions in a cluster (or clusters) and infer possible functionalities.  For example, consider the two examples of RGE subgraph clusters in Fig~\ref{fig:CElegans_K30_RGE}.  In the top subgraph in Fig.~\ref{fig:CElegans_K30_RGE}(a), we see a subgraph cluster with circuit elements associated with egg-laying, serotonin release and processing, and chemical sensing that include the interneurons ADAL and ADAR. ADA's function is not well-known. The connections to this circuit suggest that ADA may have a functional role related to chemical sensing, serotonin processing, and egg-laying.  In the bottom subgraph in Fig.~\ref{fig:CElegans_K30_RGE}(b) we see the CANL and CANR neurons that are essential for the survival of \emph{C elegans,} and whose functions are largely unknown.  One can start with the connections to the mechanosensor ALA, and trace the connections to other subgraph circuit elements such as PVDR and PVDL involved in posterior harsh touch response to develop inferences regarding the functions of CANL and CANR. RGE clustering separates large graphs into smaller subgraphs with concentrated flows and connections among neurons of relatively large importance, thereby allowing one to further investigate the functions of groups of neurons in organism function.

We provide these examples of DPE, PRE, and RGE clustering of the \emph{C elegans} hermaphrodite chemical synapse network to underscore the ability of our Flow Laplacian Edge Clustering framework to analyze different data.  We only provide a few examples of each method computed with $K=30$ due to space reasons.  As we have seen, edge clustering allows one to take a large neuron connectome, and break it up into different subgraphs that emphasize different types of directed influence and concentrated flow properties, depending upon the type of edge clustering selected.  In contrast to conventional vertex clustering that places neurons into different groups, edge clustering provides subgraph clusters that show how neurons from different groups link together to achieve different types of directed and concentrated flows. This will allow scientists to focus on smaller subgraphs/circuits while investigating neuron and circuit functions, and also provides a purely graph-based method for further understanding the functions of groups of neurons.


\paragraph{US Inter-State Migration Patterns \cite{InterStateMigration_USCB2019}}

The 2019 US inter-state migration flow data for the 50 states plus Washington D.C. \cite{InterStateMigration_USCB2019} can be represented as a 51-vertex/2,351-edge weighted digraph. The states are vertices in the graph with directed edges connecting pairs of states. Edge direction indicates the direction of migration from the starting state to the destination state and the edge weight is equal to the number of people who migrated (always a positive number); the edge's color and thickness are proportional to this edge weight.  What if we focus on the important subgraphs showing directed flows in which certain states are primarly producers of people migrating to other states while other states are mainly receptors of people?

PRE clustering generates subgraph clusters that highlight such migratory movement. Fig.~\ref{fig:USMigration_Cypress}(a) shows only one cluster generated from the application of our PRE clustering algorithm with $K=6$. Fig.~\ref{fig:USMigration_6Clusters_LPRE} show all the 6 clusters that we obtain. States/vertices with large amounts of migration (e.g., California, Texas, Florida, New York), have correspondingly large $\nu$ values.  That is why in one subgraph, California is mainly a producer, while in another it is mainly a receptor due to the PRE edge affinity. PRE clustering partitions the migration graph into $K=6$ subgraphs with each subgraph featuring important producer and receptor states with respect to human migration. 

On the other hand, RGE clustering generates `regions' within which there is higher concentrated migratory movement, regardless of the direction of movement. These regions are not determined by geographic proximity, but rather by concentrated migratory movement as captured through edge similarities guided by the RGE edge affinity. Fig.~\ref{fig:USMigration_Cypress}(b) shows only 1 cluster generated from the application of our RGE clustering algorithm with $K=6$.  Fig.~\ref{fig:USMigration_6Clusters_LRGE} show all the 6 clusters that we obtain. Recall that these 6 subgraph clusters are obtained only using the digraph weighted adjacency matrix of total migration between states; no additional data are used (i.e., no metadata or geographic information). As we can see, all of the edges associated with migration to and from California are placed in one subgraph cluster. This makes sense as California is the most populous state in the US and is associated with large migration flows, e.g., to Texas. The 5 remaining subgraph clusters feature concentrated flows generally focused on one state, i.e., New York, Texas, Florida, Minnesota, and Washington. The migration cluster centered on the midwest is noteworthy since such pockets of localized migration are not typically discussed in the public discourse. Since edges are clustered, and not vertices, states can be connected to multiple subgraph clusters. For example, one subgraph cluster involves migratory flows primarily in the Northeast and centered around New York, while New York is also involved in the subgraph cluster centered on flows involving Florida.  As we can see these subgraph clusters are not geographically confined and in fact link states across the US.


\paragraph{Florida Bay Cypress Wetlands in Dry Season \cite{Bondavalli2000JB}} 

Fig.~\ref{fig:USMigration_Cypress}(c) shows one cluster generated from the application of our DPE clustering algorithm with $K=10$ to a weighted digraph representing the Florida Bay Cypress wetlands in dry season; Figs~\ref{fig:Cypress_10Clusters_100_LDPE}-\ref{fig:Cypress_10Clusters_36_LDPE} show all the 10 clusters that we obtain. Fig.~\ref{fig:USMigration_Cypress}(d) shows one cluster generated from the application of our RGE clustering algorithm; Figs~\ref{fig:Cypress_RGE_K10A}-\ref{fig:Cypress_RGE_K10E} show all the 10 clusters. Recall that this digraph has 68 vertices with 554 weighted directed edges, with the positive edge weight proportional to the amount of carbon transfer among different organisms. 

One can extract numerous possible subgraphs from this graph.  What if we want to extract subgraphs that are meaningful and important in terms of concentrated indirect and direct carbon transfer among organisms at different trophic levels?  We can get such subgraphs through DPE clustering, and all $K=10$ such subgraphs are presented here with each subgraph providing insight into trophic levels and important directed carbon flow paths. Recall that $\nu$ values are proportional to the sum of the edge weights to which a vertex is connected thereby capturing its relative importance in carbon transfer. DPE clustering yields $K=10$ subgraph clusters, with each subgraph featuring long directed paths linking vertices with larger $\nu$ values, illustrating concentrated direct and indirect carbon transfer.  Therefore, DPE clustering groups the edges into 10 subgraphs based on an edge affinity of concentrated direct and indirect carbon transfer, and helps us better understand trophic levels.

On the other hand, RGE emphasizes concentrated flows, regardless of direction, among organisms with large $\nu$.  This then breaks up the graph into $K=10$ subgraphs of concentrated flows. Note that the largest subgraph has 231 of the 554 edges (see Fig.~\ref{fig:Cypress_RGE_K10A}(a)), meaning that it has the largest concentrated carbon flows. The other subgraphs provide insight into other concentrated flows, such as how aquatic invertebrates and small herbivorous and omnivorous fish provide both provide carbon flows to the same organisms. For instance, the RGE subgraph in Fig.~\ref{fig:Cypress_RGE_K10E}(b) (which is a a reprint of the cluster in Fig.~\ref{fig:USMigration_Cypress}(d)) reveals the important concentrated carbon flow subgraph associated with the Florida panther, an endangered animal. 


\paragraph{Dominance Relations Among Rhesus Monkeys \cite{Sade1972FP}}

The dominance relations among a 16 free-ranging adult rhesus monkey (\emph{Macaca mulatta}) population in Cayo Santiago, Puerto Rico, appears in \cite{Sade1972FP}. Each adult rhesus monkey is a unique vertex in the graph. The number of encounters in which monkey A dominates monkey B has been recorded and it is captured as a weighted directed edge from A to B in a weighted digraph. This results in the 16-vertex/105-edge weighted digraph in Fig.~\ref{fig:Rhesus}. We use two edge clustering methods to provide insight into the dominance relations. 

First, we can use DPE clustering methods to create subgraphs of concentrated dominance, allowing us to visualize hierarchical structures.  The key is that DPE clustering discourages loop backs in its subgraph clusters, and instead encourages chains of dominance for this dataset. Figs~\ref{fig:Rhesus_6Clusters_LDPE} and \ref{fig:Rhesus_6Clusters_LDPEb} show the result of applying our DPE clustering algorithm with $K=6$ to the digraph representing this dataset. Subgraph chains of dominance can be easily identified from these edge clusters. For instance, the high rank of ``066'' among the males and ``065'' and ``004'' among the females, and the low rank of ``CN'' among the males and ``KE'' among the females, are clear from these clusters, and confirms the views expressed in \cite{Sade1972FP}. Each cluster has concentrated directed paths which in this case suggests hierarchical chains of dominance.  Recall that edge affinities are stronger for edges forming a directed path through a vertex with relatively larger $\nu$ values, which in this case corresponds to rhesus monkeys with large numbers of dominance interactions. 

Second, we can use PRE clustering to obtain subgraphs of concentrated production and reception of dominance. As we have seen, vertices with relatively larger $\nu$ values have their roles as producers and receptors separated into different subgraph clusters.  In this context, rhesus monkeys that are very active (large number of dominance interactions) generally have their roles as producers of potential dominance split from their roles as receptors. The result of applying PRE clustering with $K=6$ appear in Figs.~\ref{fig:Rhesus_6Clusters_LPRE} and \ref{fig:Rhesus_6Clusters_LPREb}. For example, the monkey ``ER'' is a producer of dominance in Fig.~\ref{fig:Rhesus_6Clusters_LPRE}(d) while in Fig.~\ref{fig:Rhesus_6Clusters_LPRE}(a) it is a receptor of dominance. Similarly, in Fig.~\ref{fig:Rhesus_6Clusters_LPREb}(b) we see the important edges associated with ``004'' as a producer of dominance, while Fig.~\ref{fig:Rhesus_6Clusters_LPRE}(a) shows ``004'' as a receptor of dominance primarly asserted by ``065''.  Each edge can only be in one cluster, and so the PRE edge affinity encourages edges to be primarily associated with either production or reception of potential dominance. 


\paragraph{South Korea COVID-19 Dataset \cite{Laenen2020NEURIPS}}

The Data Science for COVID-19 (DS4C) dataset contains information about COVID-19 infections in South Korea. The 1,073-vertex/836-edge unweighted digraph representing this dataset has individuals as vertices, with directed edges indicating COVID-19 infections from individual to individual. It has 245 weakly connected components, ranging in size from 1 to 67. Fig.~\ref{fig:COVID19_3Clusters_67_LPREReg} shows the result of applying PRE clustering with $K=3$ to the largest 67-vertex/66-edge component of this digraph. With the use of the un-normalized Laplacian $L_{\PRE}$, PRE clustering manages to capture the edges which emanate from the `super' spreader while grouping the successive hops in different clusters. The unnormalized PRE Laplacian was used as there is no need to encourage equal size clusters in this case. 


\paragraph{Page Crawler-Generated URL Addresses of 100 Unique Web Pages \cite{MATLAB21}}

A 100-vertex/632-edge unweighted digraph which represents URL addresses of 100 unique web pages generated by an automatic page crawler can be accessed through \texttt{MATLAB} \cite{MATLAB21}. The page crawler starts at the URL address \texttt{https://www.mathworks.com}. It has 15 `knots' of tightly connected web pages, each having 5 vertices; each of these smaller knots is somewhat loosely connected to a larger knot of tightly connected web pages having 14 vertices. For undirected graphs, RGE edge clustering can be performed by taking each undirected edge, and converting it into two directed edges, one in each direction.  Fig.~\ref{fig:MATLAB100_LRGE} shows the edge clusters obtained by applying RGE clustering with $K=15$. Note how edges in each smaller knot are assigned to the same edge cluster.


\paragraph{Default Mode Network (DMN) of Human Brain \cite{Gordon2020PNAS}}

By replacing each undirected edge with a pair of directed edges in opposite directions, edge clustering of undirected graphs can also be carried out with our algorithms. To illustrate this, we considered the 178-vertex/1,584-edge weighted undirected graph with positive weighted edges representing the the high-level DMN of a human brain of a subject in \cite{Gordon2020PNAS}. Each vertex in the graph is a subnetwork. Subnetworks with different functionalities are highlighted with different colors (which are identical to those used in \cite{Gordon2020PNAS}): \emph{fronto-parietal} is top-down control of lower level processing systems; \emph{retrosplenial} is contextual and scene information; \emph{ventromedial} is fear and anxiety; \emph{pregenual} is reward processing; \emph{parietal} is internally oriented and social cognition; \emph{lateral DMN} are connector hubs with their connections to fronto-parietal key for understanding control of emotions. 

Figs~\ref{fig:HumanDMN_6Clusters_LRGEA}, \ref{fig:HumanDMN_6Clusters_LRGEB}, and \ref{fig:HumanDMN_6Clusters_LRGEC} show the result of applying our RGE clustering algorithm with $K = 6$.  Each cluster suggests a subgraph of meaningful concentrated activity based on the linked subnetworks. In Fig.~\ref{fig:HumanDMN_6Clusters_LRGEA}(a), we see a subgraph with top-down control of language, cognition, scene and context understanding, fear, emotion, and reward processing. Fig.~\ref{fig:HumanDMN_6Clusters_LRGEA}(b) shows a subgraph that is mostly about  internal language network processes along with their connection to top-down control, as well as integration with connector hubs linking to the DMN. Fig.~\ref{fig:HumanDMN_6Clusters_LRGEB}(a) features a subgraph that largely involves internal fronto-parietal organization processes along with their control of language functions and integration of contextual and scene information. The Fig.~\ref{fig:HumanDMN_6Clusters_LRGEB}(b) subgraph shows top-down control and integration of reward processing, scene and contextual information, and internal DMN cognition and social cognition as well as connector hubs. Fig.~\ref{fig:HumanDMN_6Clusters_LRGEC}(a) has a subgraph that is about top-down control of integrating contextual and scene information, as well as connections to connector hubs and implications for top-down control of emotion. Fig.~\ref{fig:HumanDMN_6Clusters_LRGEC}(b) shows connections between the connector hubs of the DMN and the language network, as well as internal DMN cognition and/or social cognition as well as connections to top-down control. 

In this context, RGE clustering partitions the subgraph into regions of large concentrated activity as measured by grouping together edges that share common vertices (i.e., subnetworks) with large $\nu$ values meaning vertices connected to edges with large weights. Then the coupling vertices (vertices that are connected to multiple subgraph clusters) are those that have relatively smaller $\nu$ values, meaning that their connected edges are not necessarily kept in the same region.  Therefore each RGE cluster identifies a subgraph of concentrated activity among the linked subnetworks, potentially pointing the way to identifying functionalities in the human DMN undirected graph. 


\paragraph{Minnesota Road Map \cite{MATLAB21}}

As another example of an undirected graph, take the 2,635-vertex/3,298-edge unweighted and undirected representing the road network in the state of Minnesota, USA. For edge clustering purposes, each undirected edge was replaced with a pair of directed edges in opposite directions. Then RGE clustering was performed. Similar to the Piazza Mazzini street map data, the geographic locations of the vertices are provided but not utilized in edge clustering.  Fig.~\ref{fig:MNRoadNetwork_10Clusters_LRGE} shows the result of RGE clustering with $K=10$.  We see that the area around Minneapolis-St. Paul has been partitioned into three small regions, reflecting the fact that that area has a large number of roads. 


\section{Additional Details Regarding the Edge Laplacian}

We now present some additional details on the edge Laplacian.  These details are useful for understanding how the edge Laplacian can be used to construct the flow Laplacians in edge clustering and for understanding the dual graph cuts interpretation.

\subsection{Edge Laplacian}


\begin{lemma}
\label{lem:LeEntries}
The $(p,q)$-th element of the edge Laplacian $\ul{L}_e$ in Definition~\ref{def:Le} is the following:
\begin{itemize}
  \item[\tb{(i)}] When either $e_p$ or $e_q$ is a self-edge: $(\ul{L}_e)_{pq}=0$, i.e., row and column corresponding to each self-edge are zero. 
  \item[\tb{(ii)}] When neither $e_p$ nor $e_q$ is a self-edge: the diagonal elements are  $(\ul{L}_e)_{pp}=\phi_{k_p}+\phi_{\ell_p},\, p \in \{1, \ldots, M\}$; and the off-diagonal elements (i.e., $p\neq q$) are  
\[
  (\ul{L}_e)_{pq}
    =\begin{cases}
       +\phi_{k_p},
         & \text{for $k_p=k_q$, $\ell_p\neq\ell_q$, i.e., $(e_p, e_q)$ converges to $v_{k_p}=v_{k_q}$}; \\
       +\phi_{\ell_p},
         & \text{for $\ell_p=\ell_q$, $k_p\neq k_q$, i.e., $(e_p, e_q)$ diverges from $v_{\ell_p}=v_{\ell_q}$}; \\
       -\phi_{k_p},
         & \tr{for $k_p=\ell_q$, $\ell_p\neq k_q$, i.e., $(e_p, e_q)$ flows through $v_{k_p}=v_{\ell_q}$}; \\
       -\phi_{\ell_p},
         & \text{for $\ell_p=k_q$, $k_p\neq\ell_q$, i.e., $(e_p, e_q)$ flows through $v_{\ell_p}=v_{k_q}$}; \\
       0,
         & \tr{otherwise, i.e., $(e_p, e_q)$ does not share a vertex}.
     \end{cases}
\]
\end{itemize}
\end{lemma}

Obvisouly $\ul{L}_e$ is symmetric. From Definition~\ref{def:Le} and Theorem~\ref{lem:LeEntries}, we make the following observations:
\begin{enumerate}
  \item The elements of the edge Laplacian $\ul{L}_e$ are determined by the directionality of pairs of edges and the vertex vector value at the vertex where they meet (see Figure~\ref{fig:LeEntries}). To be specific, take two edges $\{e_p, e_q\} = \{(v_{\ell_p}{\to}\, v_{k_p}), (v_{\ell_q}{\to}\, v_{k_q})\}$:
  \begin{enumerate}
    \item when they share a vertex but do not form a length-2 directed path (i.e., they converge at the shared vertex $v_{k_p}=v_{k_q}$ or they diverge from the shared vertex $v_{\ell_p}=v_{\ell_q}$), the element $(\ul{L}_e)_{pq}$ takes the vertex vector value at the shared vertex $v_i$ while inheriting the same sign, i.e., $\mr{sgn}[\ul{L}_{e,pq}] = \mr{sgn}[\phi_i]$; and
    \item when they share a vertex and form a length-2 directed path (i.e., they flow through the shared vertex $v_{k_p}=v_{\ell_q}$ or the shared vertex $v_{\ell_p}=v_{k_q}$), the element $(\ul{L}_e)_{pq}$ takes the vertex vector value at the shared vertex with the opposite sign, i.e., $\mr{sgn}[\ul{L}_{e,pq}] = -\mr{sgn}[\phi_i]$.
  \end{enumerate}  
  \item\label{item:3} Both multi-edges and loops are being accounted for: when the two edges $(e_p, e_q)$ form a multi-edge or a loop between the two vertices $\{v_{\ell_p}, v_{k_p}\}$, the element $(\ul{L}_e)_{pq}$ takes the value $+(\phi_{k_p}+\phi_{\ell_p})$ or $-(\phi_{k_p}+\phi_{\ell_p})$, respectively.
  \item The $p$-th diagonal entry of $\ul{L}_e$ associated with the edge $e_p \cong  e_{k_p\ell_p} = (v_{\ell_p} {\to}\, v_{k_p})$ is  
\begin{equation}
  (\ul{L}_e)_{pp}
    =\begin{cases}
       0,
         & \tr{when $e_p$ is a self-edge}; \\
       \phi_{k_p}+\phi_{\ell_p},
         & \tr{otherwise}.
     \end{cases}
  \label{eq:Le_pp}
\end{equation}
So, all diagonal elements of $\ul{L}_e$ associated with non-self edges are strictly positive whenever $\ul{\phi} > 0$, i.e., whenever the vertex vector is strictly positive (e.g., when $\phi_i = \sigma_i$, the social participation of vertex $v_i$).
\end{enumerate}

As an example, a 7-vertex/9-edge digraph $\mc{G}$ from \cite{Lai2010PhysicaA} is in Figure~\ref{fig:digraph_7-9}. The vertex vector values, which are taken as $\phi_i = \sigma_i$, are shown in red; the edge weight function values are shown in blue. Its edge Laplacian $\ul{L}_e$ is in Figure~\ref{fig:Le_7-9}.

Suppose $\ul{\phi}\geq 0$. Then, the following facts follow directly from Theorem~\ref{thm:DS}:
\begin{enumerate}
  \item $\ul{L}_e(\ul{\phi})$ is p.s.d. 
  \item $\ul{w}^T \ul{L}_e(\ul{\phi})\, \ul{w}=0$ iff $d_{out,i}=d_{in,i}=0,\,\forall v_i\in\mc{V}$.
  \item $\ul{L}_e(\ul{\phi})$ can be expressed as the `square' of a \emph{weighted edge incidence matrix $\ul{B}_e\in\mbb{R}^{N\times M}$} as $\ul{L}_e(\ul{\phi}) = \ul{B}_e(\ul{\phi})^T \ul{B}_e(\ul{\phi})$, where $\ul{B}_e(\ul{\phi}) = \mr{diag}[\ul{\phi}^{1/2}]\, \ul{B}$. So, for instance, associated with the digraph in Figure~\ref{fig:digraph_7-9}, we have 
\[
  \ul{B}_e
    =\begin{bmatrix}
       +\sqrt{\phi_1 } & +\sqrt{\phi_1 } & -\sqrt{\phi_1 } & 0 & 0 & 0 & 0 & 0 & 0\\
       -\sqrt{\phi_2 } & 0 & 0 & -\sqrt{\phi_2 } & 0 & 0 & 0 & 0 & 0\\
       0 & -\sqrt{\phi_3 } & 0 & 0 & -\sqrt{\phi_3 } & 0 & 0 & 0 & 0\\
       0 & 0 & +\sqrt{\phi_4 } & +\sqrt{\phi_4 } & +\sqrt{\phi_4 } & +\sqrt{\phi_4 } & 0 & 0 & 0\\
       0 & 0 & 0 & 0 & 0 & -\sqrt{\phi_5 } & +\sqrt{\phi_5 } & +\sqrt{\phi_5 } & 0\\
       0 & 0 & 0 & 0 & 0 & 0 & -\sqrt{\phi_6 } & 0 & -\sqrt{\phi_6 }\\
       0 & 0 & 0 & 0 & 0 & 0 & 0 & -\sqrt{\phi_7 } & +\sqrt{\phi_7 }
     \end{bmatrix}.
\]
  \item The diagonal element $(\ul{L}_e)_{pp}=g_p=\phi_{k_p}+\phi_{\ell_p}=0$ iff $\phi_{k_p}=\phi_{\ell_p}=0$. The only non-zero values that occur in the $p$-th row and $p$-th column of $\ul{L}_e$ correspond to edges that share a vertex with $e_p$. The values of these entries must be either $\pm\phi_{k_p}$ or $\pm\phi_{\ell_p}$, the two end-vertices of $e_p$. So, when $\phi_{k_p}=\phi_{\ell_p}=0$, the whole $p$-th row and $p$-th column of $\ul{L}_e$ are zero.
\end{enumerate}


\subsection{Dual Graph}
\label{subsec:DualGraph}


\begin{lemma}
\label{lem:DualGraph} 
Consider the dual graph $\mc{G}'(\mc{V}', \mc{E}')[\ul{\nu}', \ul{w}']$ in Definition~\ref{def:DualGraph} associated with the digraph $\mc{G}(\mc{V}, \mc{E})[\ul{\nu}, \ul{w}]$. Suppose all elements of the edge weight vector $\ul{\nu} = \{\nu_i\} \in \mbb{R}^N$ are non-zero, i.e., $\nu_i \neq 0,\, \forall i \in \{1, \ldots, M\}$. Then the following are true:
\begin{itemize}
  \item[\tb{(i)}] $\mc{G}'$ is in general a signed undirected graph with no self-edges.
  \item[\tb{(ii)}] $\mc{G}'$ has $M' = \displaystyle\sum_{v_i\in\mc{V}} \begin{pmatrix} \sigma_2 \\ 2 \end{pmatrix}$ edges and each dual edge corresponds to a digraph edge pair that shares a common vertex. Moreover, the dual graph's edge vector is $\ul{w}' \cong \ul{W}'$, where $\ul{W}'(\ul{\nu}) = \ul{L}_e(\ul{\nu}) - \mr{diag}[\mr{diag}[\ul{L}_e(\ul{\nu})]$.
  \item[\tb{(iii)}] $\mc{G}'$ has $N' = M$ vertices and each dual vertex corresponds to a digraph edge.  
\end{itemize}
\end{lemma}


\paragraph{Dual Graph vis-\`{a}-vis the Line Graph} 

Our notion of a dual graph $\mc{G}'(\mc{V}', \mc{E}')$ in Definition~\ref{def:DualGraph} can in fact be considered a `generalization' (albeit a sign difference) of another notion that one associates with a digraph, viz., the \emph{line graph} $\mc{G}''(\mc{V}'', \mc{E}'')$ \cite{Orlin1977IM}. As with the dual graph, each vertex $v''_p$ in $\mc{G}''$ represents an edge $e_p$ in $\mc{G}$. However, the edge set $\mc{E}''$ is different than $\mc{E}'$. Suppose the vertices $v''_p$ and $v''_q$ in $\mc{G}''$ represent, respectively, the two edges $e_p$ and $e_q$ in $\mc{G}$. Then $(v''_p, v''_q)$ forms an edge in $\mc{G}''$ iff $\{e_p, e_q\}$ share a vertex \emph{and} form a length-2 directed path in $\mc{G}$; if the two edges share a vertex but do \emph{not} form a length-2 directed path, then $(v''_p, v''_q)$ does not constitute an edge in $\mc{G}''$. In contrast, the vertices $v_p'$ and $v_q'$ in our dual graph $\mc{G}'$ which represent, respectively, the edges $e_p$ and $e_q$ in $\mc{G}$ constitute an edge in $\mc{G}'$ whenever they share a vertex (irrespective of whether they form a length-2 directed path or not). Indeed, one obtains the line graph $\mc{G}''$ by retaining only the `negative' edges in the dual graph $\mc{G}'$ (and changing their sign).


\section{Proofs} 
\label{sec:Proofs}


\subsection{Proof of Lemma~\ref{lem:Cost_k_MQF}}


Note that
\begin{align*}
  \tr{Cost}^{(k)}(\mc{G})
    &=\sum_{p=1}^M \sum_{q=1}^M
      \tr{Cost}^{(k)}(e_p, e_q)
     =\frac{1}{2} \sum_{p=1}^M \sum_{q=1}^M
      \phi_{pq}\,
      (x_p^{(k)} - \psi_{pq}\,x_q^{(k)})^2 
      \notag \\
    &=\frac{1}{2} \sum_{p=1}^M \sum_{q=1}^M 
      (\phi_{pq}\,x_p^{(k)^2} + \phi_{pq} \psi_{pq}^2\,x_q^{(k)^2})
        -\frac{1}{2} \sum_{p=1}^M \sum_{q=1}^M
         2\,\phi_{pq}\psi_{pq}\,x_p^{(k)}x_q^{(k)} 
      \notag \\ 
    &=\frac{1}{2} \sum_{p=1}^M \sum_{q=1}^M 
      \phi_{pq}(1+\psi_{pq}^2)\,x_p^{(k)^2}
        -\sum_{p=1}^M \sum_{q=1}^M
         \phi_{pq}\psi_{pq}\,x_p^{(k)}x_q^{(k)} 
      \notag \\
    &=\sum_{p=1}^M  
      d_{\ul{\Psi}\ul{\Phi}}(p)\,x_p^{(k)^2}
        -\sum_{p=1}^M \sum_{q=1}^M
         \phi_{pq}\psi_{pq}\,x_p^{(k)}x_q^{(k)},
\end{align*}
where $\displaystyle d_{\ul{\Psi}\ul{\Phi}}(p)=\frac{1}{2} \sum_{q=1}^M \phi_{pq}(1+\psi_{pq}^2)$ and in which we utilized the symmetry property of $\phi_{pq}$ and $\psi_{pq}$ to get $\displaystyle\sum_{p=1}^M \sum_{q=1}^M \phi_{pq}\psi_{pq}^2 x_q^{(k)^2} = \sum_{p=1}^M \sum_{q=1}^M \phi_{pq}\psi_{pq}^2 x_p^{(k)^2}$ in the third line. Recognizing the quadratic forms, now we can express $\tr{Cost}^{(k)}(\mc{G})$ as 
\[
  \tr{Cost}^{(k)}(\mc{G})
    =\ul{x}^{(k)^T}(\ul{D}_{\ul{\Psi}\ul{\Phi}} - (\ul{\Psi}\odot\ul{\Phi}))\,\ul{x}^{(k)},
\]
as claimed.
\hspace{\fill}{Q.E.D.}


\subsection{Proof of Lemma~\ref{lem:Psi}}


Consider the $(p,q)$-th element of $\ul{B}^T\ul{B}$:
\[
  (\ul{B}^T\ul{B})_{pq}
    =\sum_{v_i\in\mc{V}} (\ul{B}^T)_{pi}\,\ul{B}_{iq} 
    =\sum_{v_i\in\mc{V}} b_{ip}\, b_{iq}
    =\sum_{v_i\in\mc{V}} 
     (\delta_{i,\ell_p} - \delta_{i,k_p})\,(\delta_{i,\ell_q} - \delta_{i,k_q}).
\]
All terms within this summation are zero except when $v_i\in\{k_p, \ell_p, k_q, \ell_q\}$. So, 
\begin{multline*}
  (\ul{B}^T\ul{B})_{pq}
    =(\delta_{k_p,\ell_p} - 1)\,(\delta_{k_p,\ell_q} - \delta_{k_p,k_q})
       +(1 - \delta_{\ell_p,k_p})\,(\delta_{\ell_p,\ell_q} - \delta_{\ell_p,k_q}) \\
       +(\delta_{k_q,\ell_p} - \delta_{k_q,k_p})\,(\delta_{k_q,\ell_q} - 1)
       +(\delta_{\ell_q,\ell_p} - \delta_{\ell_q,k_p})\,(1 - \delta_{\ell_q,k_q}).
\end{multline*}
Let us consider the different cases in \eqref{eq:Psi_PRE}:
\begin{itemize}
  \item When $e_p$ and $e_q$ both point to a common vertex, we have $k_p=k_q$; when they both point away from a common vertex, we have $\ell_p = \ell_q$. In either case, we get $(\ul{B}^T\ul{B})_{pq} = +2$ so that $\tr{sgn}[(\ul{B}^T\ul{B})_{pq}] = +1$.  
  \item When $e_p$ points to and $e_q$ points away from a common vertex we have $k_p = \ell_q$; when $e_p$ points away from and $e_q$ points to a common vertex, we have $\ell_p = k_q$. In either case, we get $(\ul{B}^T\ul{B})_{pq} = -2$ so that $\tr{sgn}[(\ul{B}^T\ul{B})_{pq}] = -1$. \item When the edges $e_p$ and $e_q$ do not share a common vertex, we get $(\ul{B}^T\ul{B})_{pq} = 0$ so that $\tr{sgn}[(\ul{B}^T\ul{B})_{pq}] = 0$.  
  \item When $p = q$, we have $k_p = k_q$ and $\ell_p = \ell_q$ from which we get $(\ul{B}^T\ul{B})_{pq} = +4$ so that $\tr{sgn}[(\ul{B}^T\ul{B})_{pq}] = +1$.  
\end{itemize}
This establishes the relationship $\ul{\Psi}_{\PRE} = \mr{sgn}[\ul{B}^T \ul{B}]$. The relationships corresponding to $\ul{\Psi}_{\DPE}$ and $\ul{\Psi}_{\RGE}$ follow directly from \eqref{eq:Psi_DPE} and \eqref{eq:Psi_RGE}, respectively.
\hspace{\fill}{Q.E.D.}


\subsection{Proof of Lemma~\ref{lem:Cost_k_L}}


This follows directly from Lemma~\ref{lem:Cost_k_MQF} when one substitutes for $\ul{D}_{\phi}$ from \eqref{eq:D_phi} and use \eqref{eq:Psi_PRE}, \eqref{eq:Psi_DPE}, and \eqref{eq:Psi_RGE} for $\ul{\Psi}_{\RGE}$, $\ul{\Psi}_{\PRE}$, and $\ul{\Psi}_{\DPE}$, respectively.
\hspace{\fill}{Q.E.D.}


\subsection{Proof of Theorem~\ref{thm:DS}}


First we note that 
\begin{align*}
  (\ul{B}\, \ul{w})_i
    &=\sum_{q=1}^M b_{iq} w_q
     =\sum_{q=1}^M (\delta_{i,\ell_q}-\delta_{i,k_p})\,w_q
     =\sum_{v_i\in\mc{V}:\, i=\ell_q\neq k_q} w_q
        -\sum_{v_i\in\mc{V}:\, i=k_q\neq\ell_q} w_q \\
    &=d_{out,i}-d_{in,i}
     =d_{net,i}.
\end{align*}

To establish the edge differential at edge $e_p$, note that the $p$-th element of $\ul{L}_e \ul{w}$ is
\[
  (\ul{L}_e \ul{w})_p
    =(\ul{B}^T \mr{diag}[\ul{\nu}]\, \ul{B}\, \ul{w})_p
    =\sum_{i=1}^N (\ul{B}^T \mr{diag}[\ul{\nu}])_{pi}\, d_{net,i}.
\]
But the $(p,i)$-th element of $\ul{B}^T \mr{diag}[\ul{\nu}]$ is $\nu_i\,(\delta_{i,\ell_p}-\delta_{i,k_p})$. So, 
\[
  (\ul{L}_e\, \ul{w})_p
    =\sum_{i=1}^N \nu_i\,(\delta_{i,\ell_p}-\delta_{i,k_p})\,d_{net,i}
    =v_{\ell_p}\,d_{net,\ell_p}-v_{k_p}d_{net,k_p}.
\]

To establish the edge quadratic sum, note that  
\[
  \ul{w}^T \ul{L}_e(\ul{\nu})\, \ul{w}
    =(\ul{B}\, \ul{w})^T \mr{diag}[\ul{\nu}]\, (\ul{B}\,\ul{w})
    =\sum_{i=1}^N \nu_i\, (\ul{B}\, \ul{w})^T_i\, (\ul{B}\,\ul{w})_i 
    =\sum_{i=1}^N \nu_i\,(d_{out,i}-d_{in,i})^2.
\]

This concludes the proof.
\hspace{\fill}{Q.E.D.}


\subsection{Proof of Lemma~\ref{lem:Phi_Le}}


Compare the entries of $\ul{L}_e(\ul{\nu})$ as given in Lemma~\ref{lem:LeEntries}(ii) with those of $\ul{\Phi}(|\ul{\nu}|)$ in \eqref{eq:phi_nu}. The claim follows when one notes that these two matrices are identical in absolute value except that the diagonal entries of $\ul{\Phi}(\cdot)$ are pinned at zero. 
\hspace{\fill}{Q.E.D.}


\subsection{Proof of Theorem~\ref{thm:L_W'}}


Since $\ul{\nu} > 0$, $\mr{sgn}[\ul{B}^T \ul{B}] = \mr{sgn}[\ul{B}^T \mr{diag}[\ul{\nu}]\, \ul{B}]$. So, from Lemma~\ref{lem:Psi}, we get
\[
  \ul{\Psi}_{\PRE}
    = \mr{sgn}[\ul{L}_e];\;\;
  \ul{\Psi}_{\DPE}
    = -\mr{sgn}[\ul{L}_e];\;\;
  \ul{\Psi}_{\RGE}
    = \mr{sgn}[|\ul{L}_e|]
\]
Moreoever, for $\ul{\nu} > 0$, Lemma~\ref{lem:Phi_Le} yields
\[
  \ul{\Phi}(\ul{\nu})
    = |\ul{W}'(\ul{\nu})|,
  \;\;\tr{where}\;\;
  \ul{W}'(\ul{\nu})
    =\ul{L}_e(\ul{\nu}) - \mr{diag}[\mr{diag}[\ul{L}_e(\ul{\nu})].
\]
So, the $(p, q)$-th element of the Hadamard product $\ul{\Psi} \odot \ul{\Phi}_{\PRE}$ is given by 
\begin{alignat*}{2}
  (\ul{\Psi} \odot \ul{\Phi}_{\PRE})_{pq}
    &=\phi_{pq} \cdot (\ul{\Psi}_{\PRE})_{pq}
      &
        &=|(\ul{W}'(\ul{\nu}))_{pq}| \cdot \mr{sgn}[(\ul{L}_e)_{pq}] \\
    &=|(\ul{W}'(\ul{\nu}))_{pq}| \cdot \mr{sgn}[(\ul{W}'(\ul{\nu}))_{pq}]
      &
        &=(\ul{W}'(\ul{\nu}))_{pq},
\end{alignat*}
i.e., $\ul{\Psi} \odot \ul{\Phi}_{\PRE} = \ul{W}'(\ul{\nu})$, which also implies that 
\[
  \ul{\Psi} \odot \ul{\Phi}_{\DPE}
    =-(\ul{\Psi} \odot \ul{\Phi}_{\PRE})
    =-\ul{W}'(\ul{\nu});\;\;
  \ul{\Psi} \odot \ul{\Phi}_{\RGE}
    =|\ul{\Psi} \odot \ul{\Phi}_{\PRE}|
    =|\ul{W}'(\ul{\nu})|.     
\]
From \eqref{eq:D_phi}, we also have $\ul{D}_{\phi} = \mr{diag}[\ul{\Phi}\, \ul{1}_M] = \mr{diag}[|W'(\ul{\nu})\, \ul{1}_M]$. So we get 
\begin{alignat*}{3}
  &\ul{L}_{\PRE}(\ul{\nu})
    &
      &=\ul{D}_{\phi} - (\ul{\Psi} \odot \ul{\Phi}_{\PRE})
        &
          &=\ul{D}_{|\ul{W}'|} - \ul{W}'(\ul{\nu}); \\
  &\ul{L}_{\PRE}(\ul{\nu})
    &
      &=\ul{D}_{\phi} - (\ul{\Psi} \odot \ul{\Phi}_{\DPE})
        &
          &=\ul{D}_{|\ul{W}'|} + \ul{W}'(\ul{\nu}); \\
  &\ul{L}_{\PRE}(\ul{\nu})
    &
      &=\ul{D}_{\phi} - (\ul{\Psi} \odot \ul{\Phi}_{\RGE})
        &
          &=\ul{D}_{|\ul{W}'|} - |\ul{W}'(\ul{\nu})|.
\end{alignat*}
in which $\ul{D}_{|\ul{W}'|} = \mr{diag}[|\ul{W}'(\ul{\nu})|\, \ul{1}_M]$.
\hspace{\fill}{Q.E.D.}


\subsection{Proof of Lemma~\ref{lem:3Costs}}


By direct substitution of the expressions in Table~\ref{tab:parameters} and using the vertex vector $\ul{\nu} \in \mbb{R}^N$, where $\ul{\nu} > 0$.
\hspace{\fill}{Q.E.D.}


\subsection{Proof of Lemma~\ref{lem:3costsdualgraph} }


We establish the claim separately for each Flow Laplacian method in turn. Recall that the notation $v'_p, v'_q$ refers to vertices in the dual graph, and that $w'_{pq}$ is an edge weight in the dual graph, which is a signed and undirected graph.


\paragraph{PRE Method} 

The cost associated with the PRE method is
\[
  \tr{Cost}_{\PRE}^{(k)}(\mc{G}) =\ul{x}^{(k)^T} \ul{L}_{\PRE}\, \ul{x}^{(k)} 
    =\frac{1}{2} 
     \sum_{p=1}^M \sum_{q=1}^M 
     |w'_{pq}|\, (x_p^{(k)} - \mr{sgn}[w'_{pq}]\, x_q^{(k)})^2.
\]
We now consider the following cases: 
\begin{enumerate}
  \item \emph{When $v'_p, v'_q \in \Sigma^{(k)}$ and $w'_{pq}>0$:} 
In this case, we have $x_p^{(k)} = x_q^{(k)} = \alpha^{(k)}$ and $\mr{sgn}[w'_{pq}] = +1$. So, 
\[
  \tr{Cost}_{\PRE}^{(k)}(\mc{G})
    =\frac{1}{2} 
     \sum_{v'_p \in \Sigma^{(k)}}^M 
     \sum_{\substack{%
             v'_q \in \Sigma^{(k)} \\
             w'_{pq} > 0}}^M 
     |w'_{pq}|\, (\alpha^{(k)} - \alpha^{(k)})^2 
    =0.
\]
Thus, no clustering cost is incurred.
  \item \emph{When $v'_p, v'_q \in \overline{\Sigma}^{(k)}$:} 
In this case, $x_p^{(k)} = x_q^{(k)} = 0$, and no clustering cost is incurred.
  \item \emph{When $v'_p, v'_q \in \Sigma^{(k)}$ and $w'_{ij}<0$:} 
In this case, $x_p^{(k)} = x_q^{(k)} = \alpha^{(k)}$ and $\mr{sgn}[w'_{pq}] = -1$. So, 
\begin{align*}
  \tr{Cost}_{\PRE}^{(k)}(\mc{G})
    &=\frac{1}{2} 
      \sum_{v'_p \in \Sigma^{(k)}}^M 
      \sum_{\substack{%
              v'_q \in \Sigma^{(k)} \\
              w'_{pq} < 0}}^M 
      |w'_{pq}|\, (\alpha^{(k)} + \alpha^{(k)})^2 \\
    &=2\, {\alpha^{(k)}}^2 
      \sum_{v'_p \in \Sigma^{(k)}}^M 
      \sum_{\substack{%
              v'_q \in \Sigma^{(k)} \\
              w'_{pq} < 0}}^M 
      |w'_{pq}|
     =2\, {\alpha^{(k)}}^2\, \mr{Links}^{-}(\Sigma^{(k)}, \Sigma^{(k)}).
\end{align*}
  \item \emph{When $v'_p \in \Sigma^{(k)}$ and $v'_q \in \overline{\Sigma}^{(k)}$:} 
In this case, $x_p^{(k)} = \alpha^{(k)}$ and $x_q^{(k)} = 0$. So,
\begin{align*}
  \tr{Cost}_{\PRE}^{(k)}(\mc{G})
    &=\frac{1}{2} 
      \sum_{v'_p \in \Sigma^{(k)}}^M 
      \sum_{v'_q \in \overline{\Sigma}^{(k)}}^M 
      |w'_{pq}|\, (\alpha^{(k)} - 0)^2 \\
    &=\frac{1}{2}\, {\alpha^{(k)}}^2 
      \sum_{v'_p \in \Sigma^{(k)}}^M 
      \sum_{v'_q \in \overline{\Sigma}^{(k)}}^M 
      |w'_{pq}| 
     =\frac{1}{2} \mr{Cut}(\Sigma^{(k)}, \overline{\Sigma}^{(k)}).
\end{align*}
  \item \emph{When $v'_p \in \overline{\Sigma}^{(k)}$ and $v'_q \in \Sigma^{(k)}$:} 
In this case, $x_p^{(k)} = 0$ and $x_q^{(k)} = \alpha^{(k)}$. So,
\begin{align*}
  \tr{Cost}_{\PRE}^{(k)}(\mc{G})
    &=\frac{1}{2} 
      \sum_{v'_p \in \overline{\Sigma}^{(k)}}^M 
      \sum_{v'_q \in \Sigma^{(k)}}^M 
      |w'_{pq}|\, (0 - \mr{sgn}(w'_{pq})\alpha^{(k)})^2 \\
    &=\frac{1}{2}\, {\alpha^{(k)}}^2 
      \sum_{v'_p \in \overline{\Sigma}^{(k)}}^M 
      \sum_{v'_q \in \Sigma^{(k)}}^M 
      |w'_{pq}| 
     =\frac{1}{2}\, \mr{Cut}(\overline{\Sigma}^{(k)}, \Sigma^{(k)}).
\end{align*}
\end{enumerate}
Noting that $\mr{Cut}(\Sigma^{(k)}, \overline{\Sigma}^{(k)}) = \mr{Cut}(\overline{\Sigma}^{(k)}, \Sigma^{(k)})$, combine the above costs for all possible cases to obtain
\[
  \tr{Cost}_{\PRE}^{(k)}(\mc{G}) 
    =\ul{x}^{(k)^T} \ul{L}_{\PRE}\, \ul{x}^{(k)} = {\alpha^{(k)}}^2
     \left[
       \mr{Cut}(\Sigma^{(k)}, \overline{\Sigma}^{(k)}) 
         + 2\, \mr{Links}^{-}(\Sigma^{(k)}, \Sigma^{(k)})
     \right].
\]


\paragraph{DPE Method}

The cost associated with the DPE method is
\[
  \tr{Cost}_{\DPE}^{(k)}(\mc{G}) =\ul{x}^{(k)^T} \ul{L}_{\DPE}\, \ul{x}^{(k)} 
    =\frac{1}{2} 
     \sum_{p=1}^M \sum_{q=1}^M 
     |w'_{pq}|\, (x_p^{(k)} + \mr{sgn}[w'_{pq}]\, x_q^{(k)})^2.
\]
We now consider the following cases:
\begin{enumerate}
    \item \emph{When $v'_p, v'_q \in \Sigma^{(k)}$ and $w'_{pq}>0$:} 
In this case, we have $x_p^{(k)} = x_q^{(k)} = \alpha^{(k)}$ and $\mr{sgn}[w'_{pq}] = +1$. So, 
\begin{align*}
  \tr{Cost}_{\DPE}^{(k)}(\mc{G})
    &=\frac{1}{2} 
      \sum_{v'_p \in \Sigma^{(k)}}^M 
      \sum_{\substack{%
              v'_q \in \Sigma^{(k)} \\
              w'_{pq} > 0}}^M 
      |w'_{pq}|\, (\alpha^{(k)} + \alpha^{(k)})^2 \\
    &=2\, {\alpha^{(k)}}^2 
      \sum_{v'_p \in \Sigma^{(k)}}^M 
      \sum_{\substack{%
              v'_q \in \Sigma^{(k)} \\
              w'_{pq} > 0}}^M 
      |w'_{pq}|
     =2\, {\alpha^{(k)}}^2\, \mr{Links}^{+}(\Sigma^{(k)}, \Sigma^{(k)}).
\end{align*}
\item \emph{When $v'_p, v'_q \in \overline{\Sigma}^{(k)}$:} 
In this case, $x_p^{(k)} = x_q^{(k)} = 0$, and no clustering cost is incurred.
 \item \emph{When $v'_p, v'_q \in \Sigma^{(k)}$ and $w'_{ij}<0$:} 
In this case, $x_p^{(k)} = x_q^{(k)} = \alpha^{(k)}$ and $\mr{sgn}[w'_{pq}] = -1$. So, 
\[
  \tr{Cost}_{\DPE}^{(k)}(\mc{G})
    =\frac{1}{2} 
     \sum_{v'_p \in \Sigma^{(k)}}^M 
     \sum_{\substack{%
             v'_q \in \Sigma^{(k)} \\
             w'_{pq} < 0}}^M 
     |w'_{pq}|\, (\alpha^{(k)} - \alpha^{(k)})^2 
    =0.
\]
Thus, no clustering cost is incurred.
  \item \emph{When $v'_p \in \Sigma^{(k)}$ and $v'_q \in \overline{\Sigma}^{(k)}$:} 
In this case, $x_p^{(k)} = \alpha^{(k)}$ and $x_q^{(k)} = 0$. So,
\begin{align*}
  \tr{Cost}_{\DPE}^{(k)}(\mc{G})
    &=\frac{1}{2} 
      \sum_{v'_p \in \Sigma^{(k)}}^M 
      \sum_{v'_q \in \overline{\Sigma}^{(k)}}^M 
      |w'_{pq}|\, (\alpha^{(k)} + 0)^2 \\
    &=\frac{1}{2}\, {\alpha^{(k)}}^2 
      \sum_{v'_p \in \Sigma^{(k)}}^M 
      \sum_{v'_q \in \overline{\Sigma}^{(k)}}^M 
      |w'_{pq}| 
     =\frac{1}{2} \mr{Cut}(\Sigma^{(k)}, \overline{\Sigma}^{(k)}).
\end{align*}
  \item \emph{When $v'_p \in \overline{\Sigma}^{(k)}$ and $v'_q \in \Sigma^{(k)}$:} 
In this case, $x_p^{(k)} = 0$ and $x_q^{(k)} = \alpha^{(k)}$. So,
\begin{align*}
  \tr{Cost}_{\DPE}^{(k)}(\mc{G})
    &=\frac{1}{2} 
      \sum_{v'_p \in \overline{\Sigma}^{(k)}}^M 
      \sum_{v'_q \in \Sigma^{(k)}}^M 
      |w'_{pq}|\, (0 + \mr{sgn}(w'_{pq})\alpha^{(k)})^2 \\
    &=\frac{1}{2}\, {\alpha^{(k)}}^2 
      \sum_{v'_p \in \overline{\Sigma}^{(k)}}^M 
      \sum_{v'_q \in \Sigma^{(k)}}^M 
      |w'_{pq}| 
     =\frac{1}{2}\, \mr{Cut}(\overline{\Sigma}^{(k)}, \Sigma^{(k)}).
\end{align*}
\end{enumerate}
Noting that $\mr{Cut}(\Sigma^{(k)}, \overline{\Sigma}^{(k)}) = \mr{Cut}(\overline{\Sigma}^{(k)}, \Sigma^{(k)})$, combine the above costs for all possible cases to obtain
\[
  \tr{Cost}_{\DPE}^{(k)}(\mc{G}) 
    =\ul{x}^{(k)^T} \ul{L}_{\DPE}\, \ul{x}^{(k)} = {\alpha^{(k)}}^2
     \left[
       \mr{Cut}(\Sigma^{(k)}, \overline{\Sigma}^{(k)}) 
         + 2\, \mr{Links}^{+}(\Sigma^{(k)}, \Sigma^{(k)})
     \right].
\]


\paragraph{RGE Method}

The cost associated with the RGE method is
\[
  \tr{Cost}_{\RGE}^{(k)}(\mc{G}) =\ul{x}^{(k)^T} \ul{L}_{\RGE}\, \ul{x}^{(k)} 
    =\frac{1}{2} 
     \sum_{p=1}^M \sum_{q=1}^M 
     |w'_{pq}|\, (x_p^{(k)} - x_q^{(k)})^2.
\]
We now consider the following cases:
\begin{enumerate}
    \item \emph{When $v'_p, v'_q \in \Sigma^{(k)}$:} 
In this case, we have $x_p^{(k)} = x_q^{(k)} = \alpha^{(k)}$. So, 
\[
  \tr{Cost}_{\RGE}^{(k)}(\mc{G})
    =\frac{1}{2} 
     \sum_{v'_p \in \Sigma^{(k)}}^M 
     \sum_{v'_q \in \Sigma^{(k)}}^M 
     |w'_{pq}|\, (\alpha^{(k)} - \alpha^{(k)})^2 
    =0.
\]
Thus, no clustering cost is incurred.
\item \emph{When $v'_p, v'_q \in \overline{\Sigma}^{(k)}$:} 
In this case, $x_p^{(k)} = x_q^{(k)} = 0$, and no clustering cost is incurred.
  \item \emph{When $v'_p \in \Sigma^{(k)}$ and $v'_q \in \overline{\Sigma}^{(k)}$:} 
In this case, $x_p^{(k)} = \alpha^{(k)}$ and $x_q^{(k)} = 0$. So,
\begin{align*}
  \tr{Cost}_{\RGE}^{(k)}(\mc{G})
    &=\frac{1}{2} 
      \sum_{v'_p \in \Sigma^{(k)}}^M 
      \sum_{v'_q \in \overline{\Sigma}^{(k)}}^M 
      |w'_{pq}|\, (\alpha^{(k)} - 0)^2 \\
    &=\frac{1}{2}\, {\alpha^{(k)}}^2 
      \sum_{v'_p \in \Sigma^{(k)}}^M 
      \sum_{v'_q \in \overline{\Sigma}^{(k)}}^M 
      |w'_{pq}| 
     =\frac{1}{2} \mr{Cut}(\Sigma^{(k)}, \overline{\Sigma}^{(k)}).
\end{align*}
  \item \emph{When $v'_p \in \overline{\Sigma}^{(k)}$ and $v'_q \in \Sigma^{(k)}$:} 
In this case, $x_p^{(k)} = 0$ and $x_q^{(k)} = \alpha^{(k)}$. So,
\begin{align*}
  \tr{Cost}_{\RGE}^{(k)}(\mc{G})
    &=\frac{1}{2} 
      \sum_{v'_p \in \overline{\Sigma}^{(k)}}^M 
      \sum_{v'_q \in \Sigma^{(k)}}^M 
      |w'_{pq}|\, (0 - \alpha^{(k)})^2 \\
    &=\frac{1}{2}\, {\alpha^{(k)}}^2 
      \sum_{v'_p \in \overline{\Sigma}^{(k)}}^M 
      \sum_{v'_q \in \Sigma^{(k)}}^M 
      |w'_{pq}| 
     =\frac{1}{2}\, \mr{Cut}(\overline{\Sigma}^{(k)}, \Sigma^{(k)}).
\end{align*}
\end{enumerate}
Noting that $\mr{Cut}(\Sigma^{(k)}, \overline{\Sigma}^{(k)}) = \mr{Cut}(\overline{\Sigma}^{(k)}, \Sigma^{(k)})$, combine the above costs for all possible cases to obtain
\[
  \tr{Cost}_{\RGE}^{(k)}(\mc{G}) 
    =\ul{x}^{(k)^T} \ul{L}_{\RGE}\, \ul{x}^{(k)} = {\alpha^{(k)}}^2
     \left[
       \mr{Cut}(\Sigma^{(k)}, \overline{\Sigma}^{(k)}) 
     \right].
\]
This completes the proof.
\hspace{\fill}{Q.E.D.}


\subsection{Proof of Lemma~\ref{lem:EdgeVol}}


We will employ the following notation: 
\[
  \langle d'_{out, \ell_p}\rangle
    =\langle d_{out, \ell_p}\rangle - |w_p|;\;\;
  \langle d'_{in, k_p}\rangle 
    =\langle d_{in, k_p}\rangle - |w_p|.
\]
So, $\langle d'_{out, \ell_p}\rangle$ and $\langle d'_{in, k_p}\rangle$ take into account the absolute sum of the weights of the edges exiting vertex $v_{\ell_p}$ and the edges entering vertex $v_{k_p}$, respectively, except that they ignore the weight of the edge $e_p$. Similarly, we also use the notation
\[
  \langle d'_{\ell_p}\rangle  
    =\langle d_{\ell_p}\rangle - |w_p| 
    =\langle d'_{out, \ell_p}\rangle + \langle d_{in, \ell_p}\rangle;\;\;
  \langle d'_{k_p}\rangle 
    =\langle d_{k_p}\rangle - |w_p| 
    =\langle d'_{in, k_p}\rangle + \langle d_{out, k_p}\rangle.
\]
So, $\langle d'_{\ell_p}\rangle$ and $\langle d'_{k_p}\rangle$ denote the sums of the absolute values of weights of all edge that are connected to $v_{\ell_p}$ and $v_{k_p}$, respectively, except that they ignore the edge weight of $e_p$. 


We note that, with the vertex vector $\ul{\nu} = \{\nu_i\}$ chosen as in \eqref{eq:nuvectorchoice}, we have
\begin{equation}
  \nu_{\ell_p}
    =\frac{|w_p| + \langle d'_{out, \ell_p}\rangle + \langle d_{in, \ell_p}\rangle}
          {\Vert \ul{w}\Vert_1};\;\;
  \nu_{k_p}
    =\frac{|w_p| + \langle d_{out, k_p}\rangle + \langle d'_{in, k_p}\rangle}
          {\Vert \ul{w}\Vert_1}.
  \label{eq:nui_example}
\end{equation}

To get the expression for $f_p$, simply substitute \eqref{eq:nui_example} in Definition~\ref{def:VolEdge}(i) to yield
\begin{align}
  f_p
    &=\frac{\sigma_{\ell_p}}{2}
      \cdot
      \frac{|w_p|}{\Vert\ul{w}\Vert_1}
      \cdot
      \left(
        1 + \frac{\langle d_{in, \ell_p}\rangle}{|w_p| + \langle d'_{out, \ell_p}\rangle}
      \right)
        +\frac{\sigma_{k_p}}{2}
         \cdot
         \frac{|w_p|}{\Vert\ul{w}\Vert_1}
         \cdot
         \left(
           1 + \frac{\langle d_{out, k_p}\rangle}{|w_p| + \langle d'_{in, k_p}\rangle}
         \right)
      \notag \\
    &=\frac{\sigma_{\ell_p}}{2}
      \cdot
      \frac{|w_p|}{\Vert\ul{w}\Vert_1}
      \cdot
      \left(
        \frac{|w_p| + \langle d'_{\ell_p}\rangle}{|w_p| + \langle d'_{out, \ell_p}\rangle}
      \right)
        +\frac{\sigma_{k_p}}{2}
         \cdot
         \frac{|w_p|}{\Vert\ul{w}\Vert_1}
         \cdot
         \left(
           \frac{|w_p| + \langle d'_{k_p}\rangle}{|w_p| + \langle d'_{in, k_p}\rangle}
         \right) \notag \\
    &=\frac{\sigma_{\ell_p}}{2}
      \cdot
      \frac{|w_p|}{\Vert\ul{w}\Vert_1}
      \cdot
      \left(
        \frac{\langle d_{\ell_p}\rangle}{\langle d_{out, \ell_p}\rangle}
      \right)
        +\frac{\sigma_{k_p}}{2}
         \cdot
         \frac{|w_p|}{\Vert\ul{w}\Vert_1}
         \cdot
         \left(
           \frac{\langle d_{k_p}\rangle}{\langle d_{in, k_p}\rangle}
         \right).
  \label{eq:fp1}
\end{align} 
\begin{itemize}
  \item[(i)] Variation of $f_p$ w.r.t. $|w_p|$ (in order to show that $f_p$ is monotonically increasing with increasing $|w_p|$): From \eqref{eq:fp1}, we get 
\begin{align}
  &\frac{\partial f_p}{\partial |w_p|} 
   \notag \\
  &\;\;
     =\frac{\sigma_{\ell_p}}{2}
      \left[
        \frac{\partial}{\partial |w_p|}
        \left(
          \frac{|w_p|}{\Vert\ul{w}\Vert_1}
        \right)
        \cdot
        \left(
          \frac{|w_p| + \langle d'_{\ell_p}\rangle}{|w_p| + \langle d'_{out, \ell_p}\rangle}
        \right)
          +\frac{|w_p|}{\Vert\ul{w}\Vert_1}
           \cdot
           \frac{\partial}{\partial |w_p|}
           \left(
             \frac{|w_p| + \langle d'_{\ell_p}\rangle}{|w_p| + \langle d'_{out, \ell_p}\rangle}
           \right)
      \right] 
      \notag \\
  &\qquad
        +\frac{\sigma_{k_p}}{2}
         \left[
           \frac{\partial}{\partial |w_p|}
           \left(
             \frac{|w_p|}{\Vert\ul{w}\Vert_1}
           \right)
         \cdot
         \left(
           \frac{|w_p| + \langle d'_{k_p}\rangle}{|w_p| + \langle d'_{in, k_p}\rangle}
         \right)
           +\frac{|w_p|}{\Vert\ul{w}\Vert_1}
            \cdot
            \frac{\partial}{\partial |w_p|}
            \left(
              \frac{|w_p| + \langle d'_{k_p}\rangle}{|w_p| + \langle d'_{in, k_p}\rangle}
            \right)
         \right].
  \label{eq:fp1_diff}
\end{align}
Note that 
\[
  \frac{\partial}{\partial |w_p|}
  \left(
    \frac{|w_p|}{\Vert\ul{w}\Vert_1}
  \right)
    =\frac{\partial}{\partial |w_p|}
     \frac{|w_p|}{\sum_{q=1}^M |w_q|}
    =\frac{\Vert\ul{w}'\Vert_1}{\Vert\ul{w}\Vert_1^2},
\]
where $\Vert \ul{w}'\Vert_1 = \Vert \ul{w}\Vert_1 - |w_p| = \displaystyle \sum_{q \neq p} |w_q|$, and
\begin{alignat*}{2}
  &\frac{\partial}{\partial |w_p|}
   \left(
     \frac{|w_p| + \langle d'_{\ell_p}\rangle}{|w_p| + \langle d'_{out, \ell_p}\rangle}
   \right)
     &
       &=-\frac{\langle d_{in, \ell_p}\rangle}{(|w_p| + \langle d'_{out, \ell_p}\rangle)^2}; \\
  &\frac{\partial}{\partial |w_p|}
   \left(
     \frac{|w_p| + \langle d'_{k_p}\rangle}{|w_p| + \langle d'_{in, k_p}\rangle}
   \right)
    &
      &=-\frac{\langle d_{out, k_p}\rangle}{(|w_p| + \langle d'_{in, k_p}\rangle)^2}.
\end{alignat*}
Substitute in \eqref{eq:fp1_diff}:
\begin{align}
  \frac{\partial f_p}{\partial |w_p|} 
    &=\frac{\sigma_{\ell_p}}{2}
      \left[
        \frac{\Vert\ul{w}'\Vert_1}{\Vert\ul{w}\Vert_1^2}
        \left(
          \frac{|w_p| + \langle d'_{\ell_p}\rangle}{|w_p| + \langle d'_{out, \ell_p}\rangle}
        \right)        
          -\frac{|w_p|}{\Vert\ul{w}\Vert_1} 
           \cdot
           \frac{\langle d_{in, \ell_p}\rangle}{(|w_p| + \langle d'_{out, \ell_p}\rangle)^2}
      \right] 
      \notag \\
    &\qquad
        +\frac{\sigma_{k_p}}{2}
         \left[
           \frac{\Vert\ul{w}'\Vert_1}{\Vert\ul{w}\Vert_1^2}
           \left(
             \frac{|w_p| + \langle d'_{k_p}\rangle}{|w_p| + \langle d'_{in, k_p}\rangle}
           \right)        
             -\frac{|w_p|}{\Vert\ul{w}\Vert_1} 
              \cdot
              \frac{\langle d_{out, k_p}\rangle}{(|w_p| + \langle d'_{in, k_p}\rangle)^2}
      \right].
  \label{eq:fp11_diff}
\end{align}
From \eqref{eq:fp11_diff}, we notice that
\[
  \lim_{|w_p| \to 0} \frac{\partial f_p}{\partial |w_p|}
    =\frac{\sigma_{\ell_p}}{2}
     \cdot
     \frac{1}{\Vert\ul{w}'\Vert_1}
     \cdot
     \frac{\langle d'_{\ell_p}\rangle}{\langle d'_{out, \ell_p}\rangle}
       +\frac{\sigma_{k_p}}{2}
        \cdot
        \frac{1}{\Vert\ul{w}'\Vert_1}
        \cdot
        \frac{\langle d'_{k_p}\rangle}{\langle d'_{in, k_p}\rangle};
  \quad
  \lim_{|w_p| \to \infty} \frac{\partial f_p}{\partial |w_p|}
    =0.        
\]

We next claim that the expression within the first square brackets of \eqref{eq:fp11_diff} is non-negative:   
\begin{align*}
  &\frac{\Vert\ul{w}'\Vert_1}{\Vert\ul{w}\Vert_1^2}
   \left(
     \frac{|w_p| + \langle d'_{\ell_p}\rangle}{|w_p| + \langle d'_{out, \ell_p}\rangle}
   \right)
     \geq 
      \frac{|w_p|}{\Vert\ul{w}\Vert_1} 
      \cdot
      \frac{\langle d_{in, \ell_p}\rangle}{(|w_p| + \langle d'_{out, \ell_p}\rangle)^2} \\
  &\qquad
   \iff
   \frac{\Vert\ul{w}\Vert_1 - |w_p|}{\Vert\ul{w}\Vert_1^2}
   \left(
     \frac{\langle d_{\ell_p}\rangle}{\langle d_{out, \ell_p}\rangle}
   \right)
     \geq
      \frac{|w_p|}{\Vert\ul{w}\Vert_1} 
      \cdot
      \frac{\langle d_{\ell_p}\rangle - \langle d_{out, \ell_p}\rangle}
           {\langle d_{out, \ell_p}\rangle^2} \\
  &\qquad
   \iff
   \left(
     1 - \frac{|w_p|}{\Vert\ul{w}\Vert_1}
   \right)
   \langle d_{\ell_p}\rangle
     \geq 
      |w_p|
      \left(
        \frac{\langle d_{\ell_p}\rangle}{\langle d_{out, \ell_p}\rangle} - 1
      \right) \\
  &\qquad
   \iff
   \langle d_{\ell_p}\rangle + |w_p|
     \geq 
      \frac{|w_p|}{\langle d_{out, \ell_p}\rangle} \langle d_{\ell_p}\rangle 
        + \frac{\langle d_{\ell_p}\rangle}{\Vert\ul{w}\Vert_1} |w_p|,
\end{align*}
which is true because $|w_p| \leq \langle d_{out, \ell_p}\rangle$ and $\langle d_{\ell_p}\rangle \leq \Vert\ul{w}\Vert_1$. Similarly, the expression within the second square brackets of \eqref{eq:fp11_diff} aisre non-negative because $|w_p| \leq \langle d_{in, k_p}\rangle$ and $\langle d_{k_p}\rangle \leq \Vert\ul{w}\Vert_1$. Thus, $\partial f_p/\partial |w_p| \geq 0$, meaning that $f_p$ is monotonically increasing w.r.t. $|w_p|$.

  \item[(ii)] Variation of $f_p$ w.r.t. $\langle d_{out, \ell_p}'\rangle$ and $\langle d_{in, k_p}'\rangle$ (in order to show that $f_p$ is monotonically decreasing with increasing $\langle d_{out, \ell_p}'\rangle$ and/or $\langle d_{in, k_p}'\rangle$): We can express $f_p$ in \eqref{eq:fp1} as 
\begin{align*}
  f_p
    &= \frac{\sigma_{\ell_p}}{2}
       \cdot
       \frac{|w_p|}{\langle d'_{out, \ell_p}\rangle + \Delta'_{\ell_p}}
       \cdot
       \left(
         1 + \frac{\langle d_{in, \ell_p}\rangle}
                  {|w_p| + \langle d'_{out, \ell_p}\rangle}
       \right) \\
    &\qquad\qquad
         + \frac{\sigma_{k_p}}{2}
           \cdot
           \frac{|w_p|}{\langle d'_{in, k_p}\rangle + \Delta'_{k_p}}
           \cdot
           \left(
             1 + \frac{\langle d_{out, k_p}\rangle}
                      {|w_p| + \langle d'_{in, k_p}\rangle}
           \right),
\end{align*}
where $\Delta'_{\ell_p} = \Vert\ul{w}\Vert_1 - \langle d'_{out, \ell_p}\rangle$ and $\Delta'_{k_p} = \Vert\ul{w}\Vert_1 - \langle d'_{in, k_p}\rangle$. Now it is straightforward to show that $\partial f_p/\partial \langle d'_{out, \ell_p}\rangle \leq 0$ and $\partial f_p/\partial \langle d'_{in, k_p}\rangle \leq 0$, meaning that $f_p$ is monotonically decreasing w.r.t. $\langle d'_{out, \ell_p}\rangle$ and $\langle d'_{in, k_p}\rangle$.

  \item[(iii)] Variation of $f_p$ w.r.t. $\langle d_{in, \ell_p}\rangle$ and $\langle d_{out, k_p}\rangle$ (in order to show that $f_p$ is monotonically increasing with increasing $\langle d_{in, \ell_p}\rangle$ and/or $\langle d_{out, k_p}\rangle$): We can express $f_p$ in \eqref{eq:fp1} as
\begin{align*}
  f_p
    &= \frac{\sigma_{\ell_p}}{2}
       \cdot
       \frac{|w_p|}{\langle d_{in, \ell_p}\rangle + \Delta''_{\ell_p}}
       \cdot
       \left(
         1 + \frac{\langle d_{in, \ell_p}\rangle}
                  {|w_p| + \langle d'_{out, \ell_p}\rangle}
       \right) \\
    &\qquad\qquad
         + \frac{\sigma_{k_p}}{2}
           \cdot
           \frac{|w_p|}{\langle d_{out, k_p}\rangle + \Delta''_{k_p}}
           \cdot
           \left(
             1 + \frac{\langle d_{out, k_p}\rangle}
                      {|w_p| + \langle d'_{in, k_p}\rangle}
           \right),
\end{align*}
where $\Delta''_{\ell_p} = \Vert\ul{w}\Vert_1 - \langle d_{in, \ell_p}\rangle$ and $\Delta''_{k_p} = \Vert\ul{w}\Vert_1 - \langle d_{out, k_p}\rangle$. Now it is straightforward to show that $\partial f_p/\partial \langle d_{in, \ell_p}\rangle \geq 0$ and $\partial f_p/\partial \langle d_{out, k_p}\rangle  \geq 0$, meaning that $f_p$ is monotonically increasing w.r.t. $\langle d_{in, \ell_p}\rangle$ and $\langle d_{out, k_p}\rangle$.
\hspace{\fill}{Q.E.D.}
\end{itemize}


\subsection{Proof of Corollary~\ref{cor:EdgeVol_binary}}


This follows directly from the expression for $f_p$ in Lemma~\ref{lem:EdgeVol} when one substitutes $w_p = 1,\; \forall p = \{1, \ldots, M\}$. 


\subsection{Proof of Lemma~\ref{lem:LeEntries}}


Consider the $(p,q)$-th element of $\ul{L}_e$:
\[
  (\ul{L}_e)_{pq}
    =(\ul{B}^T \ul{\phi}\, \ul{B})_{pq}
    =\sum_{v_i\in\mc{V}} \phi_i\, (\ul{B}^T)_{pi}\, \ul{B}_{iq} 
    =\sum_{v_i\in\mc{V}} \phi_i\,
     (\delta_{i,\ell_p}-\delta_{i,k_p})
     (\delta_{i,\ell_q}-\delta_{i,k_q}).
\]
All terms within this summation are zero except when $v_i\in\{k_p, \ell_p, k_q, \ell_q\}$. So, 
\begin{multline*}
  (\ul{L}_e)_{pq}
    =\phi_{k_p}\,(\delta_{k_p,\ell_p}-1)\,(\delta_{k_p,\ell_q}-\delta_{k_p,k_q})
       +\phi_{\ell_p}\,(1-\delta_{\ell_p,k_p})\,(\delta_{\ell_p,\ell_q}-\delta_{\ell_p,k_q})
     \\
       +\phi_{k_q}\,(\delta_{k_q,\ell_p}-\delta_{k_q,k_p})\,(\delta_{k_q,\ell_q}-1)
       +\phi_{\ell_q}\,(\delta_{\ell_q,\ell_p}-\delta_{\ell_q,k_p})\,(1-\delta_{\ell_q,k_q}).
\end{multline*}
The claim now follows by direct substitution.
\hspace{\fill}{Q.E.D.}


\subsection{Proof of Lemma~\ref{lem:DualGraph}}


\begin{itemize}
  \item[\tb{(i)}]\label{item1} The fact that $\mc{G}'$ is an undirected graph with no self-edges is obvious because its adjacency matrix $\ul{W}'(\ul{\nu}) = \ul{L}_e(\ul{\nu}) - \mr{diag}[\mr{diag}[\ul{L}_e(\ul{\nu})]]$ is symmetric and has a zero diagonal. TO show that it is in general signed, consider two distinct dual vertices $v'_p, v_q',\, p\neq q$, in the dual graph $\mc{G}'$. There exists a dual edge $(v'_p, v_q')$ in the dual graph $\mc{G}'$ iff $(\ul{W}'(\ul{\nu}))_{pq} = (\ul{L}_e(\ul{\nu}))_{pq} \neq 0$. Since all elements of $\ul{\nu}$ are non-zero, this in turn occurs iff the two edges $e_p, e_q$ in the digraph $\mc{G}$ share a common vertex, say, $v_i \in \mc{V}$. Furthermore, the edge weight of the dual edge $(v'_p, v_q') \in \mc{E}'$ is $(\ul{W}'(\ul{\nu}))_{pq} = (\ul{L}_e(\ul{\nu}))_{pq}$ which assumes the value $+\nu_i$ if $e_p$ and $e_q$ are both poiting to or both pointing away from the shared vertex or the value $-\nu_i$ if $e_p$ and $e_q$ form a length-2 directed path through the shared vertex. Hence this dual graph $\mc{G}'$ is in general a signed graph in that its edge function may take on both signs. 
  \item[\tb{(ii)}] From (i) above, it is clear that a dual edge correponds to a pair of digraph edges that share a common vertex. Conversely, each pair of digraph edges that share a common vertex corresponds to a dual edge. So, the number of dual edges is equal to the number of distinct pairs of digraph edges each sharing a common vertex, i.e., $M'=\displaystyle\sum_{v_i\in\mc{V}} \begin{pmatrix} \sigma_i \\ 2 \end{pmatrix}$, where $\sigma_i$ denotes the social participation of vertex $v_i\in\mc{V}$. The fact that $\ul{w}' \cong \ul{W}'$, where $\ul{W}'(\ul{\nu}) = \ul{L}_e(\ul{\nu}) - \mr{diag}[\mr{diag}[\ul{L}_e(\ul{\nu})]$ constitutes the edge weight vector of the dual graph is obvious.
  \item[\tb{(iii)}] Each row (or column) of $\ul{W}'$ corresponds to a dual vertex because $\ul{W}'$ is the adjacency matrix of the dual graph; it also corresponds to a digraph edge because of the how $\ul{W}'$ is constructed from $\ul{L}_e$. So, the number of dual vertices is equal to the number of digraph edges, i.e., $N' = M$. 
\hspace{\fill}{Q.E.D.}
\end{itemize}


\paragraph*{Acknowledgments.}

This work is based on research supported by the University of Miami U-LINK Initiative on Interdisciplinary Inquiry and National Science Foundation Grant \#2123635. The advice of Mr. Chad Allie of The MATLAB Inc., Plano, Texas, in preparing the US migration data plots is greately appreciated.


\newpage


\setcounter{figure}{0}
\setcounter{table}{0}

\section*{Figures}


\begin{figure}[H]
  \centering
  \includegraphics[width=\textwidth]{%
    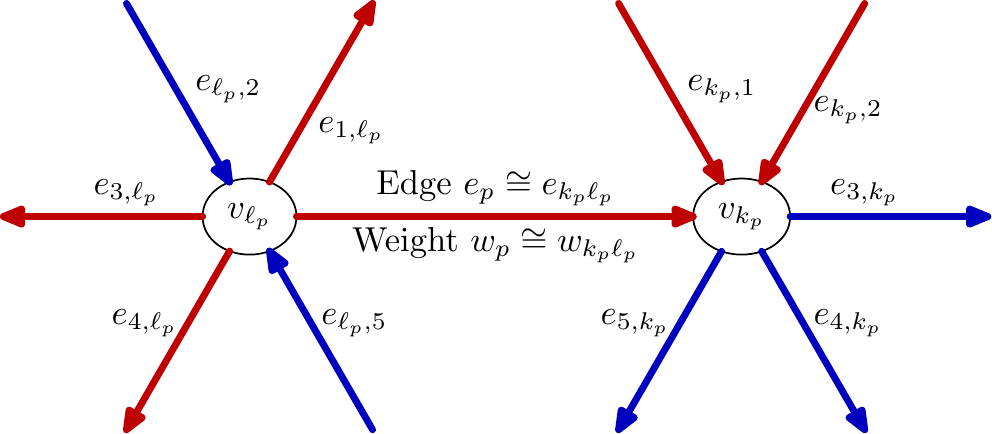}
  \caption{For edge $e_p\cong w_{k_p\ell_p}$, 
\begin{math}
  \langle d_{out, \ell_p}\rangle 
    = |w_p| + |w_{1, \ell_p}| + |w_{3, \ell_p}| + |w_{4, \ell_p}|
\end{math} 
and 
\begin{math}
  \langle d_{in, k_p}\rangle
    = |w_p| + |w_{k_p, 1}| + |w_{k_p, 2}|
\end{math}. 
These are shown in red. The volume of the edge $e_p$ is then taken as  
\begin{math}
  \mr{Vol}[e_p]
    =f_p 
    = 0.5 |w_p|
      \left(
        \sigma_{\ell_p} \nu_{\ell_p}/\langle d_{out, \ell_p}\rangle 
          + \sigma_{k_p} \nu_{k_p}/\langle d_{in, k_p}\rangle
      \right)
\end{math}. 
Thus, $f_p$ accounts for both 
\tb{(1)}~the combined importance of the two vertices $v_{\ell_p}$ and $v_{k_p}$ at the ends of the edge $e_p$; and 
\tb{(2)}~the magnitude of the edge weight $w_p$ and the directionality of the edge in the sense that it captures the role that the edge $e_p$ plays in conveying influence away from $v_{\ell_p}$ and in conveying influence toward $v_{k_p}$.} 
  \label{fig:h_p}
\end{figure}

\newpage

\begin{figure}[H]
  \centering
  \subfigure[%
    Factors affecting edge volume: Edge volume of edge $e_p \cong e_{k_p, \ell_p}$ depends on several factors: the absolute value $|w_p|$ of its weight, the role it plays in carrying the flow out of its starting vertex $v_{\ell_p}$, and the role it plays in carrying the flow into its termination vertex $v_{k_p}$. Note that $\langle d'_{out, \ell_p}\rangle
    =\langle d_{out, \ell_p}\rangle - |w_p|$ and that
  $\langle d'_{in, k_p}\rangle 
    =\langle d_{in, k_p}\rangle - |w_p|$]{%
    \label{fig:EdgeVolume_Factors}
    \includegraphics[width = 0.75\textwidth]{%
      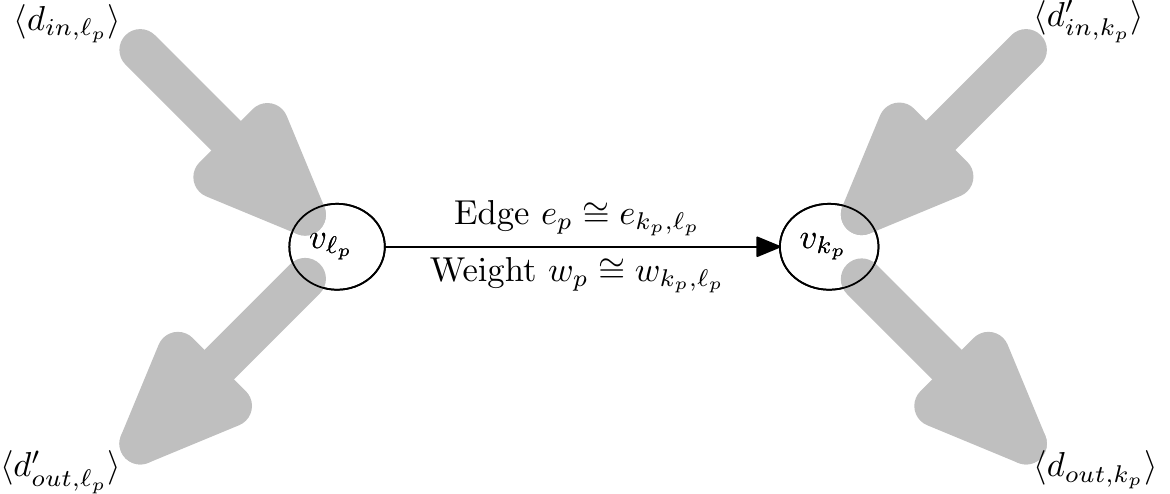}} \\
  \subfigure[%
    Minimum edge volume for binary graph: Edge volume attains its minimum when all edges of $v_{\ell_p}$ are directed out and all edges of $v_{k_p}$ are directed in.]{%
    \label{fig:EdgeVolume_Minimum}
    \includegraphics[width = 0.75\textwidth]{%
      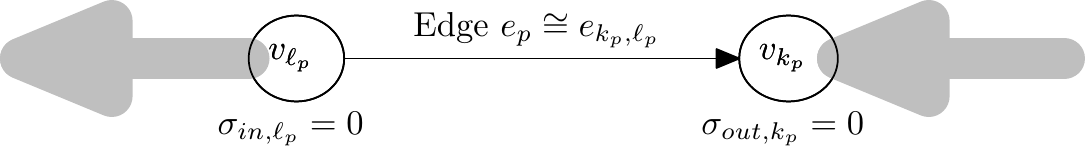}} \\
  \subfigure[%
    Maximum edge volume for binary graph: Edge volume attains its maximum when $\{v_{\ell_p}, v_{k_p}\}$ constitue the only vertices with $e_p$ being the only directed edge from $v_{\ell_p}$ to $v_{k_p}$ while all the remaining directed edges are from $v_{k_p}$ to $v_{\ell_p}$.]{%
    \label{fig:EdgeVolume_Maximum}
    \includegraphics[width = 0.75\textwidth]{%
      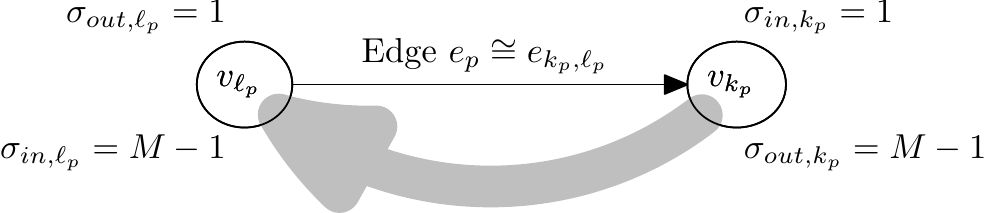}}
  \caption{%
  Edge volume of edge $e_p \cong e_{k_p, \ell_p}$.}
  \label{fig:EdgeVolume}
\end{figure}

\newpage

\begin{figure}[H]
  \centering
  \subfigure[%
    PRE clustering.]{%
    \label{fig:Benson_2Clusters_LPRE}
    \includegraphics[height = 0.325\textwidth, width = 0.31\textwidth]{%
      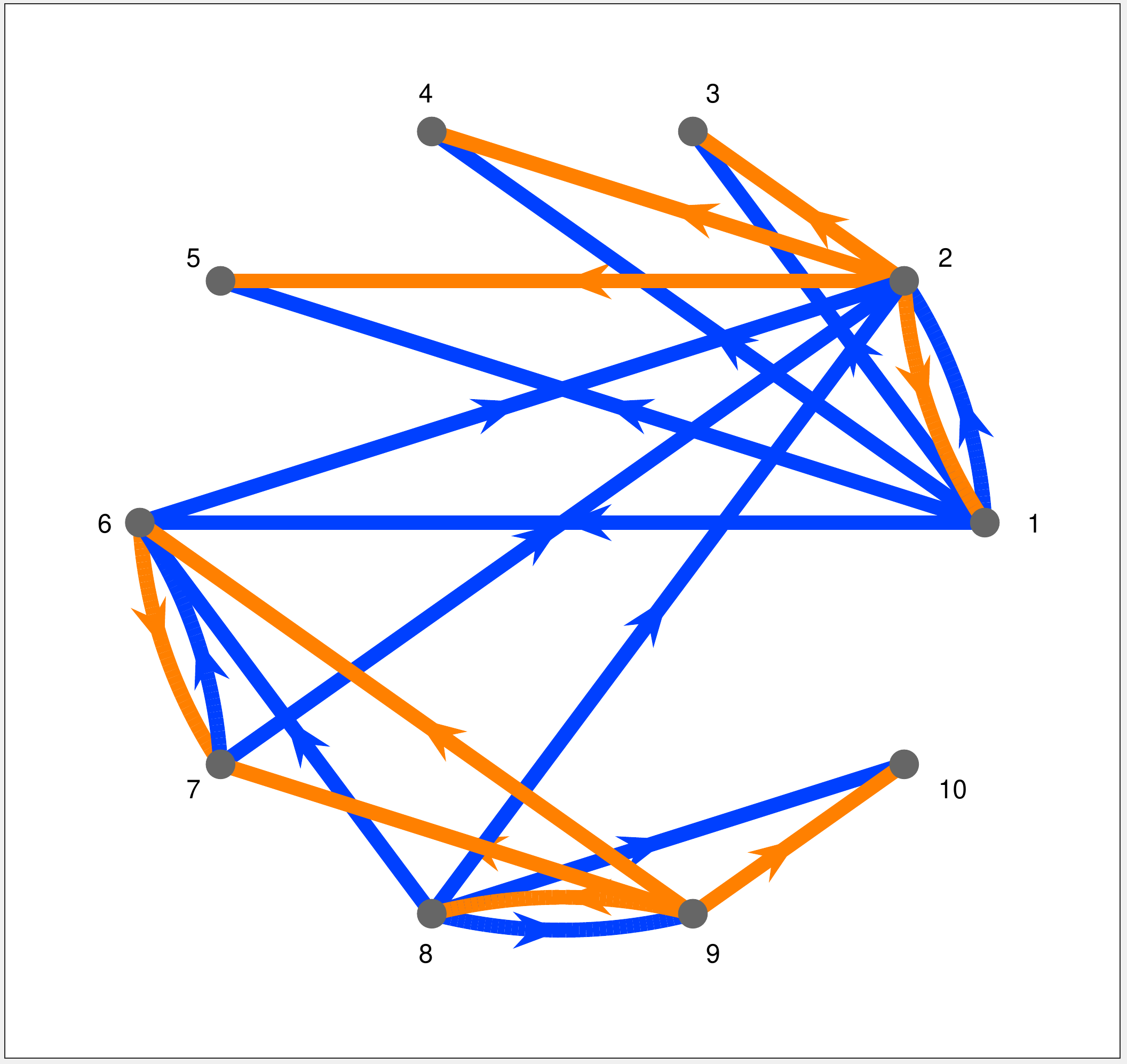}}
  \hfil
  \subfigure[%
    DPE clustering.]{%
    \label{fig:Benson_2Clusters_LDPE}
    \includegraphics[height = 0.325\textwidth, width = 0.31\textwidth]{%
      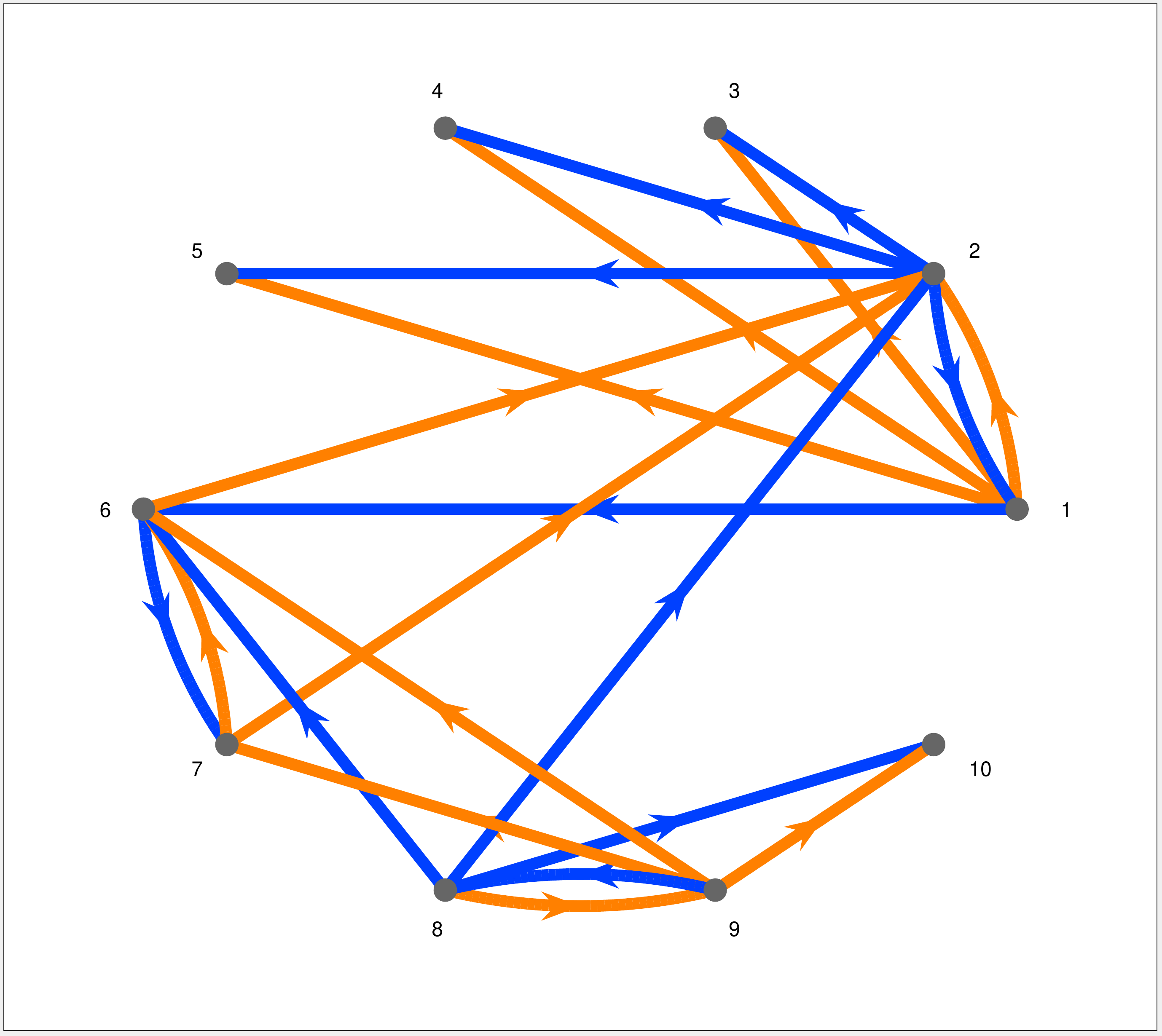}} 
  \hfil
  \subfigure[%
    RGE clustering.]{%
    \label{fig:Benson_2Clusters_LRGE}
    \includegraphics[height = 0.325\textwidth, width = 0.31\textwidth]{%
      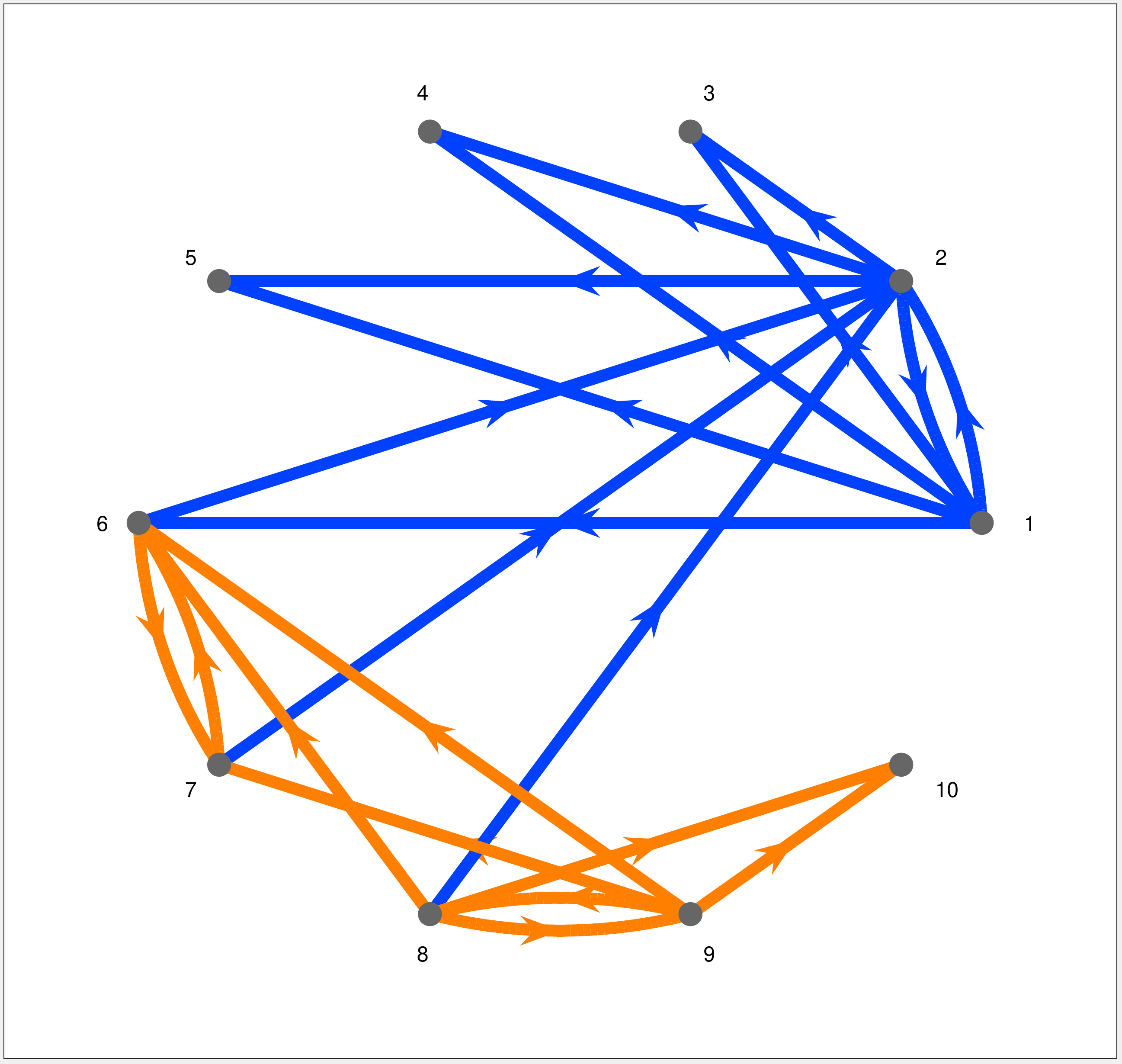}} \\
  \subfigure[%
    PRE clustering.]{%
    \label{fig:Lai_2Clusters_LPRE}
    \includegraphics[height = 0.325\textwidth, width = 0.31\textwidth]{%
      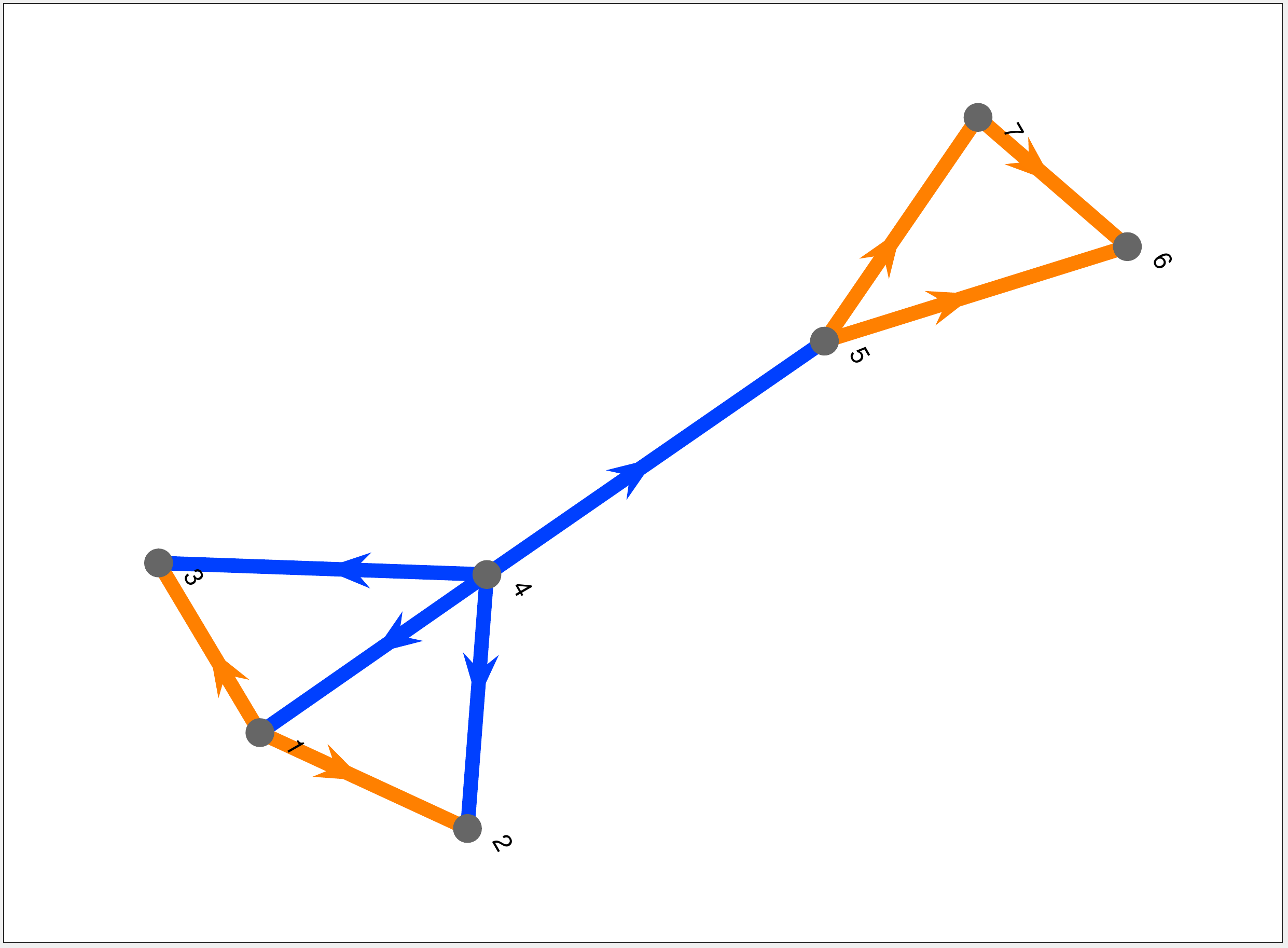}} 
  \hfil
  \subfigure[%
    DPE clustering.]{%
    \label{fig:Lai_2Clusters_LDPE}
    \includegraphics[height = 0.325\textwidth, width = 0.31\textwidth]{%
      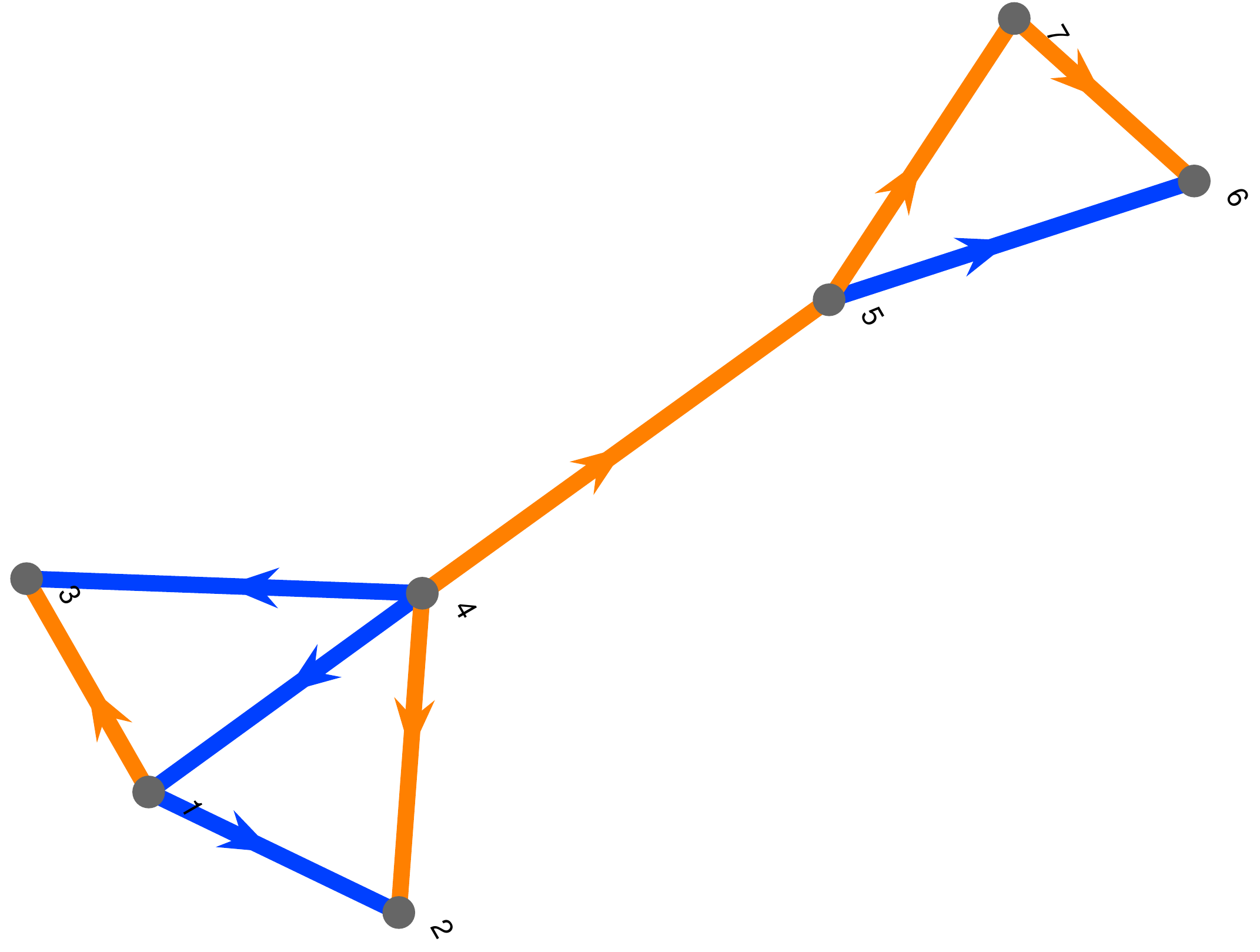}} 
  \hfil
  \subfigure[%
    RGE clustering.]{%
    \label{fig:Lai_2Clusters_LRGE}
    \includegraphics[height = 0.325\textwidth, width = 0.31\textwidth]{%
      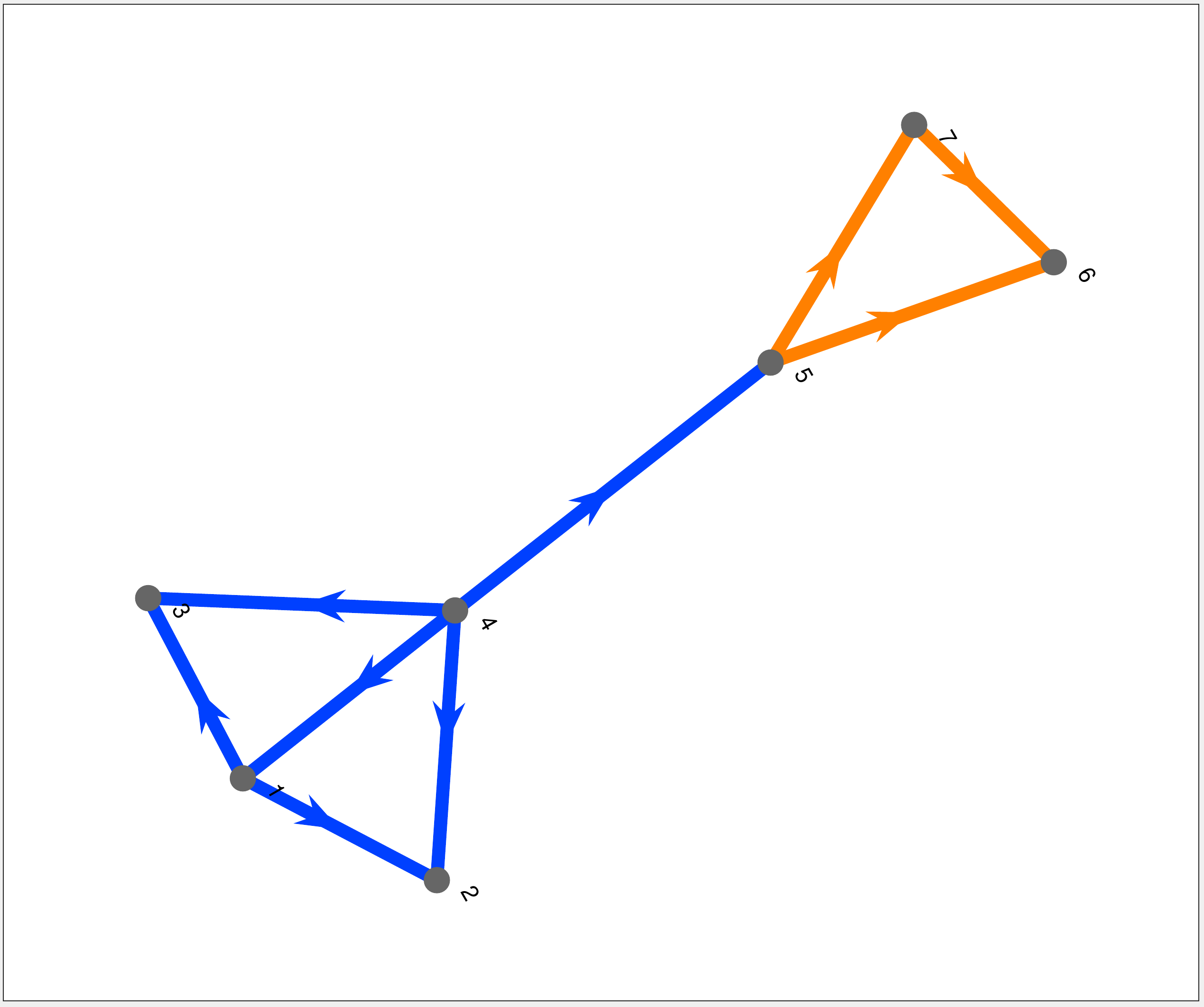}}
  \caption{%
  \tb{PRE, DPE, and RGE clustering of two synthetic unweighted digraphs} illustrate the different types of edge affinities that each algorithm extracts:  
  \tb{(a), (b), (c)} in the first row depicts the 10-vertex/21-edge digraph in \cite{Benson2016Science}; 
  \tb{(d), (e), (f)} in the second row depicts the 7-vertex/9-edge digraph in \cite{Lai2010PhysicaA}. 
  With PRE clustering in (a), for the blue cluster vertices 1 and 8 are producers while vertex 2 is a receptor; for the orange cluster vertices 2 and 9 are producers while vertex 1 is now a receptor. With PRE clustering in (d), for the blue cluster vertex 4 is a producer (it has the largest social participation); for the orange cluster, vertices 1 and 5 are producers while vertex 6 is a receptor. 
  With DPE clustering in (b), both the orange and blue clusters have length 4 directed paths, each involving vertex 2. With DPE clustering in (e), the orange cluster has a length 3 path while the blue cluster has a length 2 path, with both paths involving vertex 4 which has the largest social participation. 
  With RGE clustering in (c), vertices 6, 7, and 8 are coupling vertices between the two regions while in (f) the graph is partitioned into two regions with all of the edges of vertex 4, which has the largest social participation, in one region.}
  \label{fig:Cockroach_2Clusters_SocialScaling}
\end{figure}

\newpage

\begin{figure}[H]
  \centering
  \subfigure[PRE clustering.]{%
    \includegraphics[width=2.40in, height=2.10in]{%
      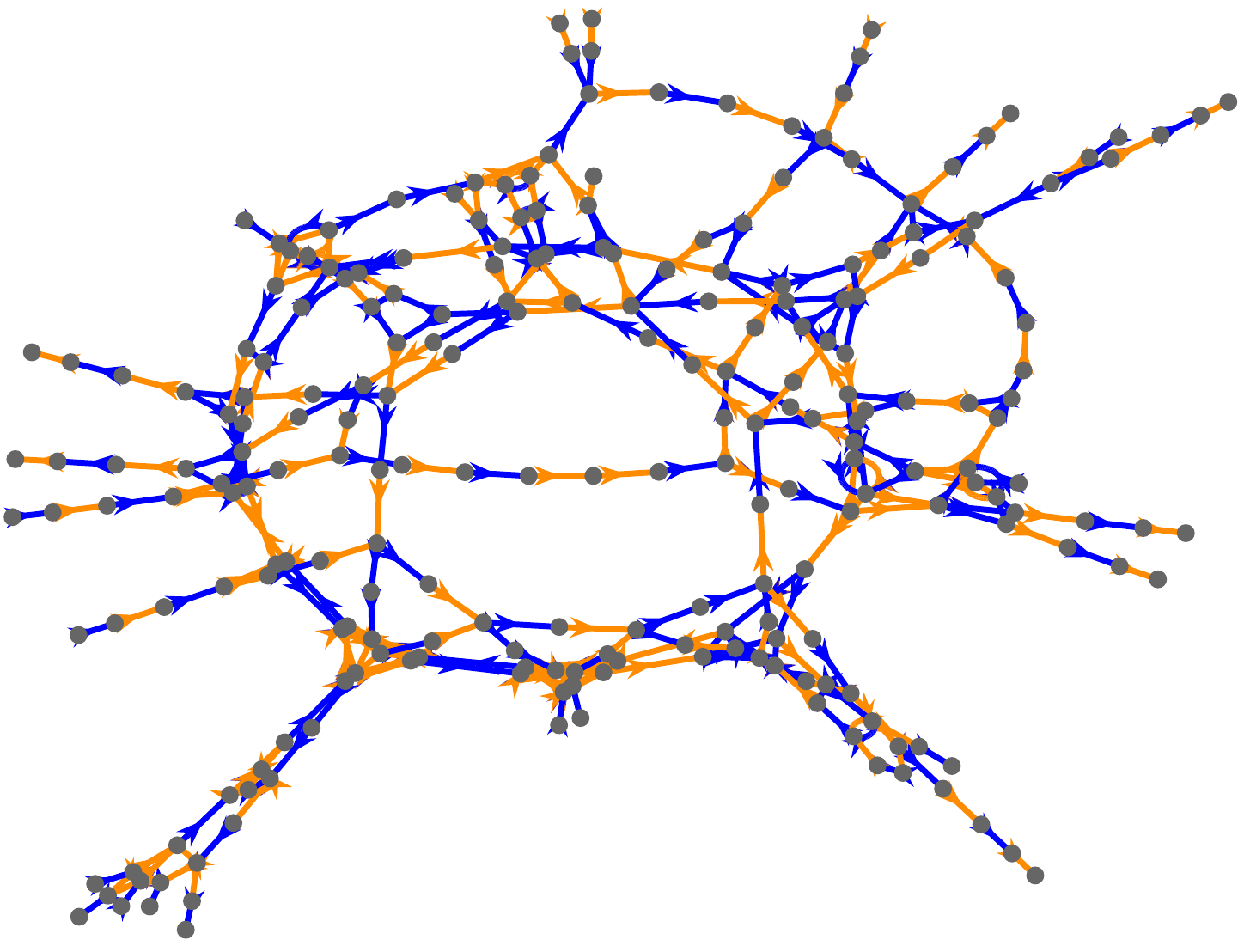}}
  \hfil
  \subfigure[PRE clustering overlaid on geo-located vertices.]{%
    \includegraphics[width=2.80in]{%
      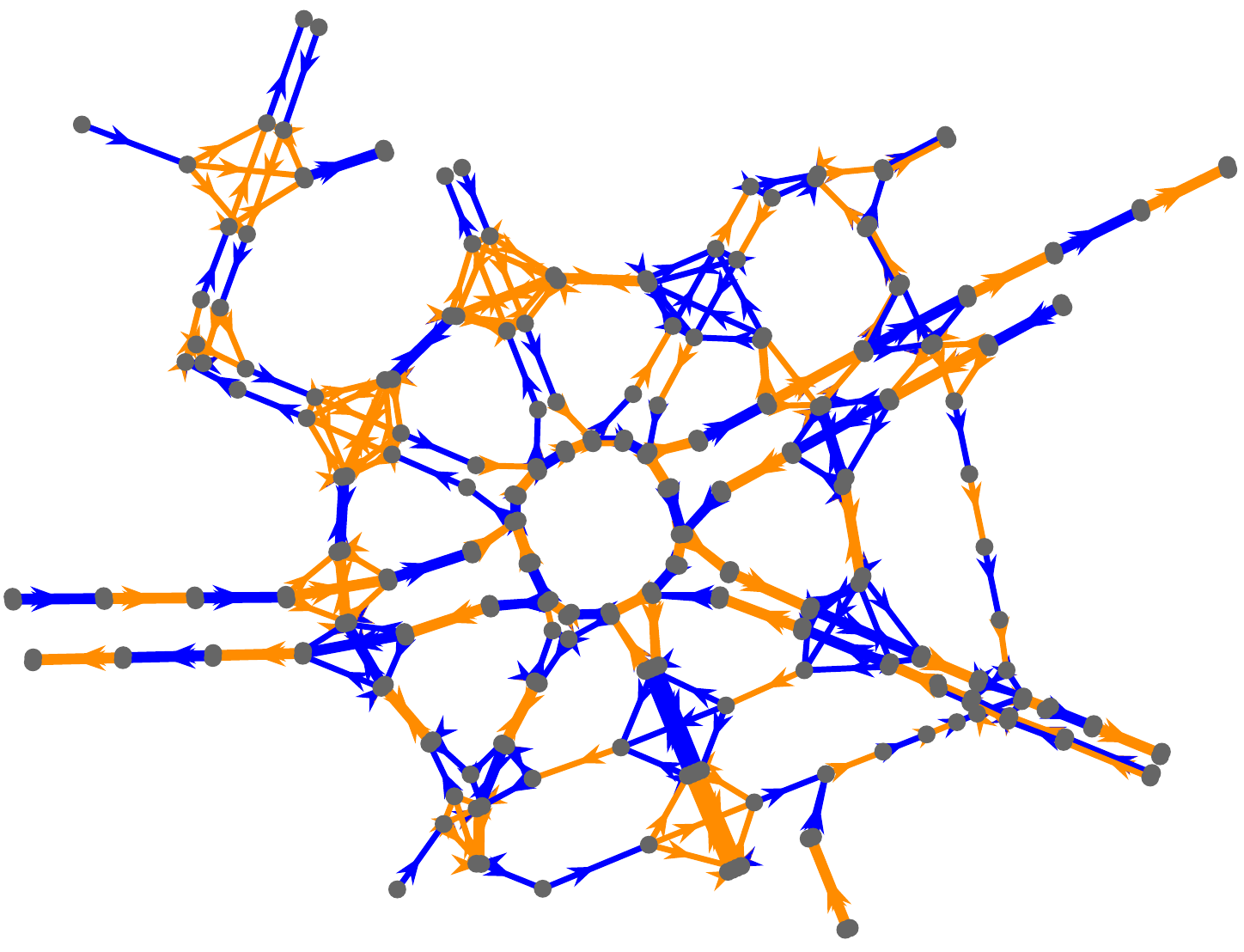}} \\   
  \subfigure[DPE clustering.]{%
    \includegraphics[width=2.40in, height=2.10in]{%
      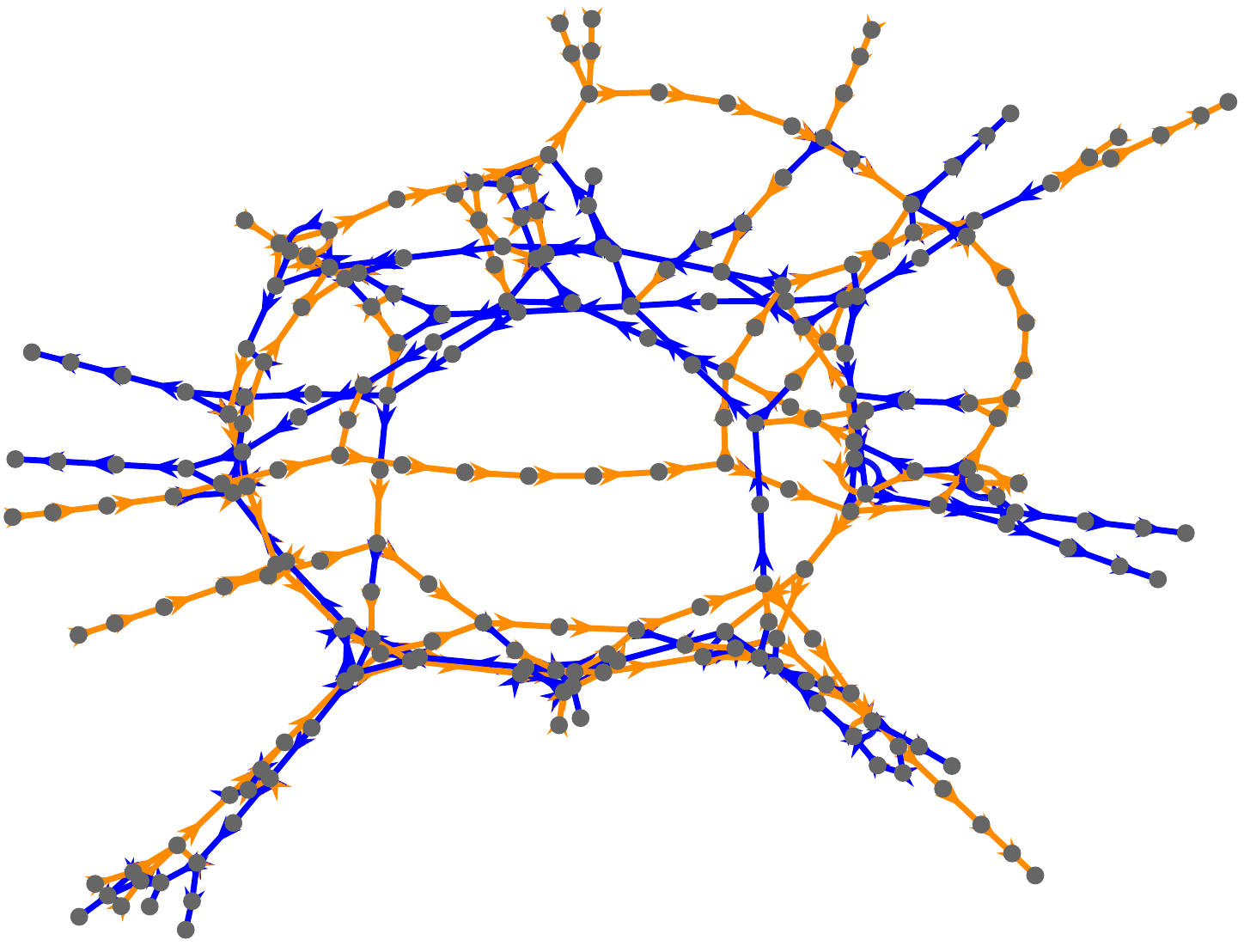}}
  \hfil
  \subfigure[DPE clustering overlaid on geo-located vertices.]{%
    \includegraphics[width=2.80in]{%
      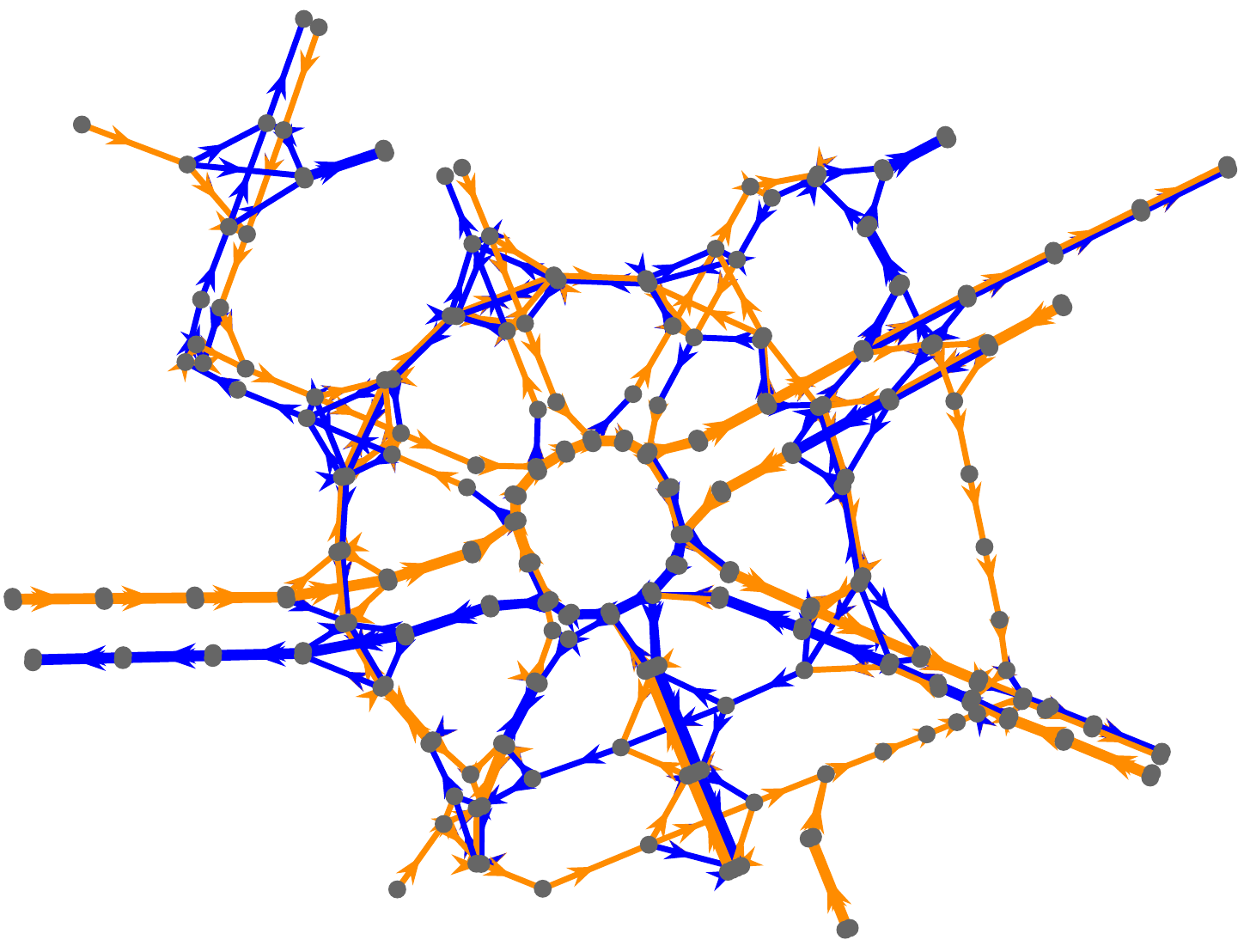}} \\    
  \subfigure[RGE clustering.]{%
    \includegraphics[width=2.40in, height=2.10in]{%
      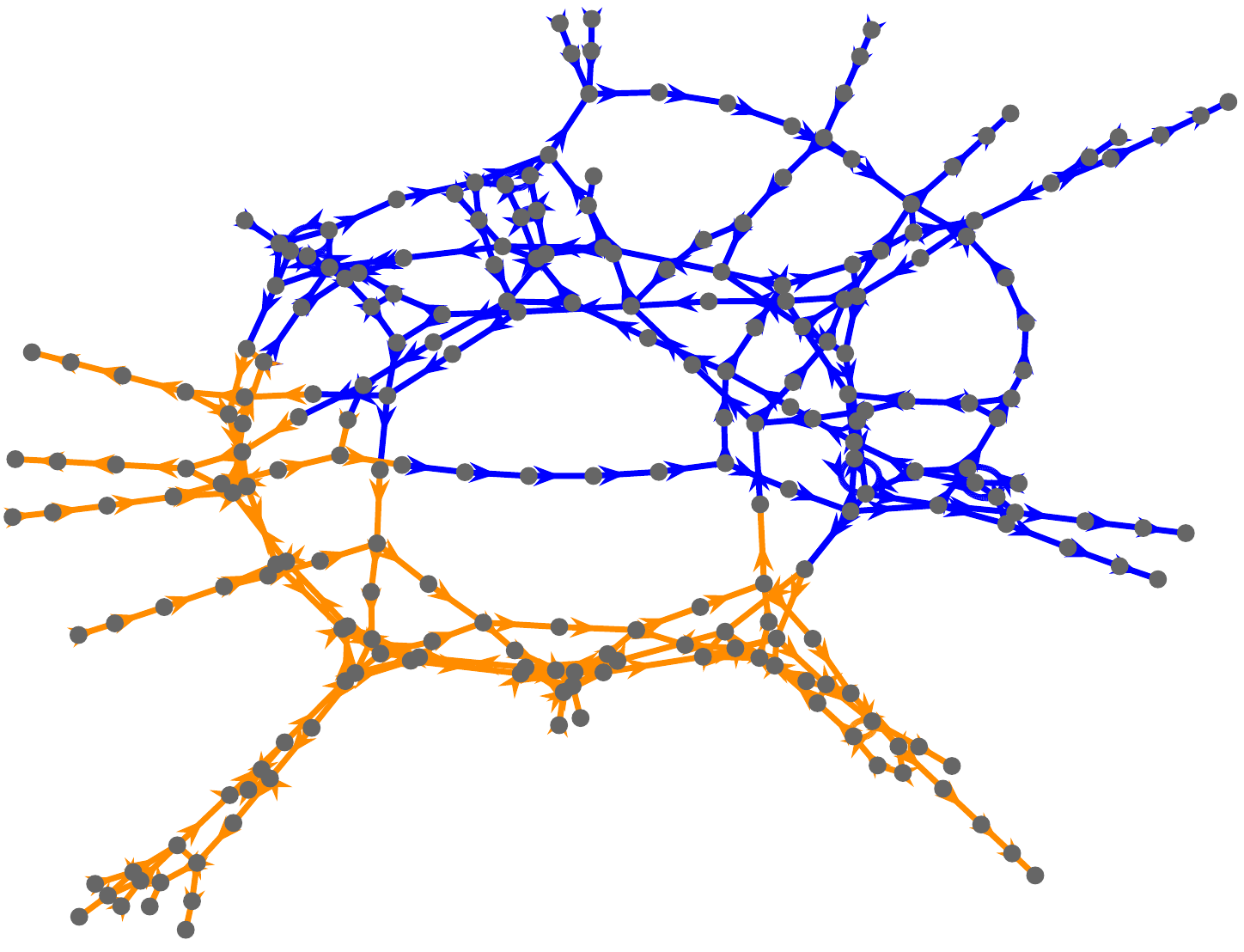}}
  \hfil
  \subfigure[RGE clustering overlaid on geo-located vertices.]{%
    \includegraphics[width=2.80in]{%
      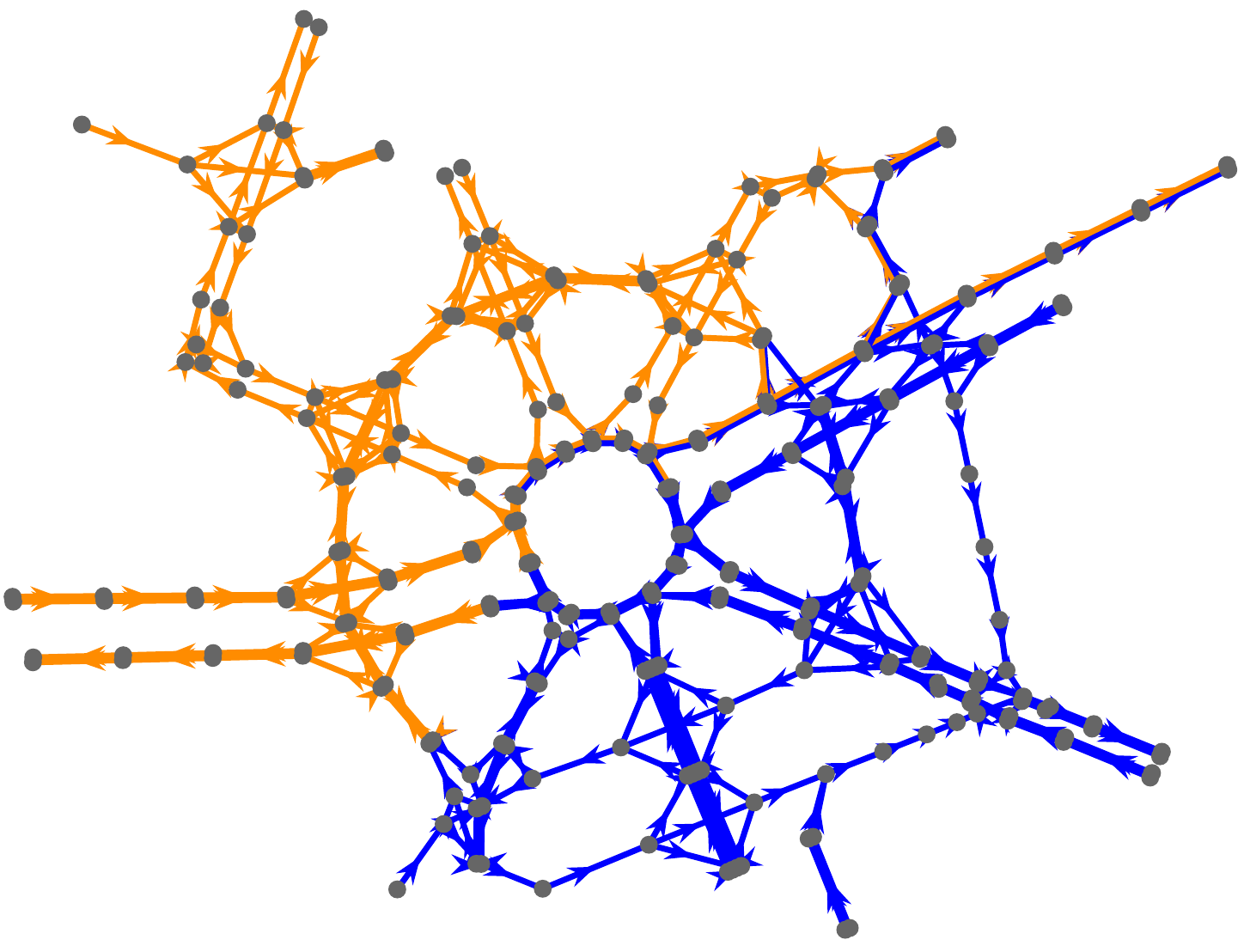}}
\end{figure}
\clearpage 
\captionof{figure}{%
\tb{Edge clustering ($K = 2$) of the road network in the vicinity of the Piazza Mazzini Square} in Rome, Italy, represented as a 236-vertex/349-edge unweighted digraph \cite{Sardellitti2017_arXiv}. Edge direction indicates traffic flow. Edge clustering yields different motifs involving intersections and roads. 
\tb{PRE clustering in (a),} which when overlaid on the vertex geographic coordinates as in \tb{(b)}, show how diamond-like edge cluster motifs of intersections having outgoing roads connected to intersections having incoming roads are being extracted. These diamonds indicate complex interchanges and points of concern for pedestrians.  Since directed paths are discouraged, segments of one way streets are in alternating groups.
\tb{DPE clustering in (c),} which when overlaid on the vertex geographic coordinates as in \tb{(d)}, shows how subgraphs of long directed roads where one may drive a long distance following the color associated with one cluster are being extracted.  These clusters show how to drive to the center, or avoid the center with long directed paths.  They also show potential bottlenecks. 
\tb{RGE clustering in (e),} which when overlaid on the vertex geographic coordinates as in \tb{(f)}, shows how the graph is being partitioned into $K=2$ distinct regions.}
\label{fig:PiazzaMazzini_2Clusters}

\newpage

\begin{figure}[H]
  \centering
  \includegraphics[width=\textwidth]{%
      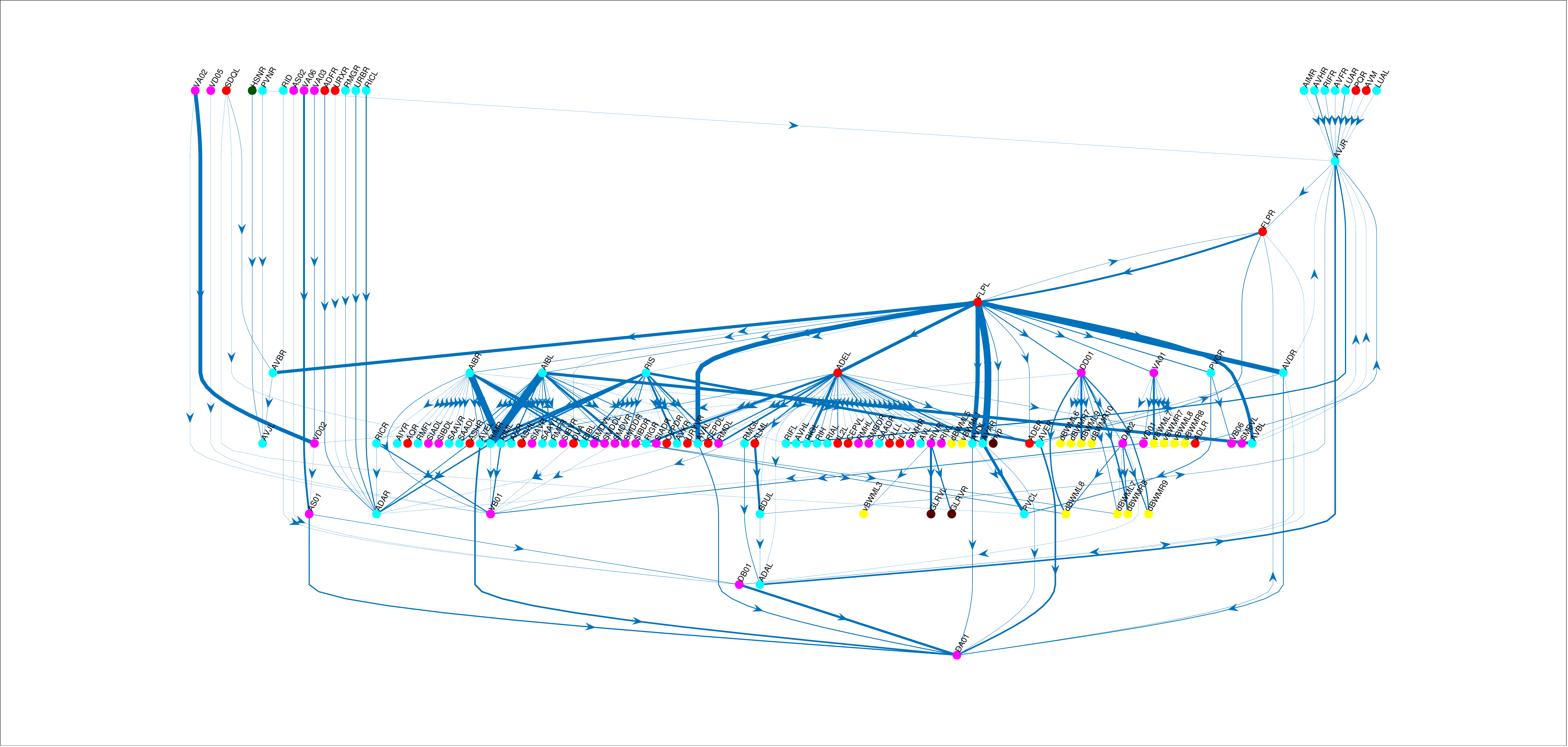}
  \caption{%
  \tb{DPE clustering ($K = 30$) of \emph{C elegans} hermaphrodite chemical synapse connectome} represented as a weighted digraph \cite{Cook2019Nature, WormWiring_Dataset}. Digraph vertices represent neurons; edges represent synaptic connections; edge thickness represents synaptic strength; and vertex colors represent the different functionalities of neurons. DPE clustering was performed using only the weighted adjacency matrix. The neurons' weighted degree is used as a proxy for the relative importances $\nu$; DPE clustering emphasizes clusters that contain direct and indirect paths linking neurons with larger $\nu$. The FLP sensory neurons associated with anterior harsh touch response only appear in two of the DPE clusters. The first one which shows the prominence of FLPR appears in Fig.~\ref{fig:CElegans_DPE_K30_N110_M183}; here we show the other DPE-based FLP cluster with 123 neurons and 198 synaptic edges that highlight directed paths prominently involving the FLPL sensory neuron. Similar to the other DPE clustering example with FLP neurons, this example subgraph also features circuit elements associated with anterior harsh touch response and backward locomotion. For instance, we see FLPL connecting with ADEL, ADER, and BDUL which are associated with anterior harsh touch response; FLPR is also indirectly connected as well 
\cite{Li2011NatureC}.  In addition, we see direct FLPL connections to the command neurons AVAL, AVAR, AVDL, AVDR, as well as FLPL connections to AIBR, AIBL, with subsequent connections to AVEL, RIMR, and RIML, and also inclusion of the motor neuron DA01 in the paths, all of which indicates circuits for backward locomotion \cite{Hallinen2021eLife}.  Note that DA01 feeds back to FLPR, possibly indicating that backward locomotion signaling is feeding back to the anterior harsh touch response sensing elements. These DPE subgraph circuits, that emphasize direct and indirect paths connecting neurons with large $\nu$, link widely varying neurons (sensory, inter, motor, muscle) and provide a topological basis for further speculative explorations of neuron and subcircuit functions.}
\label{fig:CElegans_DPE_K30_N123_M198}
\end{figure}

\newpage

\begin{figure}[H]
  \centering
  \includegraphics[width = \textwidth]{%
      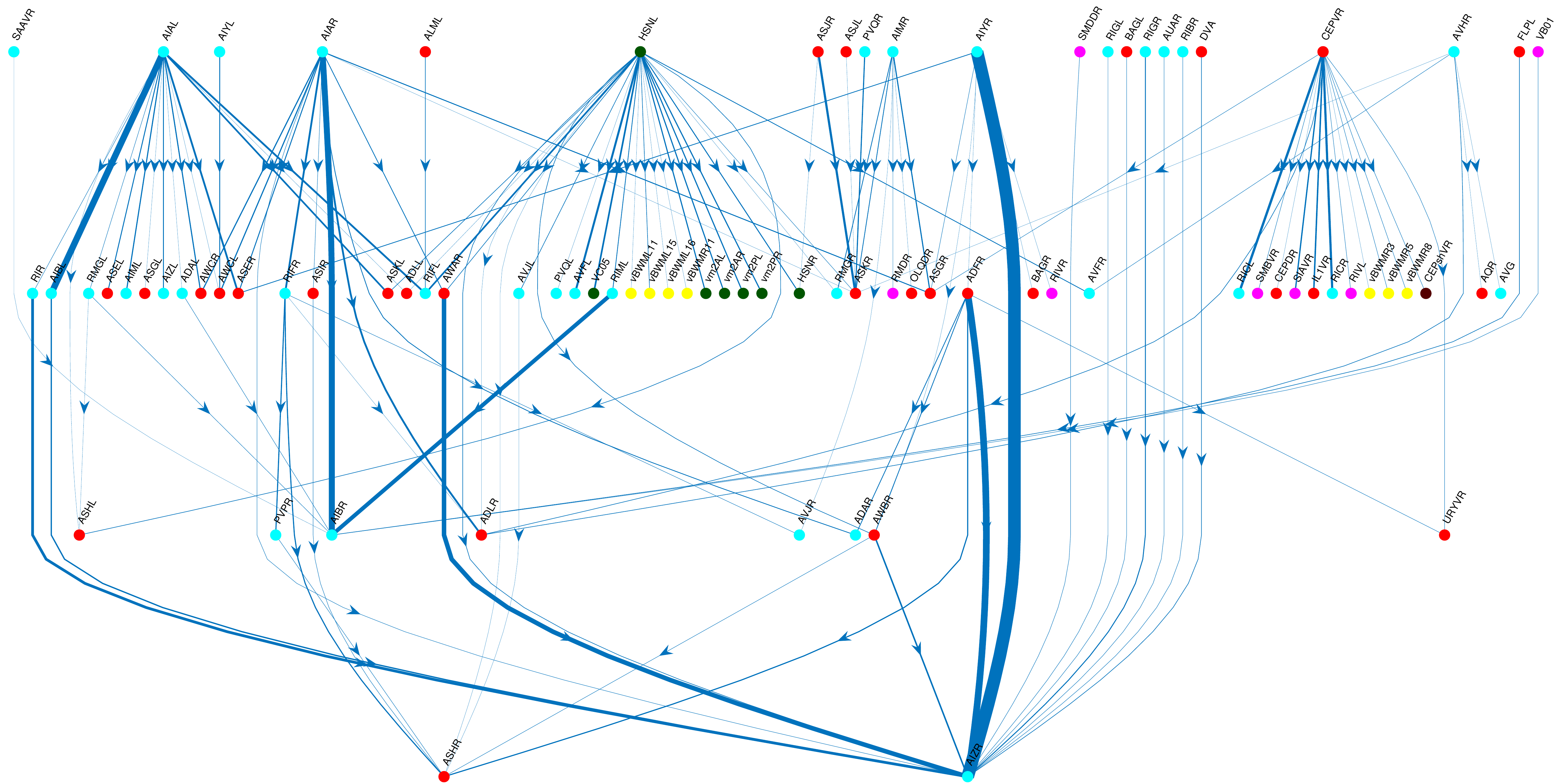}
  \caption{%
  \tb{PRE clustering ($K = 30$) of \emph{C elegans} hermaphrodite chemical synapse connectome} represented as a weighted digraph \cite{Cook2019Nature, WormWiring_Dataset}. Edge direction and thickness indicate the synaptic connection and strength, respectively. PRE clustering generates subgraphs where neurons (vertices) act either mainly as producers or receptors of synapses (directed edges). 
  Here we see a 85-vertex/122-edge PRE subgraph cluster featuring serotonin production and sensing as well as other elements of sensing and sensory integration and connections to egg laying. It shows interneuron AIZR, which is known to integrate sensory information, as a major receptor of synapses from: AIYR, a major interneuron; ADFR, the only serotonin sensory neuron in hermaphrodites; AWAR, a sensory neuron for chemotaxis; AWBR, an odor, electrical, and light sensory neuron; and RIR, a ring interneuron. The interneuron AIBR, which is also known to integrate information, is also a major receptor. AIB and AIZ are known to promote turns, and here AIBR, AIZR, AIBL and AIZL are all receptors of synapses, with AIBR, and AIZR being major receptors. AIA and AIY are known to inhibit turns; here AIAR, AIYR, AIAL, and AIYL are all producers of synapses, with AIYR, AIAR, and AIAL being major producers. Thus, we see the key neurons to which AIA and AIY direct synapses and the key neurons from which AIB and AIZ receive synapses, information that can potentially lead to better understanding of how turns are promoted and inhibited. We also see that the important sensory neuron ASKR as a major receptor, with inputs from: ASJ, a sensory neuron; PVQR, an interneuron; AIMR, a neuron possibly involved in serotonin regulation; CEPVR, a sensory neuron involved in mechanosensation; AIAR, an interneuron; and HSNL, a serotonin producer. We see HSNL as a major producer of synapses with major synapses to the VC05, and various vm2 and vbwm vertices associated with the egg-laying circuit \cite{Ravi2018JN, Collins2016eLife}. Interestingly, HSNL is connected to AWAR which is subsequently connected to AIZR. Perhaps AWAR, which is essential to sexual attraction in males, has additional functions in hermaphrodites given its connection to HSNL which is essential for egg laying and serotonin production, and to AIZR which integrates information.}
  \label{fig:CElegans_PRE_K30_N85_M122}
\end{figure}

\newpage

\begin{figure}[H]
  \centering
  \includegraphics[width = \textwidth]{%
      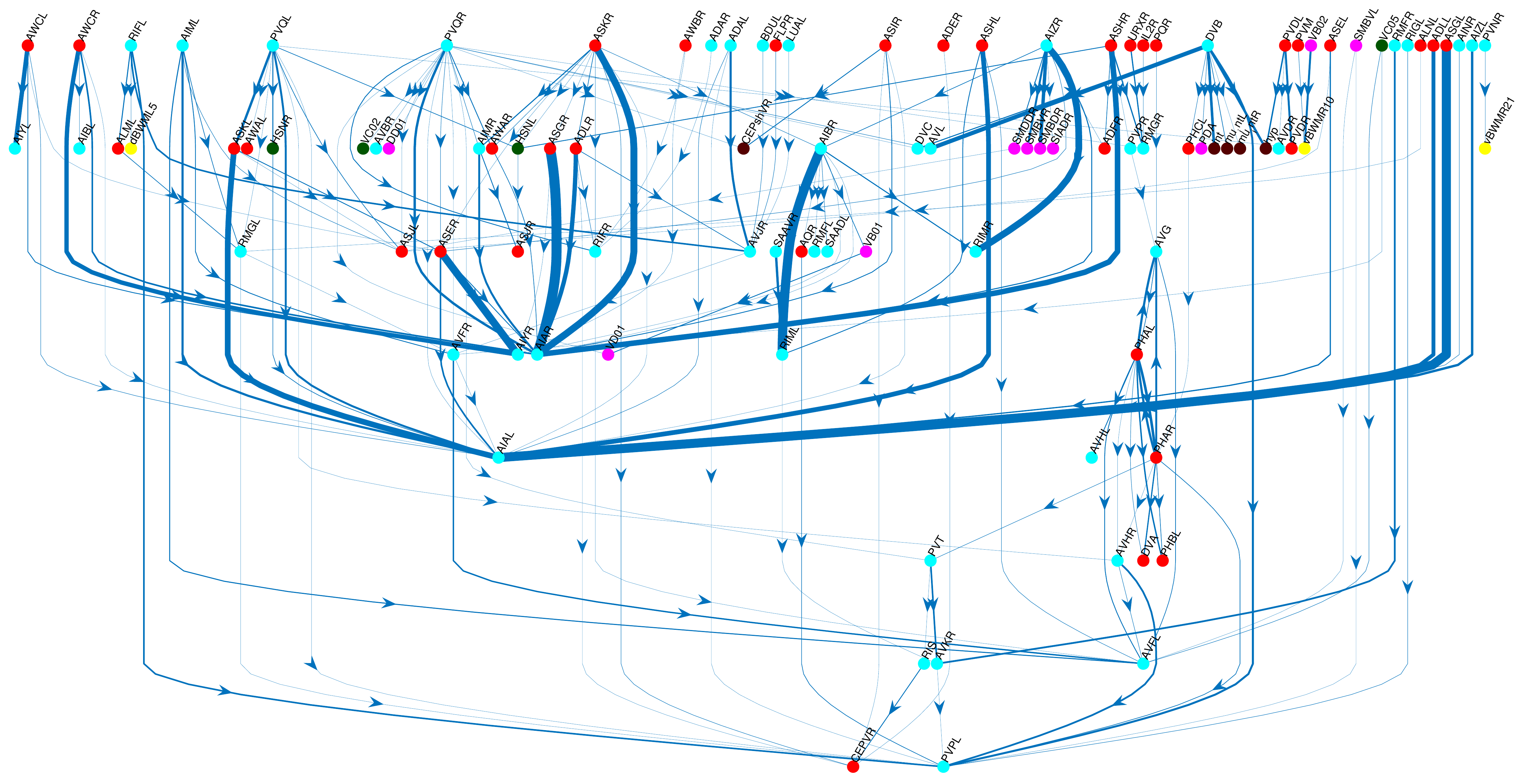}
  \caption{%
  \tb{PRE clustering ($K = 30$) of \emph{C elegans} hermaphrodite chemical synapse connectome} represented as a weighted digraph \cite{Cook2019Nature, WormWiring_Dataset}. Edge direction and thickness indicate the synaptic connection and strength, respectively. 
  Here we see a 103-vertex/179-edge PRE subgraph cluster. This edge cluster contains two other disconnected smaller components; for clarity, they are not shown. Here we notice right away that many of the major producers in the subgraph in Fig.~\ref{fig:CElegans_PRE_K30_N85_M122} take the role of major receptors; similarly many major receptors in Fig.~\ref{fig:CElegans_PRE_K30_N85_M122} are now major producers. For example, the subgraph in Fig.~\ref{fig:CElegans_PRE_K30_N85_M122} shows AIZR as a major receptor, but now it is a major receptor. Similarly, AIBR and ASHR are now major producers instead of receptors. On the other hand, the main producers in Fig.~\ref{fig:CElegans_PRE_K30_N85_M122} such as AIAL, AIAR, AIYR, and CEPVER, are now receptors. However, some major producers here (e.g., PVQR, PVQL, AIML, and DVB) here are not major receptors in Fig.~\ref{fig:CElegans_PRE_K30_N85_M122}.  HSNL is neither a producer nor receptor here; its major role as a receptor is in another subgraph not included here. Similar to the subgraph in Fig.~\ref{fig:CElegans_PRE_K30_N85_M122}, this subgraph can also be used to understand how AIB and AIZ promote turns while AIA and AIY inhibit turns by examining the concentrated production and reception of synapses.  We also see here that HSNL is receiving synapses from the major sensory neurons ASKR, ASHR, as well as PVQL. These directed synapses must affect the serotonin production and egg-laying processes controlled by HSNL.}
  \label{fig:CElegans_PRE_K30_N158_M325_trimmed_N103_M179}
\end{figure}

\newpage

\begin{figure}[H]
  \centering
  \subfigure[A RGE clustering ($K = 30$) produced 64-vertex/170-edge edge cluster.]{%
  \includegraphics[width = \textwidth, height = 3.5in]{%
      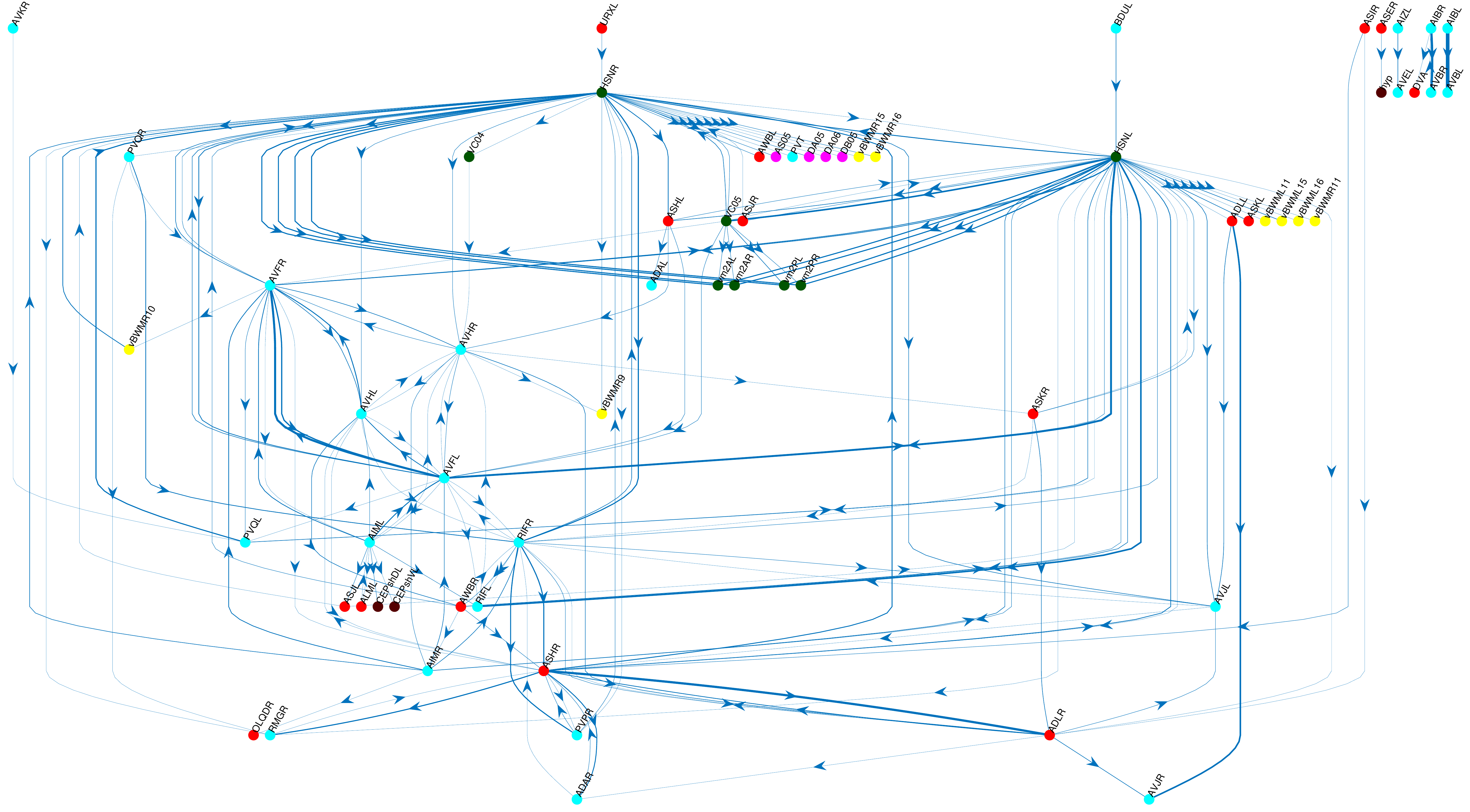}} \\
  \subfigure[A RGE clustering ($K = 30$) produced 46-vertex/70-edge edge cluster.]{%
  \includegraphics[width = \textwidth, height = 3.5in]{%
      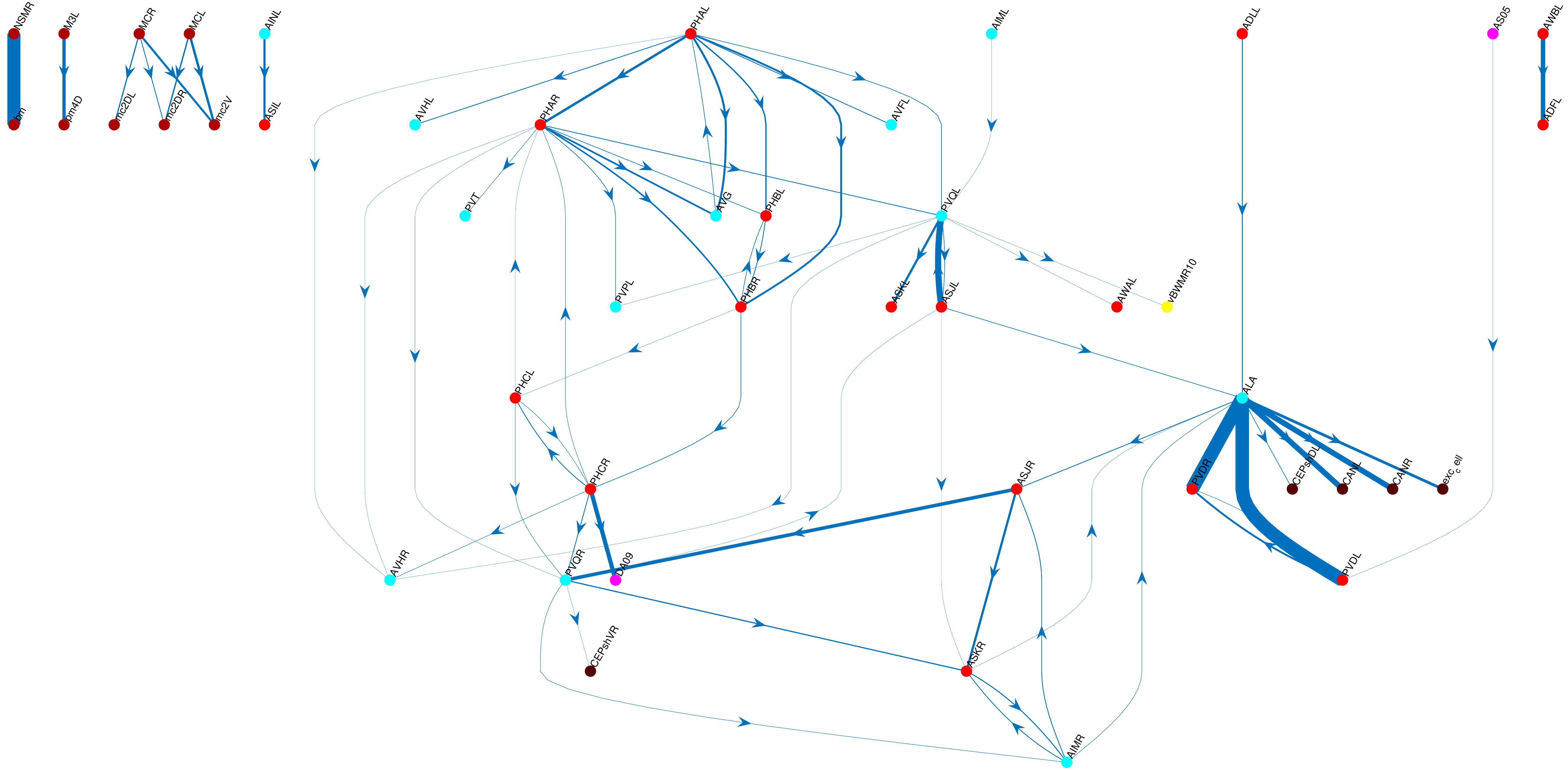}} 
\end{figure}
\clearpage 
\captionof{figure}{%
\tb{RGE clustering ($K = 30$) of \emph{C elegans} hermaphrodite chemical synapse connectome} represented as a weighted digraph \cite{Cook2019Nature, WormWiring_Dataset}. Edge direction indicates the synapse direction; edge thickness indicate weight. RGE clustering generates subgraph clusters that emphasize synaptic connections regardless of direction among neurons with large $\nu$, set here to the absolute weighted degree. 
  \tb{(a)} shows a 64 neuron and 170 synaptic edge circuit associated with egg-laying and serotonin and other chemical sensing and processing.  The HSN neurons are the driving neurons for egg laying, and the presence of connections to VC04, VC05, vm2, and vbwm neurons suggest connections to egg-laying \cite{Ravi2018JN, Collins2016eLife}.  Given that HSN releases serotonin, and given the connections to the AVF and AIM interneurons, this circuit also processes the released serotonin, possibly accomplished by the numerous interneurons such as AVJ, PVQ, and AVH. The ADA interneurons' function is largely unknown. They are connected to the processes in this circuit, including the important ASH sensory neurons, suggesting that the ADA neurons might have functions related to serotonin or chemical sensing, and potentially egg-laying. 
  \tb{(b)} The CANL and CANR neurons are essential for the survival of \emph{C elegans,} yet their function is largely unknown.  Here we see an RGE subgraph cluster with 46 neurons 70 synaptic edges focused on the strong synaptic connections involving the CAN neurons as well as ALA, PVDR, PVDL, and CEPsh glia, which are associated with sensory organs. ALA is a mechanosensor, and PVDR and PVDL are essential for posterior harsh touch response. The circuit also features strong connections among the PHA and PHB sensory neurons involved in chemorepulsion, as well as the PHC sensory neurons involved in temperature avoidance, and the PVQ interneurons.  This RGE subgraph is based on strong synaptic connections regardless of direction among neurons with larger $\nu$, and so researchers looking for graph-based clues as to the function of CANL and CANR and associated neurons can possibly start with this circuit.}
\label{fig:CElegans_K30_RGE}

\newpage

\begin{figure}[H]
  \centering
  \subfigure[Migration mainly into Florida and out of Texas.]{%
  \includegraphics[width=0.48\textwidth, height=1.5in]{%
      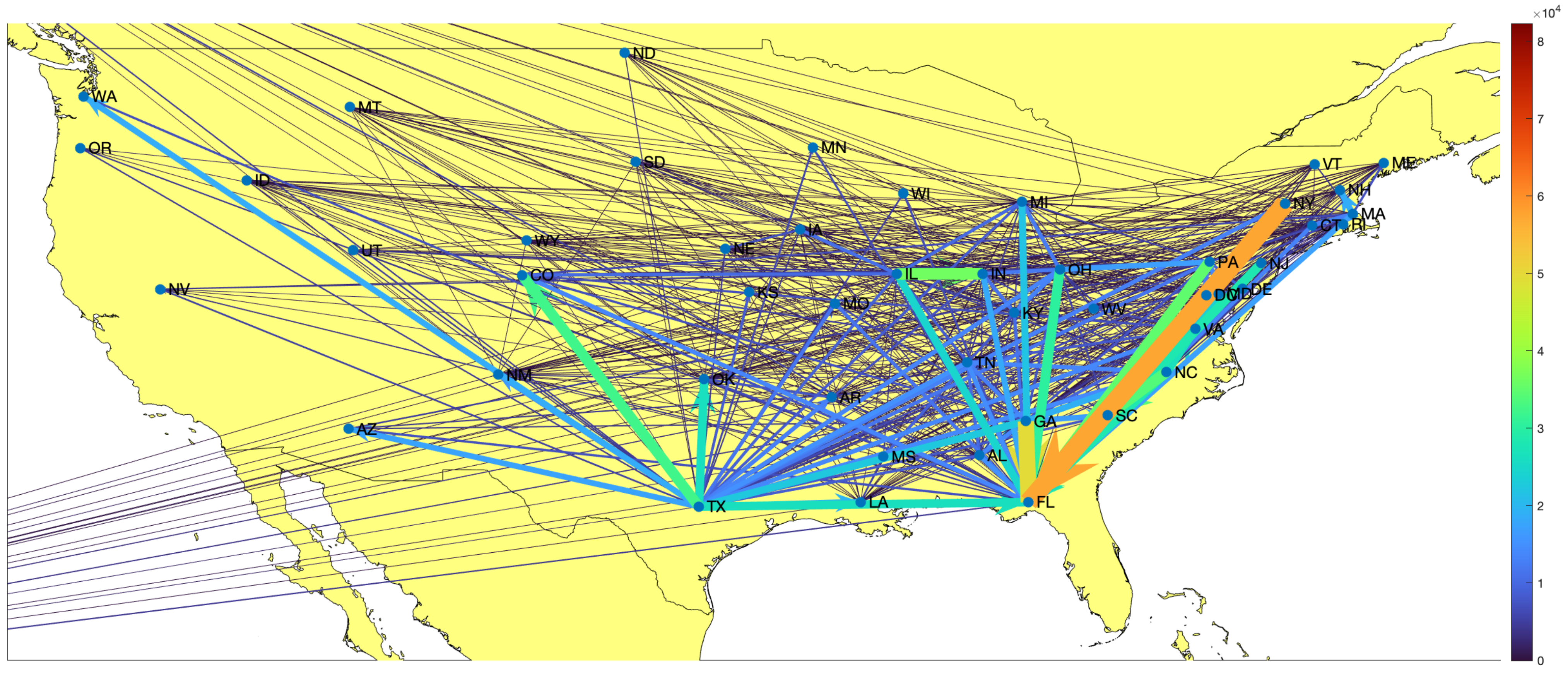}} 
  \subfigure[Migration mainly into Texas and out of Florida.]{%
  \includegraphics[width=0.48\textwidth, height=1.5in]{%
      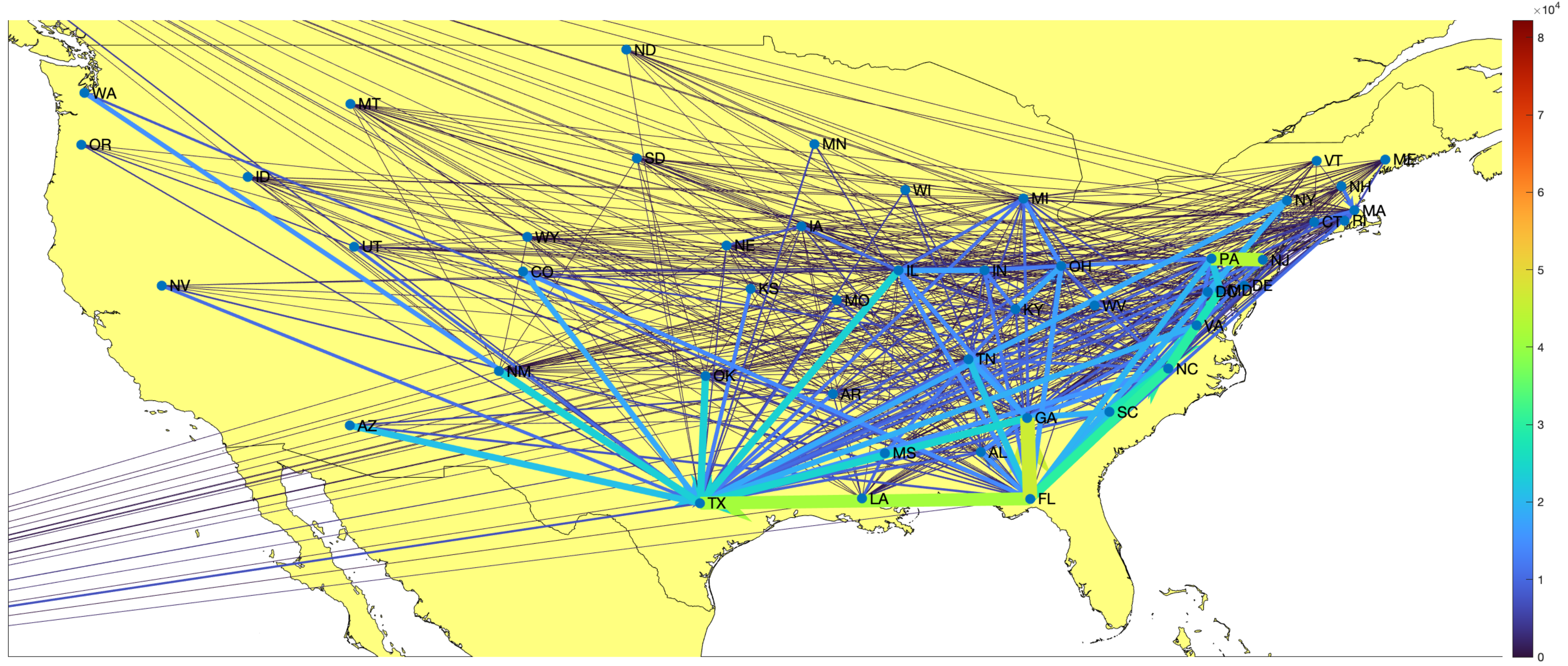}} \\
  \vspace*{-0.1in}
  \subfigure[Migration mainly out of New York.]{%
  \includegraphics[width=0.48\textwidth, height=1.5in]{%
      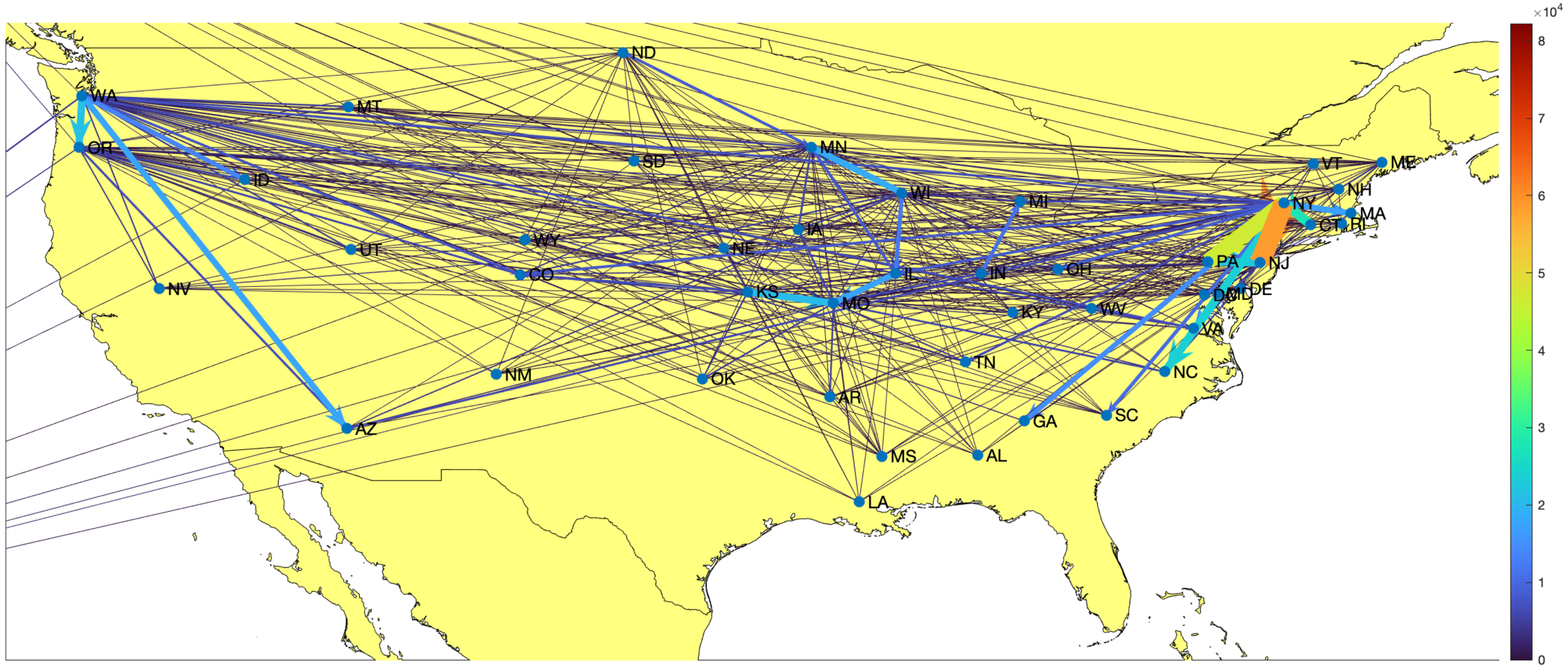}} 
  \subfigure[Migration mainly into New York.]{%
  \includegraphics[width=0.48\textwidth, height=1.5in]{%
      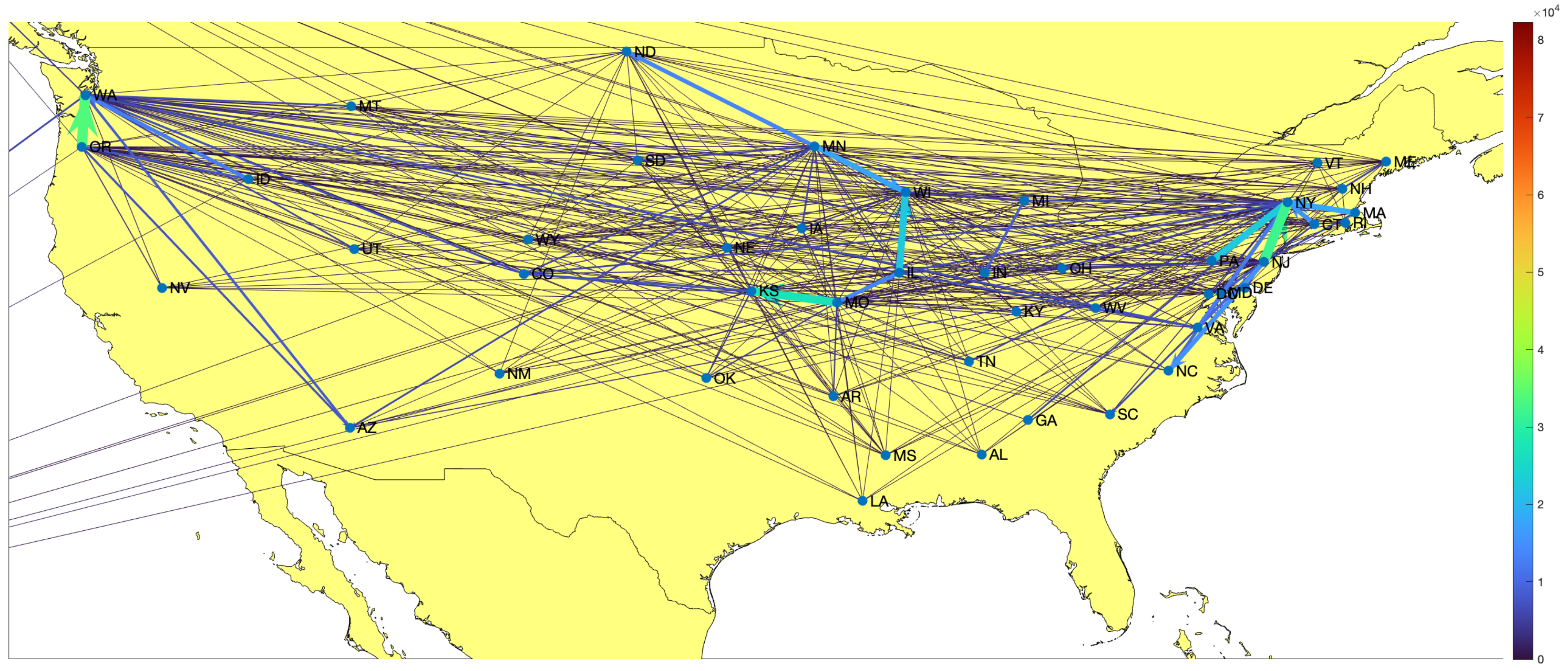}} \\ 
  \vspace*{-0.1in}
  \subfigure[Migration mainly into California.]{%
  \includegraphics[width=0.48\textwidth, height=1.5in]{%
      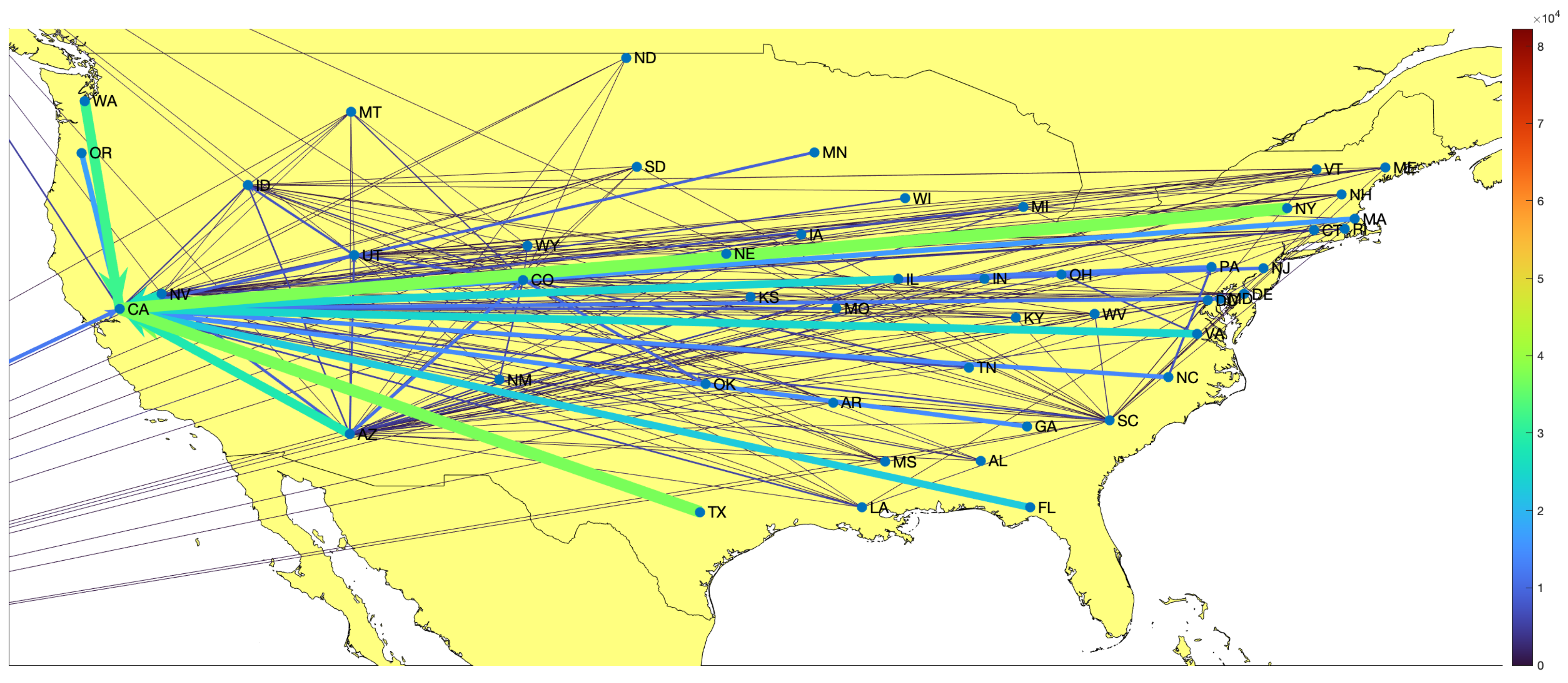}} 
  \subfigure[Migration mainly out of California.]{%
  \includegraphics[width=0.48\textwidth, height=1.5in]{%
      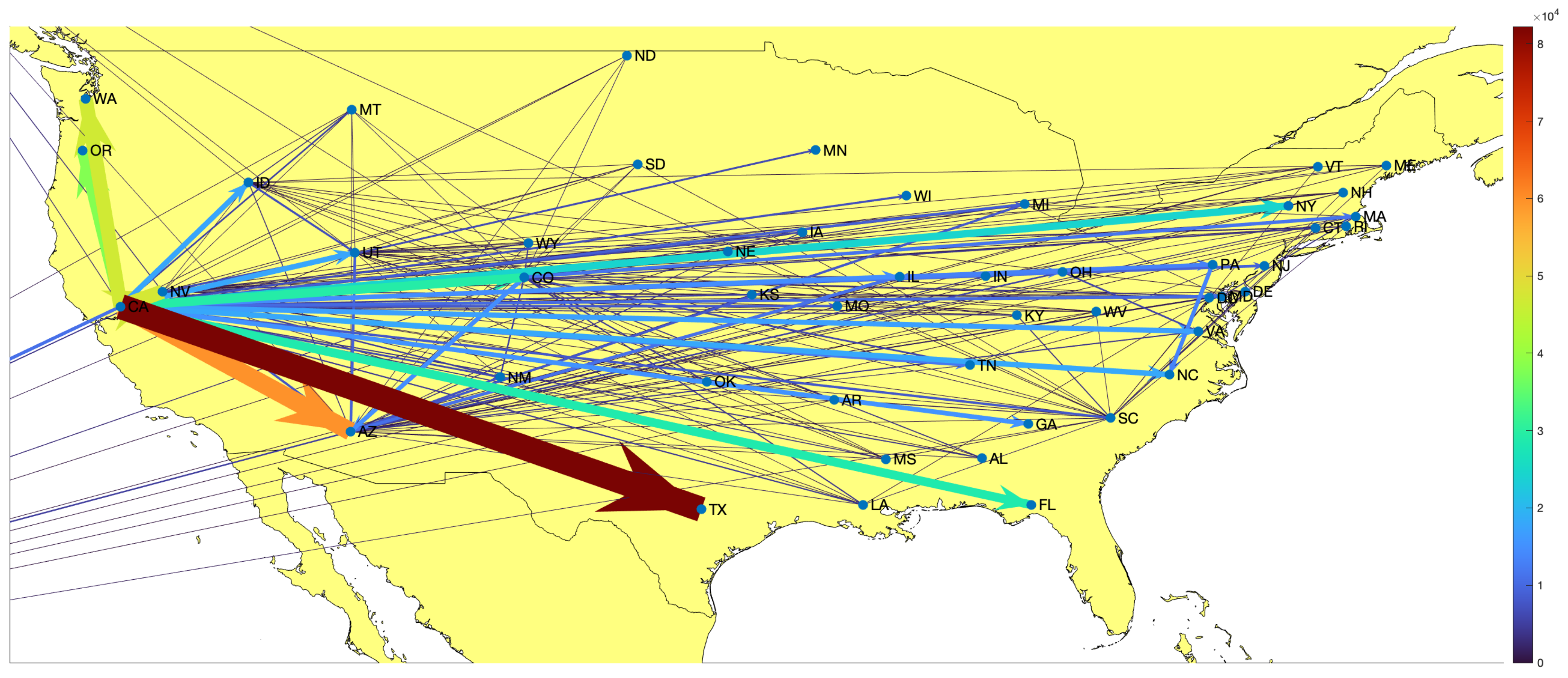}} 
  \caption{%
  \tb{PRE clustering of 2019 US inter-state migration flow data for the 50 states plus Washington D.C.} represented as a weighted digraph  \cite{InterStateMigration_USCB2019}. Edge direction indicates the direction of migration; edge color and edge thickness indicate volume. PRE clustering generates subgraph clusters that show focused migration highlighting states where people primarily emigrate from and other states where people primarily immigrate to. 
  Fig.~\ref{fig:USMigration_Cypress}(a) depicts one edge cluster (with only edges with volumes at least 5\% of the maximun shown) generated from the application of our PRE clustering algorithm with $K=6$. Here we show all the 6 edge clusters, each with all its edges:  
  \tb{(a)} shows movement mainly out of Texas (TX) and into Florida (FL) particularly from New York (NY); 
  \tb{(b)} shows movement mainly into TX and out of FL; 
  \tb{(c)} shows movement mainly out of NY; 
  \tb{(d)} shows movement mainly into NY;  
  \tb{(e)} shows movement mainly into California (CA); and 
  \tb{(f)} shows movement mainly out of CA. 
  Note that the edges moving out of the top-left and bottom-left of frame correspond to Alaska (AK) and Hawaii (HI) (which are out of frame), respectively. PRE helps us visualize the conventional wisdom regarding migration such as the outsize importance of CA as well as FL, NY, and TX.}
\label{fig:USMigration_6Clusters_LPRE}
\end{figure}

\newpage

\begin{figure}[H]
  \centering
  \subfigure[California-centered region.]{%
  \includegraphics[width=0.48\textwidth, height=1.5in]{%
      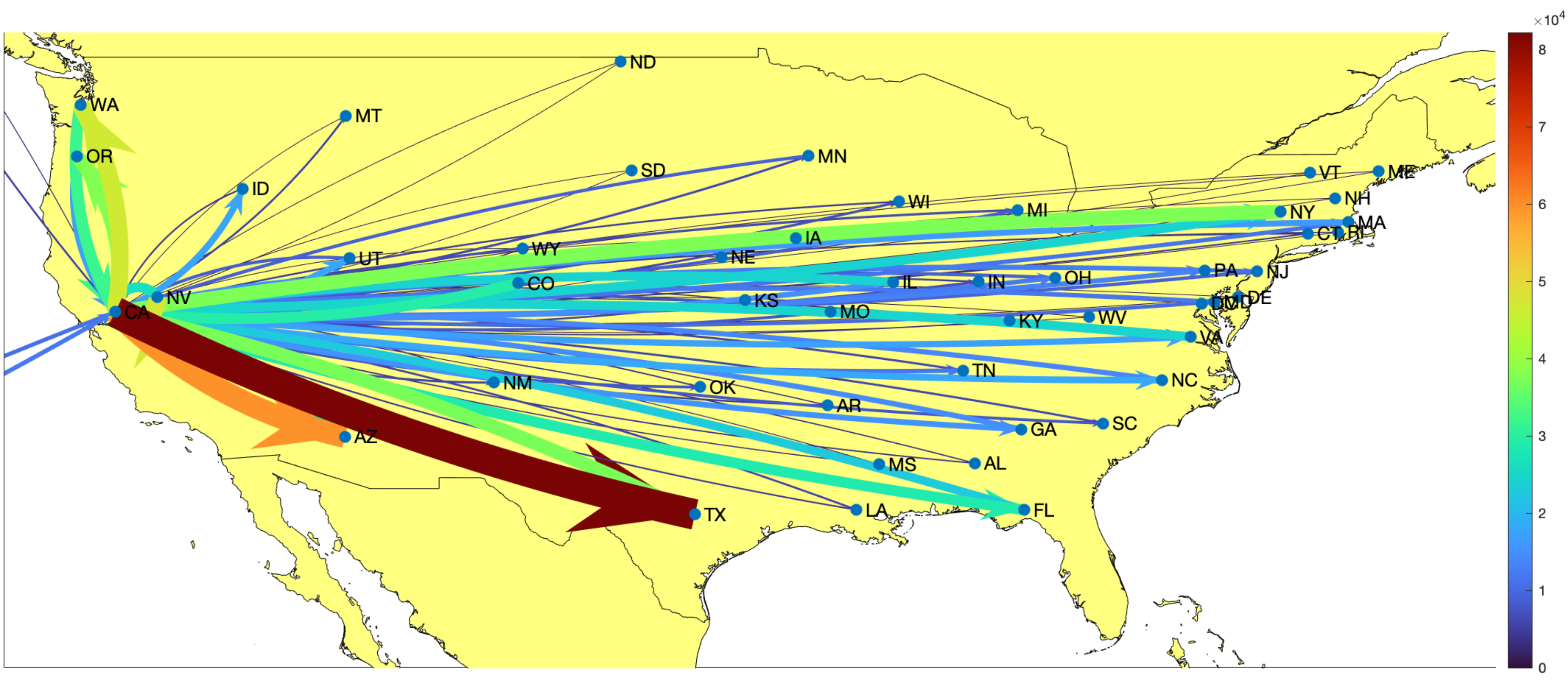}} 
  \subfigure[Florida-centered region.]{%
  \includegraphics[width=0.48\textwidth, height=1.5in]{%
      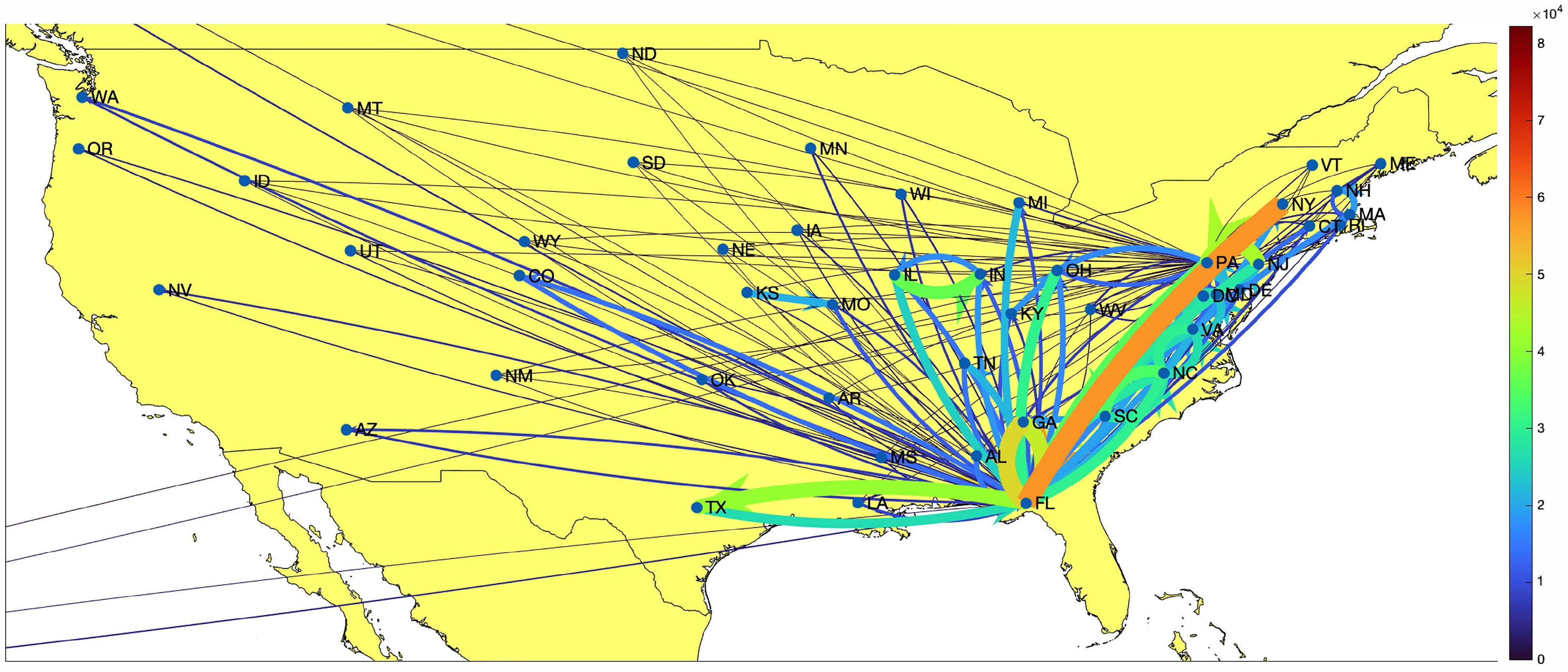}} \\
  \vspace*{-0.1in}
  \subfigure[Midwest-centered region.]{%
  \includegraphics[width=0.48\textwidth, height=1.5in]{%
      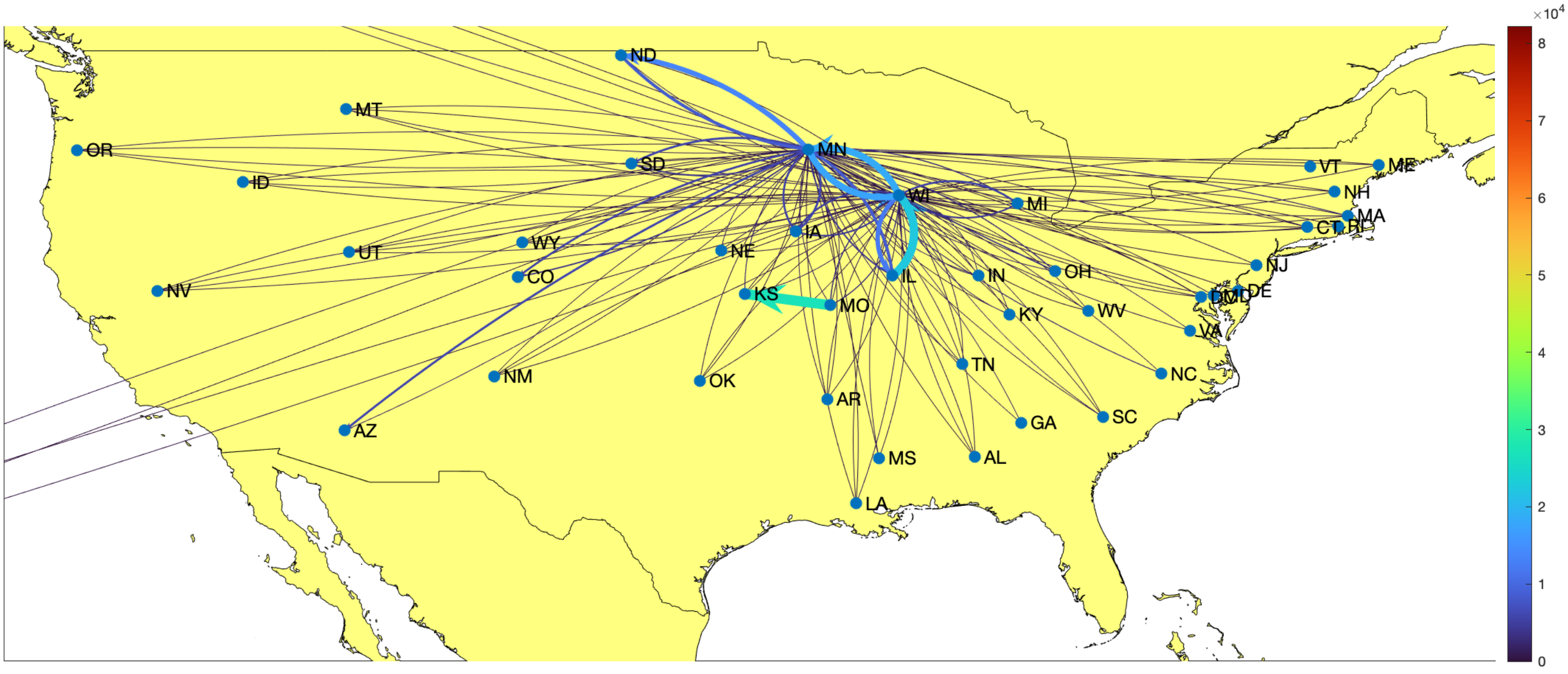}} 
  \subfigure[New York-centered region.]{%
  \includegraphics[width=0.48\textwidth, height=1.5in]{%
      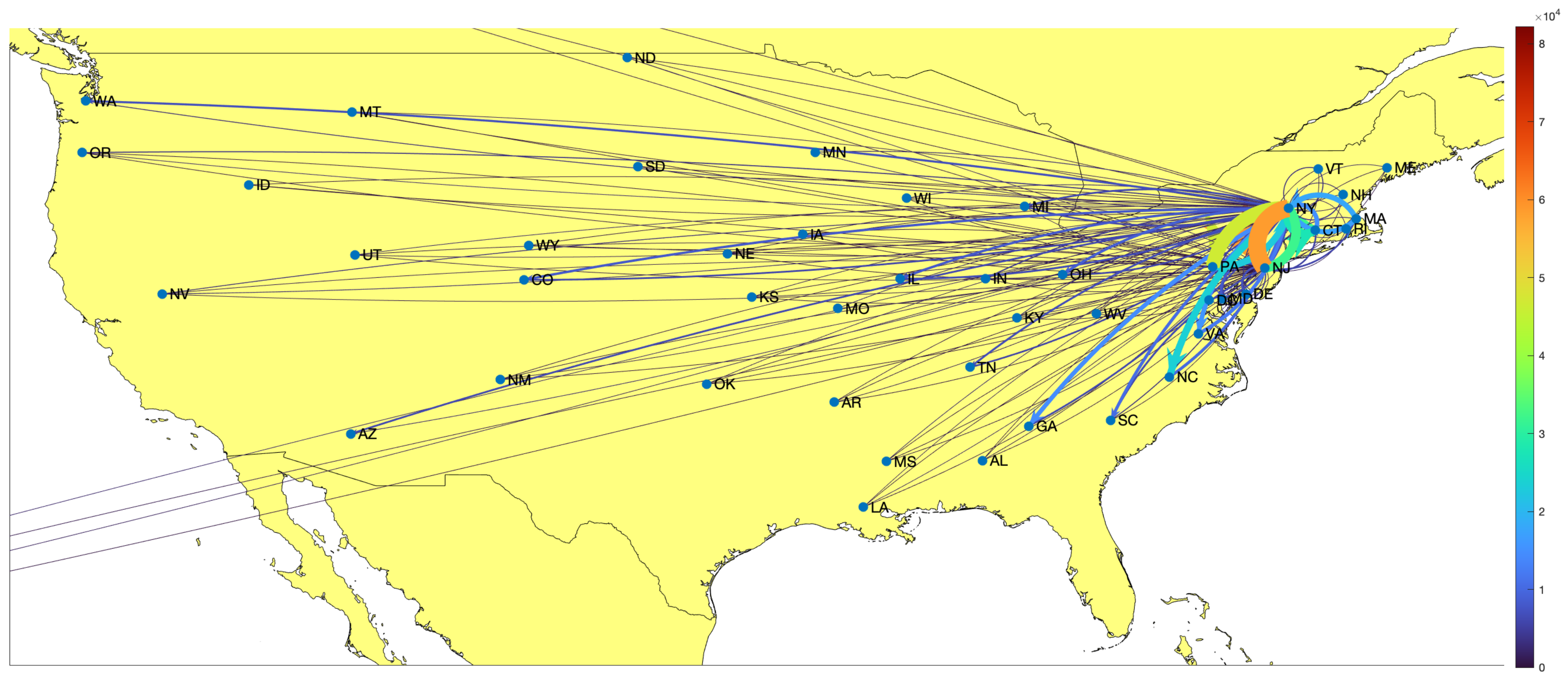}} \\ 
  \vspace*{-0.1in}
  \subfigure[Texas-centered region.]{%
  \includegraphics[width=0.48\textwidth, height=1.5in]{%
      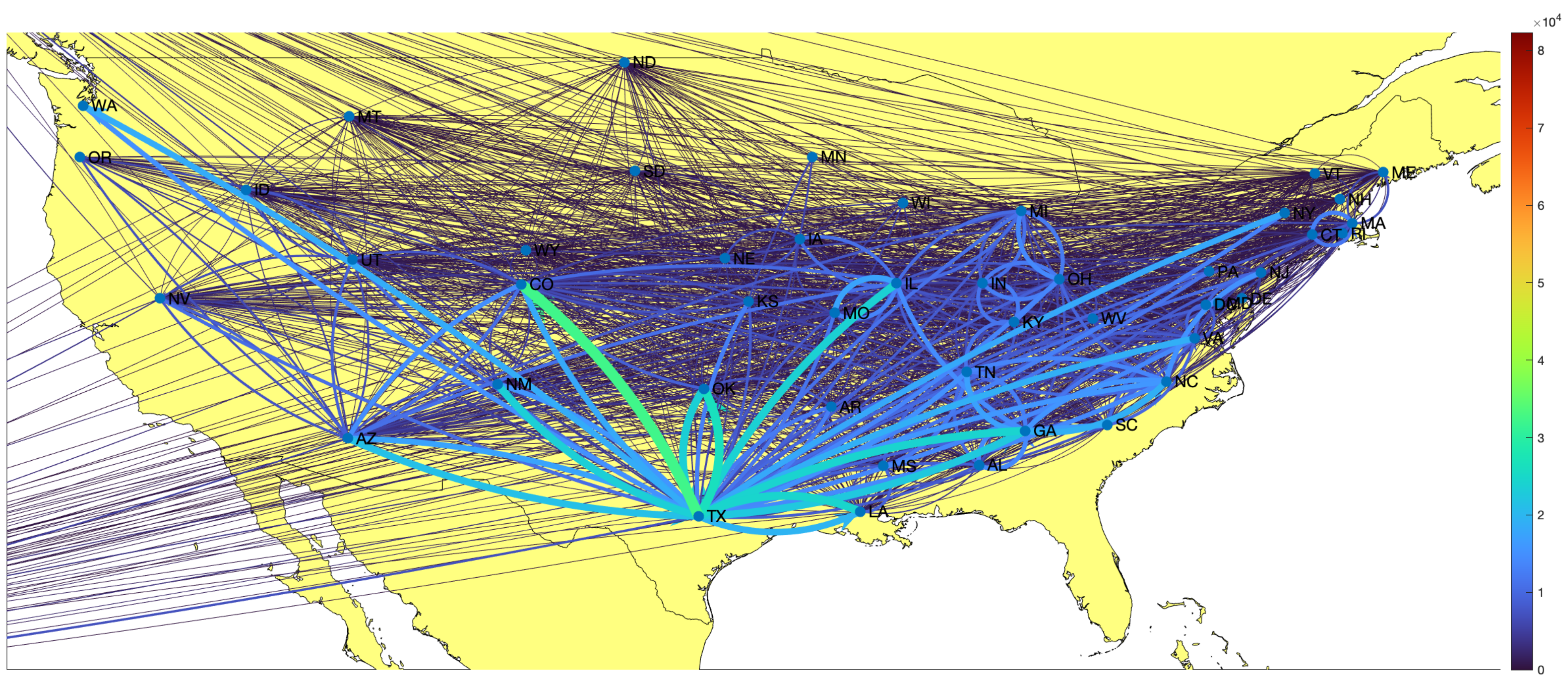}} 
  \subfigure[Washington-centered region.]{%
  \includegraphics[width=0.48\textwidth, height=1.5in]{%
      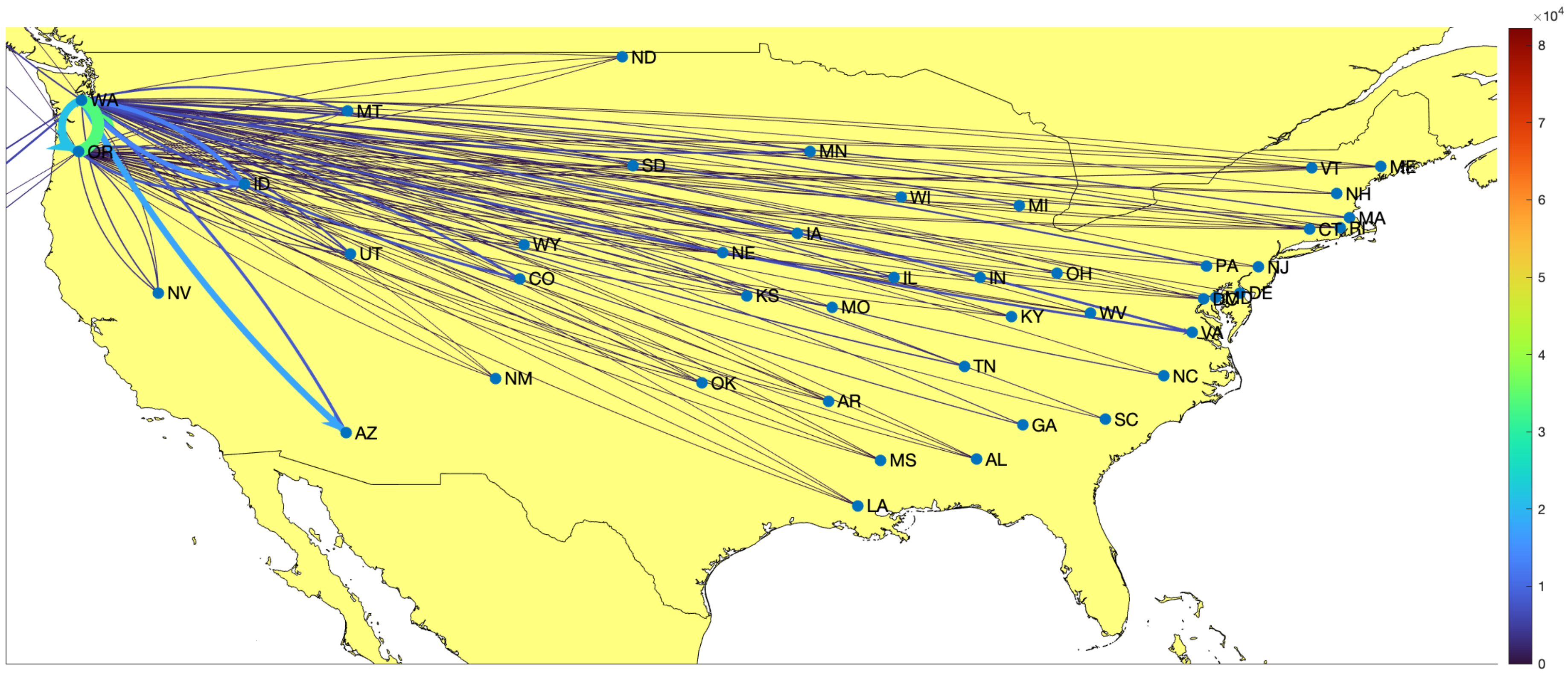}} 
  \caption{%
  \tb{RGE clustering ($K = 6$) of 2019 US inter-state migration flow data for the 50 states plus Washington D.C.} represented as a weighted digraph  \cite{InterStateMigration_USCB2019}. Edge direction indicates the direction of migration; edge color and edge thickness indicate volume. RGE clustering generates subgraph `regions' within which migratory movement is concentrated among states. 
  Fig.~\ref{fig:USMigration_Cypress}(b) depicts one edge cluster (with only edges with volumes at least 5\% of the maximun shown) generated from the application of our RGE clustering algorithm. Here we show all the 6 edge clusters, each with all its edges. 
  \tb{(a)} A cluster consisting solely of the edges connected to CA;  
  \tb{(b)} A cluster centered on FL and showing concentrated flows in the southeast and east coast; 
  \tb{(c)} This reveals a cluster of concentrated migration flows in the midwest centered on Minnesota (MN); such flows are typically not in public discourse; 
  \tb{(d)} A cluster centered on New York along with flows mainly in the northeast; 
  \tb{(e)} A cluster centered on TX; and 
  \tb{(f)} A cluster centered on Washington (WA) and flows in the Pacific Northwest with a noteworthy flow from WA to Arizona (AZ). 
  Note that the edges moving out of the top-left and bottom-left of frame correspond to AK and HI, respectively. RGE clustering reveals both well-known and less well-known migration patterns.}
\label{fig:USMigration_6Clusters_LRGE}
\end{figure}

\newpage

\begin{figure}[H]
  \centering
  \includegraphics[width=\textwidth]{%
      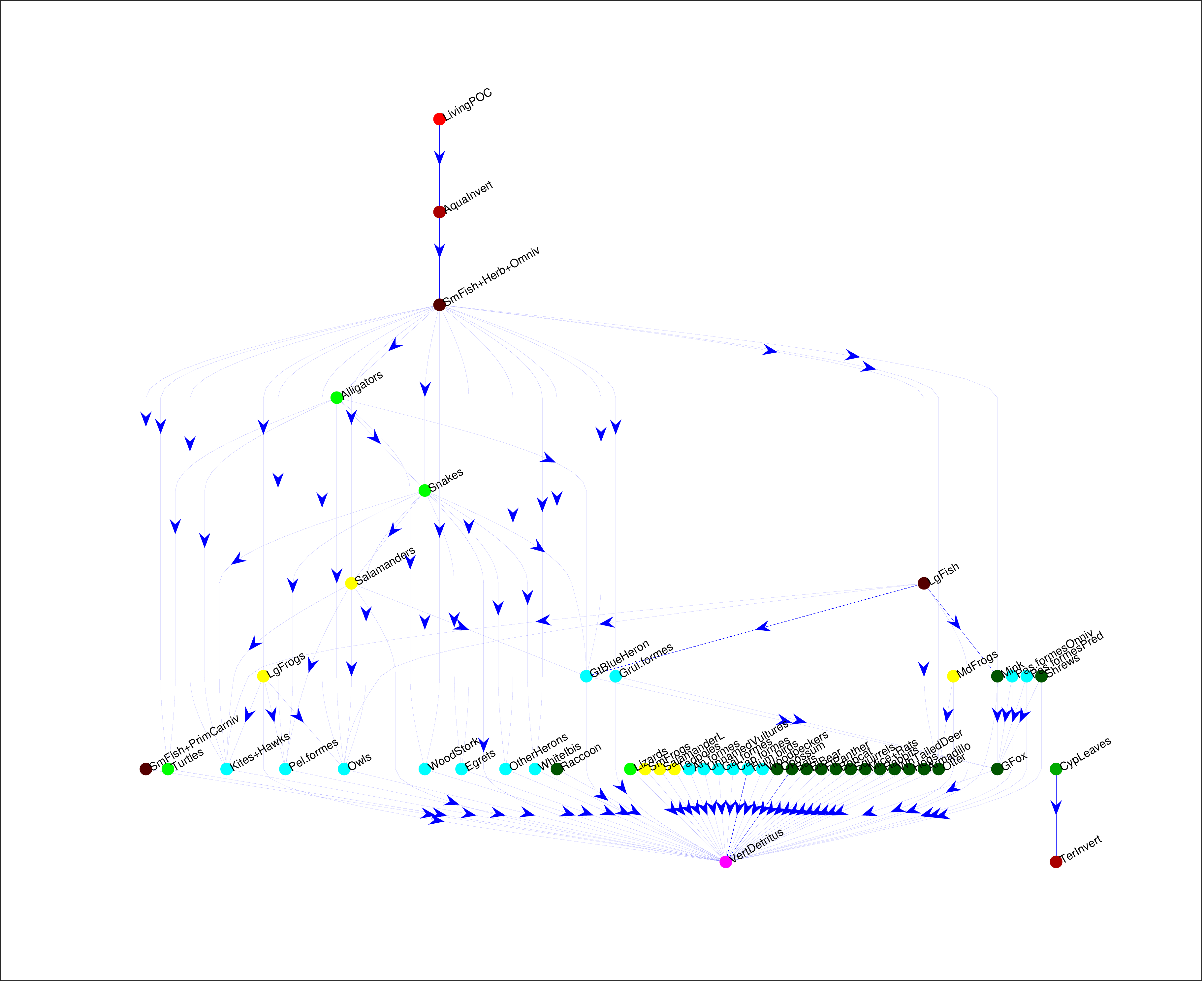}
  \caption{%
  \tb{DPE clustering ($K = 10$) of the Florida Bay Cypress Wetlands dry season food web} represented as a 68-vertex/554-edge weighted digraph \cite{Bondavalli2000JB}. Each vertex represents an organism and each directed edge (with edge width proportional to edge weight) represents carbon flow which reflects biomass exchange. Vertex colors indicate membership in one of 9 organism compartments. DPE clustering yields meaningful subgraph clusters of focused carbon transfer. Fig.~\ref{fig:USMigration_Cypress}(a) depicts one subgraph cluster generated from the application of our DPE clustering algorithm. Here we see a 100-edge cluster of focused carbon transfer from living POC to aquatic invertebrates and then to herbivorous and omnivorous small fish, which is then transferred to a variety of organisms, including large fish and alligators, ultimately ending with the final carbon transfer to vertebrate detritus. Figs~\ref{fig:Cypress_10Clusters_100_LDPE}-\ref{fig:Cypress_10Clusters_36_LDPE} show all the clusters obtained with DPE clustering ($K = 10$).}
\label{fig:Cypress_10Clusters_100_LDPE}
\end{figure}

\newpage

\begin{figure}[H]
  \centering
  \includegraphics[width=\textwidth]{%
      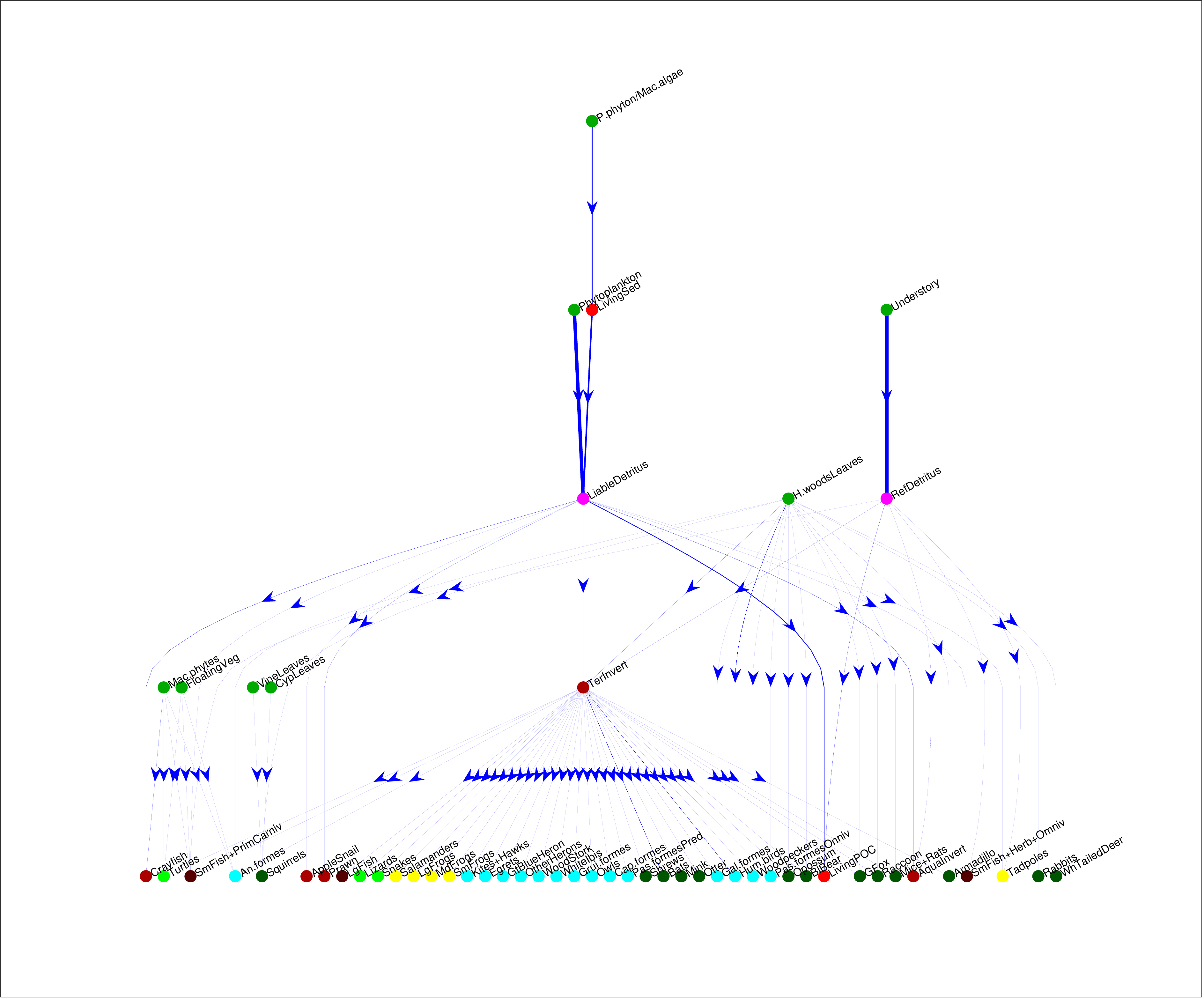}
  \caption{%
  \tb{DPE clustering ($K = 10$) of the Florida Bay Cypress Wetlands dry season food web} represented as a 68-vertex/554-edge weighted digraph \cite{Bondavalli2000JB}. Each vertex represents an organism and each directed edge (with edge width proportional to edge weight) represents carbon flow which reflects biomass exchange. Vertex colors indicate membership in one of 9 organism compartments. Here we see a 77-edge cluster highlighting a larger transfer from phytoplankton and living sediment into liable detritus which then flows to largely carnivorous small fish.  We also see how terrestrial invertebrates provide carbon to a large number of organisms, e.g., large fish, lizards, snakes, salamanders, frogs, etc. Figs~\ref{fig:Cypress_10Clusters_100_LDPE}-\ref{fig:Cypress_10Clusters_36_LDPE} show all the clusters obtained when DPE clustering is applied with $K=10$.}
\label{fig:Cypress_10Clusters_77_LDPE}
\end{figure}

\newpage

\begin{figure}[H]
  \centering
  \includegraphics[width=\textwidth]{%
      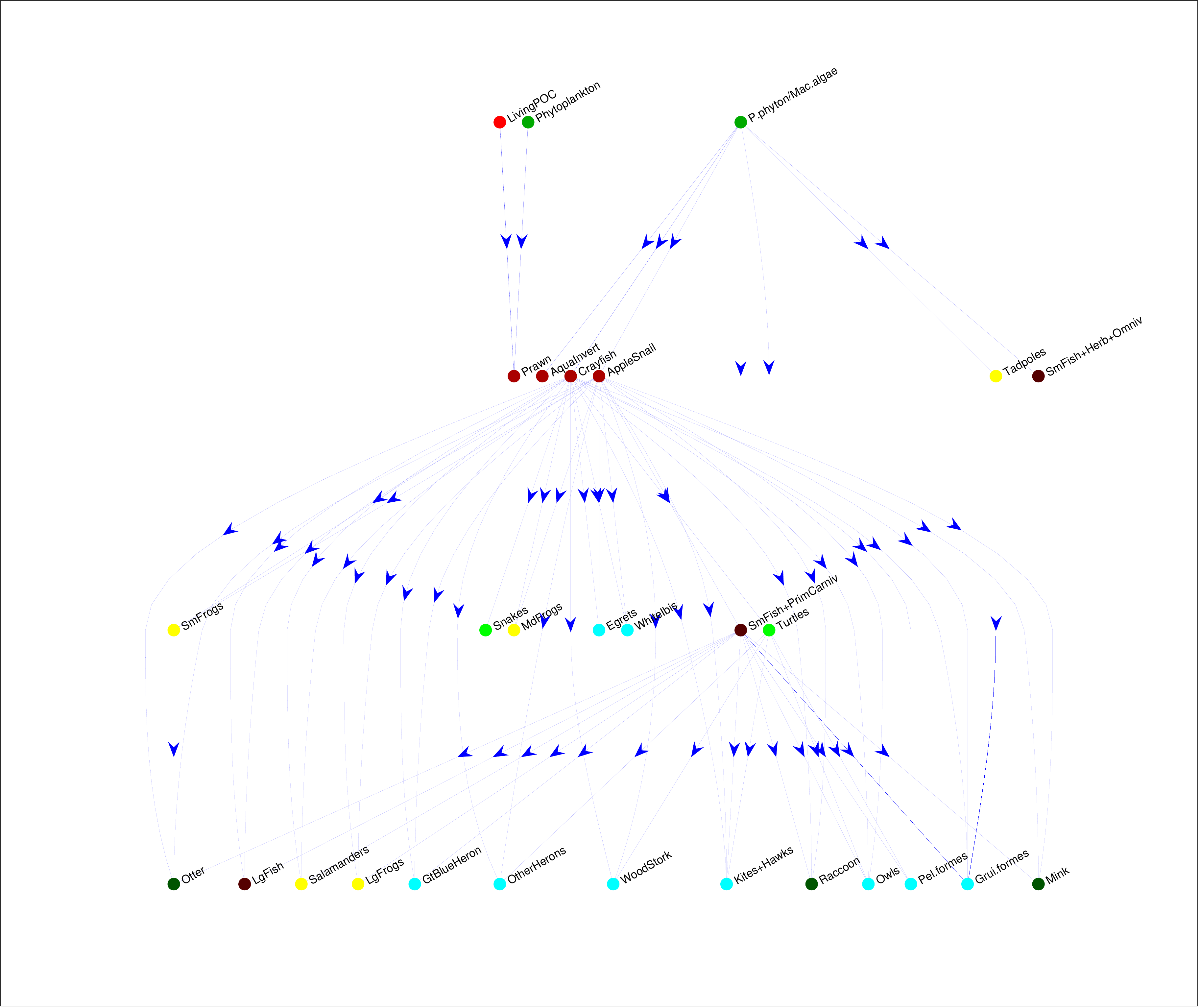}
  \caption{%
  \tb{DPE clustering ($K = 10$) of the Florida Bay Cypress Wetlands dry season food web} represented as a 68-vertex/554-edge weighted digraph \cite{Bondavalli2000JB}. Each vertex represents an organism and each directed edge (with edge width proportional to edge weight) represents carbon flow which reflects biomass exchange. Vertex colors indicate membership in one of 9 organism compartments. Here we see a 63-edge cluster highlighting how crayfish and apple snails provide carbon to a number of organisms (e.g., salamanders, large frogs, great blue heron, wood stork, in some cases partially through turtles and small fish that are primarily carnivorous. We also see how periphyton/macroalgae play an important role in the beginning of the carbon transfer. Figs~\ref{fig:Cypress_10Clusters_100_LDPE}-\ref{fig:Cypress_10Clusters_36_LDPE} show all the clusters obtained when DPE clustering is applied with $K=10$.}
\label{fig:Cypress_10Clusters_63_LDPE}
\end{figure}

\newpage

\begin{figure}[H]
  \centering
  \includegraphics[width=\textwidth]{%
      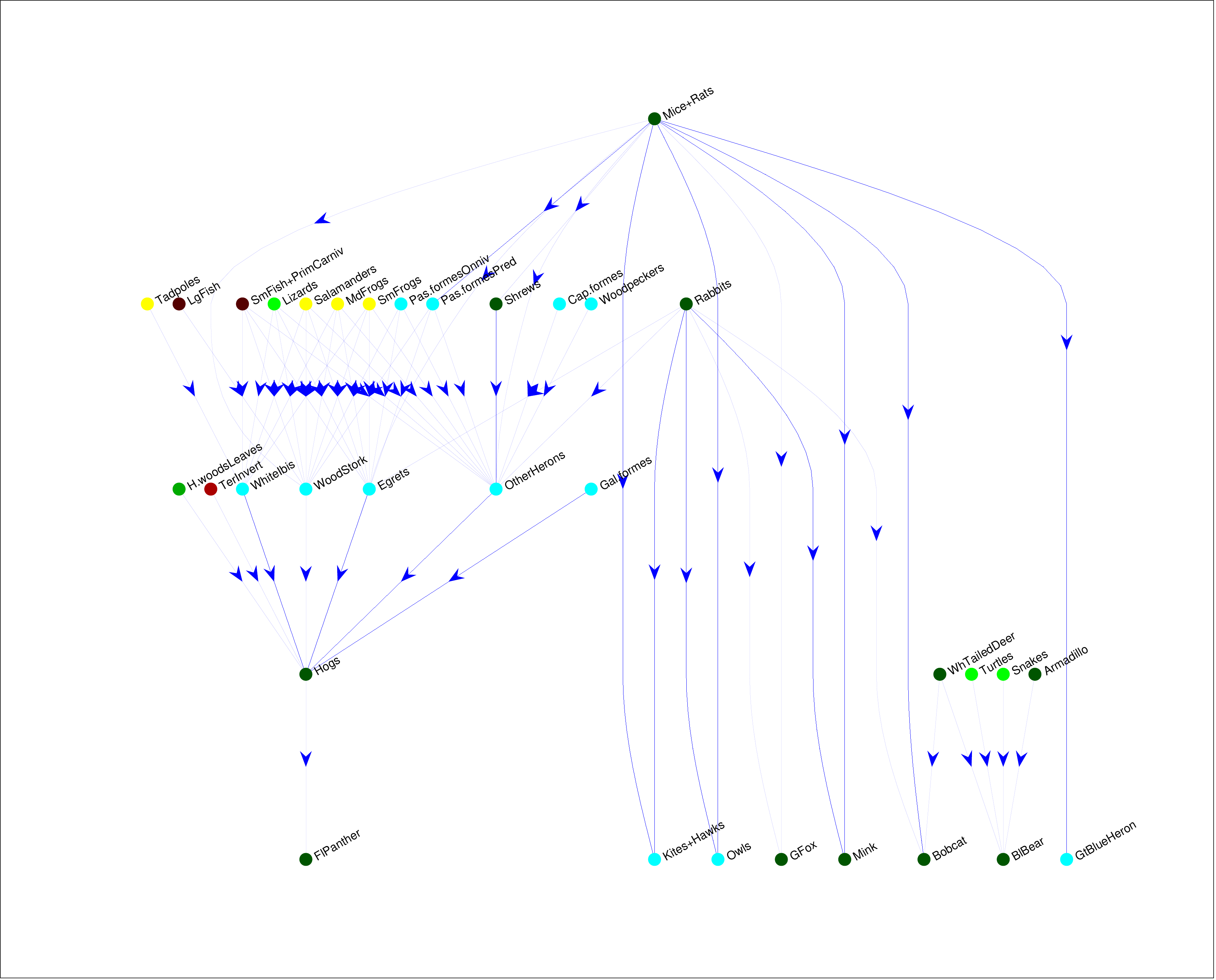}
  \caption{%
  \tb{DPE clustering ($K = 10$) of the Florida Bay Cypress Wetlands dry season food web} represented as a 68-vertex/554-edge weighted digraph \cite{Bondavalli2000JB}. Each vertex represents an organism and each directed edge (with edge width proportional to edge weight) represents carbon flow which reflects biomass exchange. Vertex colors indicate membership in one of 9 organism compartments. Here we see a reprint of the 62-edge cluster in Fig.~\ref{fig:USMigration_Cypress}(d) highlighting how mice and rats provide carbon directly and indirectly to a number of organisms, e.g., through shrews, herons, hogs, and the Florida panthers.  Connected to this, rabbits provide carbon directly and indirectly to owls, bobcats, and other animals.  We also see focused transfer to wood storks, egrets, herons, and black bears. Figs~\ref{fig:Cypress_10Clusters_100_LDPE}-\ref{fig:Cypress_10Clusters_36_LDPE} show all the clusters obtained when DPE clustering is applied with $K=10$.}
\label{fig:Cypress_10Clusters_62_LDPE}
\end{figure}

\newpage

\begin{figure}[H]
  \centering
  \includegraphics[width=\textwidth]{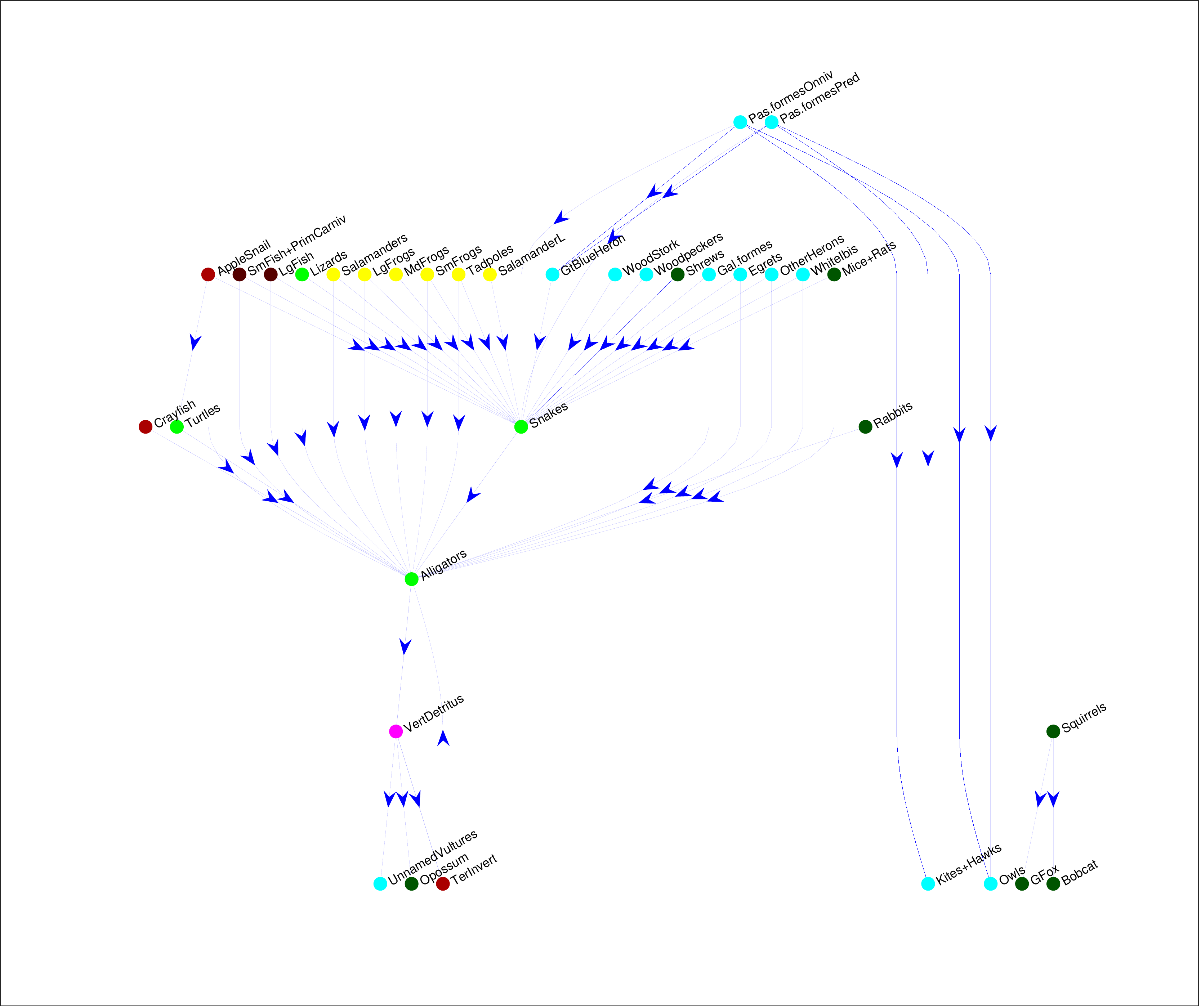}
  \caption{%
  \tb{DPE clustering ($K = 10$) of the Florida Bay Cypress Wetlands dry season food web} represented as a 68-vertex/554-edge weighted digraph \cite{Bondavalli2000JB}. Each vertex represents an organism and each directed edge (with edge width proportional to edge weight) represents carbon flow which reflects biomass exchange. Vertex colors indicate membership in one of 9 organism compartments. Here we see a reprint of the 53-edge cluster in Fig.~\ref{fig:USMigration_Cypress}(c) highlighting focused carbon transfer to both snakes and alligators, and see vultures obtaining carbon from vertebrate detritus as part of this cluster.  Figs~\ref{fig:Cypress_10Clusters_100_LDPE}-\ref{fig:Cypress_10Clusters_36_LDPE} show all the clusters obtained when DPE clustering is applied with $K=10$.}
\label{fig:Cypress_10Clusters_53_LDPE}
\end{figure}

\newpage

\begin{figure}[H]
  \centering
  \includegraphics[width=\textwidth]{%
      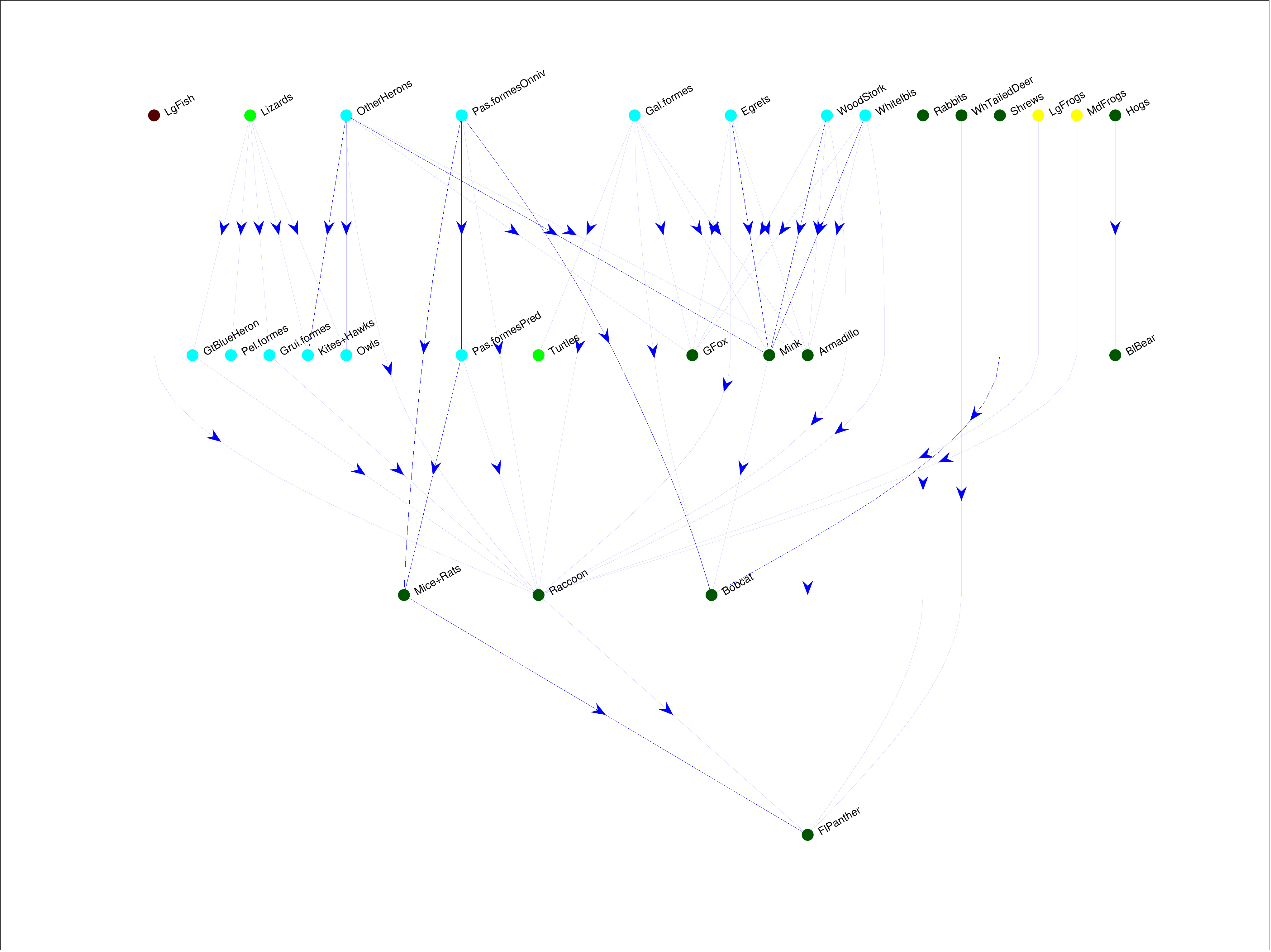}
  \caption{%
  \tb{DPE clustering ($K = 10$) of the Florida Bay Cypress Wetlands dry season food web} represented as a 68-vertex/554-edge weighted digraph \cite{Bondavalli2000JB}. Each vertex represents an organism and each directed edge (with edge width proportional to edge weight) represents carbon flow which reflects biomass exchange. Vertex colors indicate membership in one of 9 organism compartments. Here we see a 48-edge cluster highlighting direct and indirect transfer to mice and rats, raccoons, bobcats, and Florida panther, with other organisms such as lizards, egrets, and wood storks providing carbon to a number of organisms. For example, as part of this, we see focused transfers to the grey fox, mink, and armadillo.  Figs~\ref{fig:Cypress_10Clusters_100_LDPE}-\ref{fig:Cypress_10Clusters_36_LDPE} show all the clusters obtained when DPE clustering is applied with $K=10$.}
\label{fig:Cypress_10Clusters_48_LDPE}
\end{figure}

\newpage

\begin{figure}[H]
  \centering
  \includegraphics[width=\textwidth]{%
      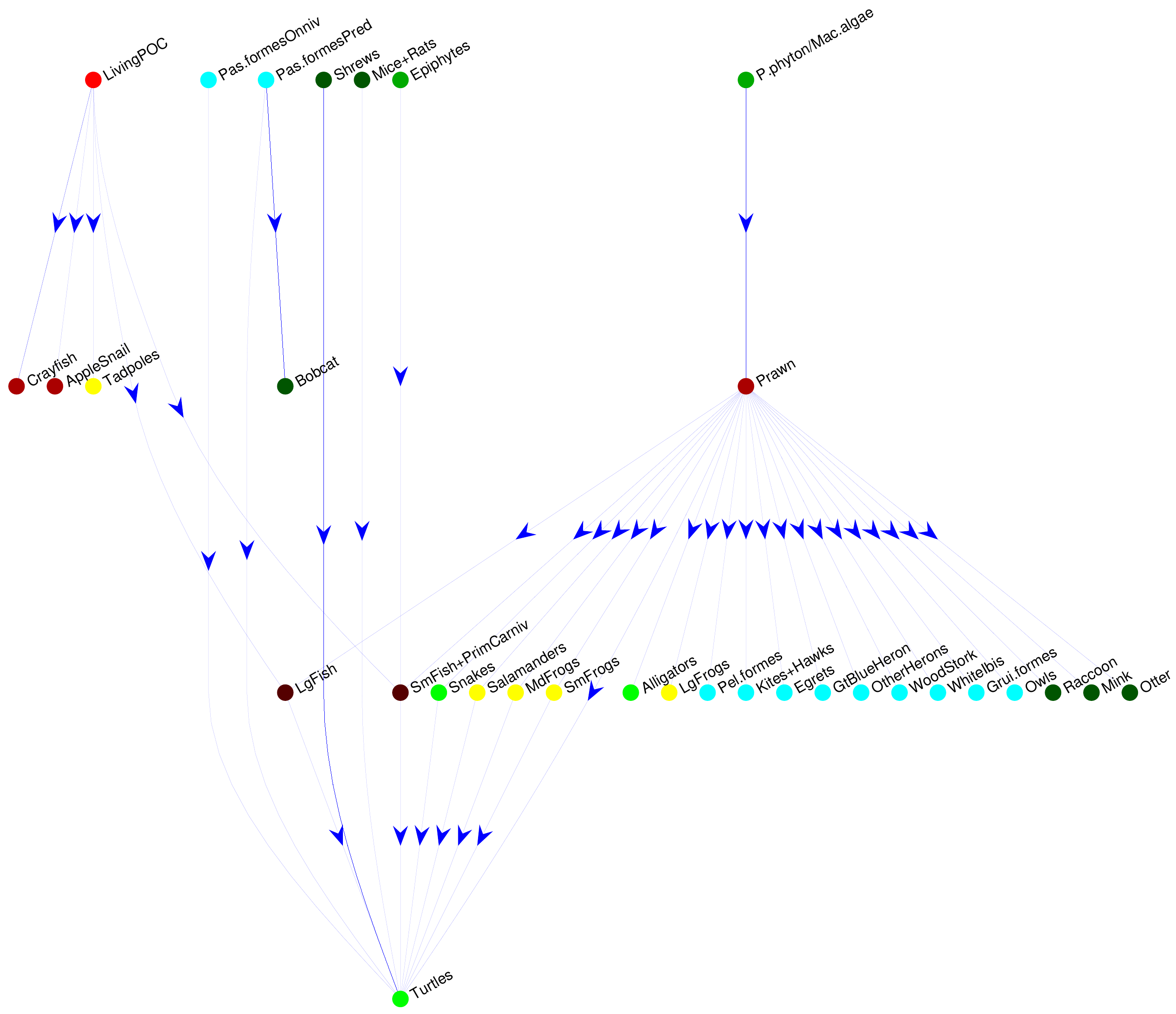}
  \caption{%
  \tb{DPE clustering ($K = 10$) of the Florida Bay Cypress Wetlands dry season food web} represented as a 68-vertex/554-edge weighted digraph \cite{Bondavalli2000JB}. Each vertex represents an organism and each directed edge (with edge width proportional to edge weight) represents carbon flow which reflects biomass exchange. Vertex colors indicate membership in one of 9 organism compartments. Here we see a 39-edge cluster where prawns provide carbon to a number of organisms as well as direct and infirect transfer to turtles, particularly from small fish that are primarily carnivorous.   Figs~\ref{fig:Cypress_10Clusters_100_LDPE}-\ref{fig:Cypress_10Clusters_36_LDPE} show all the clusters obtained when DPE clustering is applied with $K=10$.}
\label{fig:Cypress_10Clusters_39_LDPE}
\end{figure}

\newpage

\begin{figure}[H]
  \centering
  \includegraphics[width=\textwidth]{%
      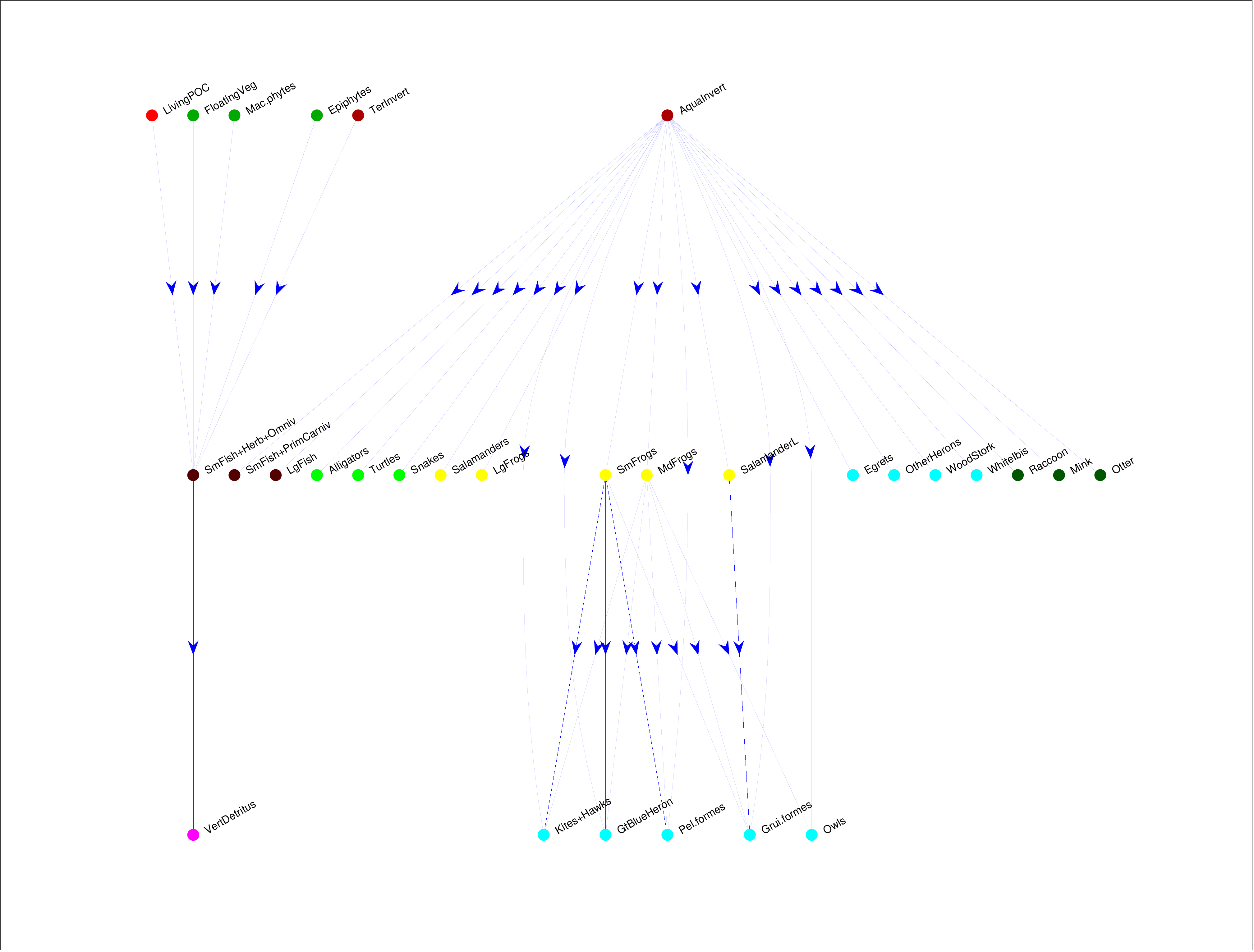}
  \caption{%
  \tb{DPE clustering ($K = 10$) of the Florida Bay Cypress Wetlands dry season food web} represented as a 68-vertex/554-edge weighted digraph \cite{Bondavalli2000JB}. Each vertex represents an organism and each directed edge (with edge width proportional to edge weight) represents carbon flow which reflects biomass exchange. Vertex colors indicate membership in one of 9 organism compartments. Here we see a 38-edge cluster highlighting how aquatic invertebrates provide carbon to a number of organisms, together with transfer to kites and hawks, great blue herons, owls, etc. Figs~\ref{fig:Cypress_10Clusters_100_LDPE}-\ref{fig:Cypress_10Clusters_36_LDPE} show all the clusters obtained when DPE clustering is applied with $K=10$.}
\label{fig:Cypress_10Clusters_38_LDPE}
\end{figure}

\newpage

\begin{figure}[H]
  \centering
  \includegraphics[width=\textwidth]{%
      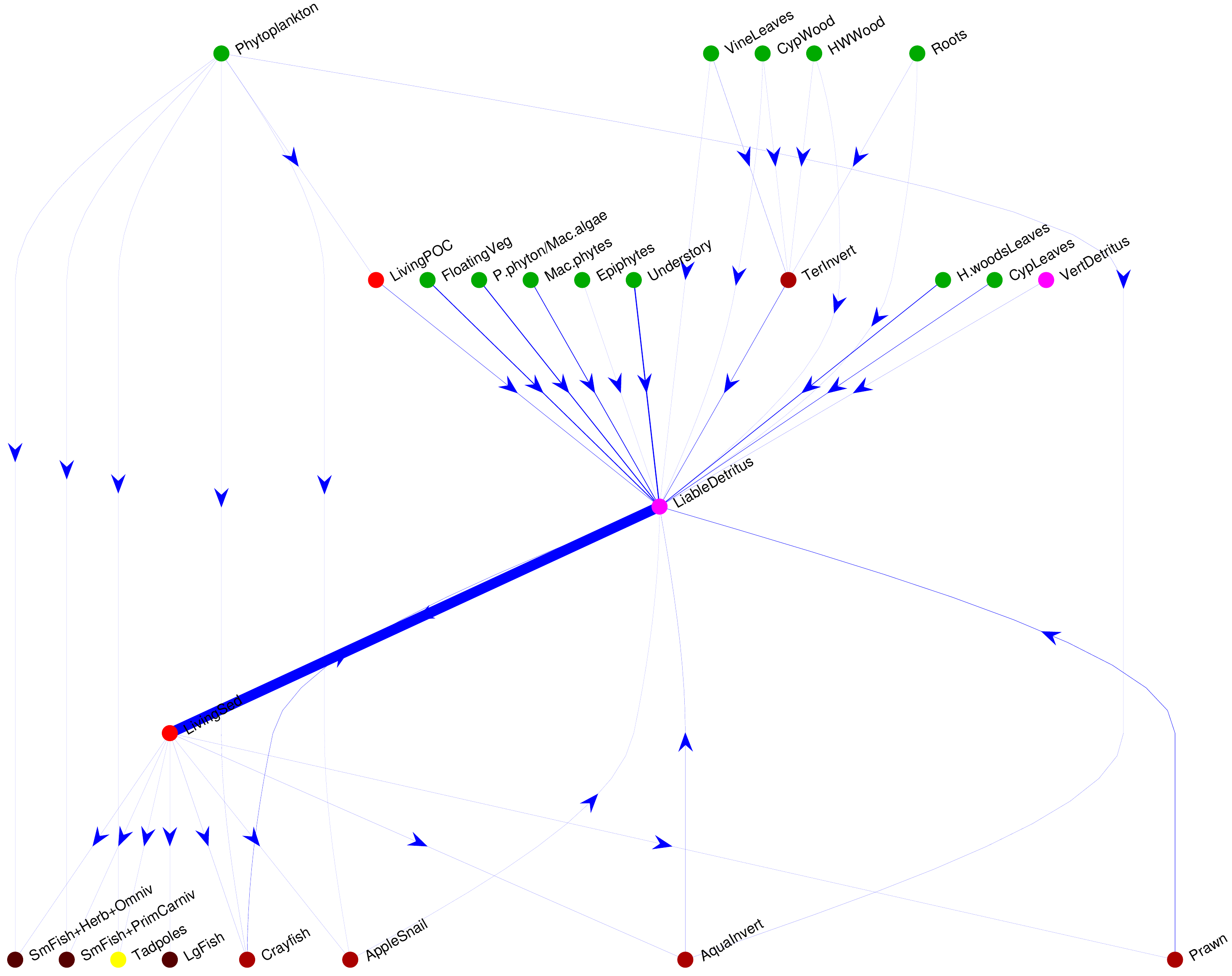}
  \caption{%
  \tb{DPE clustering ($K = 10$) of the Florida Bay Cypress Wetlands dry season food web} represented as a 68-vertex/554-edge weighted digraph \cite{Bondavalli2000JB}. Each vertex represents an organism and each directed edge (with edge width proportional to edge weight) represents carbon flow which reflects biomass exchange. Vertex colors indicate membership in one of 9 organism compartments. Here we see another 39-edge cluster showing  focused direct and indirect transfer that highlights the connections and importance of phytoplankton, liable detritus, and living sediment.  Figs~\ref{fig:Cypress_10Clusters_100_LDPE}-\ref{fig:Cypress_10Clusters_36_LDPE} show all the clusters obtained when DPE clustering is applied with $K=10$.}
\label{fig:Cypress_10Clusters_38b_LDPE}
\end{figure}

\newpage

\begin{figure}[H]
  \centering
  \includegraphics[width=\textwidth]{%
      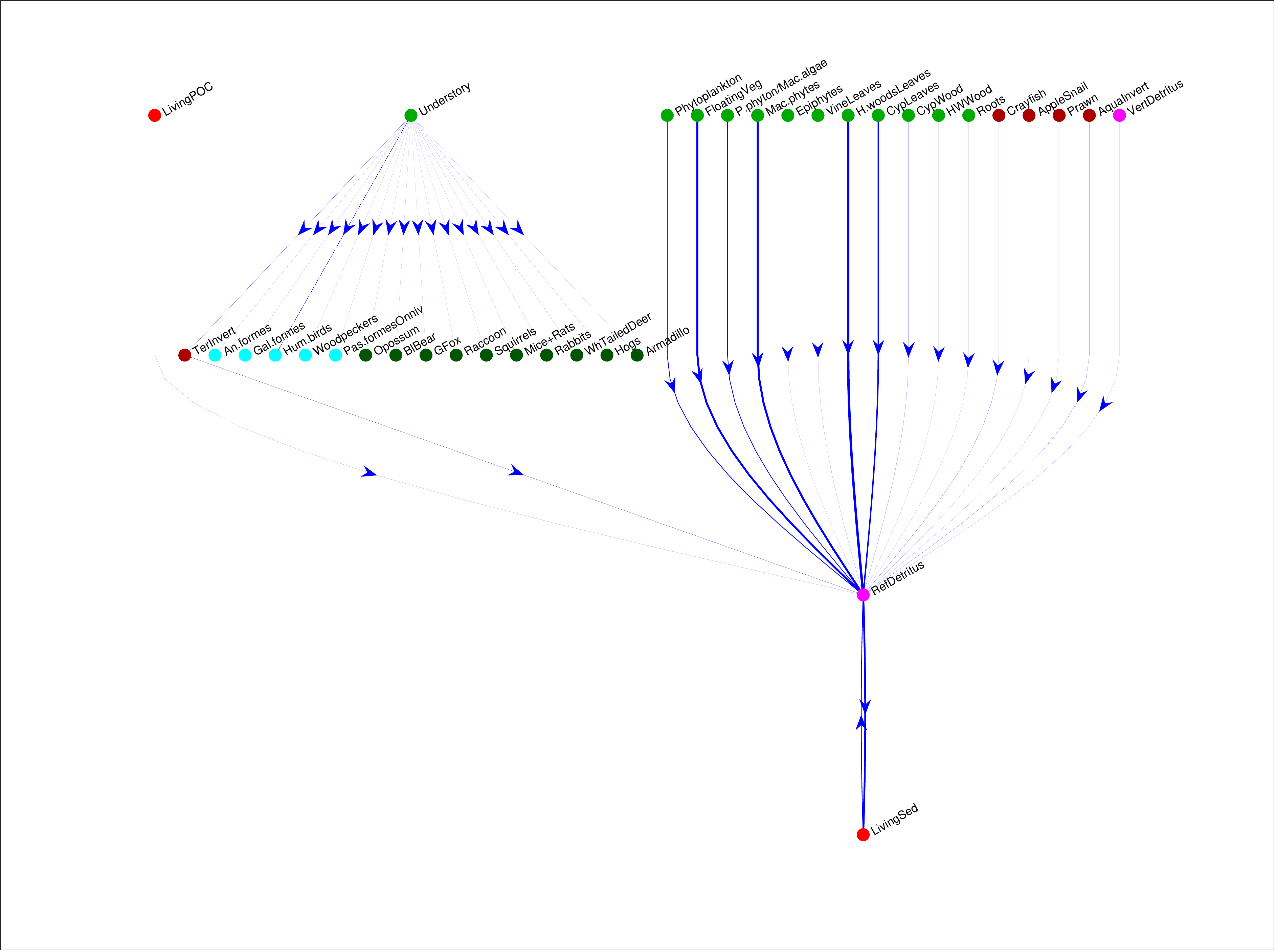}
  \caption{%
  \tb{DPE clustering ($K = 10$) of the Florida Bay Cypress Wetlands dry season food web} represented as a 68-vertex/554-edge weighted digraph \cite{Bondavalli2000JB}. Each vertex represents an organism and each directed edge (with edge width proportional to edge weight) represents carbon flow which reflects biomass exchange. Vertex colors indicate membership in one of 9 organism compartments. Here we see a 36-edge cluster highlighting focused production of refractory detritus, as well as how the understory produces carbon for a number of organisms. Figs~\ref{fig:Cypress_10Clusters_100_LDPE}-\ref{fig:Cypress_10Clusters_36_LDPE} show all the clusters obtained when DPE clustering is applied with $K=10$.}
\label{fig:Cypress_10Clusters_36_LDPE}
\end{figure}

\newpage

\begin{figure}[H]
  \centering
  \subfigure[231-edge cluster.]{%
  \includegraphics[width=3.75in]{%
      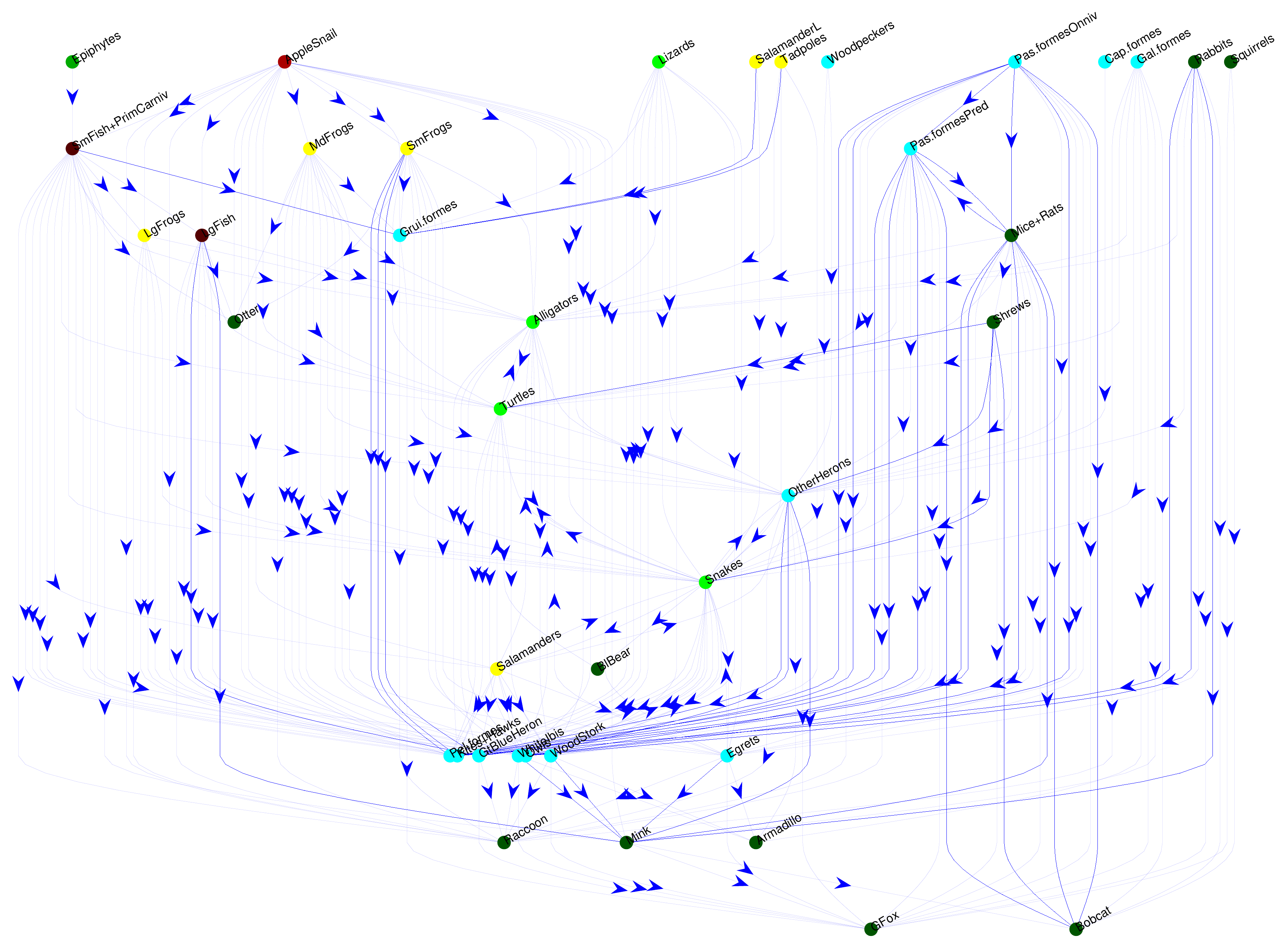}} \\
  \subfigure[56-edge cluster.]{%
  \includegraphics[width=3.75in]{%
      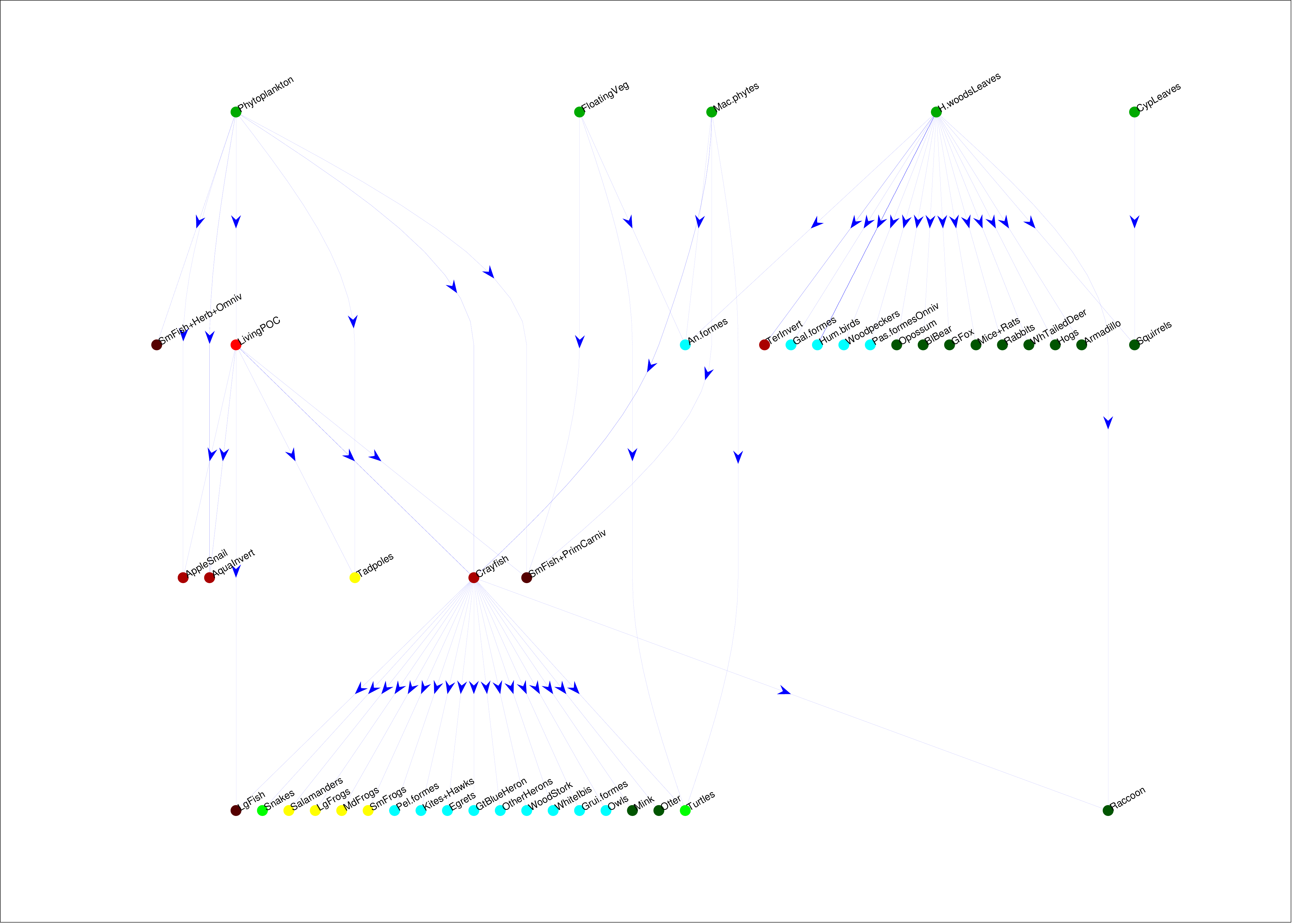}} 
  \caption{%
  \tb{RGE clustering ($K = 10$) of the Florida Bay Cypress Wetlands dry season food web} represented as a 68-vertex/554-edge weighted digraph \cite{Bondavalli2000JB} with $K=10$. Each vertex represents an organism and each directed edge (with edge width proportional to edge weight) represents carbon flow which reflects biomass exchange. Vertex colors indicate membership in one of 9 organism compartments.
  \tb{(a)} Here we see the largest edge cluster consisting of 231 edges showing concentrated carbon flows among a number of organisms.  This large edge cluster has the most concentrated flows of carbon. 
  \tb{(b)} A 56 edge cluster showing phytoplankton and living POC with outgoing flows, particularly to crayfish which provides carbon to a large number of organisms.  Also see Hardwood leaves producing carbon to a large number of organisms.}
\label{fig:Cypress_RGE_K10A}
\end{figure}

\newpage

\begin{figure}[H]
  \centering
  \subfigure[51-edge cluster.]{%
  \includegraphics[width=3.75in]{%
      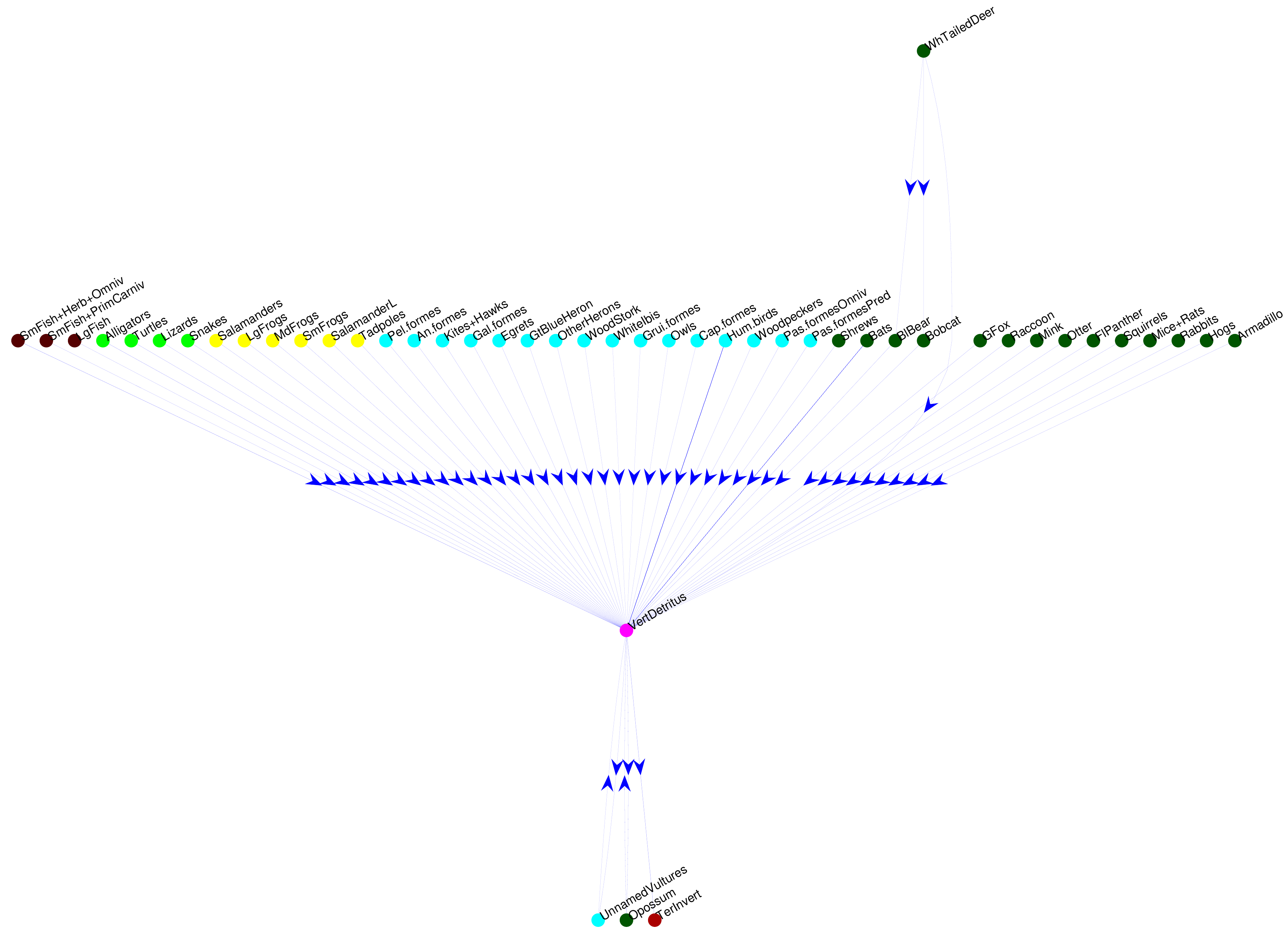}} \\
  \subfigure[47-edge cluster.]{%
  \includegraphics[width=3.75in]{%
      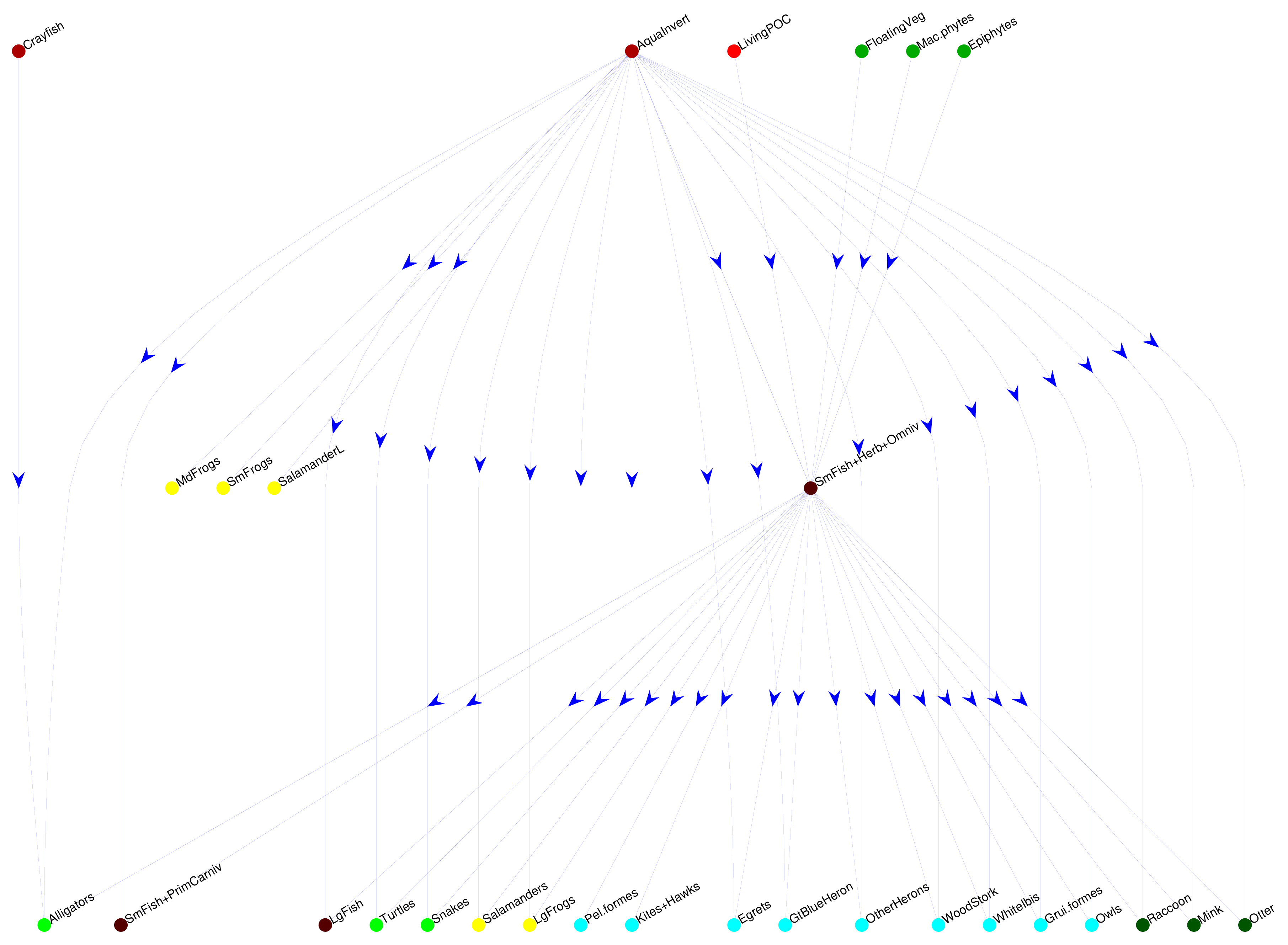}} 
  \caption{%
  \tb{RGE clustering ($K = 10$) of the Florida Bay Cypress Wetlands dry season food web} represented as a 68-vertex/554-edge weighted digraph \cite{Bondavalli2000JB}. Each vertex represents an organism and each directed edge (with edge width proportional to edge weight) represents carbon flow which reflects biomass exchange. Vertex colors indicate membership in one of 9 organism compartments.
  \tb{(a)} Here we see a 51 edge cluster that mainly shows vertebrate detritus is formed, with this detritus supplying carbon to vultures, as well as vultures becoming detritus.
  \tb{(b)} This is a 47 edge cluster that highlights how both aquatic invertebrates and small fish that are both herbivorous and omnivorous provide carbon flows to a number of the same organisms, such as alligators, turtles, egrets, and raccoons.}
\label{fig:Cypress_RGE_K10B}
\end{figure}

\newpage

\begin{figure}[H]
  \centering
  \subfigure[43-edge cluster.]{%
  \includegraphics[width=3.75in]{%
      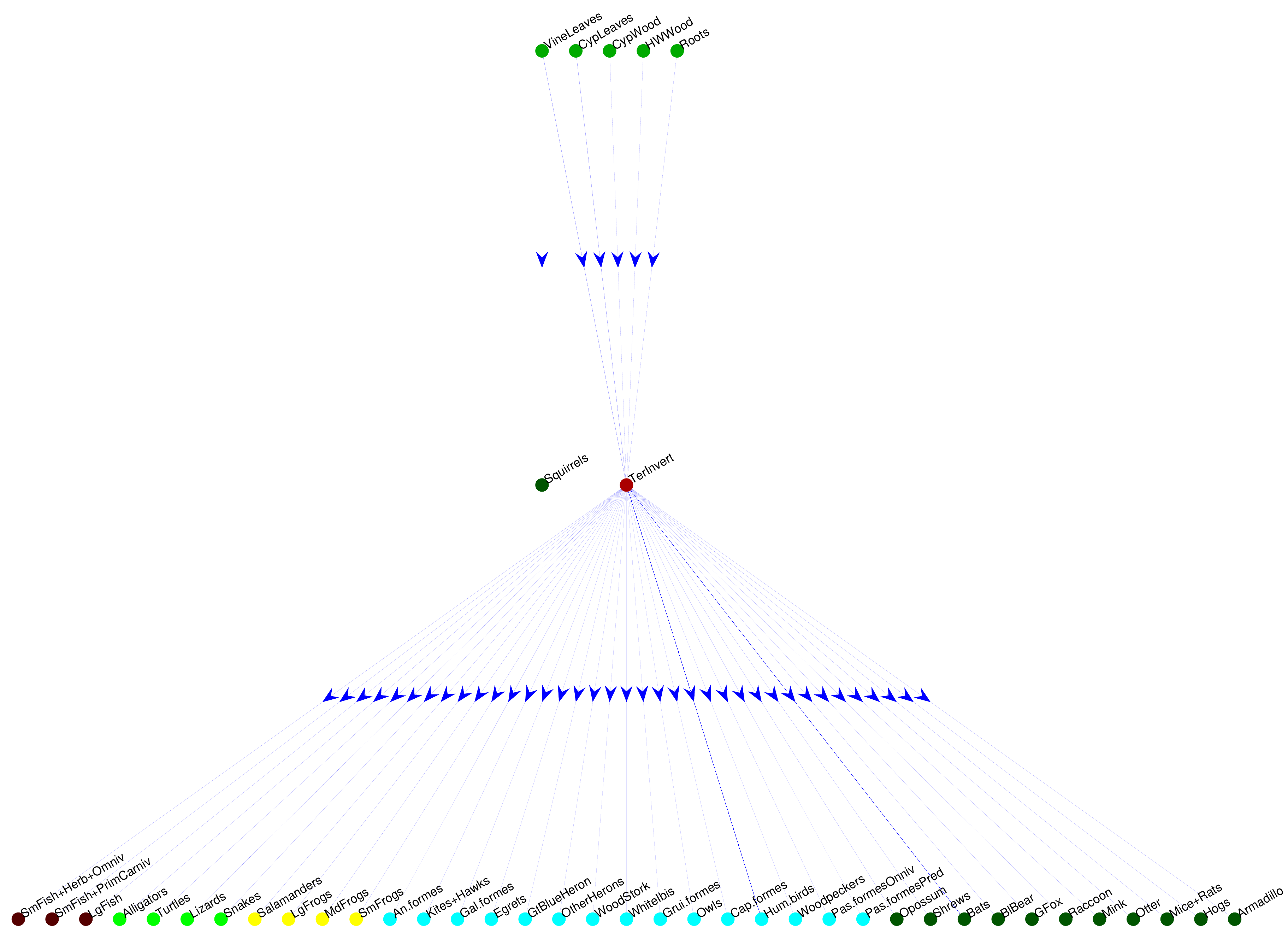}} \\
  \subfigure[40-edge cluster.]{%
  \includegraphics[width=3.75in]{%
      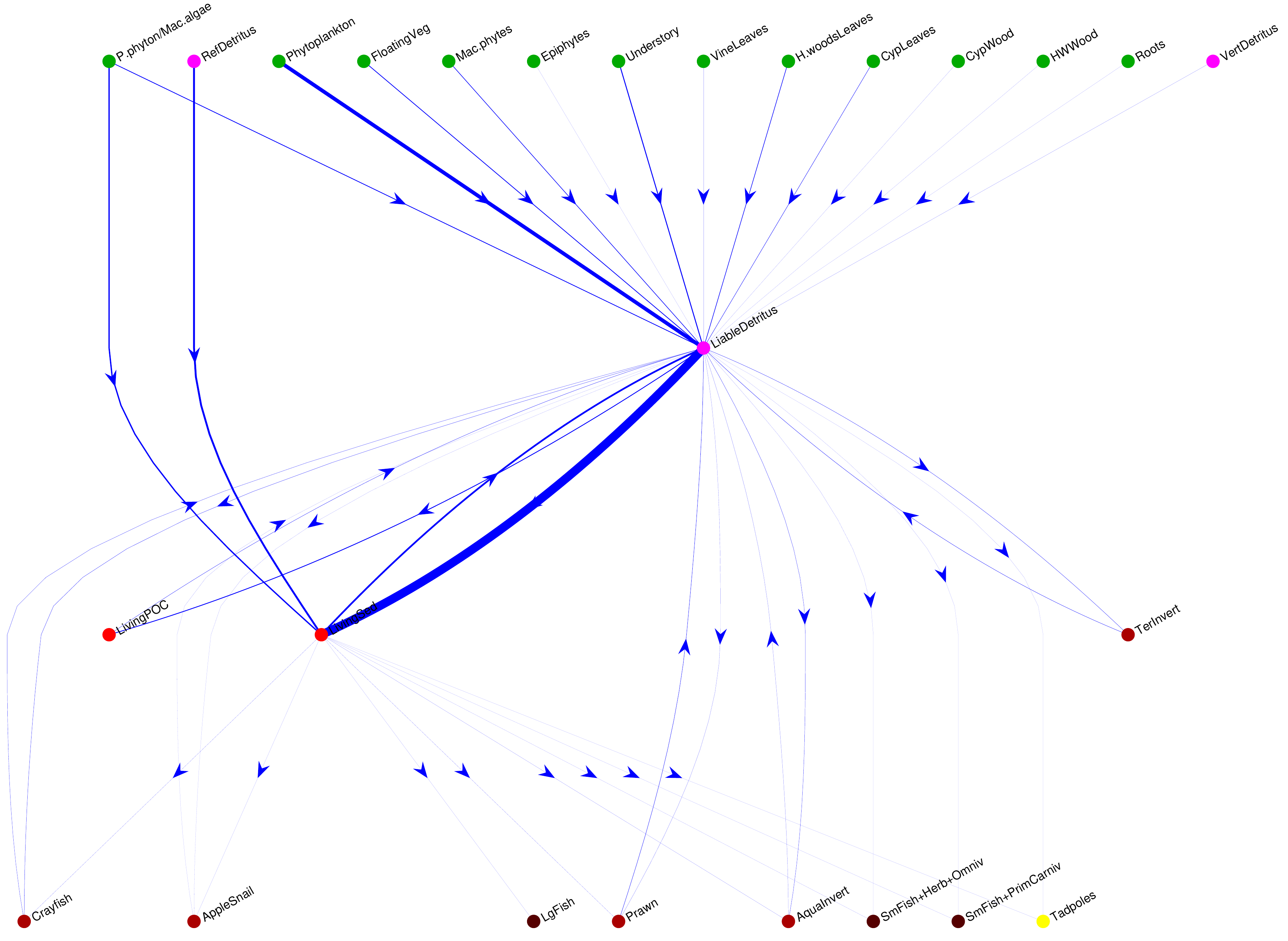}} 
  \caption{%
  \tb{RGE clustering ($K = 10$) of the Florida Bay Cypress Wetlands dry season food web} represented as a 68-vertex/554-edge weighted digraph \cite{Bondavalli2000JB}. Each vertex represents an organism and each directed edge (with edge width proportional to edge weight) represents carbon flow which reflects biomass exchange. Vertex colors indicate membership in one of 9 organism compartments.
  \tb{(a)} Here we see a 43 edge cluster highlighting carbon flows to terrestrial invertebrates which in turn provide carbon to a number of organisms. 
  \tb{(b)} This 40 edge cluster shows how liable detritus is formed, and how it also provides carbon, including a large flow to living sediment.  Living sediment in turn provides carbon to a number of organisms, e.g., crayfish, apple snail, prawns.}
\label{fig:Cypress_RGE_K10C}
\end{figure}

\newpage

\begin{figure}[H]
  \centering
  \subfigure[31-edge cluster.]{%
  \includegraphics[width=3.75in]{%
      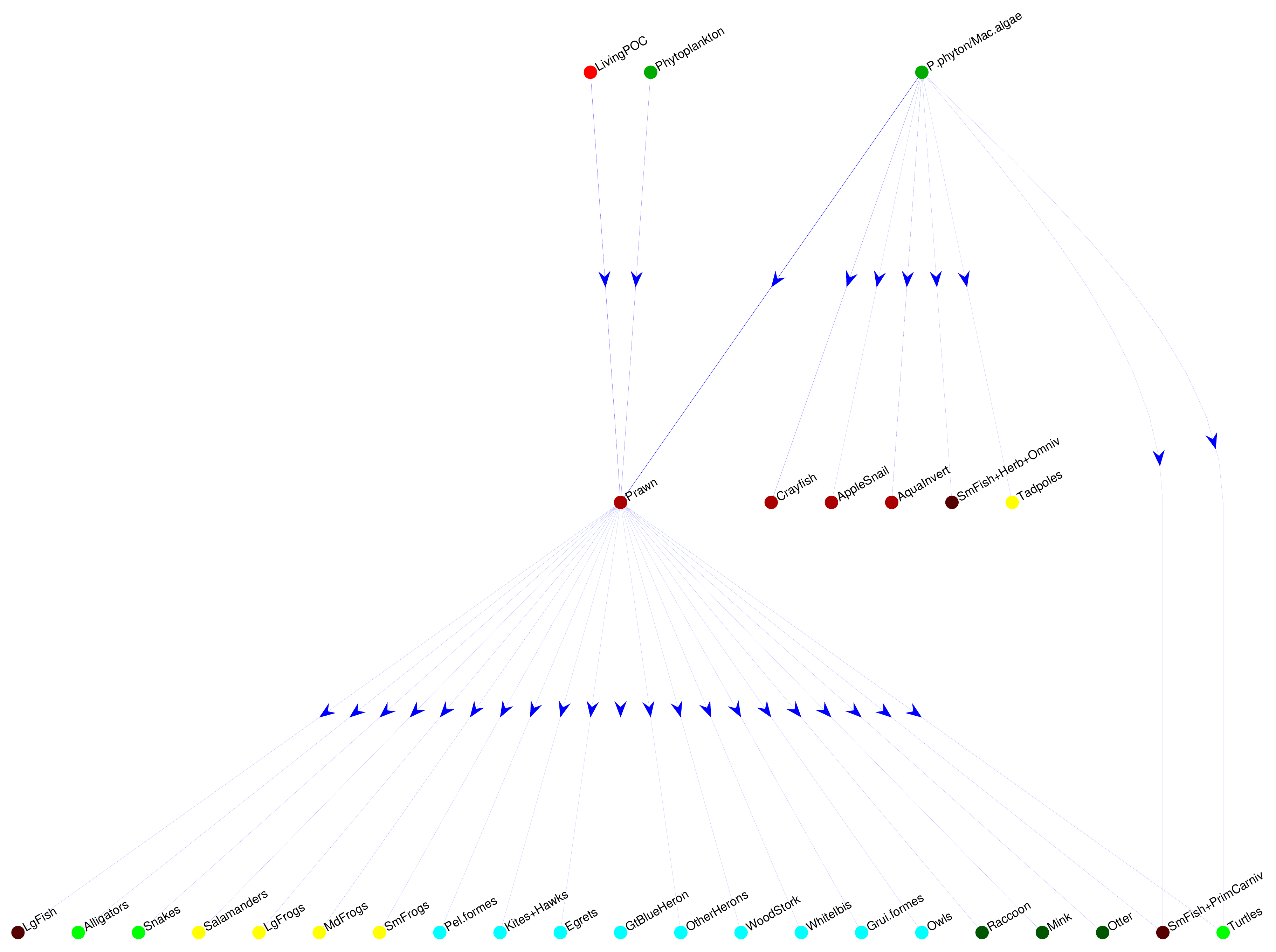}} \\
  \subfigure[25-edge cluster.]{%
  \includegraphics[width=3.75in]{%
      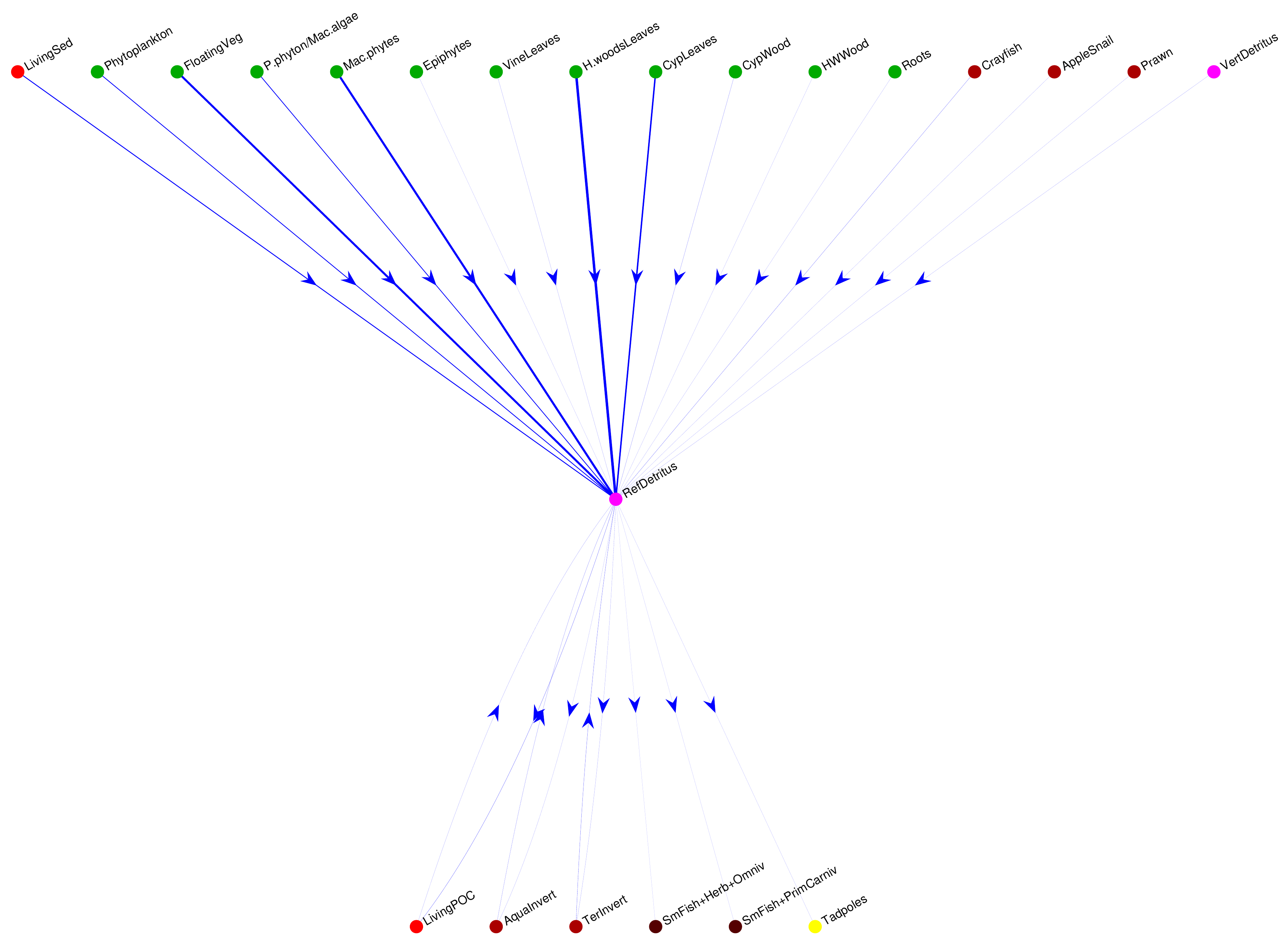}} 
  \caption{%
  \tb{RGE clustering ($K = 10$) of the Florida Bay Cypress Wetlands dry season food web} represented as a 68-vertex/554-edge weighted digraph \cite{Bondavalli2000JB}. Each vertex represents an organism and each directed edge (with edge width proportional to edge weight) represents carbon flow which reflects biomass exchange. Vertex colors indicate membership in one of 9 organism compartments.
  \tb{(a)} Here we have a 31 edge cluster showing carbon flows from periphyton/macroalgae including a flow to prawns which in turn provide carbon to a number of organisms.
  \tb{(b)} This 25 edge cluster primarily shows how carbon flows lead to the production of refractory detritus, and how this detritus also provides carbon, e.g., to tadpoles.}
\label{fig:Cypress_RGE_K10D}
\end{figure}

\newpage

\begin{figure}[H]
  \centering
  \subfigure[17-edge cluster.]{%
  \includegraphics[width=3.75in]{%
      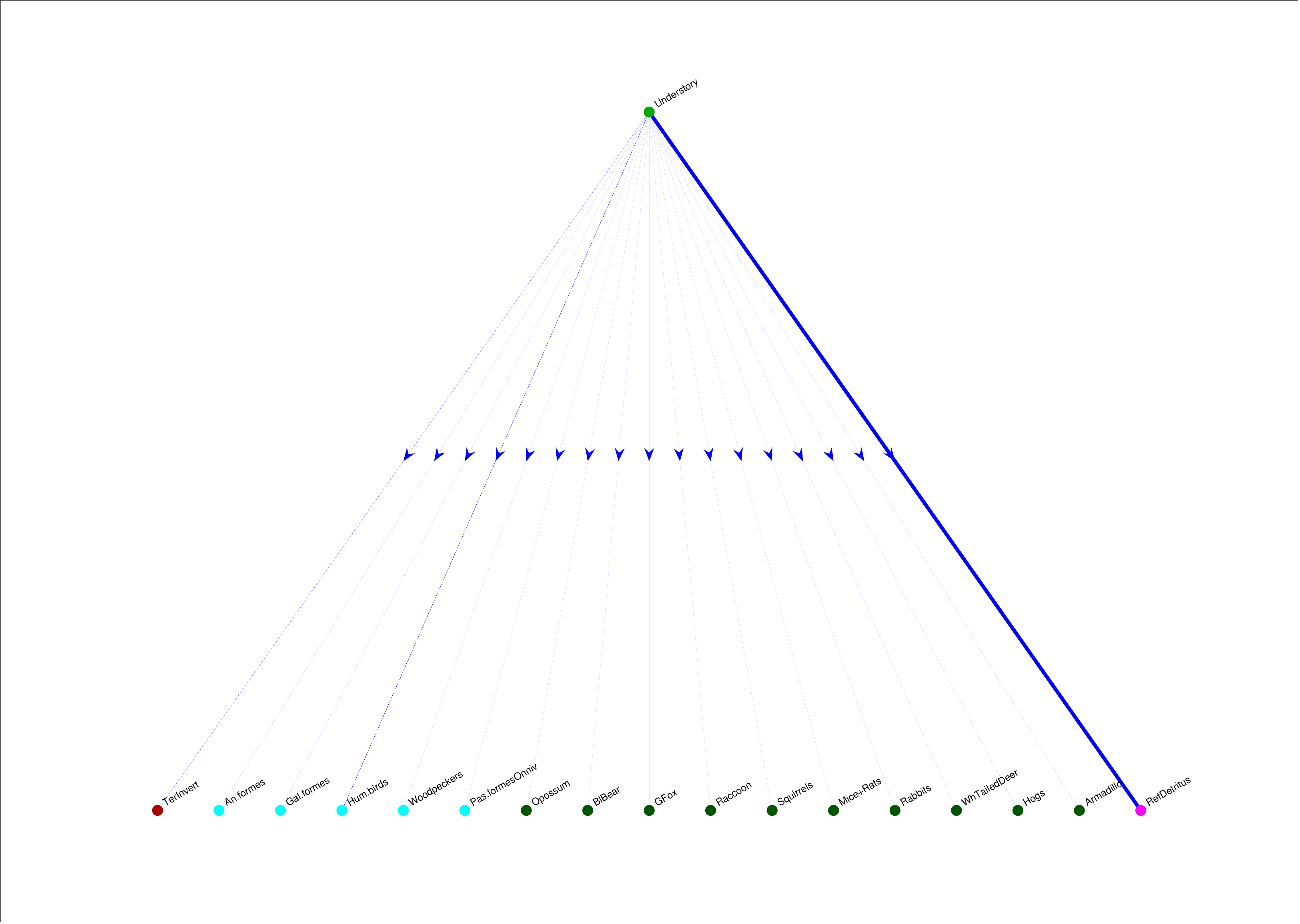}} \\
  \subfigure[13-edge cluster.]{%
  \includegraphics[width=3.75in]{%
      ./Figs/Cypress_RGE_K10_N13_M13}} 
  \caption{%
  \tb{RGE clustering ($K = 10$) of the Florida Bay Cypress Wetlands dry season food web} represented as a 68-vertex/554-edge weighted digraph \cite{Bondavalli2000JB}. Each vertex represents an organism and each directed edge (with edge width proportional to edge weight) represents carbon flow which reflects biomass exchange. Vertex colors indicate membership in one of 9 organism compartments. 
  \tb{(a)} This 17 edge cluster shows how understory provides carbon to a number of organisms.
  \tb{(b)} Fig.~\ref{fig:Cypress_RGE_K10E}(b) is a reprint of the cluster in Fig.~\ref{fig:USMigration_Cypress}(d). It shows the critical carbon flows associated with the endangered Florida panther, including carbon flows to hogs which in turn provide carbon to the panther and black bears.}
\label{fig:Cypress_RGE_K10E}
\end{figure}

\newpage

\begin{figure}[H]
  \centering
  \includegraphics[width=5.5in]{%
      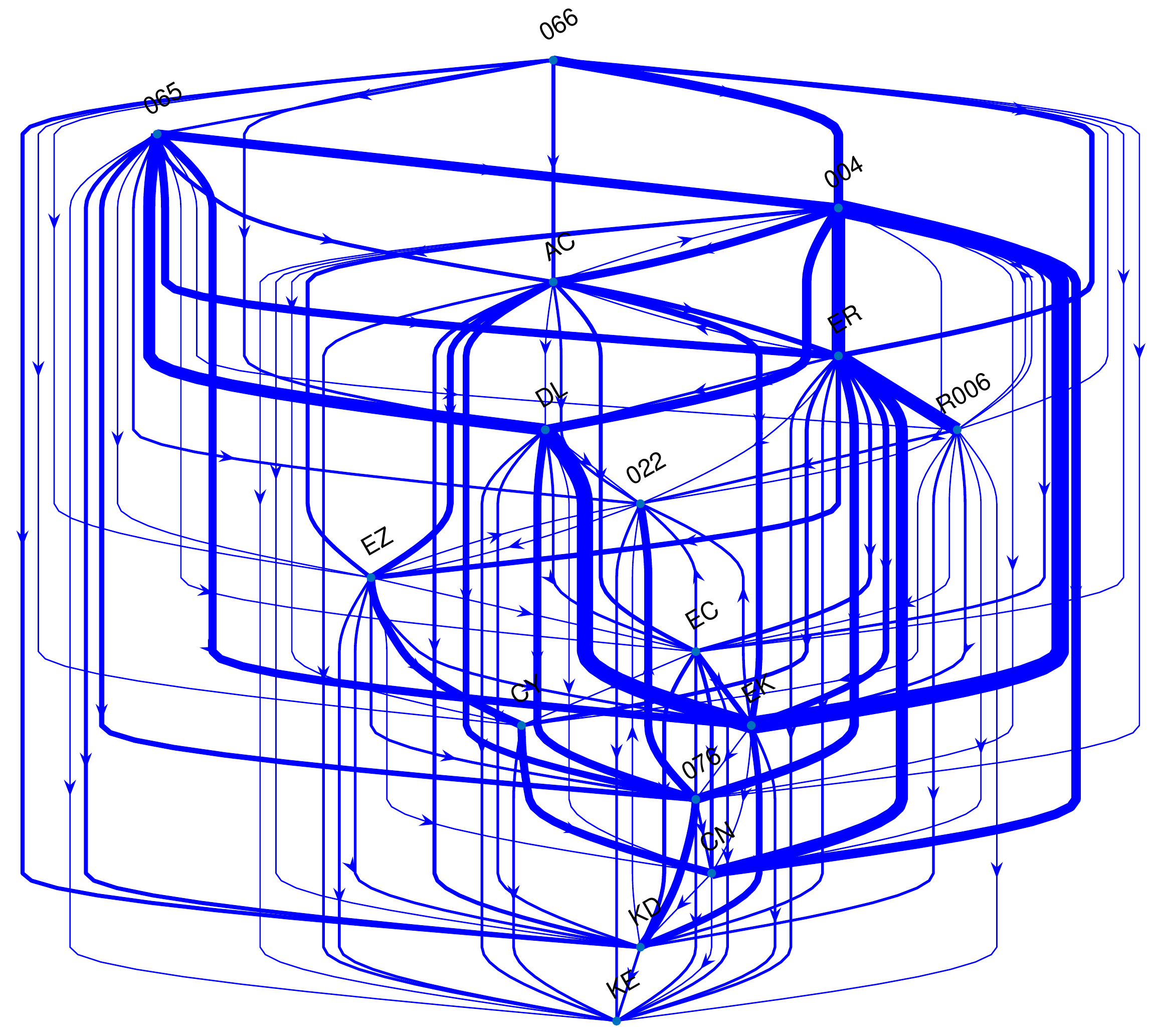}
  \caption{%
  The dominance relationhips of a \tb{rhesus monkey population} represented as a 16-vertex/105-edge weighted digraph \cite{Sade1972FP}. The start vertex and end-vertex of each directed edge indicates the monkey who is dominating and the monkey who is being dominated, respectively. Edge width is proportional to edge weight (which reflects number of encounters between two monkeys). Blue and pink vertices identify the males and females, respectively. 
DPE clustering of this digraph appears in Figs~\ref{fig:Rhesus_6Clusters_LDPE} and \ref{fig:Rhesus_6Clusters_LDPEb}; and 
PRE clustering appears in Figs~\ref{fig:Rhesus_6Clusters_LPRE} and \ref{fig:Rhesus_6Clusters_LPREb}.}
\label{fig:Rhesus}
\end{figure}

\newpage

\begin{figure}[H]
  \centering
  \subfigure[29-edge cluster.]{%
  \includegraphics[width=2.75in]{%
      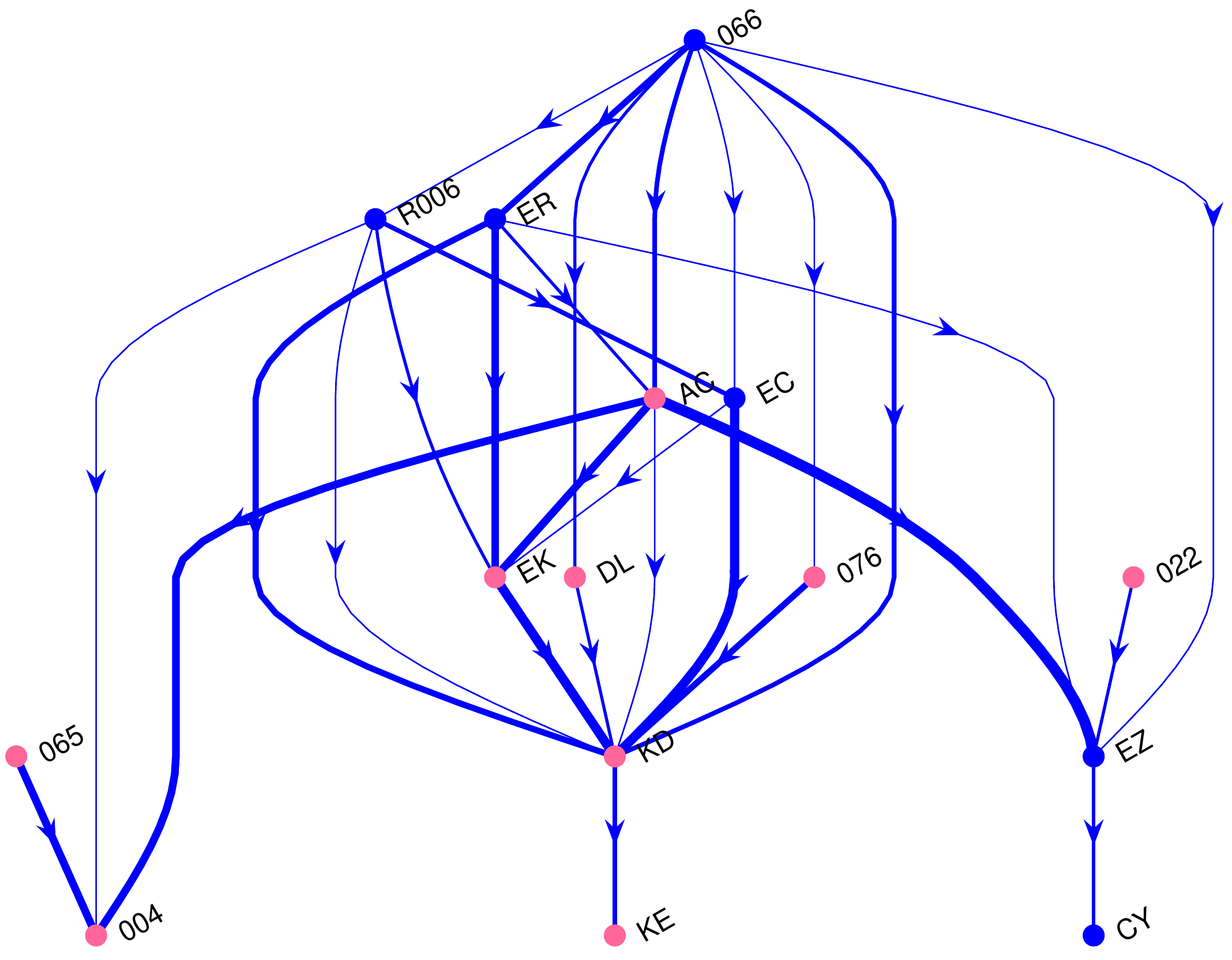}}
  \hfil
  \subfigure[19-edge cluster.]{%
  \includegraphics[width=2.75in]{%
      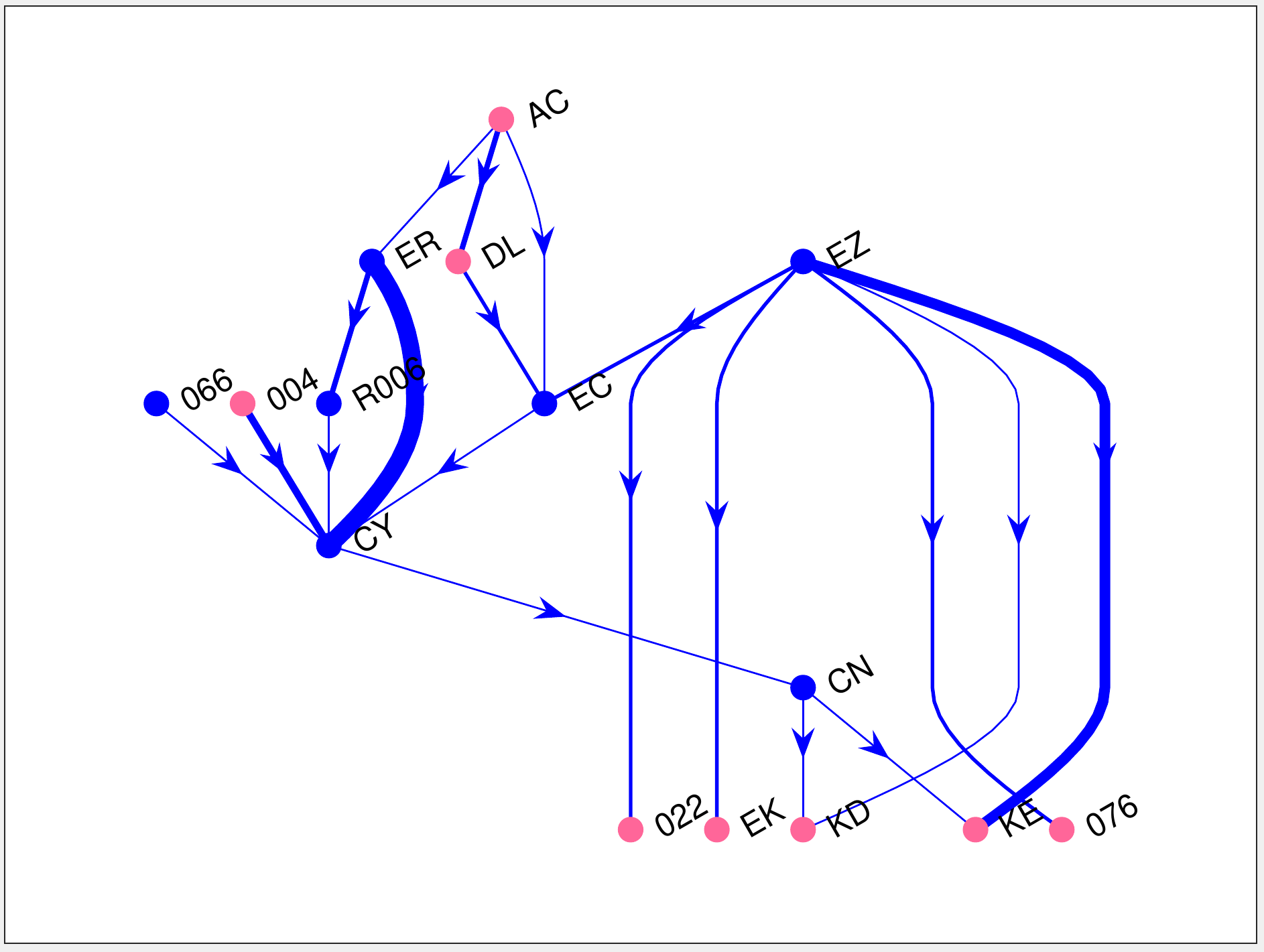}} \\
  \subfigure[16-edge cluster.]{%
  \includegraphics[width=2.75in]{%
      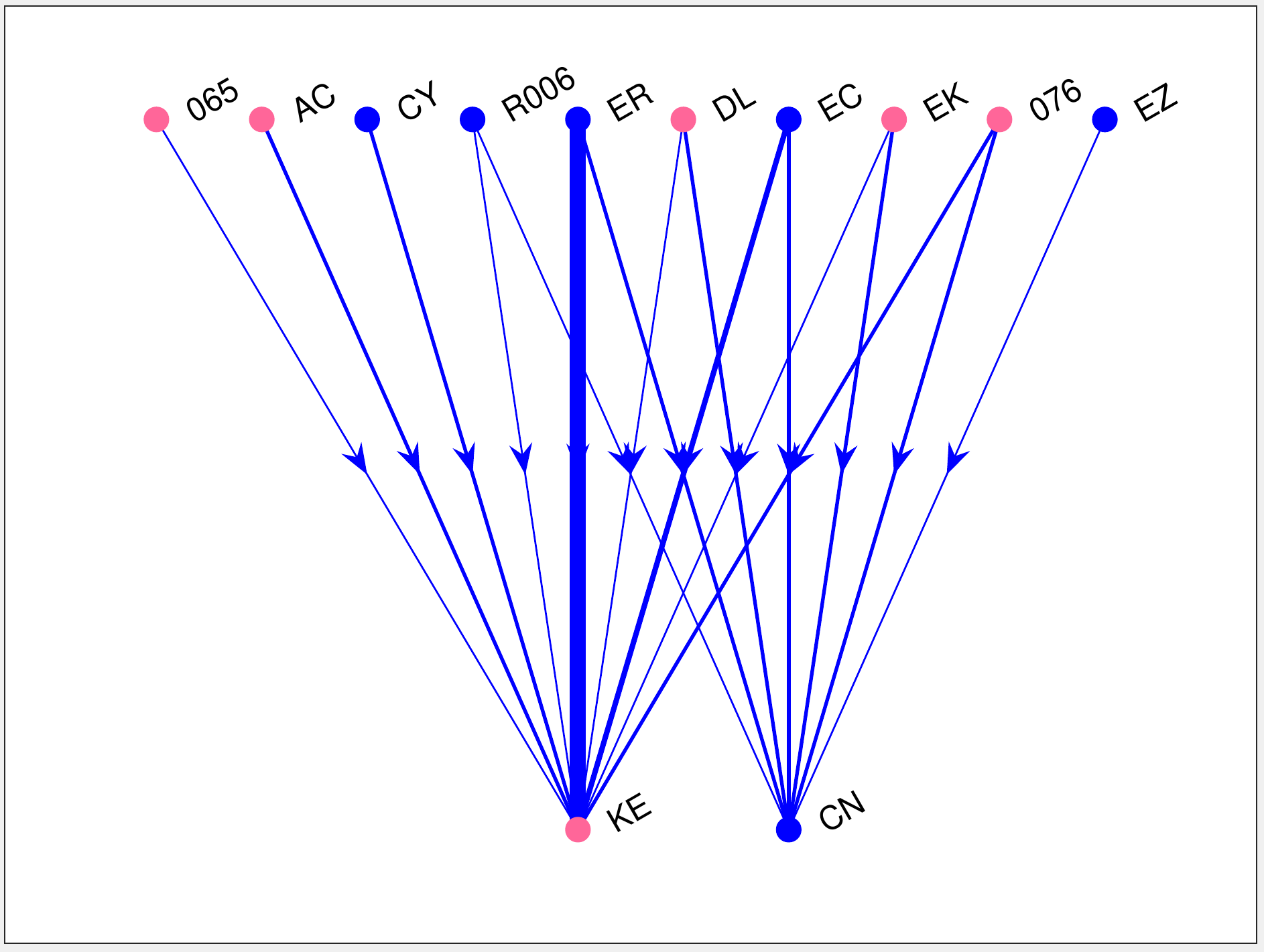}}
  \hfil
  \subfigure[Another 16-edge cluster.]{%
  \includegraphics[width=2.75in]{%
      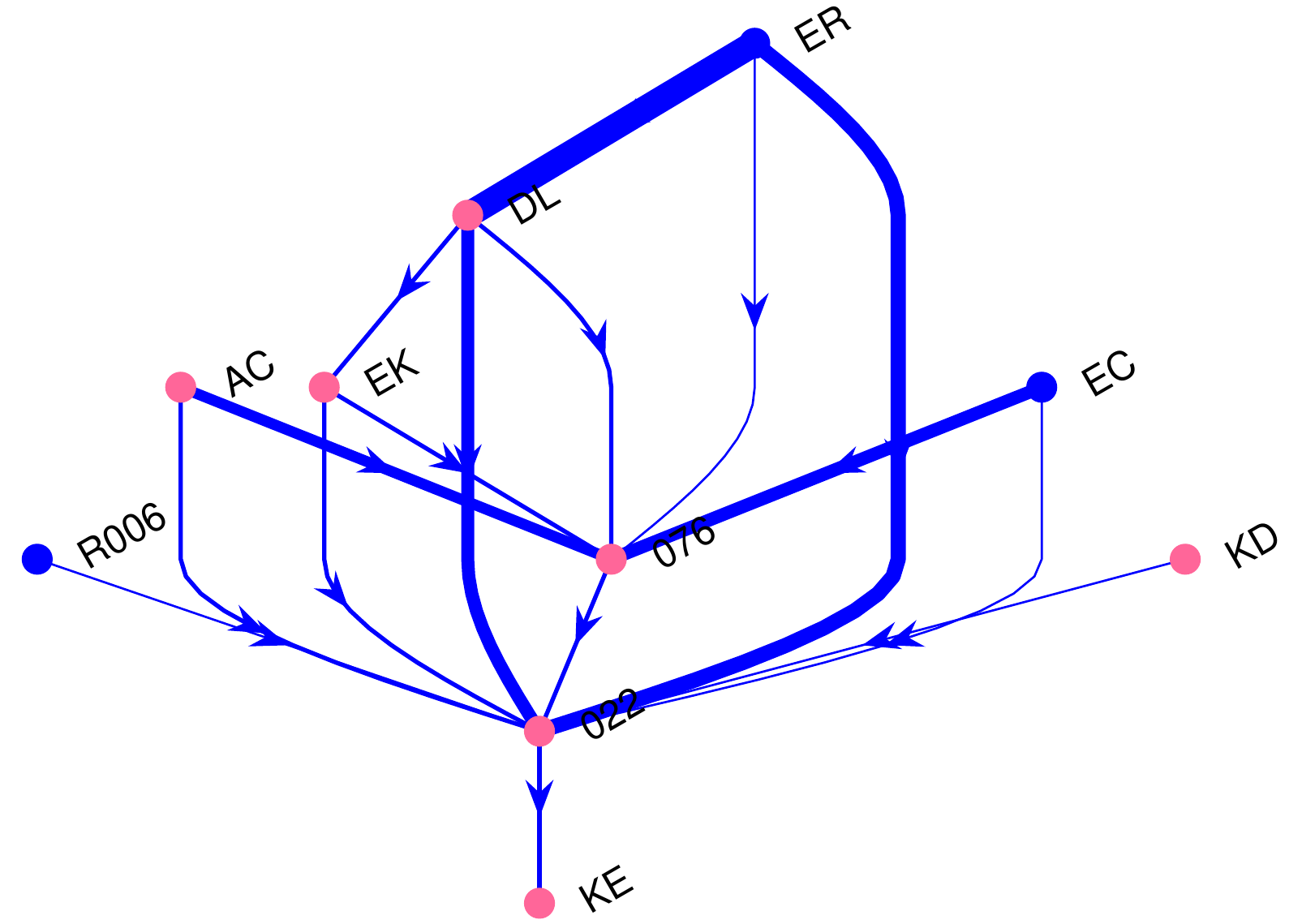}}
  \caption{%
  \tb{DPE clustering} ($K = 6$) of the dominance relationhips of a \tb{rhesus monkey population} represented as a weighted digraph \cite{Sade1972FP}. The start vertex and end-vertex of each directed edge indicates the monkey who is dominating and the monkey who is being dominated, respectively. Edge width is proportional to edge weight (which reflects number of encounters between two monkeys). Blue and pink vertices identify the males and females, respectively. DPE clustering reveals subgraph clusters of concentrated chains of hierarchical dominance. This figure shows the largest 4 clusters obtained when DPE clustering is applied with $K=6$. The high rank of the male ``066'' is apparent. The low rank of ``CN'' among the males and ``KE'' among the females is clear from these clusters. Fig.~\ref{fig:Rhesus_6Clusters_LDPEb} shows the remaining clusters.}
\label{fig:Rhesus_6Clusters_LDPE}
\end{figure}

\newpage

\begin{figure}[H]
  \centering
  \subfigure[14-edge cluster.]{%
  \includegraphics[width=2.75in]{%
      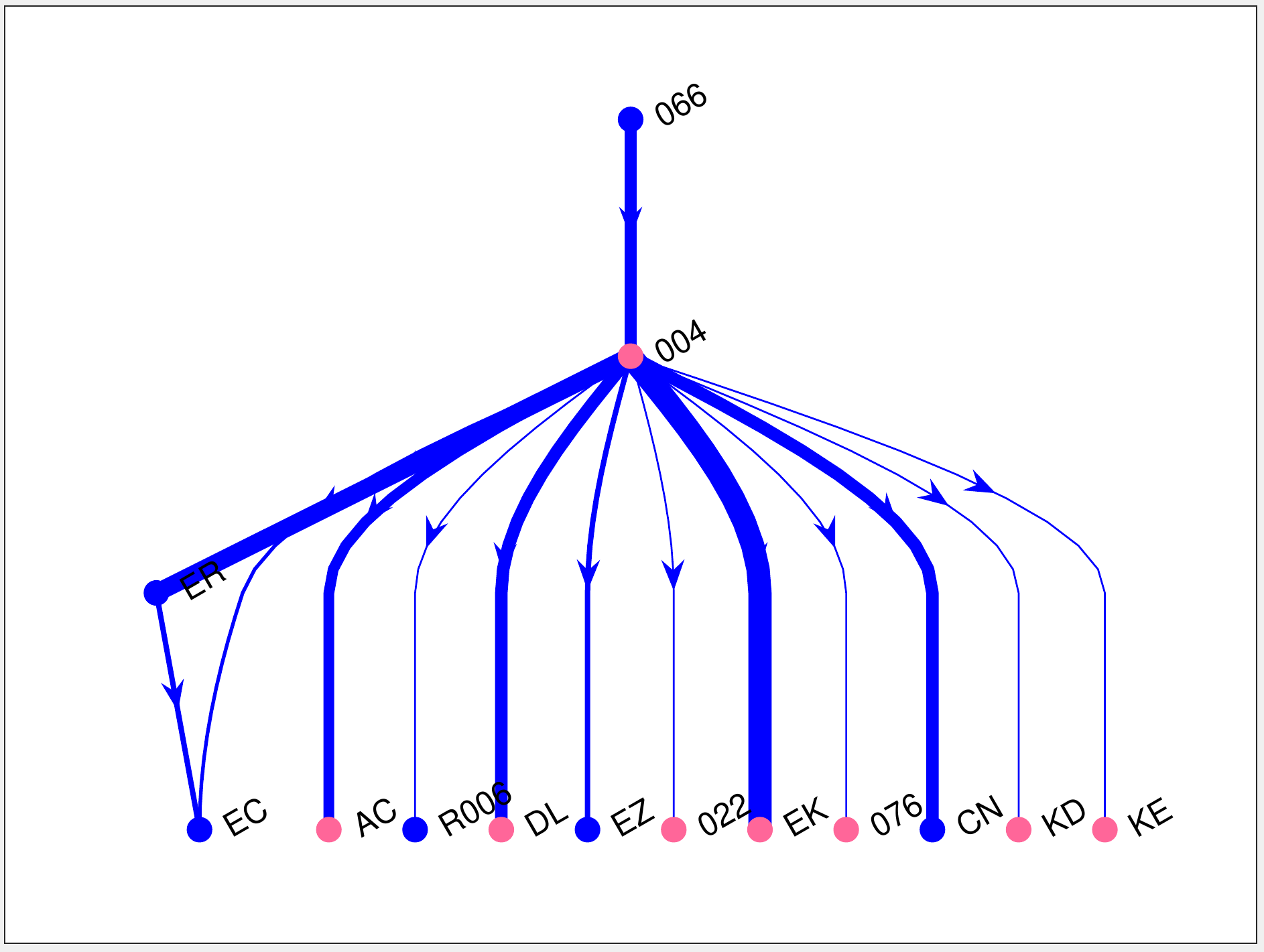}}
  \hfil
  \subfigure[11-edge cluster.]{%
  \includegraphics[width=2.75in]{%
      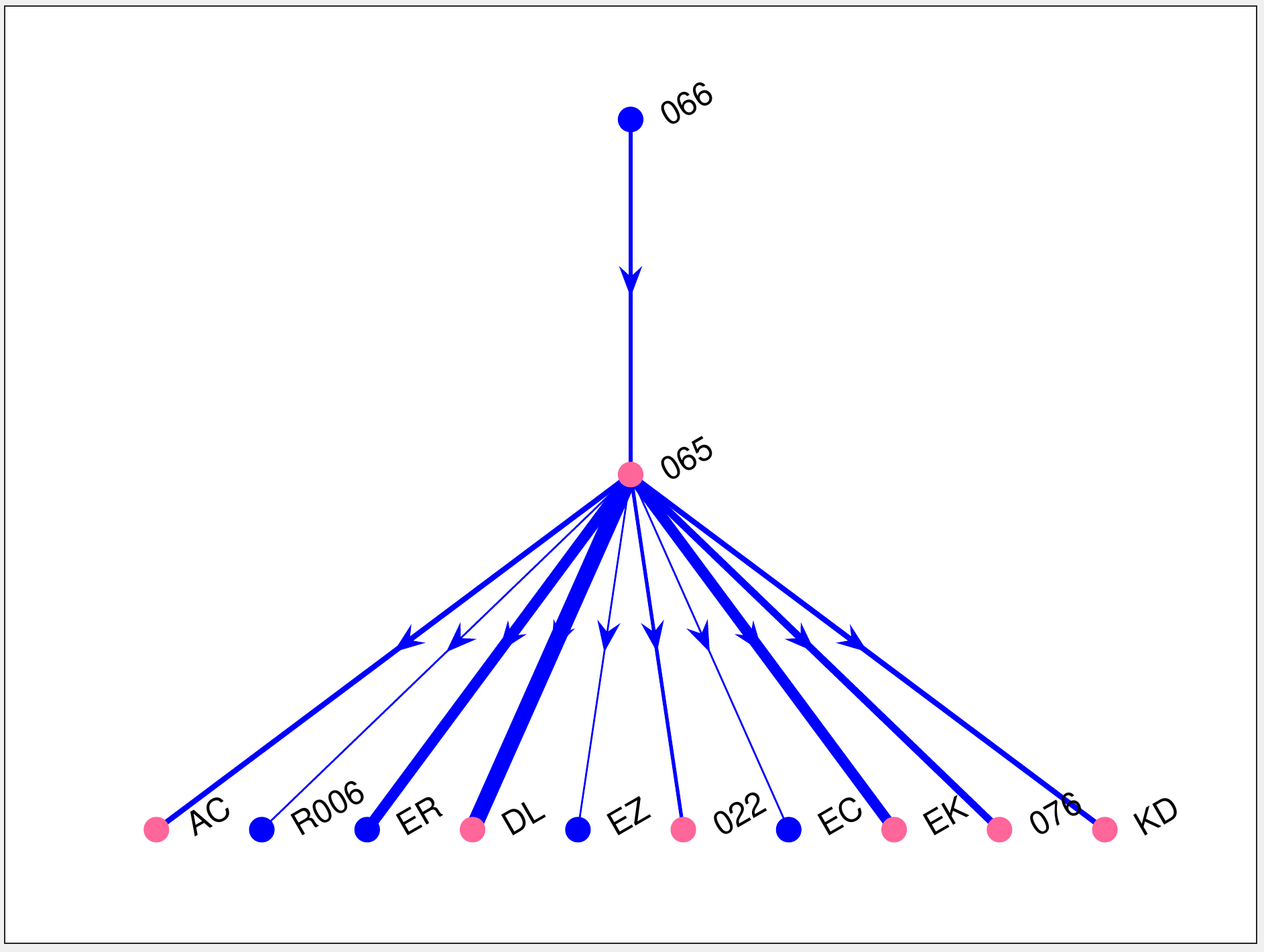}} 
  \caption{%
  (continued from Fig.~\ref{fig:Rhesus_6Clusters_LDPE}) 
  \tb{DPE clustering} ($K = 6$) of the dominance relationhips of a \tb{rhesus monkey population} represented as a weighted digraph \cite{Sade1972FP}. The start vertex and end-vertex of each directed edge indicates the monkey who is dominating and the monkey who is being dominated, respectively. Edge width is proportional to edge weight (which reflects number of encounters between two monkeys). Blue and pink vertices identify the males and females, respectively. DPE clustering reveals subgraph clusters of dominance. This figure shows the smallest 2 clusters obtained when DPE clustering is applied with $K=6$. The dominance of ``066'' among the males and the dominance of ``065'' and ``004'' among the females are clear from these two clusters. Figs~\ref{fig:Rhesus_6Clusters_LDPE} shows the remaining clusters.}
\label{fig:Rhesus_6Clusters_LDPEb}
\end{figure}

\newpage

\begin{figure}[H]
  \centering
  \subfigure[28-edge cluster.]{%
  \includegraphics[width=2.75in]{%
      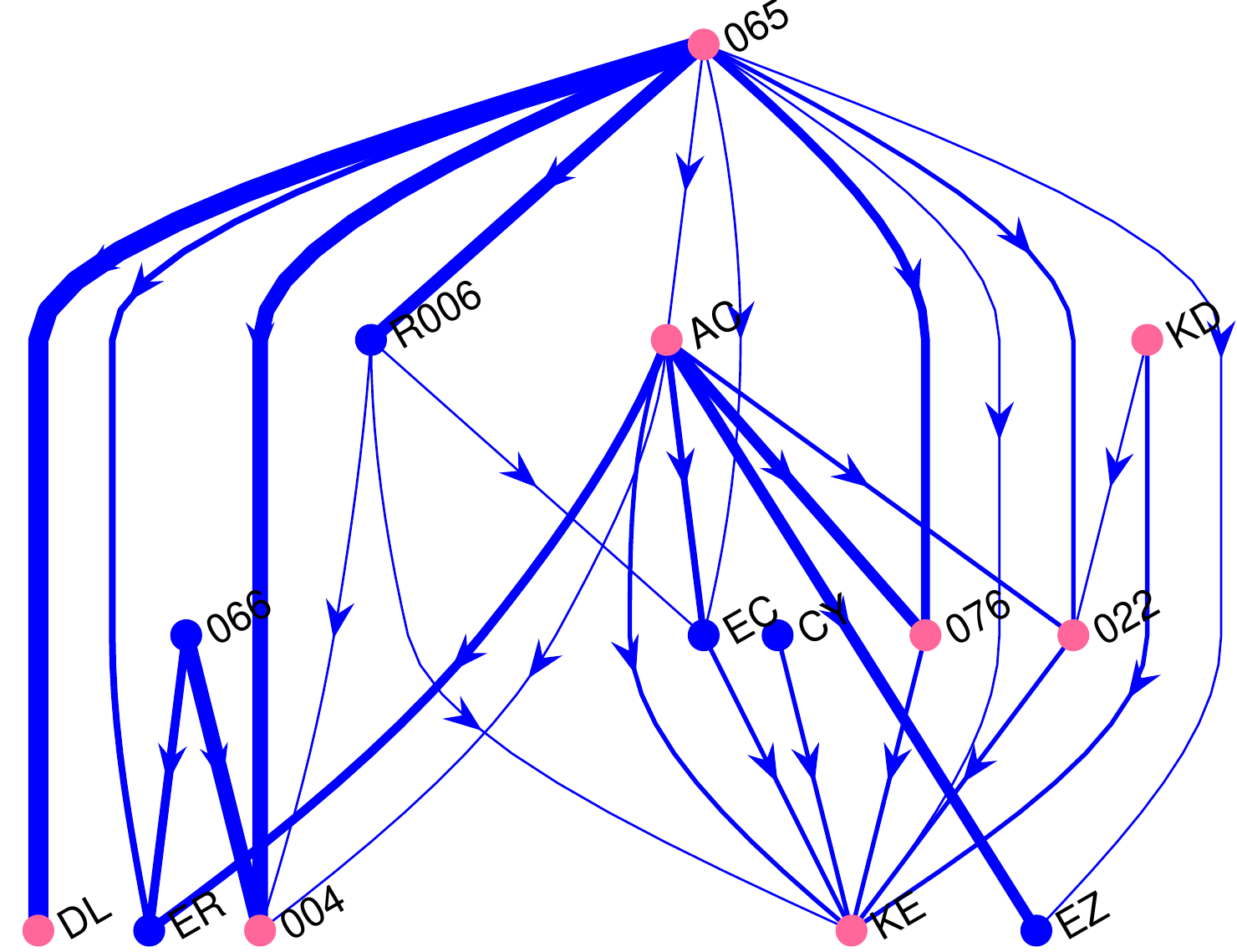}}
  \hfil
  \subfigure[20-edge cluster.]{%
  \includegraphics[width=2.75in]{%
      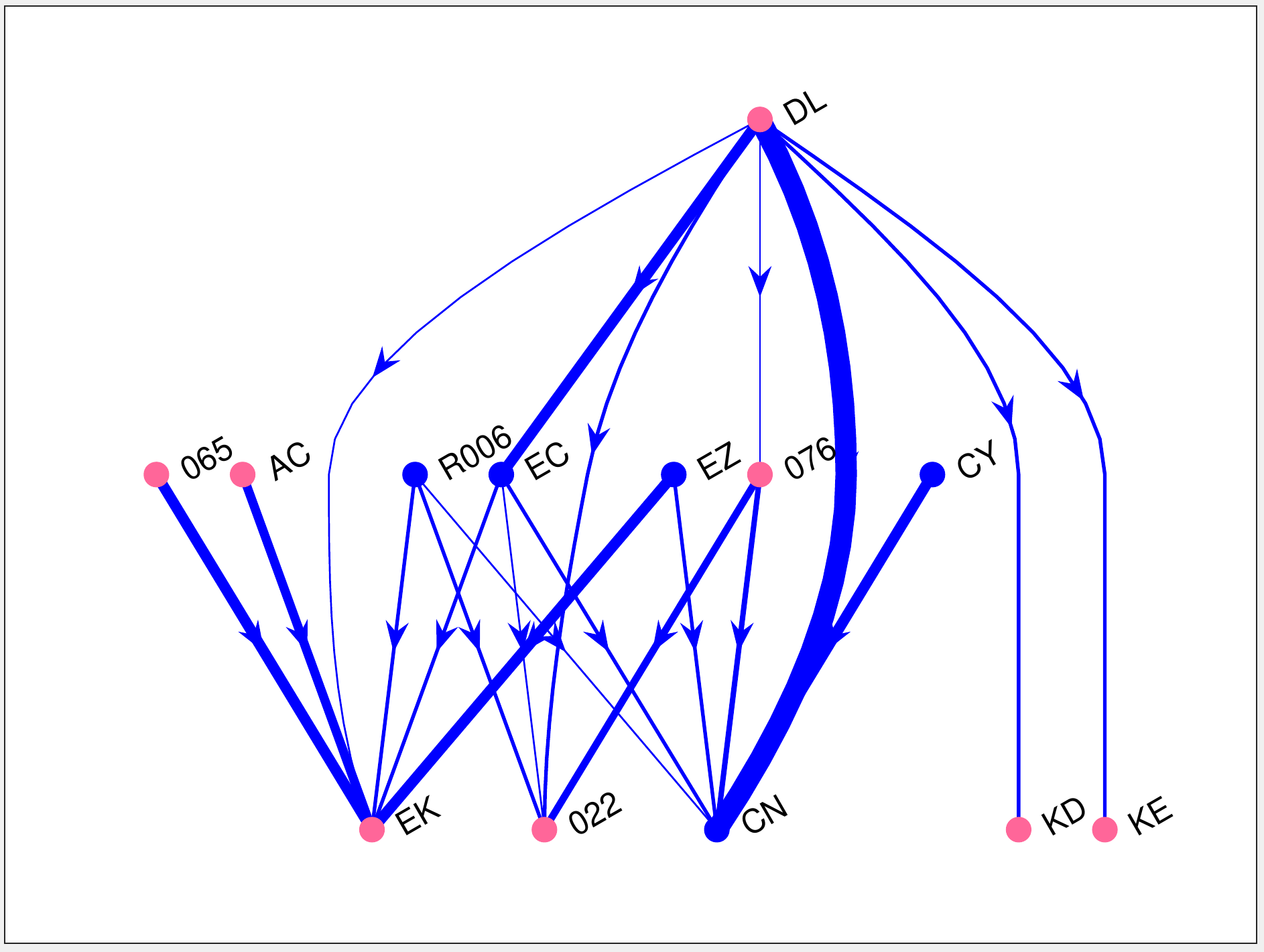}} \\
  \subfigure[18-edge cluster.]{%
  \includegraphics[width=2.75in]{%
      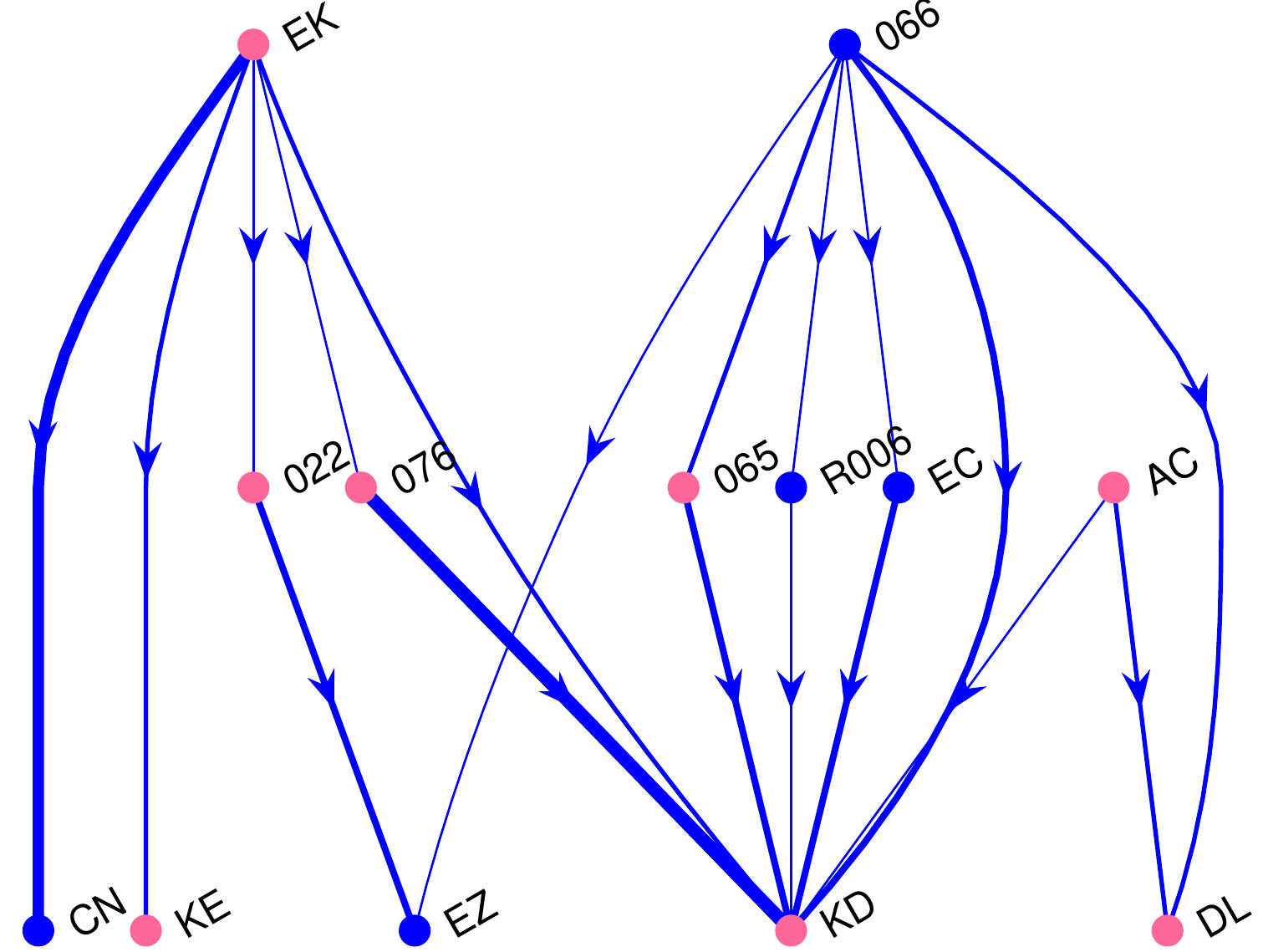}}
  \hfil
  \subfigure[13-edge cluster.]{%
  \includegraphics[width=2.75in]{%
      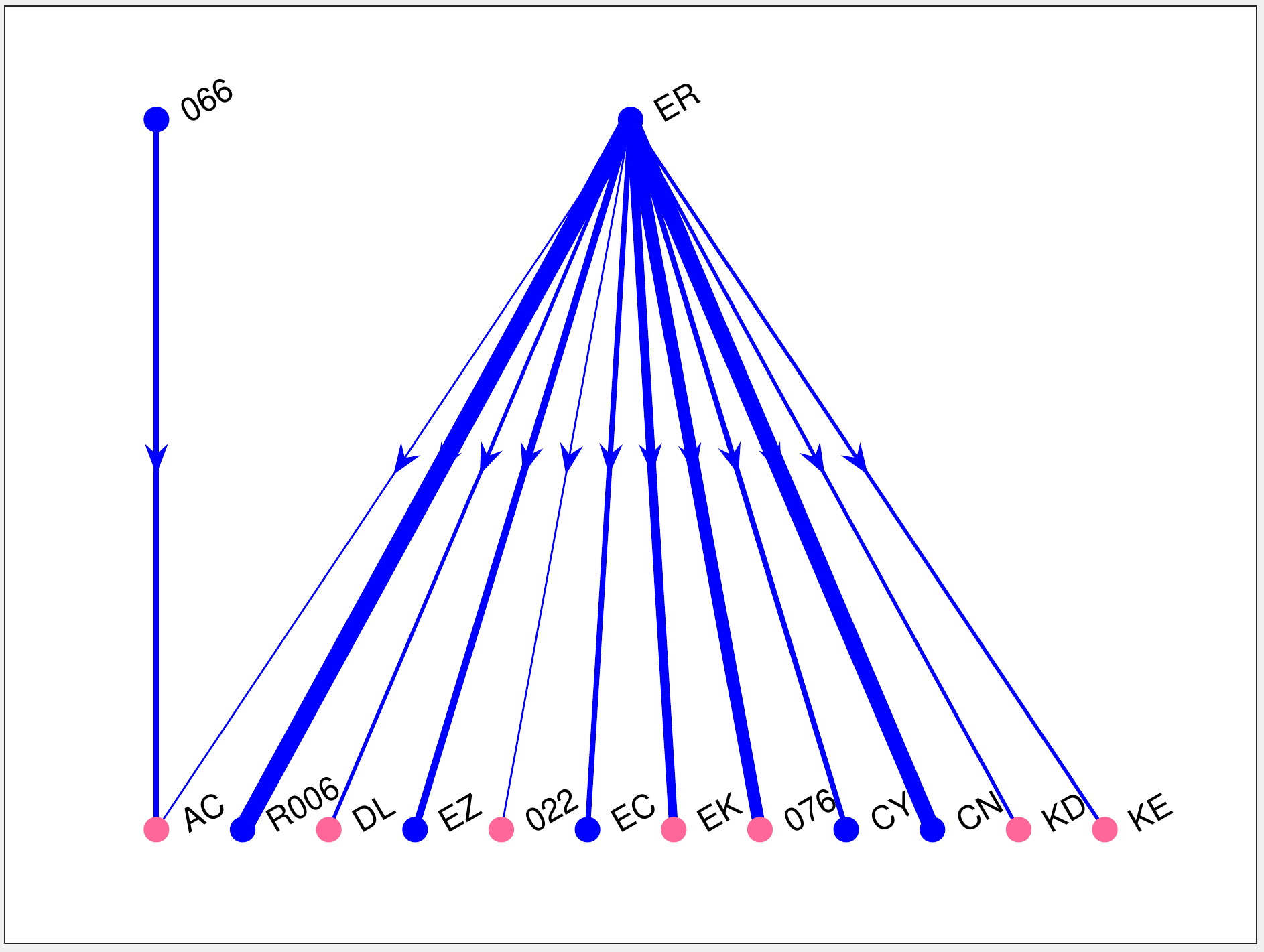}}
  \caption{%
  \tb{PRE clustering} ($K = 6$) of the dominance relationhips of a \tb{rhesus monkey population} represented as a weighted digraph \cite{Sade1972FP}. The start vertex and end-vertex of each directed edge indicates the monkey who is dominating and the monkey who is being dominated, respectively. Edge width is proportional to edge weight (which reflects number of encounters between two monkeys). Blue and pink vertices identify the males and females, respectively. This figure shows the largest 4 clusters obtained when PRE clustering is applied with $K=6$; Figs~\ref{fig:Rhesus_6Clusters_LPREb} shows the remaining clusters. The PRE clustering results in subgraphs with concentrated production and reception of dominance.  For example, here we see ``ER'' both producing dominance (a) and receiving dominance (d), while ``004'' produces dominance (b) and receives dominance (a), primarily asserted by ``0065''. }
\label{fig:Rhesus_6Clusters_LPRE}
\end{figure}

\newpage

\begin{figure}[H]
  \centering
  \subfigure[Another 13-edge cluster.]{%
  \includegraphics[width=2.75in]{%
      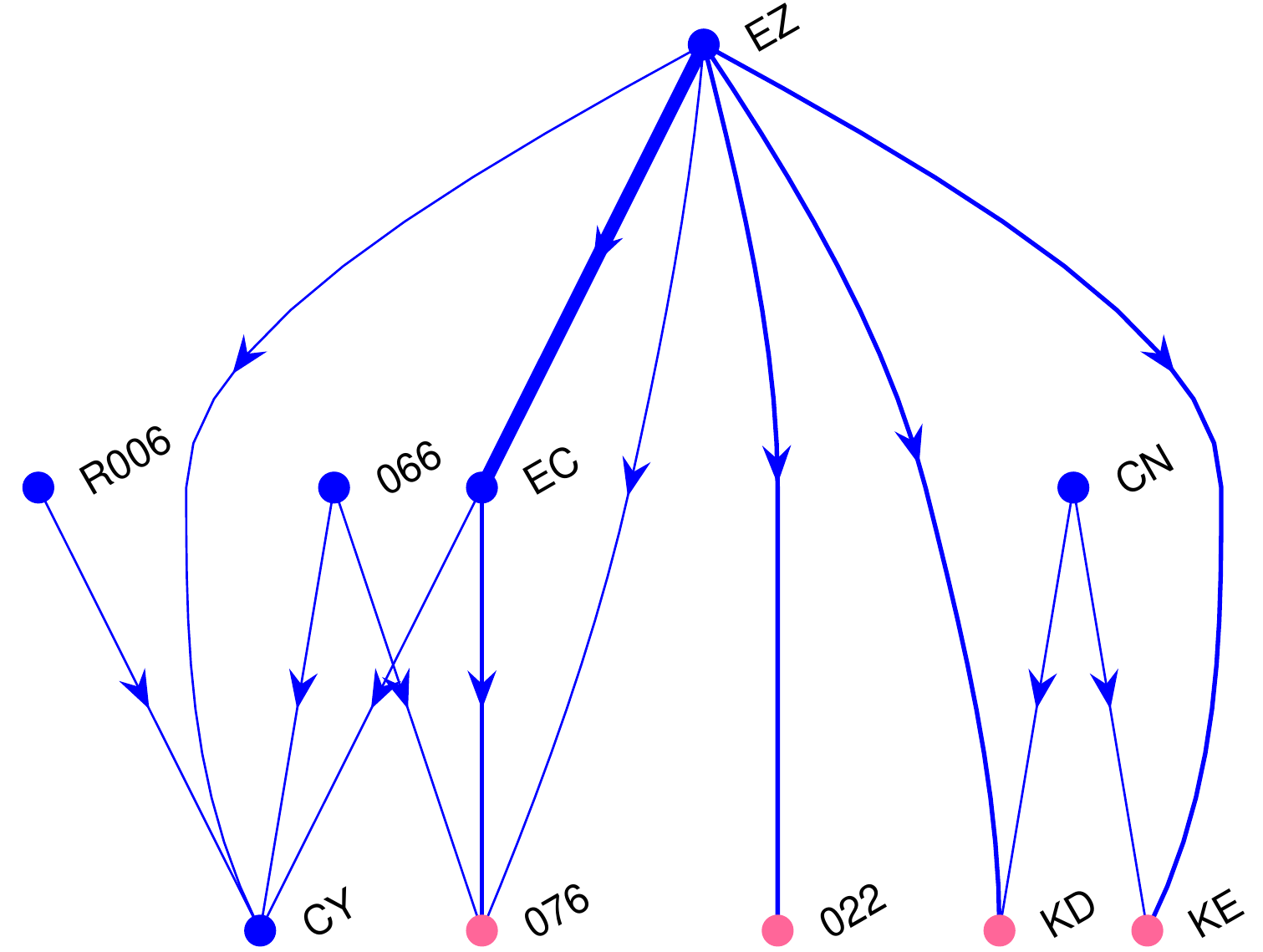}}
  \hfil
  \subfigure[Yet another 13-edge cluster.]{%
  \includegraphics[width=2.75in]{%
      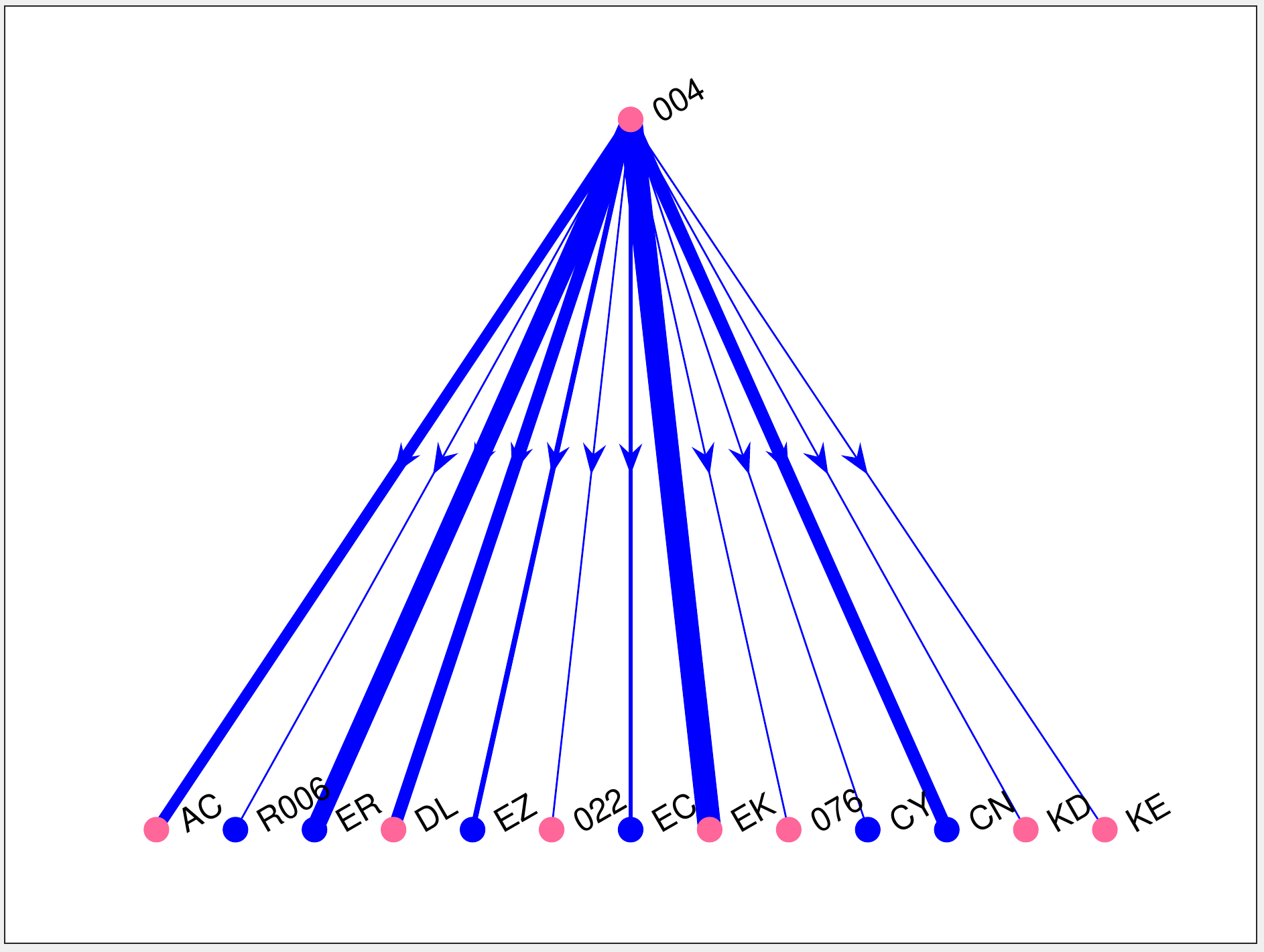}} 
  \caption{%
  (continued from Fig.~\ref{fig:Rhesus_6Clusters_LPRE}) 
  \tb{PRE clustering} ($K = 6$) of the dominance relationhips of a \tb{rhesus monkey population} represented as a weighted digraph \cite{Sade1972FP}. The start vertex and end-vertex of each directed edge indicates the monkey who is dominating and the monkey who is being dominated, respectively. Edge width is proportional to edge weight (which reflects number of encounters between two monkeys). Blue and pink vertices identify the males and females, respectively. This figure shows the smallest 2 clusters obtained when PRE clustering is applied with $K=6$; We see ``004'' and ``EZ'' as producing dominance in these clusters. Figs~\ref{fig:Rhesus_6Clusters_LPRE} shows the remaining clusters.}
\label{fig:Rhesus_6Clusters_LPREb}
\end{figure}

\newpage

\begin{figure}[H]
  \centering
  \includegraphics[width=5.5in]{%
      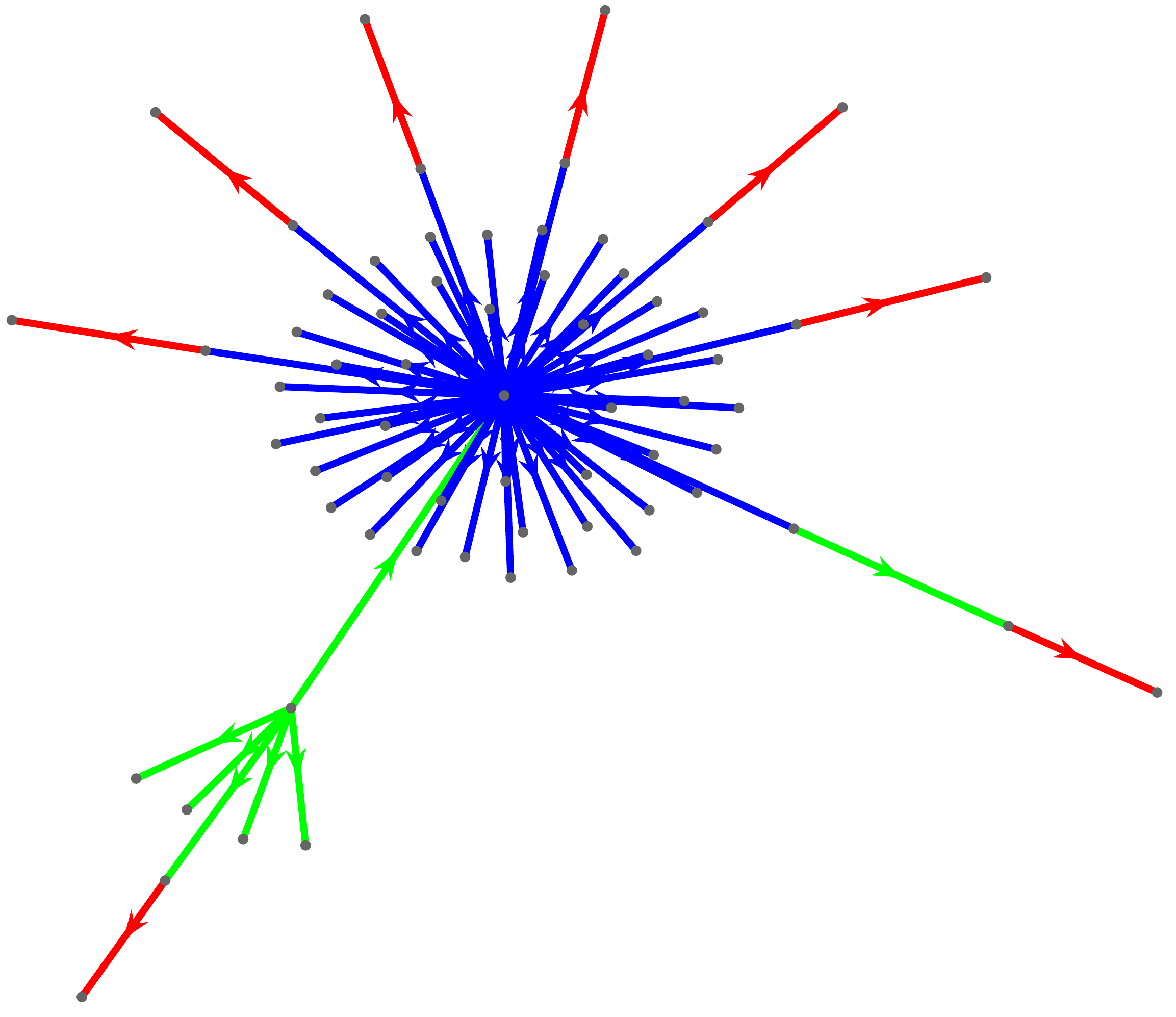}
  \caption{%
  \tb{PRE clustering} ($K = 3$) of a dataset of \tb{COVID-19 infections among South Korean citizens} represented as an unweighted digraph \cite{Laenen2020NEURIPS}. Clustering is carried out only the largest component (66 edges). The dataset consists of 245 disconnected components ranging in size from 1 edge to 66 edges. The start vertex and end-vertex of each directed edge indicates the person spreading the virus and the person being infected by it, respectively.  This figure shows the PRE clustering ($K = 3$) of the 66 edges of the largest digraph component in the dataset.  In this example, one does not need to encourage clusters with equal volumes, and so the unnormalized $L_{\PRE}$ is used. We can see that the edge clustering captures the two superspreader events; the two superspreading events are in distinct clusters while edges within a superspreader event are grouped together.}
\label{fig:COVID19_3Clusters_67_LPREReg}
\end{figure}

\newpage

\begin{figure}[H]
  \centering
  \includegraphics[width=5.5in]{%
      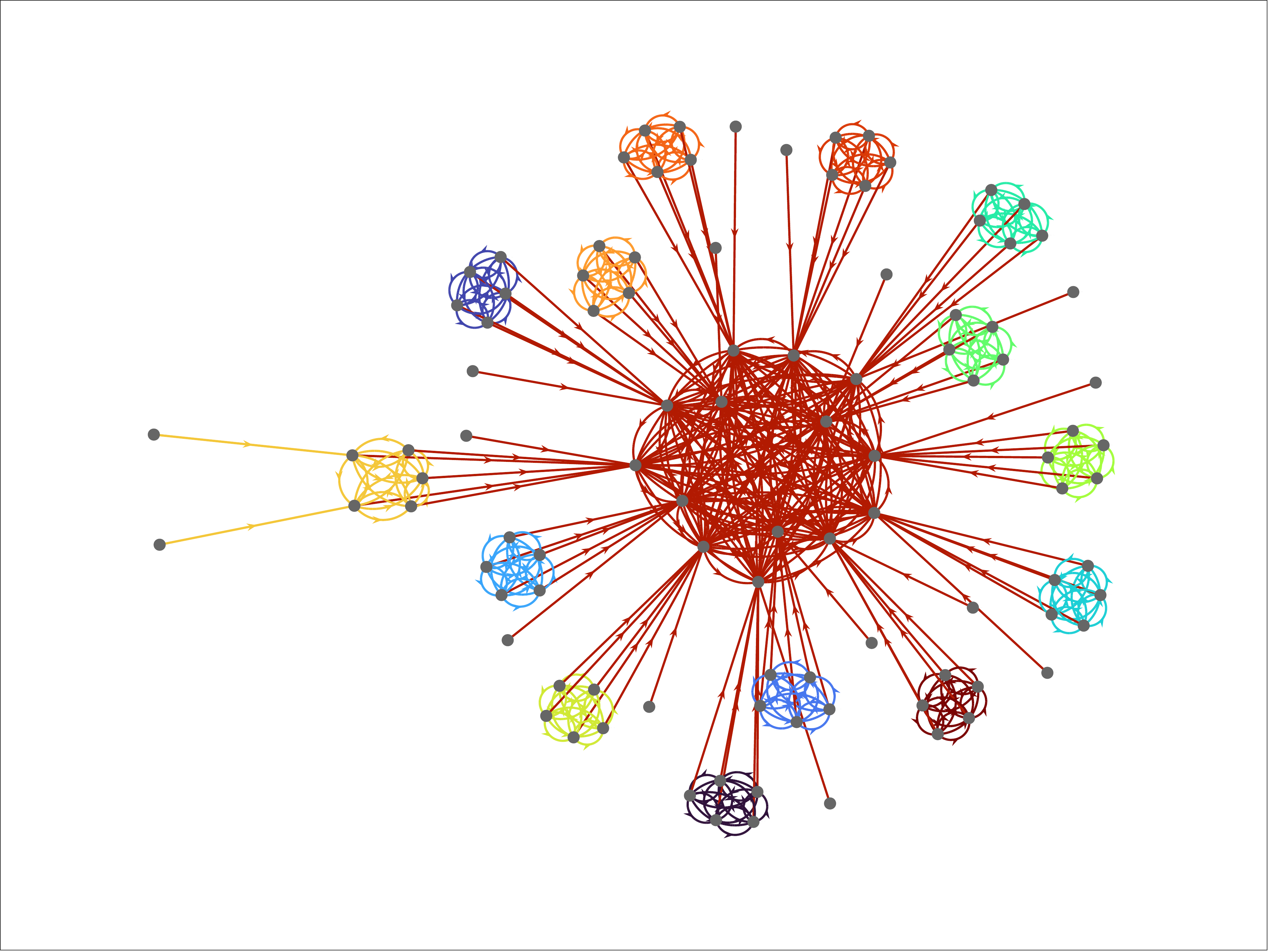}
  \caption{%
  \tb{RGE clustering} ($K = 15$) of the URL addresses of 100 unique \tb{web crawler-generated web pages} starting at \texttt{https://www.mathworks.com} \cite{MATLAB21} represented as a 100-vertex/548-edge unweighted digraph. Note how edges in each smaller knot are assigned to the same edge cluster.}
\label{fig:MATLAB100_LRGE}
\end{figure}

\newpage

\begin{figure}[H]
  \centering
  \subfigure[850-edge cluster.]{%
  \includegraphics[width=0.45\textwidth, height=0.45\textwidth]{%
      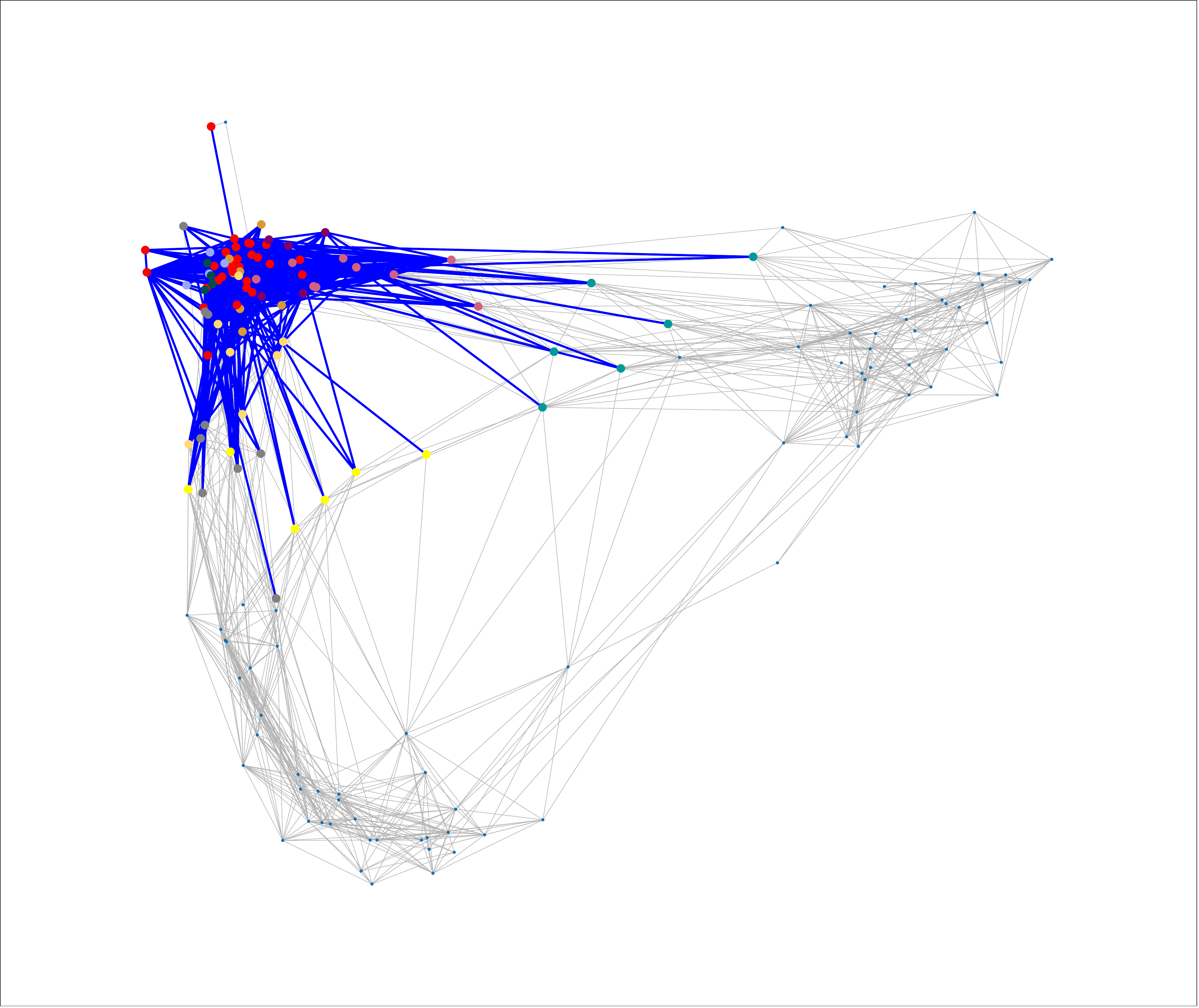}} 
  \quad
  \subfigure[283-edge cluster.]{%
  \includegraphics[width=0.45\textwidth, height=0.45\textwidth]{%
      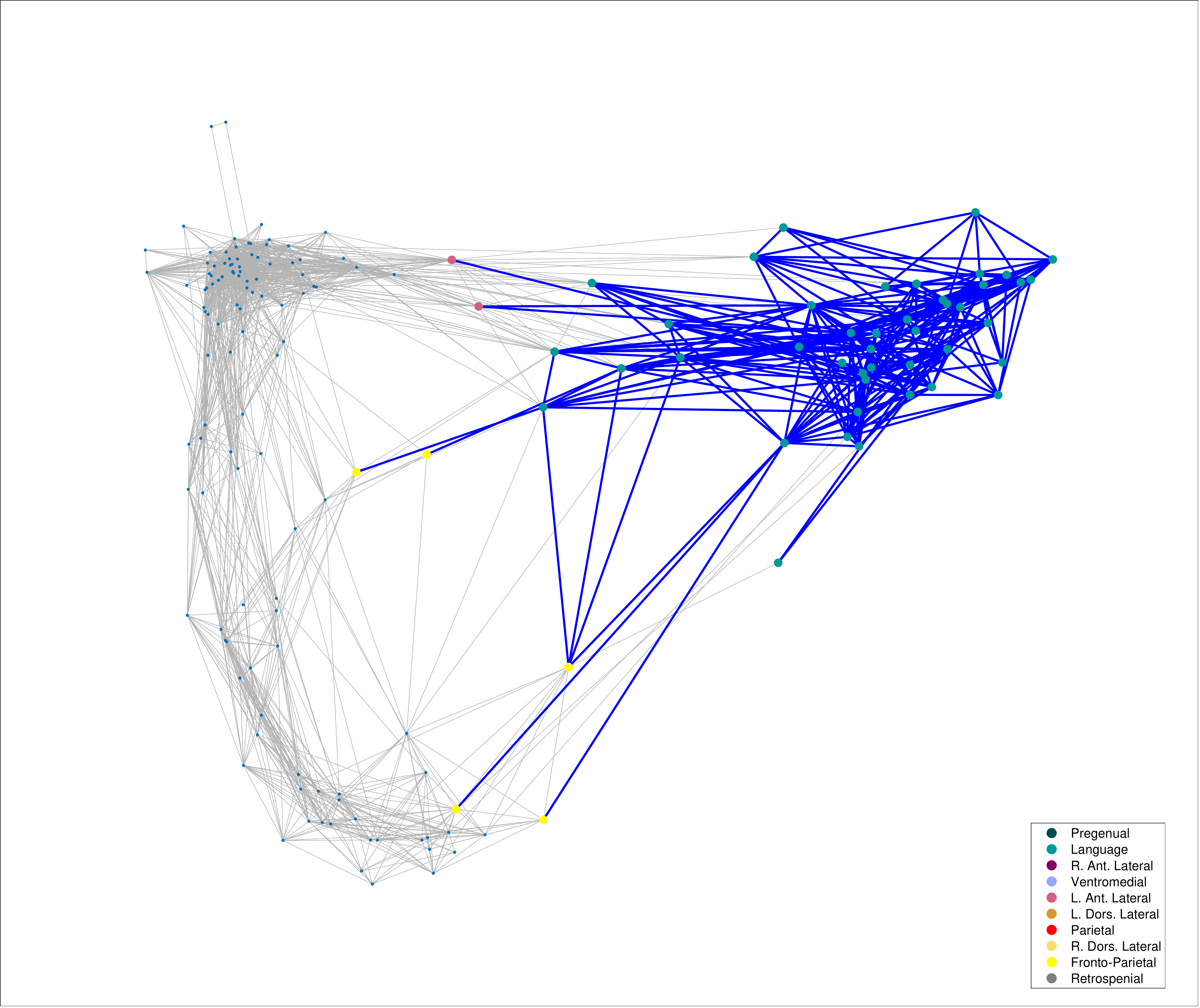}} 
  \caption{%
  \tb{RGE clustering} ($K = 6$) of the high-level \tb{default mode network (DMN) of a human brain} represented as a weighted undirected graph \cite{Gordon2020PNAS}. Different colors highlight subnetworks with different functionalities: \emph{fronto-parietal} is top-down control of lower level processing systems; \emph{retrosplenial} is contextual and scene information; \emph{ventromedial} is fear and anxiety; \emph{pregenual} is reward processing; \emph{parietal} is internally oriented and social cognition; \emph{lateral DMN} are connector hubs. Replacing each edge by an oppositely directed edge pair, RGE clustering generates subgraph clusters that show concentrated activity among DMN subnetworks. Here we show two of the RGE clusters generated with $K = 6$ (the others are in Figs~ \ref{fig:HumanDMN_6Clusters_LRGEB} and \ref{fig:HumanDMN_6Clusters_LRGEC}): 
  \textbf{(a)}~top-down control of  language, cognition, scene and context understanding, as well as fear, emotion, and reward processing; 
  \textbf{(b)}~internal language network processes with connections to top-down control and connector hubs for DMN.}
\label{fig:HumanDMN_6Clusters_LRGEA}
\end{figure}

\newpage

\begin{figure}[H]
  \centering
  \subfigure[276-edge cluster.]{%
  \includegraphics[width=0.45\textwidth, height=0.45\textwidth]{%
      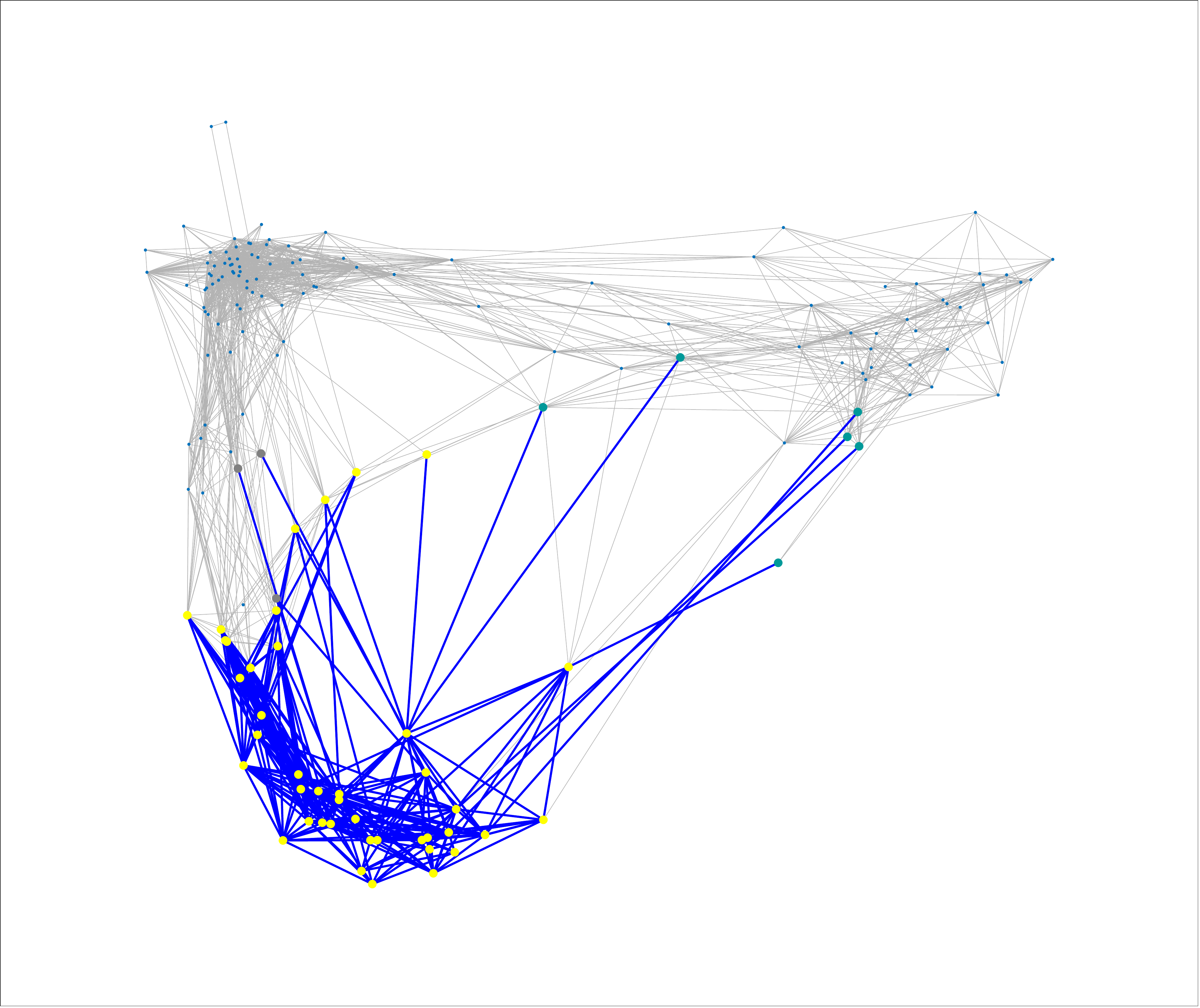}} 
  \quad
  \subfigure[64-edge cluster.]{%
  \includegraphics[width=0.45\textwidth, height=0.45\textwidth]{%
      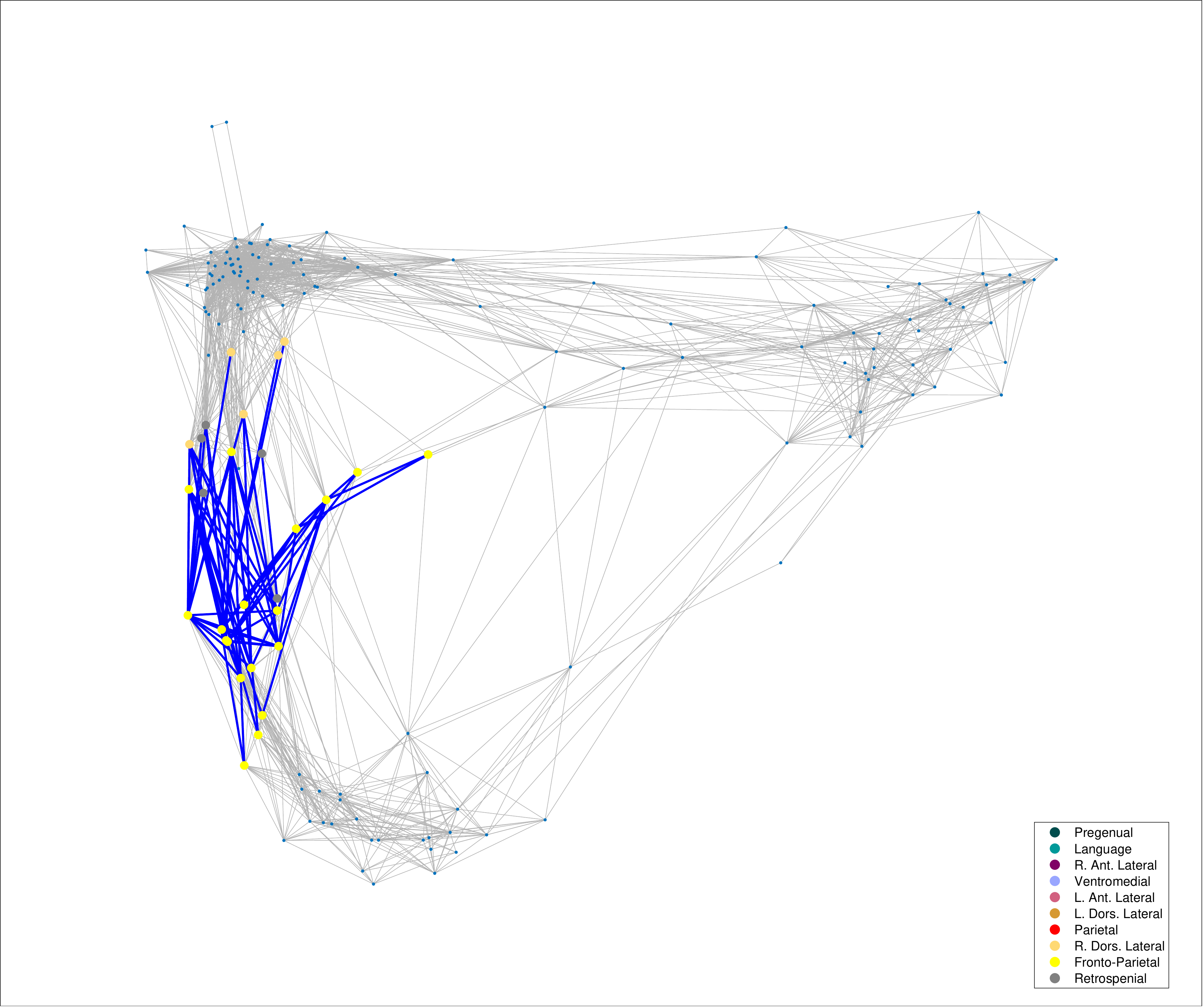}} 
  \caption{%
  (continued from Fig.~\ref{fig:HumanDMN_6Clusters_LRGEA}) 
  \tb{RGE clustering} ($K = 6$) of the high-level \tb{default mode network (DMN) of a human brain} represented as a weighted undirected graph \cite{Gordon2020PNAS}. Different colors highlight subnetworks with different functionalities: \emph{fronto-parietal} is top-down control of lower level processing systems; \emph{retrosplenial} is contextual and scene information; \emph{ventromedial} is fear and anxiety; \emph{pregenual} is reward processing; \emph{parietal} is internally oriented and social cognition; \emph{lateral DMN} are connector hubs. Replacing each edge by an oppositely directed edge pair, RGE clustering generates subgraph clusters that show concentrated activity among DMN subnetworks. Here we show two of the RGE clusters generated with $K = 6$ (the others are in Figs~ \ref{fig:HumanDMN_6Clusters_LRGEA} and \ref{fig:HumanDMN_6Clusters_LRGEC}): \textbf{(a)}~fronto-parietal processes and control of language and integration of contextual and scene information; 
  \textbf{(b)}~top-down control and integration of reward processing, scene and contextual information, and internal DMN cognition and social cognition as well as connector hubs.}
\label{fig:HumanDMN_6Clusters_LRGEB}
\end{figure}

\newpage

\begin{figure}[H]
  \centering
  \subfigure[63-edge cluster.]{%
  \includegraphics[width=0.45\textwidth, height=0.45\textwidth]{%
      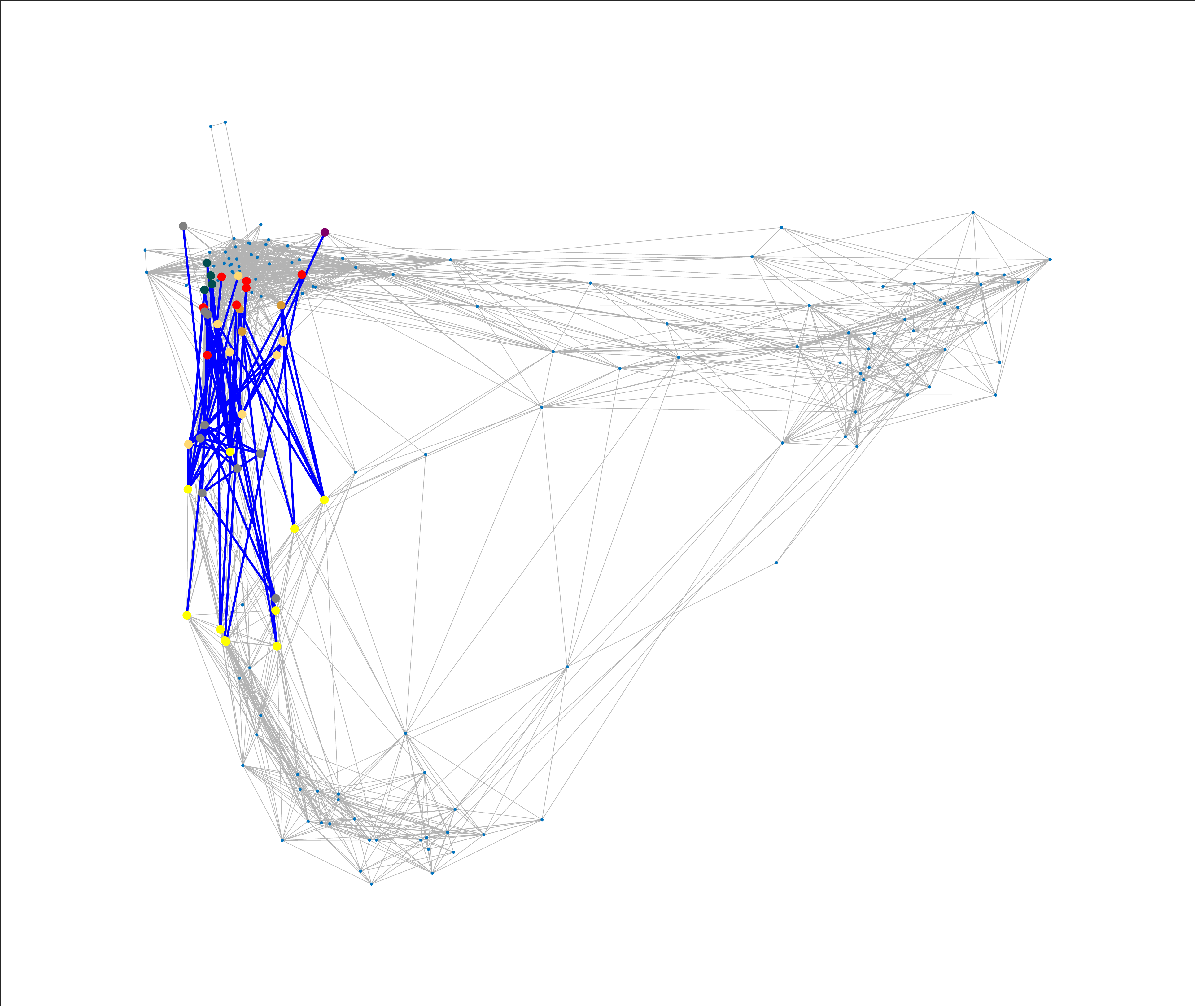}} 
  \quad
  \subfigure[45-edge cluster.]{%
  \includegraphics[width=0.45\textwidth, height=0.45\textwidth]{%
      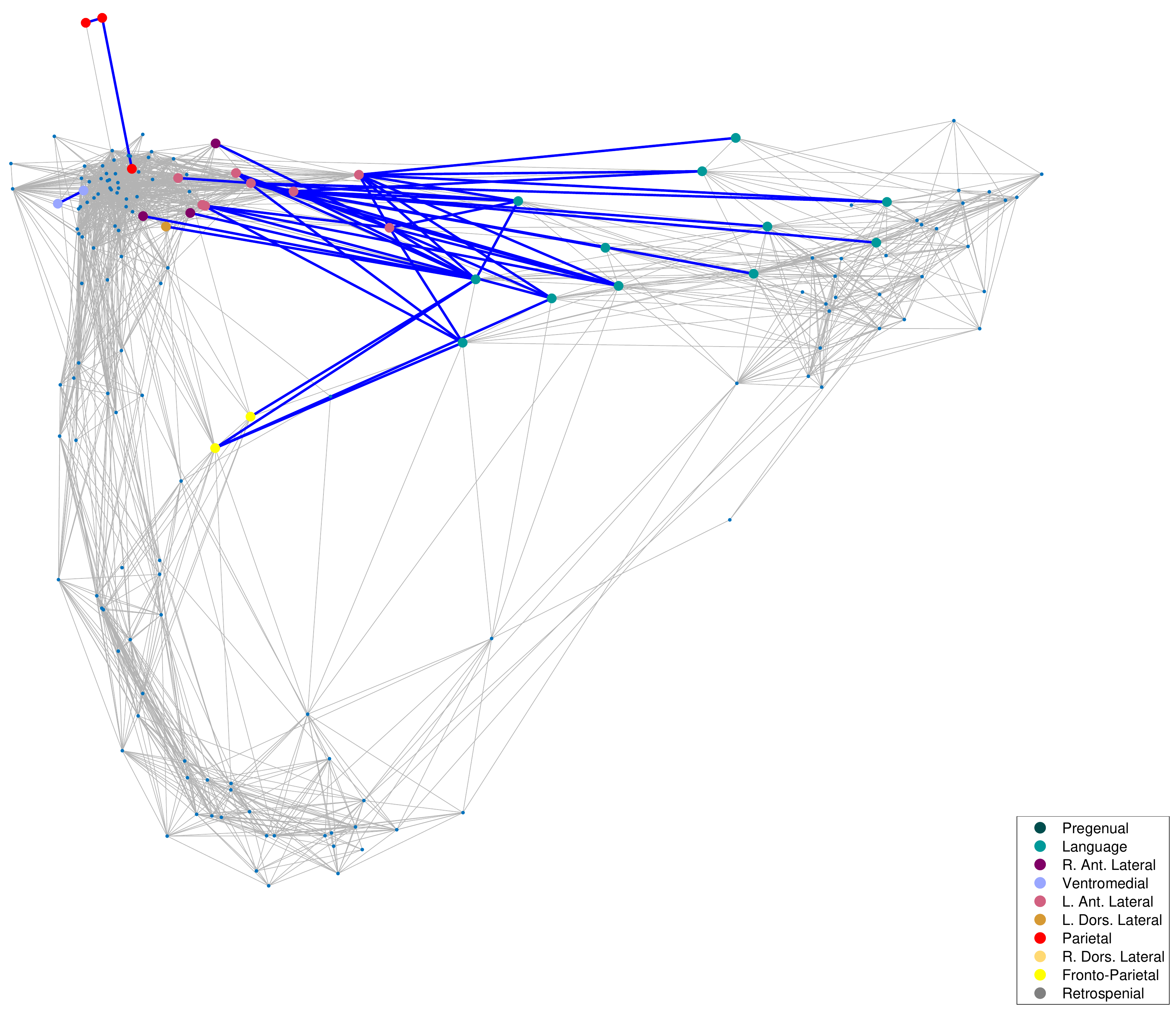}} 
  \caption{%
  (continued from Fig.~\ref{fig:HumanDMN_6Clusters_LRGEB}) 
  \tb{RGE clustering} ($K = 6$) of the high-level \tb{default mode network (DMN) of a human brain} represented as a weighted undirected graph \cite{Gordon2020PNAS}. Different colors highlight subnetworks with different functionalities: \emph{fronto-parietal} is top-down control of lower level processing systems; \emph{retrosplenial} is contextual and scene information; \emph{ventromedial} is fear and anxiety; \emph{pregenual} is reward processing; \emph{parietal} is internally oriented and social cognition; \emph{lateral DMN} are connector hubs. Replacing each edge by an oppositely directed edge pair, RGE clustering generates subgraph clusters that show concentrated activity among DMN subnetworks. Here we show two of the RGE clusters generated with $K = 6$ (the others are in Figs~ \ref{fig:HumanDMN_6Clusters_LRGEA} and \ref{fig:HumanDMN_6Clusters_LRGEB}): \textbf{(a)}~top-down control of scene and contexual information as well as possibly emotion with connector hubs; 
\textbf{(b)}~connections between connector hubs and the language network as well as internal DMN cognition and/or social cognition and connections to top-down control.}
\label{fig:HumanDMN_6Clusters_LRGEC}
\end{figure}

\newpage

\begin{figure}[H]
  \centering
  \includegraphics[width=5.5in]{%
      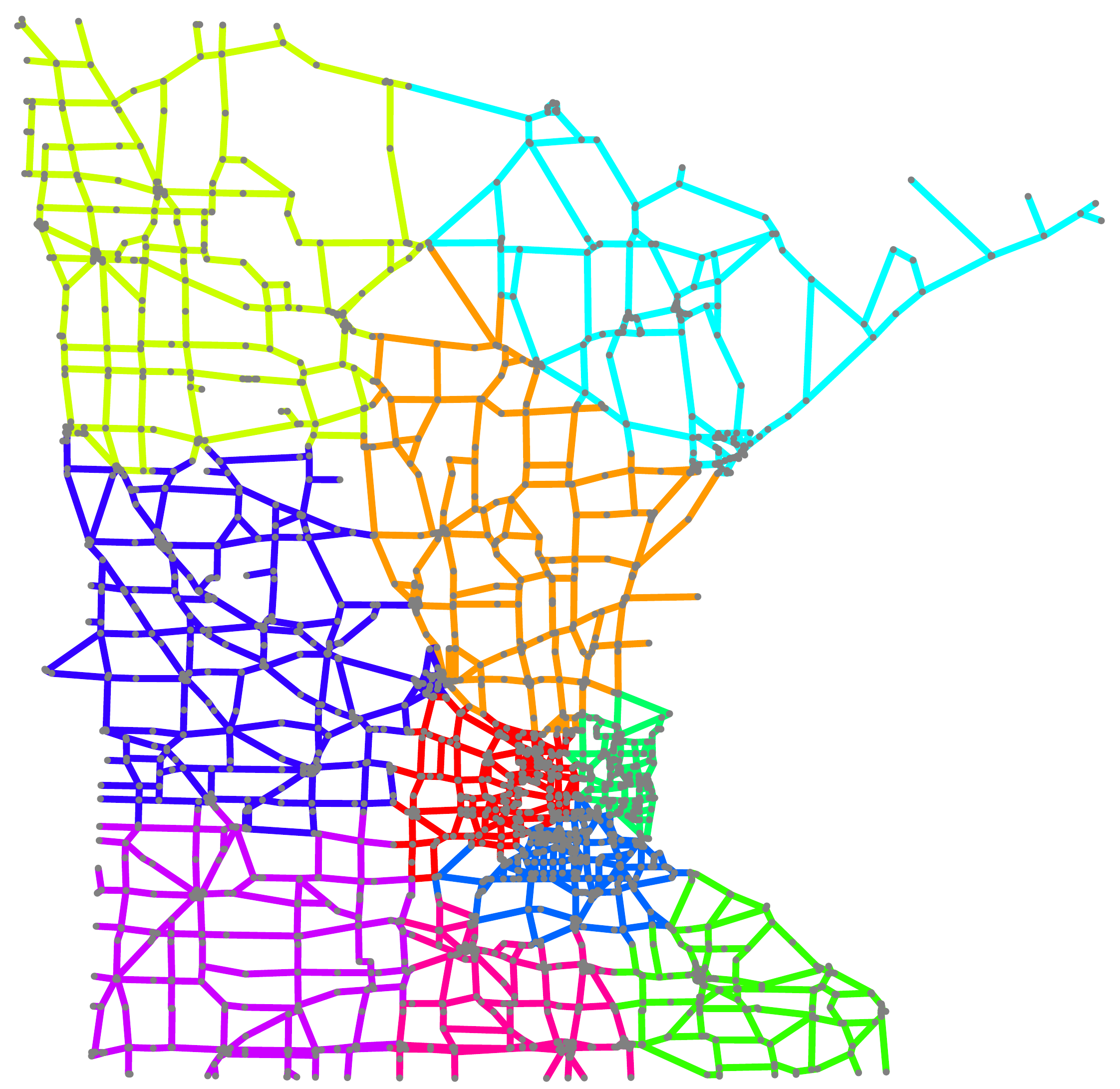}
  \caption{%
  \tb{RGE clustering} ($K = 10$) of the \tb{road network in the state of Minnesota, USA,} represented as a 2,635-vertex/3,298-edge unweighted undirected graph. Note that the Minneapolis-St. Paul area, which has many more roads than the rest of the state, is associated with 3 smaller regional clusters.}
\label{fig:MNRoadNetwork_10Clusters_LRGE}
\end{figure}

\newpage

\begin{figure}[H]
  \centering
  \subfigure[%
    The left two figures show two edges $\{e_p, e_q\}$ sharing a vertex but not forming a length-2 directed path: when the shared vertex is $v_{k_p}=v_{k_q}$ so that the edges are converging, $(L_e)_{pq}=+\phi_{k_p}=+\phi_{k_q}$; when the shared vertex is $v_{\ell_p}=v_{\ell_q}$ so that the edges are diverging, $(L_e)_{pq}=+\phi_{\ell_p}=+\phi_{\ell_q}$. 
    The right two figures show two edges $\{e_p, e_q\}$ sharing a vertex and forming a length-2 directed path: when the shared vertex is $v_{k_p}=v_{\ell_q}$, $(L_e)_{pq}=-\phi_{k_p}=-\phi_{\ell_q}$; when the shared vertex is $v_{k_q}=v_{\ell_p}$, $(L_e)_{pq}=-\phi_{k_q}=-\phi_{\ell_p}$.]{%
    \label{fig:LeEntries_Lines}
    \includegraphics[width=0.95\textwidth]{%
      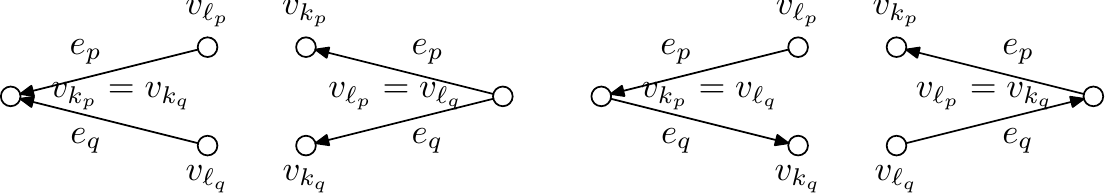}} \\
  \subfigure[%
    When as in the left figure two edges $\{e_p, e_q\}$ form a multi-edge between the vertices $v_{\ell_p}=v_{\ell_q}$ and $v_{k_p}=v_{k_q}$, $(L_e)_{pq}=+(\phi_{k_p}+\phi_{\ell_p})$; when as in the right figure two edges $\{e_p, e_q\}$ form a loop between the vertices $v_{\ell_p}=v_{k_q}$ and $v_{k_p}=v_{\ell_q}$, $(L_e)_{pq}=-(\phi_{k_p}+\phi_{\ell_p})$.]{%
    \label{fig:LeEntries_Loops}
    \includegraphics[width=0.95\textwidth]{%
      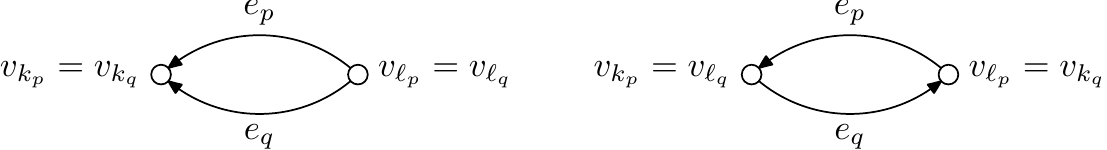}} \\
    \caption{The elements of the edge Laplacian $L_e$ is determined by pairs of edges that share a vertex, in particular, the directionality of the edges and the vertex signal value at the shared vertex.} 
  \label{fig:LeEntries}
\end{figure}

\newpage

\begin{figure}[H]
  \centering
  \subfigure[%
    A digraph $\mc{G}$.]{%
    \label{fig:digraph_7-9}
    \includegraphics[width=1.50in]{%
      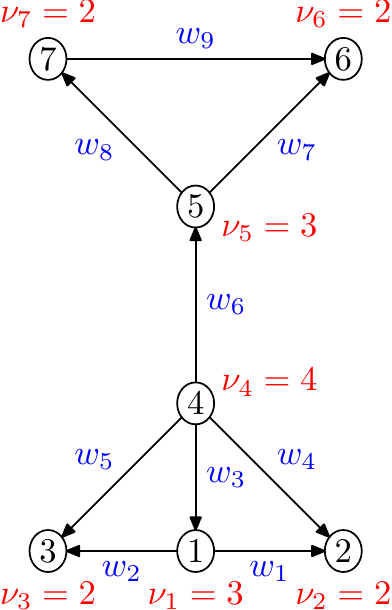}}
  \hfil
  \subfigure[%
    Edge Laplacian of $\mc{G}$.]{%
    \label{fig:Le_7-9}
    \includegraphics[width=4.25in]{%
      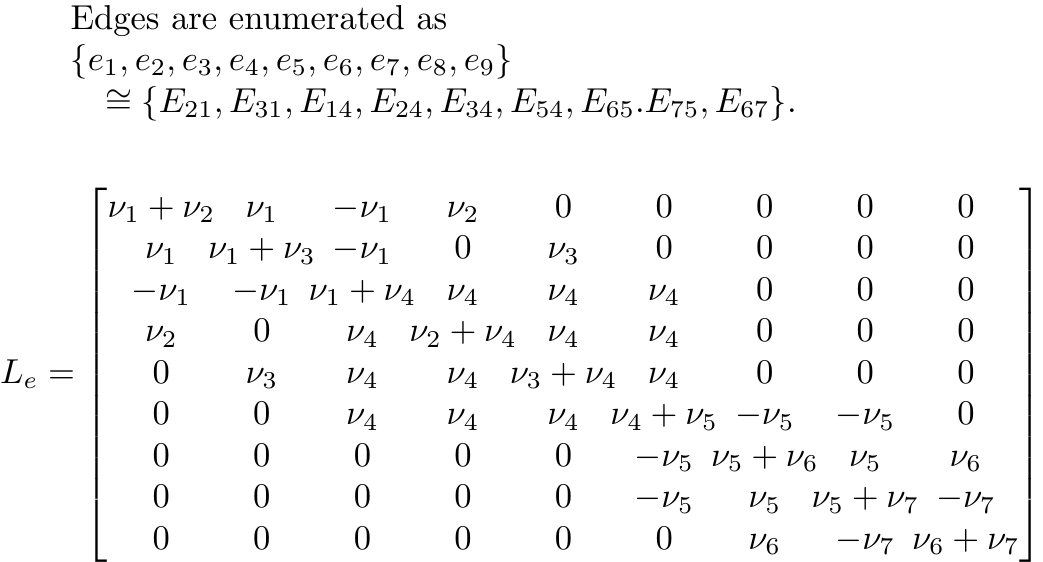}} \\
  \subfigure[%
    The dual graph $\mc{G}'$ of $\mc{G}$.]{%
    \label{fig:dualgraph_7-9}
    \includegraphics[width=1.50in]{%
      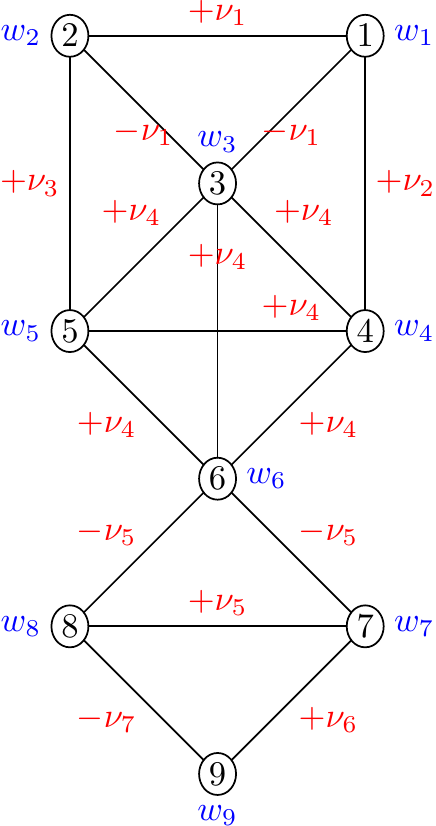}} 
  \hfil
  \subfigure[%
    The adjacency matrix $W'$ of $\mc{G}'$.]{%
    \label{fig:Wprime_7-9}
    \includegraphics[width=4.25in]{%
      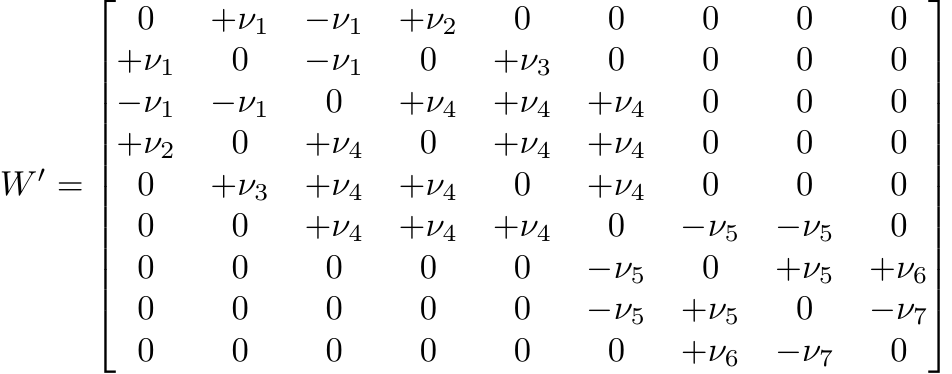}} 
  \caption{A 7-vertex/9-edge digraph $\mc{G}$ from \cite{Lai2010PhysicaA} is in Figure~\ref{fig:digraph_7-9}. The vertex vector values are in red (numerical values refer to the social participation function) and the edge function values are in blue. Its edge Laplacian $L_e$ is in Figure~\ref{fig:Le_7-9}, its 9-vertex/16-edge dual graph $\mc{G}'$ is in Figure~\ref{fig:dualgraph_7-9}, and the adjacency matrix of $\mc{G}'$ is in Figure~\ref{fig:Wprime_7-9}. Each vertex in $\mc{G}'$ represents an edge in $\mc{G}$; each edge in $\mc{G}'$ represenst two edges in $\mc{G}$ that share a vertex. When they form a length-2 directed path in $\mc{G}$, the edge in $\mc{G}'$ inherits the vertex vector value at the shared vertex in $\mc{G}$ with no change in sign; when they do not form a length-2 directed path in $\mc{G}$, the edge in $\mc{G}'$ inherits the vertex vector at the shared vertex in $\mc{G}$ but with the opposite sign.}
  \label{fig:Lai2010PhysicaA_7_9}
\end{figure}

\newpage


\section*{Tables}


\begin{table}[H]
  \centering
  \renewcommand{\arraystretch}{1.2}\addtolength{\tabcolsep}{-0pt}
  \begin{tabular}{l p{4.5in}}
    \hline
    \hfil\tb{Symbol}
      & \hfil\tb{Meaning} \\
    \hline\hline
    $\mc{X}$, $\overline{\mc{X}}$, $\mr{card}[\mc{X}]$ 
      & A set, its complement, and its cardinality. \\
    $\mbb{N}$, $\mbb{N}_+$
      & Integers and non-negative integers. \\
    $\mbb{R}$, $\mbb{R}^N$, $\mbb{R}^{N\times M}$
      & The reals, $N\times 1$-sized real-valued column vectors, and $N\times M$-sized real-valued matrices. \\
    $\mbb{R}_+$, $\mbb{R}_+^N$, $\mbb{R}_+^{N\times M}$
      & The non-negative reals, $N\times 1$-sized non-negative real-valued column vectors, and  $N\times M$-sized non-negative real-valued matrices. \\
    $|x|$, $\mr{sgn}[x]$
      & Absolute value and sign of $x\in\mbb{R}$. \\
    $\delta_{ij}$
      & Kronecker delta, i.e., $\delta_{ij} = 1$ for $i=j$, and $\delta_{ij} = 0$ otherwise. \\
    \hline
    $\ul{x} = \{x_i\} \in \mbb{R}^N$ 
      & $N\times 1$ column vector populated with the elements $x_i,\, i \in \{1, \ldots, N\}$. \\
    $|\ul{x}|$, $\mr{sgn}[\ul{x}]$
      & For $\ul{x} \in \mbb{R}^N$, $|\ul{x}| = \{|x_i|\}$, $\mr{sgn}[\ul{x}] = \{\mr{sgn}[x_i]\}$. \\
    $\ul{X} = \{x_{ij}\} \in \mbb{R}^{N\times M}$ 
      & $N\times M$ matrix populated with the elements $x_{ij},\, i \in \{1, \ldots, N\},\, j \in \{1,\ldots, M\}$. \\
    $\ul{X}^T$, $|\ul{X}|$, $\mr{sgn}[\ul{X}]$
      & For $\ul{x} \in \mbb{R}^{N\times M}$, $\ul{X}^T$ is its transpose, $|\ul{X}| = \{|x_{ij}|\}$, and $\mr{sgn}[\ul{x}] = \{\mr{sgn}[x_{ij}]\}$. \\
    $\mr{Tr}[\ul{X}]$
      & For the square matrix $\ul{X} = \{x_{ij}\} \in\mbb{R}^{N \times N}$, sum of its diagonal elements, i.e., $\mr{Tr}[\ul{X}] = \sum_{i=1}^N x_{ii}$. \\
    $\ul{1}_N$
      & $N \times 1$ column vector of all $1$\,s. \\
    $\mr{diag}[\ul{x}] \in \mbb{R}^{N \times N}$
      & For $\ul{x} \in \mbb{R}^N$, the diagonal matrix with $\ul{x}$ on its diagonal. \\
    $\mr{diag}[\ul{X}] \in \mbb{R}^N$
      & For $\ul{X} \in \mbb{R}^{N \times N}$, the $N \times 1$ column vector generated from the main dagonal of $\ul{X}$. \\
    $\ul{X} \geq 0$, $\ul{X} > 0$
      & All elements of $\ul{X}$ are non-negative (i.e., $x_{ij} \geq 0,\;\forall i, j$, all elements of $\ul{X}$ are positive (i.e., $x_{ij} > 0,\;\forall i, j$. \\
    $\ul{X} \odot \ul{Y}$
      & Hadamard product (i.e., element-wise product) of matrices $\ul{X}, \ul{Y} \in \mbb{R}^{N \times M}$. \\
    \hline
    $\{\ul{v}_i, \lambda_i\}$
      & $i$-th unit eigenpairs of the square symmetric matrix $\ul{X} \in \mbb{R}^{N \times N}$, i.e., $\ul{X}\, \ul{v}_p = \lambda_p \ul{v}_p,\, p \in \{1, \ldots, N\}$, where $\Vert \ul{v} \Vert_p = 1$ and the eigenvalues $\lambda_p$ are ordered as $\lambda_{\min} \equiv \lambda_1 \leq \cdots \leq \lambda_N \equiv \lambda_{\max}$. \\      
    \hline
  \end{tabular}
  \caption{Table of notation: Basic notions.}
  \label{tab:notation}
\end{table}

\newpage

\begin{table}[H]
  \centering
  \renewcommand{\arraystretch}{1.2}\addtolength{\tabcolsep}{-0pt}
  \begin{tabular}{l p{4.5in}}
    \hline
    \hfil\tb{Symbol}
      & \hfil\tb{Meaning} \\
    \hline\hline
    $N, M \in \mbb{N}_+$
      & Number of vertices (nodes) and number of edges. \\
    $\mc{V} = \{v_1, \ldots, v_N\}$
      & Set of $N$ vertices. $v_i$ is the $i$-th vertex. \\
    $\mc{E} = \{e_1, \ldots, e_M\}$  
      & Set of $M$ directed edges. $e_q$ is the $q$-th edge \emph{from} vertex $v_{\ell_q}$ \emph{to} $v_{k_q}$. \\
    $\mc{G}(\mc{V}, \mc{E})$
      & Digraph with $\mc{V}$ as its set of vertices and $\mc{E}$ as its set of edges. \\
    $e_q \cong e_{k_q,\ell_q}$
      & Identifies that $e_q$ and $e_{k_q,\ell_q}$ is the same edge. This enumeration scheme which pairs a particular edge $e_q$ with $e_{k_q,\ell_q}$ is assumed fixed. \\
    \hline
    $\ul{B} \in \mbb{N}^{N\times M}$
      & \emph{Unweighted incidence matrix} of $\mc{G}=\mc{G}(\mc{V}, \mc{E})$. $\ul{B} = \{b_{ip}\}$, where $b_{ip} = +1$ if edge $e_p$ (with weight $w_p$) has vertex $v_i$ as its source vertex, $-1$ if vertex $v_i$ is its destination vertex, and $0$ otherwise. \\
    $\ul{A} \in \mbb{N}_+^{N\times N}$
      & \emph{Unweighted adjacency matrix} of $\mc{G}=\mc{G}(\mc{V}, \mc{E})$. $A = \{a_{ij}\}$, where $a_{ij} = 1$ if $e_{ij} = (v_j {\to}\, v_i) \in \mc{E}$ and $a_{ij} = 0$ otherwise. \\
    $\ul{\nu} \in \mbb{R}^N$
      & Vertex vector $\ul{\nu} = [\nu_1, \ldots, \nu_N]^T$, where $\nu_i: \mc{V} \mapsto \mbb{R}$ is a real-valued function defined on $\mc{V}$. \\
    $\ul{w} \in \mbb{R}_+^M$ 
      & Edge weight vector $\ul{w} = [w_1, \ldots, w_M]^T$, where $w_q: \mc{E} \mapsto \mbb{R}_+$ is the non-negative real-valued edge weight of edge $e_q\cong e_{k_q,\ell_q}$. \\
    $\ul{W} \in \mbb{R}_+^{N\times N}$ 
      & \emph{Weighted adjacency matrix} of $\mc{G}(\mc{V}, \mc{E})$. $\ul{W} = \{w_{ij}\}$, where $w_{ij} = w_q$ if $e_q \cong (v_j \to v_i)$ and $w_{ij} = 0$ otherwise. \\
    $\ul{w} \cong \ul{W}$
      & Identifies that $\ul{W}$ is populated with the edge weights in $\ul{w}$ s.t. $w_{ij} = w_q$ if $e_q \cong (v_j \to v_i)$ and $w_{ij} = 0$ otherwise. \\
    $\mc{G}(\mc{V}, \mc{E})[\ul{\nu}, \ul{w}]$ 
      & Digraph $\mc{G}(\mc{V}, \mc{E})$ with vertex vector $\ul{\nu}$ and edge weight vector $\ul{w}$. \\
    \hline
    $\sigma_i \in \mbb{N}_+$
      & \emph{Social participation} of vertex $v_i \in \mc{V}$, i.e., total integer number of edges it is connected to. \\
    $d_{out,i}, d_{in,i} \in \mbb{R}$
      & (Weighted) out-degree and weighted in-degree of vertex $v_i \in \mc{V}$, i.e., $d_{out,i} = \sum_j w_{ji}$ and $d_{in,i} = \sum_j w_{ij}$. \\  
    $d_i \in \mbb{R}$
      & (Weighted) total degree of vertex $v_i \in \mc{V}$, i.e., $d_i = d_{out,i} + d_{in,i}$. \\
     $\langle d_{out,i}\rangle, \langle d_{in,i}\rangle \in \mbb{R}_+$
       & Sum of the absolute values of outgoing edge weights and sum of the absolute values of incoming edge weights of vertex $v_i \in \mc{V}$, i.e., $\langle d_{out,i}\rangle = \sum_j |w_{ji}|$ and $\langle d_{in,i}\rangle = \sum_j |w_{ij}|$. \\
     $\langle d_i \rangle \in \mbb{R}_+$
       & $\langle d_i \rangle = \langle d_{out,i}\rangle +  \langle d_{in,i}\rangle$. \\
     $\ul{d}_{out}, \ul{d}_{in} \in \mbb{R}^N$
       & (Weighted) out-degree vector and (weighted) in-degree vector of vertex $v_i \in \mc{V}$, i.e., $\ul{d}_{out} = \ul{W}\,\ul{1}_N\in\mbb{R}^N$ and $\ul{d}_{in} = \ul{W}^T\ul{1}_N\in\mbb{R}^N$. \\
     $\ul{D}_{out}$, $\ul{D}_{in} \in \mbb{R}^{N \times N}$
       & $N \times N$ \emph{out-degree matrix} and \emph{in-degree matrix} whose diagonals are populated with the degree vectors $\mr{diag}[\ul{d}_{out}]$ and $\mr{diag}[\ul{d}_{in}]$, i.e., $\ul{D}_{out} = \mr{diag}[\ul{d}_{out}] \in \mbb{R}^{N\times N}$ and $\ul{D}_{in} = \mr{diag}[\ul{d}_{in}] \in \mbb{R}^{N\times N}$. \\
    \hline
  \end{tabular}
  \caption{Table of notation: Graph-related notions.}
  \label{tab:notation2}
\end{table}

\newpage

\begin{table}[H]
  \centering
  \renewcommand{\arraystretch}{1.2}\addtolength{\tabcolsep}{-0pt}
  \begin{tabular}{l p{3.71in} p{0.65in}}
    \hline
    \hfil\tb{Symbol}
      & \hfil\tb{Meaning} 
      & \tb{Refer To} \\
    \hline\hline
    $K \in \mbb{N}_+$
      & Number of clusters the set $\mc{E}$ of digraph edges is to be partitioned. \\
    $\{\Sigma^{(k)}, \ldots, \Sigma^{(K)}\}$
      & Set of $K$ edge clusters. $\Sigma^{(k)}$ is the $k$-th cluster and $\bigcup_{k=1}^K \Sigma^{(k)} = \mc{E}$ and $\Sigma^{(k)} \cap \Sigma^{(\ell)} = \emptyset$ for $k \neq \ell$. \\
    $M^{(k)} \in \mbb{N}_+$
      & Cardinality of edge cluster $\Sigma^{(k)}$, i.e., $M^{(k)} = |\Sigma^{(k)}|$. \\
    $X = \{x_p^{(k)}\} \in \mbb{R}^{M \times K}$
      & Partition matrix, where $x_p^{(k)} = \alpha^{(k)}$ if edge $e_p \in \Sigma^{(k)}$, $x_p^{(k)} = 0$ otherwise, and $\alpha^{(k)} \in \mbb{R}_+,\, k \in \{1, \ldots, K\}$. 
      & \eqref{eq:X}, \eqref{eq:Indicator} \\
    $\ul{x}^{(k)} \in \mbb{R}^M$
      & Indicator vector for edge membership in cluster $\Sigma^{(k)}$. $k$-th column of $\ul{X}$, i.e., $\ul{X} = [\ul{x}^{(1)}, \ldots, \ul{x}^{(K)}]$. 
      & \\
    \hline
    $\tr{Cost}^{(k)}(e_p, e_q) \in \mbb{R}_+$
      & Cost associated with edge pair $\{e_p, e_q\}$ in cluster $\Sigma^{(k)}$. \newline $\tr{Cost}^{(k)}(e_p, e_q) = (1/2)\, \phi_{pq}\, (x_p^{(k)} - \psi_{pq}\,x_q^{(k)})^2$.
      & Def.~\ref{def:Cost_k_pq} \\
    $\ul{\Psi} = \{\psi_{pq}\} \in \mbb{R}^{M \times M}$
      & Symmetric matrix that captures the edge functional affinity. $\psi_{pq}$ captures the edge functional affinity between the edge pair $\{e_p, e_q\}$. 
      & \\
    $\ul{\Phi} = \{\phi_{pq}\} \in \mbb{R}_+^{M \times M}$
      & Symmetric non-negative matrix of weights associated with the squared pairwise edge label comparisons. $\phi_{pq} \geq 0$ is the weight associated with edge pair $\{e_p, e_q\}$; $\phi_{pp} = 0$, i.e., $\ul{\Phi}$ has a zero diagonal. We pick $\ul{\Phi}$ to be a function of a non-negative vertex vector $\ul{\nu} \geq 0$. 
      & \eqref{eq:phi_nu} \\
    $\ul{D}_{\ul{\Psi}\ul{\Phi}} \in \mbb{R}^{M \times M}$
      & Diagonal matrix $\ul{D}_{\ul{\Psi}\ul{\Phi}} = \mr{diag}[\ul{d}_{\ul{\Psi}\ul{\Phi}}]$ whose diagonal is populated with  $\ul{d}_{\ul{\Psi}\ul{\Phi}} = [d_{\ul{\Psi}\ul{\Phi}}(1), \ldots, d_{\ul{\Psi}\ul{\Phi}}(M)]^T \in \mbb{R}^M$, where $d_{\ul{\Psi}\ul{\Phi}}(p) = (1/2) \sum_{q=1}^M \phi_{pq} (1 + \psi_{pq}^2)$. 
      & \eqref{eq:D_Psi_Phi} \\
    $\ul{\Psi}_{\PRE}$, $\ul{\Psi}_{\DPE}$, $\ul{\Psi}_{\RGE}$
      & $M \times M$ matrices that capture edge functional affinities. For PRE, $\ul{\Psi}_{\PRE} = \mr{sgn}[\ul{B}^T \ul{B}]$; for DPE, $\ul{\Psi}_{\DPE} = -\ul{\Psi}_{\PRE}$; and for RGE, $\ul{\Psi}_{\RGE} = |\ul{\Psi}_{\PRE}|$. 
      & Lem.~\ref{lem:Psi} \\
    $\ul{D}_{\ul{\Phi}} \in \mbb{R}^{M \times M}$
      & Diagonal matrix $\ul{D}_{\ul{\Psi}\ul{\Phi}} = \mr{diag}[\ul{d}_{\ul{\Phi}}]$, where $\ul{d}_{\ul{\Phi}} = \ul{\Phi}\, \ul{1}_M \in \mbb{R}^M$. With $\ul{\Psi}$ as in Lemma~\ref{lem:Psi}, $\ul{D}_{\ul{\Phi}} = \ul{D}_{\ul{\Psi}\ul{\Phi}}$.  
      & \eqref{eq:D_phi} \\
    $\ul{L}_{\PRE}$, $\ul{L}_{\DPE}$, $\ul{L}_{\RGE}$
      & $M \times M$ `Laplacian' matrices associated with edge functional affinities. For PRE,  $\ul{L}_{\PRE} = \ul{D}_{\ul{\Phi}} - (\ul{\Psi}_{\PRE}\odot\ul{\Phi})$; for DPE,  $\ul{L}_{\PRE} = \ul{D}_{\ul{\Phi}} + (\ul{\Psi}_{\PRE}\odot\ul{\Phi})$; and for RGE, $\ul{L}_{\PRE} = \ul{D}_{\ul{\Phi}} - |\ul{\Psi}_{\PRE}\odot\ul{\Phi}|$.
      & Lem.~\ref{lem:Cost_k_L} \\
    $\tr{Cost}^{(k)}(\mc{G}) \in \mbb{R}_+$
      & Cost associated with cluster $\Sigma^{(k)}$. With $\ul{L}$ being  $\ul{L}_{\PRE}$, $\ul{L}_{\DPE}$, or $\ul{L}_{\RGE}$ depending on edge functional affinity, $\tr{Cost}^{(k)}(\mc{G}) = \ul{x}^{(k)^T} \ul{L}\, \ul{x}^{(k)}$, 
      & Lem.~\ref{lem:Cost_k_L} \\
    $\ul{L}_v(\ul{\nu}) \in \mbb{R}^{N \times N}$
      & Vertex Laplacian. $\ul{L}_v(\ul{\nu}) = \ul{B}\, \mr{diag}[\ul{w}]\, \ul{B}^T$. \\
    \hline
  \end{tabular}
  \caption{Table of notation: Edge clustering-related notions.}
  \label{tab:notation3}
\end{table}

\newpage

\begin{table}[H]
  \centering
  \renewcommand{\arraystretch}{1.2}\addtolength{\tabcolsep}{-0pt}
  \begin{tabular}{l p{3.71in} p{0.65in}}
    \hline
    \hfil\tb{Symbol}
      & \hfil\tb{Meaning} 
      & \tb{Refer To} \\
    \hline\hline
    $\ul{L}_e(\ul{\nu}) \in \mbb{R}^{M \times M}$
      & Edge Laplacian. $\ul{L}_e(\ul{\nu}) = \ul{B}^T \, \mr{diag}[\ul{\nu}]\, \ul{B}$. 
      & Def.~\ref{def:Le} \\
    $\mc{G}'(\mc{V}', \mc{E}')[\ul{\nu}', \ul{w}']$
      & Undirected dual graph of digraph $\mc{G}(\mc{V}, \mc{E})[\ul{\nu}, \ul{w}]$. 
      & Def.~\ref{def:DualGraph} \\
    $\ul{W}' \in \mbb{R}^{M \times M}$
      & Weighted adjacency matrix of dual graph $\mc{G}'(\mc{V}', \mc{E}')$. $\ul{W}' = \ul{L}_e - \mr{diag}[\mr{diag}[\ul{L}_e]]$ and, when $\ul{\nu} \geq 0$, $\ul{\Phi}(\ul{\nu}) = |\ul{W}'(\ul{\nu})|$. 
      & Lem.~\ref{lem:Phi_Le} \\
    $\mr{Cut}(\mathcal{A}, \mathcal{B})$
      & Absolute cut between dual graph vertex sets $\mc{A}, \mc{B} \subseteq \mc{V}'$. \newline 
      $\mr{Cut}(\mathcal{A}, \mathcal{B}) = \sum_{v'_p \in \mc{A}} \sum_{v'_q \in \mc{B}} |w'_{pq}|$.
     & Def.~\ref{def:Cuts} \\
    $\mr{Links}^+(\mc{A}, \mc{B})$
      & Sum of positive links between vertex sets $\mc{A}, \mc{B} \in \mc{V}'$ of dual graph $\mc{G}'(\mc{V}', \mc{E}')$. \newline
      $\mr{Links}^+(\mc{A}, \mc{B}) = \sum_{v'_p \in \mc{A}} \sum_{v'_q \in \mc{B}}
      \max\{0, +w'_{pq}\}$.
      & Def.~\ref{def:Cuts} \\
    $\mr{Links}^-(\mc{A}, \mc{B})$
      & Sum of negative links between dual graph vertex sets $\mc{A}, \mc{B} \in \mc{V}'$ of dual graph $\mc{G}'(\mc{V}', \mc{E}')$. \newline
      $\mr{Links}^-(\mc{A}, \mc{B}) = \sum_{v'_p \in \mc{A}} \sum_{v'_q \in \mc{B}}
      \max\{0, -w'_{pq}\}$.
      & Def.~\ref{def:Cuts} \\
    $\mr{Vol}(e_p) = f_p \in \mbb{R}_+$
      & Volume of edge $e_p = (v_{\ell_p} {\to}\, v_{k_p})$. 
      & Def.~\ref{def:VolEdge} \\
    $\ul{F} \in \mbb{R}^{M \times M}$
      & Diagonal matrix $\ul{F} = \mr{diag}[\ul{f}]$ whose diagonal is populated with $\ul{f} = [f_1, \ldots, f_M]^T \in \mbb{R}^M$, where $f_p = \mr{Vol}(e_p)$. 
      & Def.~\ref{def:VolEdge} \\
    $\mr{Vol}(\Sigma^{(k)})$
      & Volume of cluster $\Sigma^{(k)} \subseteq\mc{E} = \sum_{e_p \in \Sigma^{(k)}} f_p$.
      & Def.~\ref{def:VolEdge} \\
    $\tr{NCost}^{(k)}(\mc{G}) \in \mbb{R}_+$
      & Normalized cost associated with cluster $\Sigma^{(k)}$. \newline
      $\tr{NCost}^{(k)}(\mc{G}) = \ul{x}^{(k)^T} \ul{L}\, \ul{x}^{(k)}/\ul{x}^{(k)^T} \ul{F}\, \ul{x}^{(k)}$.
      & Def.~\ref{def:NCuts} \\
    $\tr{NCost}(\mc{G}) \in \mbb{R}_+$
      & Total normalized cost for clustering all edges. \newline
      $\tr{NCost}(\mc{G}) = \sum_{k=1}^K \tr{NCost}^{(k)}(\mc{G})$.
      & Def.~\ref{def:NCuts} \\
    $\widetilde{\ul{L}}_{\PRE}$, $\widetilde{\ul{L}}_{\DPE}$, $\widetilde{\ul{L}}_{\RGE}$
      & $M \times M$ normalized Laplacian matrices associated with edge functional affinities. 
      & Def.~\ref{def:normalizedFlowLaplacians} \\
    \hline
  \end{tabular}
  \caption{Table of notation: More edge clustering-related notions.}
  \label{tab:notation4}
\end{table}

\newpage

\begin{table}[htpb]
  \centering
  \renewcommand{\arraystretch}{1.2}\addtolength{\tabcolsep}{-0pt}
  \begin{tabular}{c p{1.55in}ll}
  
    \multicolumn{4}{l}{\tb{When $\ul{\Phi}$ is as in Definition~\ref{def:Cost_k_pq}:}} \\
    \hline
    \textbf{Edge}
      & \hfil$\ul{\Phi} \in \mbb{R}^{M\times M}$
      & \hfil$\ul{\Psi} \in \mbb{R}^{M\times M}$ 
      & \hfil$\ul{L} \in \mbb{R}^{M\times M}$ \\
    \textbf{Functional}
      & 
      & \hfil(Lemma~\ref{lem:Psi})
      & \hfil(Lemma~\ref{lem:Cost_k_L}) \\
    \textbf{Affinity}
      &&& \\
    \hline
    PRE
      & $\phi_{pq} \geq 0$, when $\{e_p, e_q\}$ share a common vertex and $p \neq q$, and $\phi_{pq} = 0$ otherwise. 
      & $\ul{\Psi}_{\PRE}
           =\tr{sgn}[B^TB]$ 
      & $\ul{L}_{\PRE} = \ul{D}_{\ul{\Phi}} - (\ul{\Phi} \odot \ul{\Psi}_{\PRE})$ \\
    DPE
      & \hfil -- do --
      & $\ul{\Psi}_{\DPE}
           =-\tr{sgn}[B^TB]$ 
      & $\ul{L}_{\DPE} = \ul{D}_{\ul{\Phi}} + (\ul{\Phi} \odot \ul{\Psi}_{\PRE})$ \\
    RGE
      & \hfil -- do --
      & $\ul{\Psi}_{\RGE}
           =\tr{sgn}[|B^TB|]$ 
      & $\ul{L}_{\RGE} = \ul{D}_{\ul{\Phi}} - |\ul{\Phi} \odot \ul{\Psi}_{\PRE}|$ \\

    \hline
    \\

    \multicolumn{4}{l}{\tb{When $\ul{\Phi}$ is as in \eqref{eq:phi_nu} with vertex vector $\ul{\nu} > 0$:}} \\
    \hline     
    {}
      & \hfil$\ul{\Phi} \in \mbb{R}^{M\times M}$ 
      & \hfil$\ul{\Psi} \in \mbb{R}^{M\times M}$ 
      & \hfil$\ul{L} \in \mbb{R}^{M\times M}$ \\
    {}
      & 
      & \hfil(proof of Theorem~\ref{thm:L_W'})
      & \hfil(proof of Theorem~\ref{thm:L_W'}) \\
    {}
      & 
      && \\
    \hline
    PRE
      & $\phi_{pq} = \nu_i > 0$, when $\{e_p, e_q\}$ share a common vertex and $p \neq q$, and $\phi_{pq} = 0$ otherwise.
      & $\ul{\Psi}_{\PRE}
           =\tr{sgn}[\ul{L}_e(\ul{\nu})]$ 
      & $\ul{L}_{\PRE} = \ul{D}_{|\ul{W}'|} - \ul{W}'(\ul{\nu})$ \\
    DPE
      & \hfil -- do --
      & $\ul{\Psi}_{\DPE}
           =-\tr{sgn}[\ul{L}_e(\ul{\nu})]$ 
      & $\ul{L}_{\DPE} = \ul{D}_{|\ul{W}'|} + \ul{W}'(\ul{\nu})$ \\
    RGE
      & \hfil -- do --
      & $\ul{\Psi}_{\RGE}
           =\tr{sgn}[|\ul{L}_e(\ul{\nu})|]$ 
      & $\ul{L}_{\RGE} = \ul{D}_{|\ul{W}'|} - |\ul{W}'(\ul{\nu})|$ \\
    \hline
  \end{tabular}
  \caption{Parameters associated with the three different edge functional affinities. When $\ul{\Phi} \in \mbb{R}^{M\times M}$ is as in Definition~\ref{def:Cost_k_pq}, $\ul{D}_{\ul{\Phi}} = \mr{diag}[\ul{\Phi}\, \ul{1}_M]$. When $\ul{\Phi} \in \mbb{R}^{M\times M}$ is as in \eqref{eq:phi_nu} with vertex vector $\ul{\nu} > 0$, $\ul{\Phi}(\ul{\nu}) = |\ul{W}'(\ul{\nu})|$ so that $\ul{D}_{\ul{\Phi}} = \ul{D}_{|\ul{W}'|}$ (see proof of Theorem~\ref{thm:L_W'}).}
  \label{tab:parameters}
\end{table}

\newpage

\begin{table}[!htbp]
  \floatname{algorithm}{Procedure}
  \begin{algorithmic}[1]
    
    \Procedure{EdgeClustering\_WithLe~}{Digraph $\mc{G}$, Vertex Vector $\ul{\nu}$, \# of Clusters $K$}
    
    \% Edge clustering algorithm using the edge Laplacian $\ul{L}_e$ and used in the simulations.
    
    \% User-specified input vertex vector $\ul{\nu}$ contains the user-defined relative importance of each vertex, e.g., use \eqref{eq:nuvectorchoice}.
    
    \% User must specify a numbering of the edges in the graph, and must specify the fixed mapping between edge number and its associated edge weight in the adjacency matrix.
    
    \State Compute edge Laplacian [Definition~\ref{def:Le}]: \newline
    \indent 
    $\ul{L}_e(\ul{\nu}) \gets \ul{B}^T \mr{diag}[\ul{\nu}]\, \ul{B}$.
    
    \State Compute dual graph adjacency matrix [Lemma~\ref{lem:Phi_Le}]: \newline
    \indent 
    $W'(\ul{\nu}) \gets \ul{L}_e(\ul{\nu}) - \mr{diag}[\mr{diag}[\ul{L}_e(\ul{\nu})]]$; \newline
    \indent 
    $\ul{D}_{|\ul{W}'|}(\ul{\nu}) \gets \mr{diag}[|\ul{W}'(\ul{\nu})|\, \ul{1}_M]$. 
    
    \State Compute Flow Laplacian $\ul{L}$ depending on edge affinity [Theorem~\ref{thm:L_W'}]: \newline
    \indent 
    $\ul{L}_{\PRE}(\ul{\nu}) \gets \ul{D}_{|\ul{W}'|}  - \ul{W}'(\ul{\nu})$, 
    $\ul{L}_{\DPE} \gets \ul{D}_{|\ul{W}'|} + \ul{W}'(\ul{\nu})$, or 
    $\ul{L}_{\RGE} \gets \ul{D}_{|\ul{W}'|} - |\ul{W}'(\ul{\nu})|$. 
    
    \State Generate diagonal edge volume matrix $\ul{F} = \{f_p\}$ [Definition~\ref{def:VolEdge}]: \newline
    \indent
    $f_p \gets 
     0.5\,|w_p|
       \left(
         \displaystyle
         \frac{\sigma_{\ell_p}\nu_{\ell_p}}{\langle d_{out, \ell_p} \rangle} 
                 + \frac{\sigma_{k_p}\nu_{k_p}}{\langle d_{in, k_p} \rangle}
       \right)$.

    \State Compute normalized Laplacian [Definition~\ref{def:normalizedFlowLaplacians}]: \newline
    \indent
    $\ul{\widetilde{L}} \gets \ul{F}^{-1/2} \ul{L}\, \ul{F}^{-1/2}$. 
    
    \State Generate $\widehat{\ul{Y}}$, the $M \times K$ matrix containing the $K$ eigenvectors corresponding to the $K$ smallest eigenvalues of $\ul{\widetilde{L}}$:
    \For {$q \gets 1:K$}
        \State $\ul{\widetilde{u}}_q \gets$ Eigenvector corresponding to $q$-th smallest eigenvalue of $\ul{\widetilde{L}}$;
        \State $\widehat{\ul{Y}} \gets \begin{bmatrix} \widehat{\ul{Y}}, & \ul{\widetilde{u}}_q\end{bmatrix}$.
    \EndFor
    
    \State Row Normalize $\widehat{\ul{Y}}$ to generate the $M \times K$ matrix $\mr{FeatureRecords}$:
    \For {$i \gets 1:M$}
        \State $\mr{FeatureRecords}(i,:) \gets \widehat{\ul{Y}}(i,:)/\sqrt{\sum_{j=1}^K \widehat{Y}_{ij}^2}$.        
    \EndFor
    
    \% Row $i$ of $\mr{FeatureRecords}$ is a $1 \times K$ unit feature vector associated with edge $e_i$.
   
    \State Clustering these $M$ rows/feature vectors of $\mr{FeatureRecords}$ into $K$ groups: \newline
    \indent 
    $\begin{bmatrix} C_1, & \ldots, C_K\end{bmatrix} 
     \gets 
     \mr{Cluster}(\mr{FeatureRecords}, K)$. 
    
    \% $C_i$ contains the indices of edges belonging to the $i$-th cluster. 

    \State Return $\begin{bmatrix} C_1, & \ldots, & C_K\end{bmatrix}$.
  
  \EndProcedure
  
  \end{algorithmic}
  \caption{Pseudocode for the Flow Laplacian Edge Clustering Method used in all simulations.}
  \label{Pro:EdgeClustering}
\end{table}

\pagebreak

\begin{table}[!htbp]
  \floatname{algorithm}{Procedure}
  \begin{algorithmic}[1]
    
    \Procedure{EdgeClustering\_General~}{Digraph $\mc{G}$, Matrix $\ul{\Phi}$, \# of Clusters $K$}
    
    \% General Edge Clustering algorithm using $\ul{\Phi}$, the matrix containing the weights for the pairwise comparisons of edge labels in the edge clustering cost function.
    
    \% User-specified input matrix $\ul{\Phi}$ must be symmetric and non-negative [see Definition~\ref{def:Cost_k_pq}].  
    
    \% User must specify a numbering of the edges in the graph, and must specify the fixed mapping between edge number and its associated edge weight in the adjacency matrix.
    
    \State Compute pertinent edge affinity matrix $\ul{\Psi}$ [Lemma~\ref{lem:Psi}]: \newline
    \indent
    $\ul{\Psi}_{\PRE} \gets \mr{sgn}[\ul{B}^T\ul{B}]$, 
    $\ul{\Psi}_{\DPE} \gets -\ul{\Psi}_{\PRE} = -\mr{sgn}[\ul{B}^T\ul{B}]$, 
    $\ul{\Psi}_{\RGE} \gets |\ul{\Psi}_{\PRE}| = \mr{sgn}[|\ul{B}^T\ul{B}|]$.
    
    \State Compute matrix $\ul{D}_{\Phi}$ [equation~\eqref{eq:D_phi}]: \newline
    \indent
    $\ul{D}_{\ul{\Phi}} = \mr{diag}[\ul{\Phi}\,\ul{1}_M]$.
    
    \State Compute Flow Laplacian $\ul{L}$ depending on edge affinity [Lemma~\ref{lem:Cost_k_L}] \newline
    \indent 
    $\ul{L}_{\PRE} \gets \ul{D}_{\ul{\Phi}} - (\ul{\Psi}_{\PRE}\odot\ul{\Phi})$, 
    $\ul{L}_{\DPE} \gets \ul{D}_{\ul{\Phi}} + (\ul{\Psi}_{\PRE}\odot\ul{\Phi})$, 
    $\ul{L}_{\RGE} \gets \ul{D}_{\ul{\Phi}} - |\ul{\Psi}_{\PRE}\odot\ul{\Phi}|$.
    
    \State Generate diagonal edge volume matrix $\ul{F} = \{f_p\}$ [Definition~\ref{def:VolEdge}]: \newline
    \indent
    $f_p \gets 
     0.5\,|w_p|
       \left(
         \displaystyle
         \frac{\sigma_{\ell_p}\nu_{\ell_p}}{\langle d_{out, \ell_p} \rangle} 
                 + \frac{\sigma_{k_p}\nu_{k_p}}{\langle d_{in, k_p} \rangle}
       \right)$.

    \State Compute normalized Laplacian [Definition~\ref{def:normalizedFlowLaplacians}]: \newline
    \indent
    $\ul{\widetilde{L}} \gets \ul{F}^{-1/2} \ul{L}\, \ul{F}^{-1/2}$. 
    
    \State Generate $\widehat{\ul{Y}}$, the $M \times K$ matrix containing the $K$ eigenvectors corresponding to the $K$ smallest eigenvalues of $\ul{\widetilde{L}}$:
    \For {$q \gets 1:K$}
        \State $\ul{\widetilde{u}}_q \gets$ Eigenvector corresponding to $q$-th smallest eigenvalue of $\ul{\widetilde{L}}$;
        \State $\widehat{\ul{Y}} \gets \begin{bmatrix} \widehat{\ul{Y}}, & \ul{\widetilde{u}}_q\end{bmatrix}$. 
    \EndFor
    
    \State Row Normalize $\widehat{\ul{Y}}$ to generate the the $M \times K$ matrix $\mr{FeatureRecords}$:
    \For {$i \gets 1:M$}
        \State $\mr{FeatureRecords}(i,:) \gets \widehat{\ul{Y}}(i,:)/\sqrt{ \sum_{j=1}^K \widehat{Y}_{ij}^2 }$.
    \EndFor
    
    \% Row $i$ of $\mr{FeatureRecords}$ is a $1 \times K$ unit feature vector associated with the edge $e_i$.
   
    \State Cluster these $M$ rows/feature vectors of $\mr{FeatureRecords}$ into $K$ groups: \newline
    $\begin{bmatrix} C_1, & \ldots, C_K\end{bmatrix} 
     \gets 
     \mr{Cluster}(\mr{FeatureRecords}, K)$.
    
    \% $C_i$ contains the indices of edges belonging to the $i$-th cluster. 
    
    \State Return $\begin{bmatrix} C_1, & \ldots, & C_K\end{bmatrix}$.
  
  \EndProcedure
  
  \end{algorithmic}
  \caption{Pseudocode for the Flow Laplacian Edge Clustering Method with a more general user-specified $\ul{\Phi}$.}
  \label{Pro:GenericEdgeClustering}
\end{table}


\end{document}